\documentclass[12pt,twoside,openright]{report}
\usepackage[top=21mm,bottom=20mm,left=30mm,right=25mm]{geometry} 
\usepackage{fancyhdr}

\fancypagestyle{plain}{%
\fancyhead{} 
}

\usepackage[parfill]{parskip}    
\usepackage{graphicx}
\usepackage{amssymb}
\usepackage{amsmath}
\usepackage{epstopdf}
\usepackage{setspace}
\usepackage[small]{caption}
\usepackage{tabularx}
\usepackage{longtable}
\usepackage{cleveref}

\crefname{section}{\S\!}{\S\S}
\Crefname{section}{\S\!}{\S\S}

\DeclareMathOperator*{\argmin}{arg~min}

\title{Modelling the Role of Nitric Oxide in Cerebral Autoregulation}
\author{Mark Catherall}

\begin{document}

\begin{center}
\thispagestyle{empty}
\LARGE
\rule[-0.4cm]{\textwidth}{0.1pt}\\
\textbf{Modelling the Role of Nitric Oxide in Cerebral Autoregulation}
\rule[1cm]{\textwidth}{0.1pt}\\
\vspace{0.4cm}

\Large
\textbf{Mark Catherall}\\
University College

Supervisor: \textbf{ Prof. S. J. Payne }
\vspace{40pt}

\includegraphics[width=0.25\columnwidth]{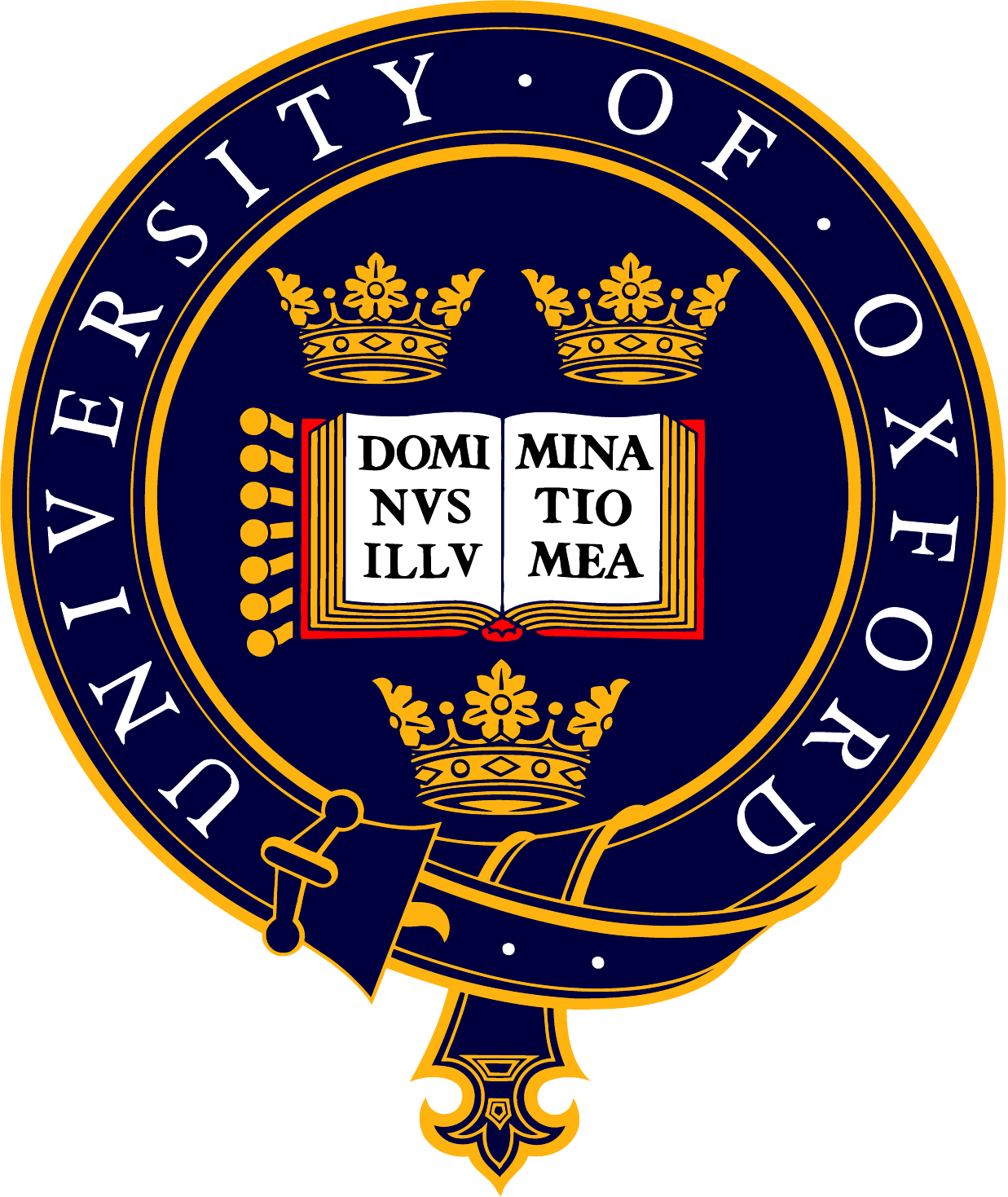}%
\vspace{20pt}

\normalsize
PUMMA Research Group \\
Department of Engineering Science \\
University of Oxford \\

\vspace{30pt}
\Large
Thesis submitted for the degree of\\
\emph{Doctor of Philosophy}\\
\textbf{October 2014}
\vspace{30pt}

~\\
~\\
~\\
\thispagestyle{empty}

\end{center}

\newcommand{\ca}{Ca$^{2+}$~}
\newcommand{\cae}{Ca$^{2+}$}
\newcommand{\cca}{[Ca$^{2+}$]~}
\newcommand{\cno}{[NO]~}
\newcommand{\cam}{Ca^{2+}}
\newcommand{\ccam}{\lbrack Ca^{2+} \rbrack}
\newcommand{\ud}{\,\mathrm{d}}
\newcommand{\diff}[2]{\frac{\ud #1}{\ud #2}}
\newcommand{\pdiff}[2]{\frac{\partial #1}{\partial #2}}
\newcommand{\conc}[1]{\lbrack#1\rbrack}
\newcommand{\beq}{\begin{equation}}
\newcommand{\eeq}{\end{equation}}
\newcommand{\e}[1]{\times10^{#1}}
\newcommand{\logt}{\log_{10}}
\newcommand{\ccgmp}{\conc{cGMP}~}

\begin{abstract}
Malfunction of the system which regulates the bloodflow in the brain is a major cause of stroke and dementia, costing many lives and many billions of pounds each year in the UK alone. This regulatory system, known as \emph{cerebral autoregulation}, has been the subject of much experimental and mathematical investigation yet our understanding of it is still quite limited. One area in which our understanding is particularly lacking is that of the role of nitric oxide, understood to be a potent vasodilator. The interactions of nitric oxide with the better understood myogenic response remain un-modelled and poorly understood. In this thesis we present a novel model of the arteriolar control mechanism, comprising a mixture of well-established and new models of individual processes, brought together for the first time. We show that this model is capable of reproducing experimentally observed behaviour very closely and go on to investigate its stability in the context of the vasculature of the whole brain. In conclusion we find that nitric oxide, although it plays a central role in determining equilibrium vessel radius, is unimportant to the dynamics of the system and its responses to variation in arterial blood pressure. We also find that the stability of the system is very sensitive to the dynamics of \ca within the muscle cell, and that self-sustaining \ca waves are not necessary to cause whole-vessel radius oscillations consistent with vasomotion.
\end{abstract}

\section*{Nomenclature}
\begin{longtable}{c l}
\hline
$A$ & area\\
ABP & arterial blood pressure \\
Ca & calcium \\
CBFV & cerebral blood flow velocity \\
cGMP & cyclic guanosine monophosphate \\
CICR & calcium induced calcium release \\
$C$ & compliance\\
$C_w$ & concentration of NO in vessel wall\\
$\Delta X$ & difference in $X$ between two places\\
$\epsilon$ & error between target and response\\
$\eta$ & dimensionless variance\\
eNOS & endothelial nitric oxide synthase \\
$f$ & viscous friction coefficient\\
$F$ & force \\
$h_w$ & NO wall transport coefficient\\
IP$_3$ & inositol 1,4,5-triphosphate \\
$k$ & constant or iteration counter\\
$K$ & rate constant\\
$\lambda$ & normalized cell length\\
$L$ & length \\
mmHg & millimetres of mercury\\
$\mu_b$ & viscosity of blood\\
$n$ & number or exponent constant\\
NO & nitric oxide \\
$P$ & pressure\\
$pX$ & $\log_{10}(X)$\\
$q$ & flow or constant\\
$Q$ & flow\\
$r$ & radius or decay rate\\
$R$ & resistance to flow\\
$\sigma$ & stress\\
\S & section\\
$s_w$ & NO production rate in vessel wall\\
SERCA & sarco-(endo)plasmic reticulum calcium ATPase \\
SR & sarcoplasmic reticulum \\
$\tau$ & shear stress \\
$\tau_x$ & time constant for variable denoted by subscript $x$\\
$t_w$ & vessel wall thickness\\
$v$ & speed\\
$V_x$ & volume of compartment $x$\\
VSMC & vascular smooth muscle cell \\
$w_c$ & cell width\\
$\omega_n$ & natural frequency\\
$\bar x$ & equilibrium value of $x$ \\
$\dot{x}$ & first time derivative of $x$ ($\diff{x}{t}$)\\
$\ddot{x}$ & second time derivative of $x$ ($\diff{^2x}{t^2}$)\\
$X_L$ & Lower value of $X$\\
$X_U$ & Upper value of $X$\\
$X^{z\pm}$ & ion of element X with charge $\pm z$ \\
\conc{$X$} & concentration of $X$ \\
\hline
\end{longtable}

\tableofcontents

\chapter{Introduction}

\section{The Problem}
Stroke and dementia are between them responsible for approximately 68,000 deaths per year in the England and Wales alone \cite{strokestats,ONS}. The cost to the UK of stroke and dementia combined is an estimated \pounds 24.6 billion per year \cite{demUK,PISC}. Stroke can be either haemorrhagic (caused by bleeding in the brain), or ischaemic (caused by insufficient blood flow to a part of the brain). There are several different types of dementia; one of them, called vascular dementia, occurs due to insufficient blood supply within the brain, just as in ischaemic stroke, only on a much smaller scale. Ischaemia damages the brain because when brain tissue does not receive enough blood it is not supplied with enough oxygen to respire (a condition known as hypoxia), this in turn leads to a build up of toxic chemicals within the cell and ultimately to the death of the cell. The volume of cells that die during an ischaemic episode is called the infarct. In the case of ischaemic stroke, the infarct is relatively large, comparable to the size of a golf ball, and this can cause loss of a specific mental function, e.g., speech or control of a limb. In the case of dementia, rather than an acute ischaemic attack that leads to a large infarct, chronic slight ischaemia causes many small infarcts to accrue over months and years. These infarcts are very small, of the order of 1mm across, and do not lead to the sudden loss of a specific mental function, but rather to a general degradation of cognitive ability, affecting the reasoning and memory of those afflicted. Dementia can then be thought of as a chronic condition brought about as the result of an ongoing series of tiny strokes, each one occurring over a short time scale.

\section{The Brain and its Blood Supply}
\label{introBrain}
The brain consumes more oxygen than any other organ (in a resting state), with around 15\% of the total cardiac output (blood flow) going to the brain \cite{anaesthesiaUK}. Brain tissue is also much more sensitive than most tissues to ischaemia; a leg can be deprived of blood flow, and hence oxygen, for up to two hours without damage, in contrast, the brain can suffer irreversible damage after only 4 minutes of oxygen deprivation at normal temperatures \cite{headway}. There are biological control systems that exist to maintain the brain's blood supply at a constant level in order that the oxygen sensitive tissue of the cerebral cortex is never without oxygen. The combination of all such systems is known as cerebral autoregulation. An impaired cerebral autoregulatory system will affect the brain's vascular response to changing conditions and so will affect the progress of ischaemic stroke and vascular dementia. If we can better understand the mechanisms of cerebral autoregulation we may be able to prevent or slow down the onset of cerebral ischaemia, furthermore, we may be able to propose improvements to post infarction treatment based on a better understanding of the brain's own response to the event.

The primary mechanisms by which the flow of blood through the brain is regulated are vasodilation and vasoconstriction -- the enlargement or narrowing of the blood vessels. Various substances interact in the tissue comprising the blood vessels to cause these changes in radius. One such substance, the role of which is not yet well understood, is nitric oxide (NO).

\section{The Value of Mathematical Modelling}
Mathematical modelling provides direction for clinical experimentation. It serves to interpolate between the observed behaviours of a system and so suggest fruitful avenues of further investigation. It can also serve as a quick and cheap test-bed for new treatment ideas -- the alternative clinical trials are very costly and time-consuming and it would be very wasteful to embark upon such a venture without the best possible guidance as to the target of the investigation. Mathematical models provide such guidance. In the limit, where a completely faithful mathematical reconstruction of a biological system were available, there would be no need for speculative clinical investigation to discover new treatments, nor for long clinical trials to evaluate efficacy and safety. Without mathematical modelling we are limited to very simple hypotheses, and so are limited in our understanding to only those things which are simple in nature. Mathematical modelling allows us to extend our understanding to complex systems and enables us to make useful predictions about such systems.

\section{Objective}
\label{introObjective}
In this thesis we will answer two principle questions; the first relates to the role of NO within the cerebral autoregulatory system, in particular some experimental work carried out by Ide et al. \cite{Ide07} (described in \cref{interactionsNOMyoMech}) produced results which could not properly be explained within the current understanding of the role of NO in cerebral autoregulation; we aim to build a mathematical model of the regulating blood vessels and use this to elucidate the results of \cite{Ide07}. The second is concerned with a phenomenon known as \emph{vasomotion} and the mechanism by which is arises. Vasomotion could play a role in protecting tissue from transient ischaemia by enhancing oxygen delivery - we will use our blood vessel model to replicate vasomotion and examine the conditions under which is arises. Both the experiments and results of \cite{Ide07} and the phenomenon of vasomotion are described in more detail in \cref{chapLitRev} in order that they may be understood in the wider biological context.

\section{Overview}
\label{overview}
This thesis is presented in several chapters, each relating to a distinct phase of our investigations into the subject, here we provide an overview of the thesis and the contents of each chapter:

In \cref{chapLitRev} we review some background physiology of which the reader should be aware to put the subsequent discussion into context. The experimental work that has led to the current understanding of the cerebral autoregulatory system is then reviewed and commented upon before we move on to reviewing the mathematical modelling and theoretical literature relevant to our thesis.

\Cref{chapVascModel} presents the equations, and origins thereof, that constitute our novel model of the cerebral arteriole. The model presented is an agglomeration of parts of previous models and original elements.

\Cref{chapArtFit} introduces the experimental data which we will use to validate our model and details the optimization process through which parameter values are found which enable the model to best reproduce the behaviour represented in the data. The results of this optimized fitting process are presented and insights are gained into the nature of the system. Also presented here is an analysis of the sensitivity of the response of the system to variations in its parameter values.

In \Cref{chapAutoreg} we integrate the fitted arteriole model into a representation of the vasculature of the whole brain. We then modify this model to better fit the balance between venous and arterial blood volume observed in vivo. Finally we examine the frequency response of the combined model and draw conclusions about the autoregulatory system as a whole, its possible pathologies and methods of diagnosis.

\Cref{chapOscill} details further modifications made to the autoregulatory model of \cref{chapAutoreg} in search of both greater physiological fidelity and, crucially, the development of instability and spontaneous oscillation representative of vasomotion. We identify features of the dynamics of the vessel which are sufficient to cause spontaneous oscillations and examine the limits and possible triggers of such oscillations.

Finally, in \Cref{chapConcs}, we summarise what we have found through this iterative modelling, fitting, and examination process. We reflect on the successes and limitations of our investigation and make suggestions for future work to address these limitations.

\chapter{Literature Review}
\label{chapLitRev}
In this chapter we will first review some of the fundamental material which underpins our current understanding of bloodflow in the brain before going on to review the more focussed research which is directly relevant to cerebral autoregulation. Finally we will review existing mathematical models of relevant elements of the system.

\section{The Cardiovascular System}
In order to function and to stay alive, all cells of the body require energy, and this energy is provided by respiration, which requires oxygen. Consequently all cells of the body require oxygen. The absorption of oxygen from the air and its distribution throughout the body is taken care of by the cardiovascular system. The heart pumps blood through the lungs wherein it absorbs oxygen, chiefly through binding to haemoglobin (Hb) molecules found in red blood cells. The oxygenated blood returns to the heart where it is pumped once again, this time through the rest of the body via the system of blood vessels known as the vasculature. As the blood passes through the vasculature, oxygen is transferred from the blood to the tissue surrounding the blood vessel. By the time the blood is returned to the heart it has lost much of the oxygen it formerly carried.

\subsection{Blood Vessels}
The vessels which carry the blood through the body form an intricate network. The network can be thought of as having two sides, both tree like in structure: the arterial side takes oxygenated blood leaving the heart through the aorta and distributes it throughout the body; the venous side collects the blood from throughout the body and returns it to the heart. Whereas the arterial side starts in one large vessel (the aorta) and divides repeatedly into smaller vessels, the venous side starts in a multitude of small vessels and, through repeated confluence, ends up in one large vessel (the vena cava). The junction of the two systems is the capillary bed, a very fine mesh of blood vessels which themselves are only just wide enough for one red blood cell to pass through. It is through the capillary bed that most of the oxygen delivery takes place. Capillaries have very thin walls (only one cell thick) to allow easy diffusion of oxygen from the blood to the surrounding tissue.

\begin{figure}[!hbtp] \centering \includegraphics[scale = 0.8]{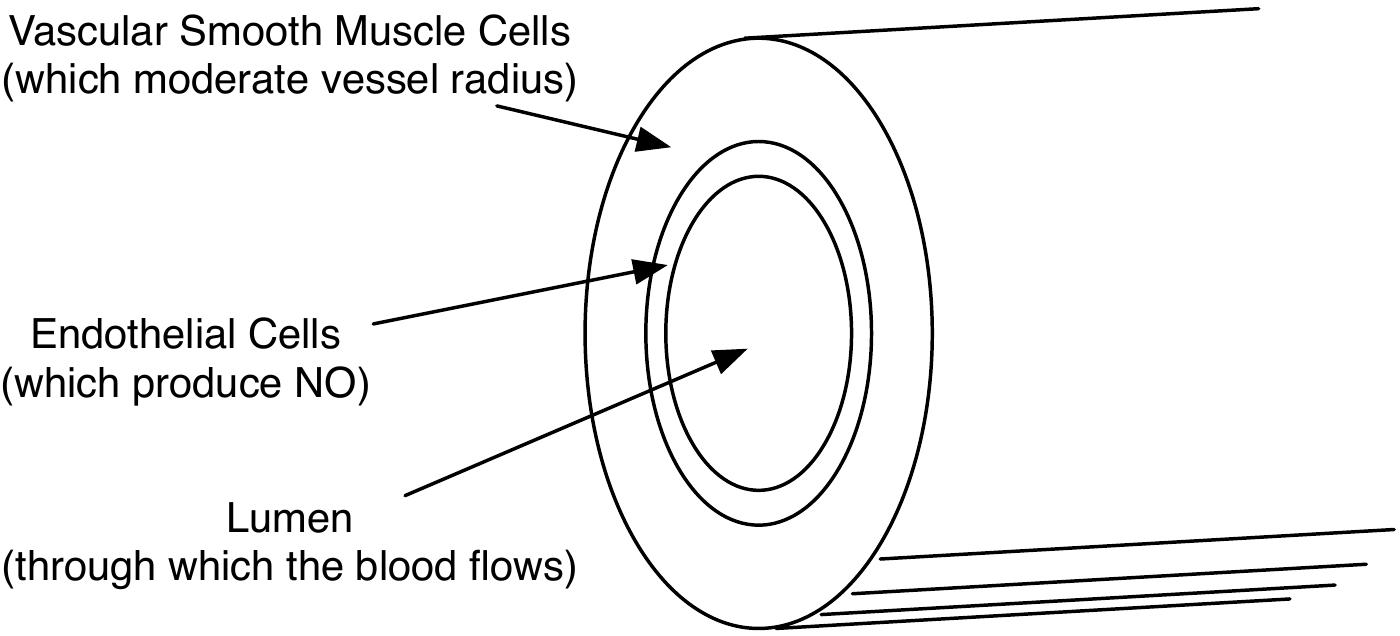} \caption{Structure of an arteriole.} \label{figarteriole} \end{figure}

The generation of blood vessels which immediately precedes the capillaries is made up of arterioles. It is predominantly these arterioles which change size in order to modulate flow through the tissue. Whereas the walls of capillaries are only one cell thick, the walls of arterioles have two layers of cells: the first is the same as that which the capillaries have, a single-cell-thick layer of endothelial cells; the second, outer, layer is a one or two cell thick layer of muscle cells. It is these muscle cells  (vascular smooth muscle cells or VSMCs) which enable the arterioles to change size and so to regulate the flow of blood through the tissues. The structure of an arteriole is shown in \cref{figarteriole}.

Because they carry blood at such different pressures, arteries and veins have very different structures. Arteries have thick, elastic walls with substantial layers of muscle to contain blood at high pressures relative to the surrounding tissue, and to smooth out highly pulsatile flow, whereas veins have thin, inelastic walls and are easily collapsed between pulses.

\subsection{Blood Pressure}
Blood flows through the vasculature because the heart raises the blood pressure sufficiently. The amount of pressure drop along a vessel is a function of the viscosity of the blood, the vessel radius, and the vessel length; taken together these constitute the overall \emph{resistance to flow} of the system. As the cardiovascular system generally acts to maintain constant blood flow throughout the vasculature, an increase in the resistance to flow is generally followed by an increase in arterial blood pressure (ABP). This can be achieved through increased heart rate and/or increased stroke volume of the heart. For example, when the blood vessels constrict due to mental stress \cite{Lacy95}, or are partially occluded by cholesterol build up \cite{Martin86}, ABP necessarily rises to achieve sufficient blood flow through the vasculature.

Blood pressure in the aorta is necessarily higher than anywhere else in the body (correcting for elevation). As blood progresses through the arterial tree, and overcomes the resistance of each vessel in turn, its pressure falls. By the time the blood returns to the heart via the vena cava it is at a similar pressure to that of the surrounding tissue. 

\subsection{Blood Flow}
As mentioned in \cref{introBrain}, the tissue of the brain is particularly sensitive to hypoxia (lack of oxygen) and so the supply of blood to the brain must be critically controlled. The flow through a vessel is determined by the difference in blood pressure from one end of the vessel to the other and the internal radius of the vessel. Smaller vessels present greater resistance and so admit less flow than larger vessels, for the same driving pressure. This relationship is normally assumed to be governed by the Poiseuille equation for flow through cylindrical vessels:
\beq R = \frac{8\mu L}{\pi r^4} = \frac{\Delta P}{Q} \label{Poiseuille} \eeq

where $R$ is the resistance of the vessel, $\mu$ is the viscosity of the fluid, $L$ is the length of the vessel, $r$ is the radius of the vessel, $\Delta P$ is the difference in fluid pressure between the ends of the vessel, and $Q$ is the volumetric flow rate through the vessel.

By varying the radius, and thus the resistance to flow, of the blood vessels in the brain, the autoregulatory system can prevent fluctuations in blood pressure from causing fluctuations in bloodflow which might otherwise lead to lack, or excess, of oxygen in the brain tissue, both of which would be damaging.

\section{Cells}
\subsection{Cell Structure}
The cells of animals all have a similar structure; the cell is contained by a membrane made from two layers of lipid (fat) molecules containing a phosphorus atom. The phosphorus atom at one end of the molecule polarises the bonds surrounding it, thus enabling affinity with other polar molecules, such as water; This end is known as the hydrophilic end (liking water). The other end is a chain (or two) of carbon atoms  surrounded by hydrogen atoms, much like in alkanes, or oils. This end of the molecule does not have polarised bonds and so does not have any electrostatic affinity with other polar molecules; this is known as the hydrophobic end.

During cell formation these phospholipids become arranged so that there are two layers, facing in opposite directions, all of the hydrophobic ends are in the middle while all the hydrophilic ends face outwards, and are in contact with the watery solution that makes up both the fluid between cells and the fluid within the cell. Because the centre of this membrane is hydrophobic, polar species, including charged ones such as metal ions cannot readily diffuse across it. This makes the membrane electrically insulating, which means that the electrical potential on one side can be different from that on the other. 

Differences in potential can arise through different concentrations of charged species on either side of the membrane (i.e., inside and outside the cell). The potential difference across the membrane is known as \emph{membrane potential} and is usually noted by $V_m$. The membrane is not entirely insulating however; embedded in the membrane are transmembrane proteins that can act as channels to allow certain species to cross the membrane, and thereby enter or leave the cell. These protein channels come in various forms, some require energy to operate, some swap one ion with another, some must be activated by the binding of some species on side or the other before they allow another species to cross the membrane, and some are activated by the membrane potential. It is these channels that allow the regulation of ionic concentrations within the cell which in turn can allow the cell to change its behaviour, or to affect other cells.

\begin{figure}[!hbtp] \centering \includegraphics[scale = 0.8]{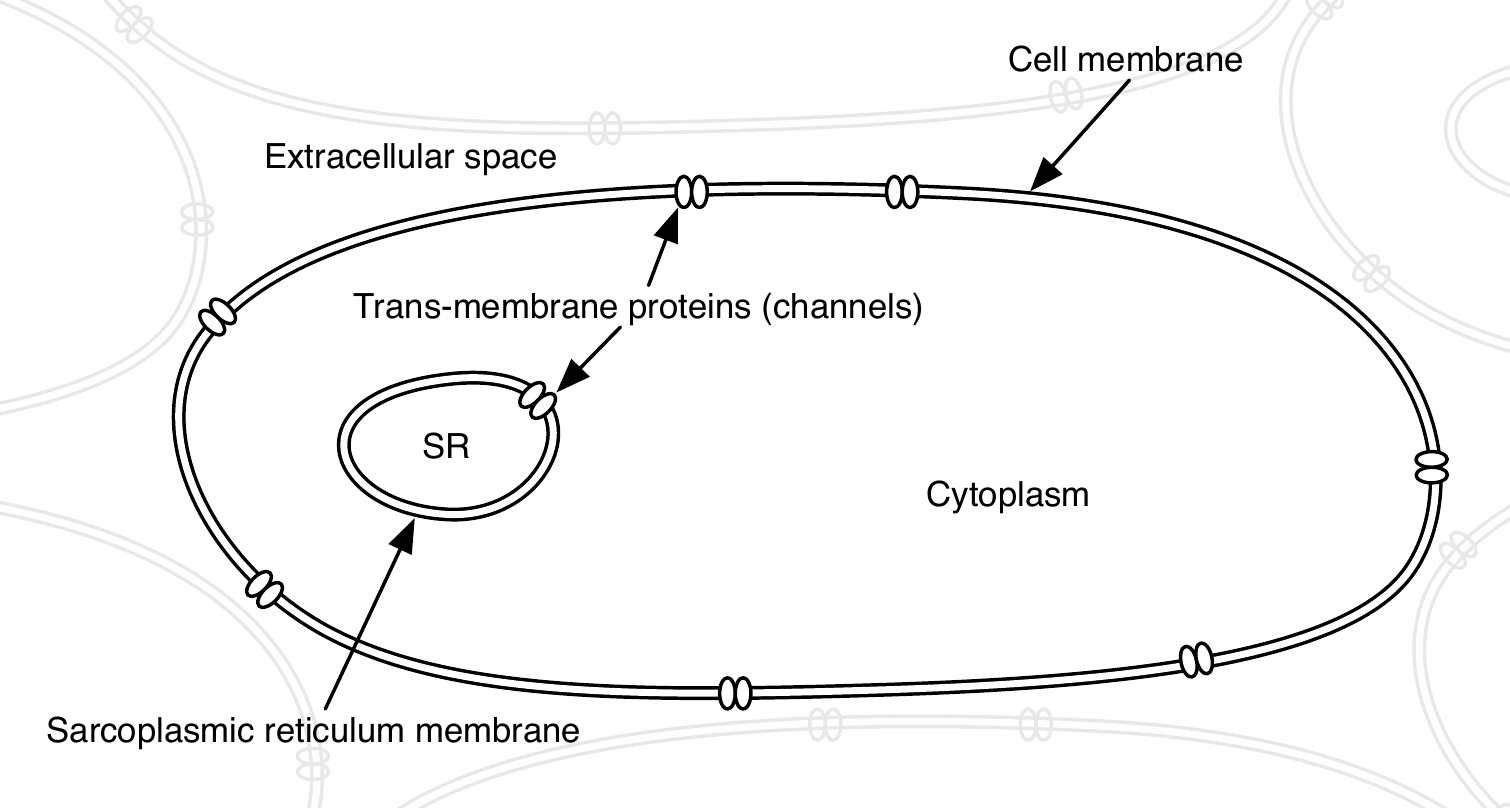} \caption{Schematic diagram of a cell showing only those structures and compartments that are relevant. SR denotes the Sarcoplasmic Reticulum.} \label{figcell} \end{figure}

The volume enclosed by the cell membrane is called the lumen, and is filled mainly with a watery solution of ions and other molecules called the cytoplasm. Floating about in the cytoplasm there are various other structures and molecular machines, depending on the type of cell. One such structure is the sarcoplasmic reticulum, or SR. This sub-compartment is contained by its own membrane, which is very similar to the cell membrane in structure and it too has transmembrane proteins that regulate the passage of polar species across the membrane. The SR contains \ca ions at a much higher concentration than in the cytoplasm, this will be important later. \Cref{figcell} shows the basic structure of an animal cell.

\subsection{Smooth Muscle Cells}
\label{musclestructure}
\begin{figure}[!hbtp] \centering \includegraphics[scale = 1]{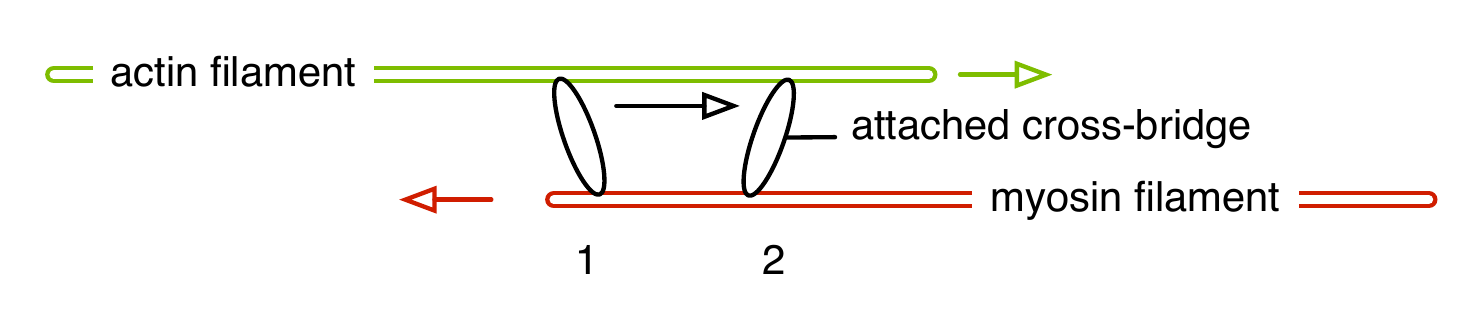} \caption{Schematic diagram of a cross bridge pulling an actin filament past a myosin filament; the cross bridge attaches in position 1, then tilts so that the actin filament is drawn past, as in position 2. The cross bridge will then detach and reposition, the repeated process is known as cross-bridge cycling.} \label{figactinmyosin} \end{figure}

Smooth muscle cells contain molecules of myosin and actin, both of which form thin strands which lie parallel to one another. Contraction is achieved through the globular heads of myosin filaments (which protrude from the sides of the filament) attaching to the actin filaments and then changing angle, pulling the actin filaments past the myosin filaments as they do so. The attached myosin head, or \emph{cross-bridge}, then detaches and tilts back to its original position where it can attach to another site on the actin filament. This process of attachment, dragging past, detachment and repositioning of the myosin heads is called cross-bridge cycling. \Cref{figactinmyosin} shows the action of cross-bridge cycling pulling the myosin filament past the actin filament.

Myosin heads cannot attach to the actin filaments until they have been activated by phosphorylation. Phosphorylation requires an enzyme called myosin light-chain kinase, or MLCK, which is in turn activated by a complex formed from \ca and a molecule called calmodulin, called the calcium-calmodulin complex. This activation chain has the consequence that the contraction of smooth muscle is chiefly modulated by the concentration of free \ca ions in the cytoplasm (which is where the actin-myosin filament structures are found). Bundles of overlapping actin and myosin filaments are joined in chains across the smooth muscle cell and attached to the cell walls such that the cycling of cross-bridges throughout the cell causes the cell to contract. Each myosin filament has many heads that can form attached cross-bridges along its length; the maximum force that can be created by this cross-bridge cycling is therefore related to the amount of overlap between actin and myosin filaments. One pair of overlapping actin and myosin filaments is known as a \emph{sarcomere}.

\section{Autoregulation}

\subsection{Myogenic Response}
\label{myogenicresp}
The automatic contraction of arterioles in response to elevated blood pressure is known as the myogenic response. First discovered by Bayliss in 1902 \cite{Bayliss02}, it was not until isolated vessel experimental techniques were developed in 1981 that the mechanisms behind the response began to be uncovered. Experiments involving stretching isolated VSMCs bathed in various solutions \cite{Harris91,Herlihy73}, and cannulated isolated arterioles \cite{Kuo91,VanBavel94,Falcone91}, have lead to the consensus that the myogenic response is driven by a force-sensitive element in-line with the contractile elements (actin and myosin filaments) of the VSMC; this is known as the \emph{wall tension hypothesis}. These experiments have repeatedly shown that the myogenic response is dependent on the availability of \cae. In cases where the fluid surrounding the experimental preparation does not contain \cae, the myogenic response is not observed. This observation is consistent with the idea that \ca is a vital part of the pathway to activation of MLCK and subsequent phosphorylation of myosin heads which leads to increased vascular tone.

\subsection{The Role of Nitric Oxide}
\label{roleOfNO}
Nitric oxide (NO) has now been identified as the formerly enigmatic \emph{Endothelium-Derived Relaxing Factor}, or EDRF as it was known. NO is produced primarily in the endothelium by an enzyme called endothelial nitric oxide synthase, or eNOS. It acts on VSMCs in the opposite way to \cae, through an intermediary molecule, cyclic guanosine mono-phosphate (cGMP) \cite{Denninger99}, acting to de-phosphorylate myosin heads and to cause a reduction in smooth muscle tone. The action of cGMP is twofold: firstly it reduces \ca concentration in the VSMC via various mechanisms \cite{Caravajal00}, primarily the activation of the \ca ATPase, an enzyme which actively pumps \ca out of the cell \cite{Rashatwar87}; and secondly it indirectly stimulates myosin light-chain phosphatase (MLCP) \cite{Lee97}, which does the opposite to MLCK and \emph{de}-phosphorylates myosin heads, thus reducing smooth muscle tone.

Inhibition of eNOS causes significant vasoconstriction and a subsequent increase in blood pressure \cite{Rees90}, suggesting that there is some baseline production of NO by eNOS. NO produced in the endothelium diffuses either into the blood or into the VSMCs surrounding the endothelium. The decay rate of NO is quite fast as it is not a particularly stable diatom. Decay or absorption of NO in whole blood is very rapid due to uptake by haemoglobin in place of oxygen. For this reason approximately 90\% \cite{Buerk01} of the NO produced in the endothelium is carried away in the blood.

It has been found using endothelial cells grown on the surface of small glass beads \cite{Buga91} that NO production is stimulated in response to shear stress at the surface of the endothelial cells in contact with the blood. This suggests that the action of NO in autoregulation could be antagonistic to that of the myogenic response. Whereas a rise in pressure would cause the myogenic response to contract the arteriole, the accompanying rise in flow and therefore shear stress would cause an increased production of NO which would act antagonistically, moderating the contraction due to the myogenic mechanism or overriding it and causing dilation.

\subsection{Interactions of NO and Myogenic Mechanisms}
\label{interactionsNOMyoMech}
After removing arterioles from rat skeletal muscle, Falcone et al.\cite{Falcone91} modulated the pressure of the fluid in the arterioles and observed the ensuing changes in the vessels' diameter. They found that over a mid range of intramural pressures the arterioles contracted in response to elevated pressure. Critically they did this before and after removing the endothelium from the vessels, thus establishing that the myogenic response is independent of the endothelium. However, they did go on to point out that this result may not hold true for cerebral arterioles. In fact Table 2 in \cite{Falcone91} summarises previous studies on arterioles isolated from different tissues and highlights the fact that studies \cite{Harder87,Harder89,Katusic87,Rubanyi88} using cerebral vessels tend to find that the myogenic response of these vessels \emph{is} dependent on an intact endothelium, although some of these studies use vessels larger than arterioles, which may be significant. The role of shear-stress-induced NO production in this case is ruled out not only by the removal of the endothelium but also by the absence of flow through the vessel during the experiment.

In his review of mechanisms underlying the myogenic response \cite{Davis99}, Davis makes the observation that in experiments on arterioles in which arterial and venous pressures are raised equally, as in \cite{Meininger87}, the contribution of the myogenic response to autoregulation is much greater than in experiments where flow increases as well. This may be due to a mitigating effect of flow-induced NO production. He also refers to the wall tension hypothesis (see \cref{myogenicresp}) which posits that the myogenic response is mediated by some tension sensing element in the VSMC in series with the contractile element. This is supported and taken further in \cite{VanBavel94} where it is concluded that wall tension not only mediates \ca release but also modulates \cca sensitivity.

In order to separate the effects of flow and intramural pressure, Kuo et al. \cite{Kuo91} isolated coronary arterioles from a pig and subjected them both to a range of intramural pressures and to a range of pressure differences between the ends (thus a range of flow rates). They found similar results to \cite{Falcone91} for the no-flow case and observed significant vasodilation in response to flow. This flow-induced dilation was not present after the introduction of N\textsuperscript{G}--monomethyl--L--arginine (L-NMMA), which is an eNOS inhibitor, thus supporting the idea that shear-stress-induced release of NO is the driver behind flow-induced vasodilation. They also found that the vasodilation in response to flow was endothelium-dependent, i.e., when the endothelium was removed, the flow-induced dilation was no longer observed. In addition to the aforementioned steady-state experiments, Kuo et al. did a series of dynamic experiments, subjecting an isolated arteriole to step changes in intramural pressure and flow and observing the response of the vessel over time.

In the study which provided the motivation for this thesis, Ide et al. \cite{Ide07} recorded the ABP and cerebral blood flow velocity (CBFV) of several patients over ten minute periods under different conditions. One of these conditions was infusion with L-NMMA. To record a baseline for comparison with similar ABP to the L-NMMA condition (L-NMMA causes significant vasoconstriction and hence elevated ABP, see \cref{roleOfNO}) the experimenters used an $\alpha$-adrenergic agonist called phenylephrine. This stimulates the release of \ca from the SR via a molecule called inositol triphosphate, or IP$_3$. The correlation between ABP and CBFV is often used as an indicator of autoregulatory function -- after all it is the function of the autoregulatory system to control CBFV in spite of fluctuations in ABP. We would expect then that this ABP-CBFV correlation would be quite different under the influence of L-NMMA (if NO is involved in autoregulation) than in the baseline condition with an equally high ABP (the phenylephrine condition). Wavelet analysis of the ABP-CBFV data shows that in fact no discernible difference can be observed between the two cases \cite{Abatay}. This surprising result raises questions about the role that NO plays in the cerebral autoregulation, which seems unaffected by its absence.

\subsection{Other Autoregulatory Mechanisms}
Whilst the NO mechanism is the primary focus of this investigation, and the myogenic response is the primary autoregulatory mechanism, the autoregulatory system includes several other mechanisms by which the flow of blood is controlled. These include but are not limited to: metabolic mechanisms, through which the consumption of resources, e.g. oxygen, in surrounding tissue feeds back to the blood vessel; neural control, whereby nervous stimulation of the VSMCs affects tone; cell-to-cell conducted response, in which responses are propagated from cell to cell along the vessel walls via gap junctions; ATP release from red blood cells, shear stress within the blood can cause ATP to be released from red blood cells which then binds to the endothelium and stimulates release of NO; and, over longer timescales, capillary recruitment - the process by which new capillaries are created in response to ongoing demand for higher perfusion. Most of these responses will not be explicitly included in this investigation, instead we will keep our model as simple as possible whilst still providing enough detail and verisimilitude to provide insight into our key questions.

\section{Vasomotion}
\label{vasomotion}
Arterioles in vivo have been found to exhibit oscillations in vascular tone under certain conditions, typically low arterial pressure and hence hypo-perfusion \cite{Slaaf87,Meyer88,Rucker00}. This oscillation in vascular tone manifests as a rhythmic contraction and dilation of the arteriole known as vasomotion.  Although vasomotion has been frequently observed experimentally, the biochemical mechanisms driving the rhythmic oscillations in smooth muscle tone are still unclear \cite{Aalkjaer11}. There is however consensus that vasomotion requires the coordination of all of the VSMCs in the vessel to contract in synch, and that this coordination is achieved through the electrical coupling of the cells. Experiments by Gustafsson \cite{Gustafsson93} and Hill \cite{Hill99} found that the membrane potentials of all VSMCs in an arteriole exhibiting vasomotion oscillate with the same frequency as that of the vasomotion, and with a slight phase lead. Whereas Gokina \cite{Gokina96} and Lamb \cite{Lamb89} found that each arterial contraction was accompanied by a burst of action potentials from the VSMCs. Peng et al. \cite{Peng01} showed that  oscillations in VSMC membrane potential are caused by oscillations in \ca concentration in the cytoplasm, and further that the synchronisation of these oscillations between VSMCs requires either an endothelium or the external introduction of cGMP. The action of cGMP in this study suggests a further link between NO and the autoregulatory system.

The purpose, if there is one, of vasomotion is also the subject of some debate; the suggestion that vasomotion may benefit oxygenation of adjacent tissue is one that has received much attention. Hapuarachchi et al. demonstrated mathematically in \cite{Hapuarachchi10} that oxygen transport from blood to surrounding tissue is enhanced in a vessel which oscillates in radius compared to a stable vessel of the same mean radius. This conclusion is corroborated by the experimental work of R\"ucker et al. \cite{Rucker00} who measured tissue oxygenation in rat hindlimbs with and without vasomotion and found that vasomotion did indeed increase the perfusion of oxygen into the adjacent muscle tissue.

The combination of findings suggesting that vasomotion occurs only in pathophysiological conditions (such as hypo-perfusion) and that vasomotion may increase the delivery of oxygen from the blood to the adjacent tissue has lead to the hypothesis that vasomotion is an evolved response to keep tissue alive during periods of vascular strain. If this is the case then vasomotion may be of particular importance when considering dementia. Brain tissue is exceptionally sensitive to hypoxia and it may be that vasomotion could prevent the death of cells during brief periods of localised ischaemia which would otherwise form an infarct and contribute to the degradation of mental function that characterises the disease.

\section{Mathematical Models in the Literature}
\label{Models in the Literature}
What follows is a review of models that have been proposed previously, and that are relevant to our investigation. We shall thus be mainly looking at models of the electrochemistry of the smooth muscle cell, models of the interaction between chemical concentrations and smooth muscle contraction, models of the mechanics of smooth muscle, and models of the production and transport of NO.

\subsection{Hodgkin and Huxley}
\begin{figure}[!hbtp] \centering \includegraphics[scale = 0.8]{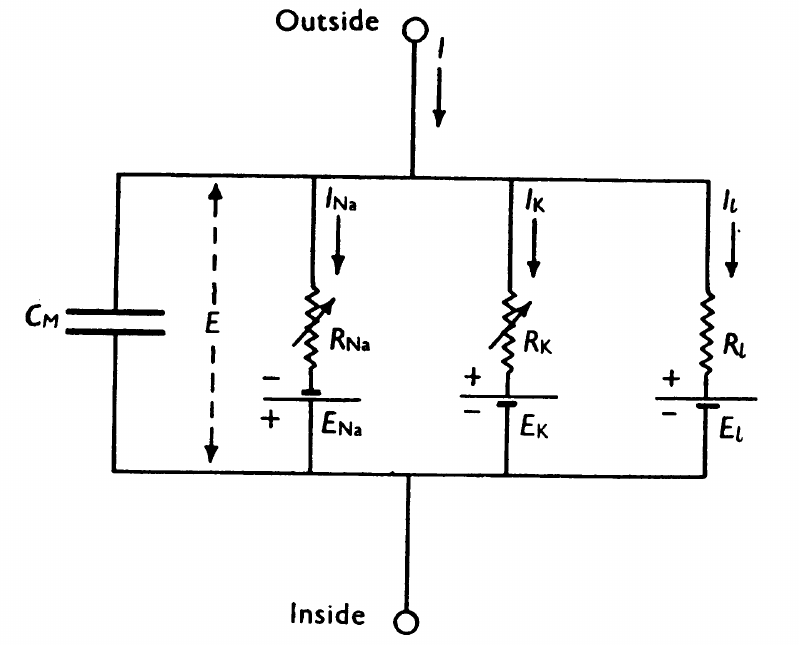} \caption{Schematic of the equivalent circuit used in the Hodgkin-Huxley model. Taken from \cite{HH52}.} \label{figHHcircuit} \end{figure}

In 1952 Hodgkin and Huxley published a series of five papers which described and validated an electrical equivalent circuit model of the activity of the cell membrane. Specifically they sought to model the generation and propagation of action potentials in the squid giant axon (a long nerve). An action potential is a short electrical pulse in which the membrane potential of a cell rapidly rises and falls. An action potential is used to transmit information along a cell, in the case of neurons, or in the case of muscle cells, to trigger a series of processes that result in muscular contraction. Although the exact electrical equivalent circuit used is adapted to suit the cell or behaviour being modelled, many models of cell electrochemistry are based on Hodgkin and Huxley's approach. \Cref{figHHcircuit} is taken from \cite{HH52} and shows the circuit that was used to model the activity of the cell membrane from the axon of a squid.

Hodgkin and Huxley were considering a travelling action potential, of the sort that transmits information through nerves and derived from their model the following:
\beq \diff{V}{t} = - \frac{1}{C_m} \left(\bar g_K n^4 \left(V - V_K \right) + \bar g_{Na} m^3 h \left(V - V_{Na} \right) + \bar g_l \left(V - V_l \right) \right) + \frac{z}{K} \eeq
where $V$ is the membrane potential, $V_K$, $V_{Na}$ and $V_l$ are the equilibrium potentials for the various species (see below), $g_i$ represents the conductance of the $i$ channel while $m$, $n$ and $h$ are parameters representing the number and activation of the channels, $z = \diff{^2V}{t^2}$ and $K$ is a variable related to the speed of propagation. If we assume that the cell in question is small, or that the speed of propagation is infinite (or at least very large compared to other speeds and processes) then $K$ tends to infinity and the last term tends to zero. What we have then is an equation which expresses the membrane potential in terms of the ionic currents into and out of the cell, the ionic currents being given in the form $I_X = g_X \left(V - V_X \right)$.

The parameters $m$, $n$ and $h$ control the number of channels open to a particular ion, these parameters vary with membrane potential and it is through this feedback of membrane potential (potential determines $m$, $n$ and $h$ which determine ionic currents which affect membrane potential) that the system is capable of generating action potentials.

The equilibrium potential $V_X$ for the species $X$ is the membrane potential which would arise if there were zero flux of that ion across the membrane, i.e., the concentrations of that ion inside and outside the cell were in equilibrium. The mathematical expression for this potential was first formulated by Walther Nernst \cite{nernst}:
\beq V_X = \frac{RT}{Fz_X} \cdot \log\!\left(\frac{\conc{X}_o}{\conc{X}_i}\right) \eeq
where $R$, $T$, and $F$ are the ideal gas constant, the absolute temperature, and the Faraday constant respectively, $z_X$ is the charge on the ion $X$, and $\conc{X}$ is the concentration of ion $X$ with the subscipts $o$ and $i$ standing for \emph{outside} the cell and \emph{inside} the cell, log here representing the natural logarithm.

\subsection{DeBoer}
\label{DeBoer}
There are several different frequencies present in the subsystems that make up the cardiovascular system, some are very obvious, such as the frequency of the heart beat and the frequency of respiration. Some are less obvious and do not arise from a single oscillatory source but from the interplay between system components and their feedback control mechanisms. For example, the heart rate is controlled in part by feedback from blood pressure sensors in the neck. Feedback from these pressure sensors (baroreceptors) comes in two forms of nervous activation, vagal and sympathetic. A model first published by DeBoer \cite{DeBoer} shows that the difference in speed between the vagal and sympathetic nervous systems, or more accurately the slowness of the sympathetic nervous system, can lead to a 0.1Hz oscillation in blood pressure. The frequency of 0.1Hz from DeBoer's model is purely emergent from the feedback system, it is not a reflection of some other frequency such as heart rate or breathing but a property of the system as a whole.

Debate continues as to whether such oscillations in actual subjects are the result of a feedback instability, as in DeBoer's model, or come from an oscillator within the vagal nervous system itself. It is sufficient for our purposes to know that, as a consequence of autoregulatory systems, there can be periodic oscillations in blood pressure and blood flow. The important feature of these oscillations is that they reflect the contributions of all parts of the autoregulatory system; as highlighted in \cite{Malpas02}, all parts of a feedback system contribute to its gain and natural frequency, so the removal of one element will change the behaviour of the system as a whole. It is for this reason that we would expect to see a change in the autoregulatory system reflected in a change in the correlation between ABP and CBFV, hence why the result from \cite{Ide07} is intriguing. 

\subsection{Yang}
In a 2003 paper \cite{Yang03a} Yang describes a model of the VSMC including three main parts: the electrochemical processes of ions entering and leaving the cell, the kinetics of phosphorylation of myosin cross bridges, and the mechanical interactions that ultimately determine the force sustained by the VSMC. This model contains several of the key elements that we are interested in modelling here.

Yang's electrochemical model is a lumped-parameter Hodgkin-Huxley type representation which accounts for changes in membrane potential via the equation:
\beq \diff{V_m}{t} = -\frac{1}{C_m} \left( I_{Ca,L} + I_K + I_{K,Ca} + I_{K_i} + I_M + I_{Na,Ca} + I_{Na,K} + I_{Ca,P} + I_B \right) \eeq 
where $V_m$ is the cell membrane potential, $C_m$ is the capacitance of the cell membrane, and $I_{X,Y}$ is generally the current attributable to the flow of an ion through a particular channel. Crucially, $I_M$ is the total stretch-sensitive current across the membrane, a current which increases with increasing strain on the VSMC.

This electrochemical model goes together with a fluid compartment model that includes one compartment for the cytoplasm, one compartment for the extracellular space and another compartment for the sarcoplasmic reticulum, which serves an important role in buffering the concentration of \cae. Through the coupling of this fluid compartment model and the electrochemical model, it is possible to keep track of all of the ionic currents between the compartments and therefore the concentrations of each species within each compartment and hence the potential of the cell membrane.

The second element of Yang's model is taken from Hai and Muphy \cite{Hai88a} and comprises a 4-state kinetic model describing the attachment and phosphorylation of actin and myosin filaments; this provides the molecular basis for smooth muscle contraction. The important feature of this representation is the way in which it links the electrochemical model and the mechanical model of the cell. The link to the electrochemical model is through the rate constants for phosphorylation of myosin filaments which is represented as having a sigmoidal dependance on the concentration of the calcium-calmodulin complex, which in turn is strongly related to the concentration of free \cae. The force generated by the myosin filaments depends on the number of them which are in particular states (attached, phosphorylated). In this way the concentration of \ca in the cytoplasm, which is an output of the electrochemical model, affects the force generated by the myosin filaments in the cell.

The final part of Yang's model is a mechanical, spring-and-damper type of model representing the passive stiffness of the whole cell, the stiffness of latched myosin filaments and the tension generated by the cycling cross-bridges actually pulling the actin and myosin strands past each other. There is also a term modulating the active force by the amount of overlap between the actin and myosin filaments. Overall this part of the model takes, as inputs, the states of the kinetic model and translates them into the length of the VSMC.

Yang has combined three models to produce a viable, and near complete, model of the smooth muscle cell, and validated several aspects of it against experimental data. One thing that is conspicuous by its absence however is any mention of NO. 

Furthermore the implementation of the myogenic response in Yang's model is flawed. Whereas this model includes a \emph{stretch}-sensitive leakage current, it is well established that the myogenic response is driven by \emph{stress}, not stretch. In fact due to the negative relationship between stress and stretch found in the autoregulating regime, a stretch-sensitive leakage current would act in the opposite sense to that which is observed -- only limiting dilation rather than causing constriction.

\subsection{Stalhand}
A slightly modified version of Yang's mechanical model of the VSMC is combined with Hai and Murphey's 4-state kinetic model in an energy based treatment of the system given by Stalhand \cite{Stalhand08}. In his work a general expression of the mechanochemical model is given and it is then shown that the model of Yang is recovered by a linearisation of the general model in the case of small strains. Stalhand's formulation is somewhat more thorough than Yang's and its generality, especially over larger strains, is an advantage when seeking to build a vessel model in which radius may increase by a factor of two.

\subsection{Abatay}
In the last chapter of his thesis \cite{Abatay}, Abatay attempted to take the model proposed by Yang \cite{Yang03a} and modify it to include the mass transport and vasodilatory effects of NO. In fact Abatay presents several different models in his thesis; when a reference is made here to `Abatay', the model presented in chapter 6 is meant. This adaptation comprises of several changes, firstly the electrochemical model is replaced by a much simplified version that only includes the \ca dynamics, ignoring the other ions; secondly a model of NO mass transport is included, and thirdly the stretch-sensitive currents are removed.

The electrochemical model used by Abatay is taken from Gonzalez-Fernandez, 1994 \cite{Gonzalez94} and has only three states, the open probability of a \ca\!\!-sensitive K$^+$ channel, the internal concentration of \ca\!\!, and the membrane potential. This model also has the feature that it oscillates in its `steady-state'. In other words, the electrochemical model does not reach a single equilibrium state but rather tends towards a stable oscillation involving the release and re-uptake of \ca\!\!. This baseline oscillation in \cca could be thought of as representing what is known as vasomotion; the spontaneous oscillation in vascular tone that is found in many mammals under certain conditions, and is of myogenic origin \cite{Nilsson03}. Myogenic means `generated by the muscle' and in this context refers to the fact that vasomotion is not a response to oscillatory nervous stimulation, rather it is a result of the feedback mechanisms inherent in muscle cells themselves. Vasomotion is thought to enhance oxygen transport under conditions of reduced flow \cite{Nilsson03,Hapuarachchi10}. It is debatable whether a model of the effects of NO should include an oscillatory electrochemical model. Certainly the use of a model which is  capable of spontaneous \cca oscillations under certain circumstances is desirable, in order to provide verisimilitude, but it should not necessarily exhibit this behaviour under normal, baseline conditions.

Abatay's NO model builds on that presented in \cite{Payne05} by also including the concentration of NO in the blood. Production of NO is assumed to be proportional to shear stress at the vessel wall, which means that in steady-state NO production is proportional to pressure gradient and vessel radius. The vasodilatory effect of NO is modelled through a sigmoidal relationship between the dephosphorylation rate of the 4-state kinetic model and the concentration of NO in the vessel wall\footnote{It is assumed that NO diffuses perfectly throughout the vessel wall so that even though it is produced in the endothelium, it is present in identical concentration in the VSMC.}. This means that [NO] has an effect converse to that of \cca\!\!; NO facilitates dephosphorylation of myosin chains while \ca facilitates phosphorylation. Whilst proportionality with shear stress is consistent with the current knowledge of NO production \cite{Buga91}, it means that NO acts as a form of positive feedback on the system: if the vessel dilates, shear stress rises (as flow velocity rises) and so more NO is produced which causes the vessel to dilate further. The system doesn't actually tend towards dilation indefinitely due to the passive stiffness of the VSMC, but an increase in ABP in the model leads to a greater increase (proportionally) in blood flow. 

A key feature which is lacking from the Abatay model is the myogenic response. In order to properly include the myogenic response however, a different electrochemical model is necessary. 

\subsection{Rowley-Payne}
Another candidate for the electrochemical model is that proposed by Payne and Rowley \cite{Rowley04}. This model is a Hodgkin-Huxley type model of the cell membrane including single currents for \ca, K$^+$ and Na$^+$ that only depend on membrane potential and a stretch-sensitive leakage current as described in \cite{Yang03a}. This stretch-sensitive current is a problem as mentioned in relation to Yang's model. Additionally this simplified model is not capable of spontaneous, vasomotion-generating oscillations under any circumstances.

\subsection{Jacobsen}
\label{Jacobsen}
Jens Jacobsen from the University of Copenhagen developed a model of the VSMC for his investigations into the effect of cGMP on calcium waves/oscillations in vascular smooth muscle \cite{Jacobsen07a}. Jacobsen was interested in the transition from a state where waves of high \cca travelled from one end of a cell to the other to a state where \cca in the whole cytoplasm oscillates as one. His work was undertaken with a view to understanding the process by which \cca waves in cells transition to whole-cell \cca oscillations and so go on to synchronised oscillations amongst all VSMCs, resulting in vasomotion. As mentioned in \cref{vasomotion}, the role of cGMP in this process may be critical to the transition from \cca waves to whole-cell \cca oscillations. Jacobsen's model of the VSMC describes a cell divided along its length into many finite elements, through which substances diffuse at a finite rate. If these spatial variations are ignored, i.e., no division of the cell and instant diffusion, then the resulting model is complex enough to enable vasomotion-generating oscillations in \cca\!\!.

\begin{figure}[!h] \centering \includegraphics[scale = 0.8]{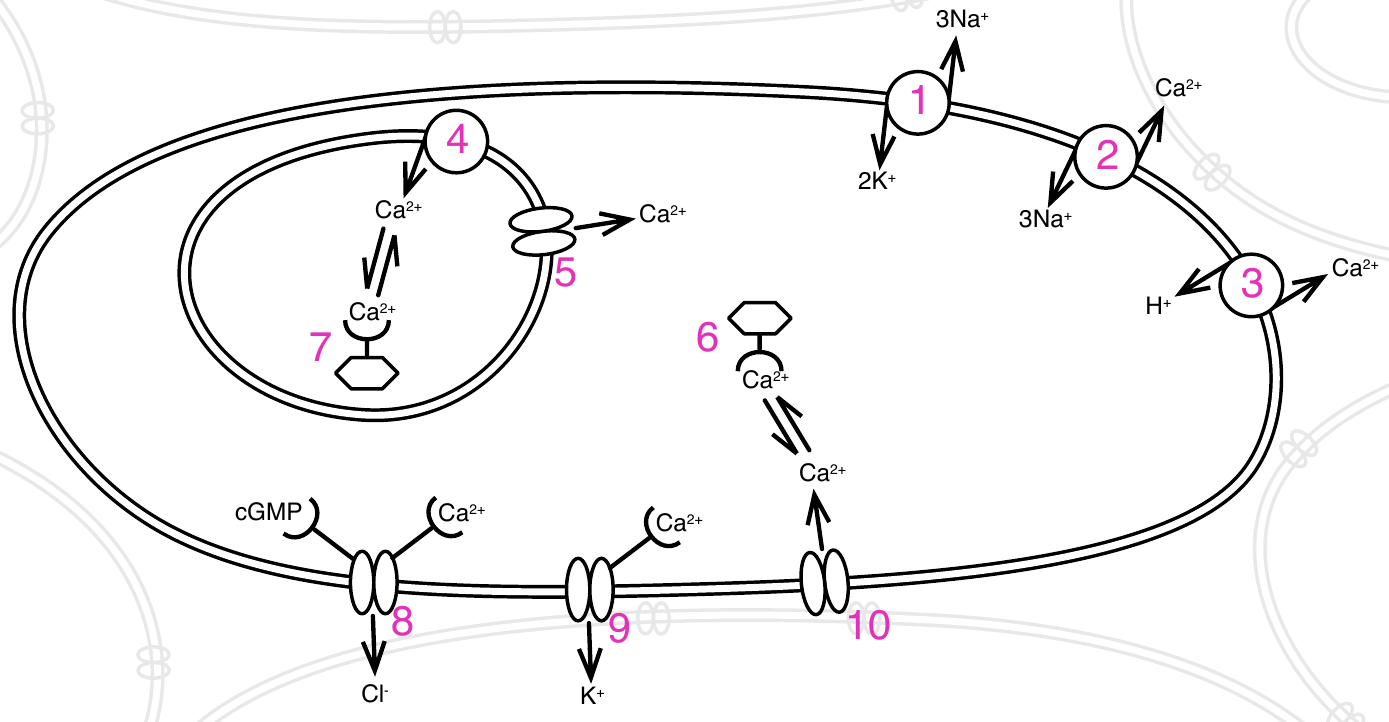} \caption{Schematic of the cell model in \cite{Jacobsen07a}, the labelled structures are: 1) Na$^+$ - K$^+$ ATPase, 2) Na$^+$ - \ca exchanger, 3) cell membrane \ca ATPase, 4) SR \ca ATPase, 5) SR \ca release channel, 6) cytosol \ca buffer, 7) SR \ca buffer, 8) cGMP sensitive, \ca dependent Cl$^-$ channel, 9) voltage dependent, \ca sensitive K$^+$ channel, 10) voltage dependent \ca channel (L-type channel). Background leakage and stretch sensitive ion channels are not shown.} \label{figJacobsenCell} \end{figure}

The model is similar to Yang's in that there are three fluid compartments, the cytoplasm, the sarcoplasmic reticulum (SR) and the extracellular space. Although the currents for all of the ions are included in the model, the actual concentrations of all but \ca are assumed to be constant within the cytoplasm because their native concentrations are so high that the small flux across the cell membrane makes very little difference. The currents are included because of their important influence on membrane potential. Another important feature is that both the cytoplasm and the SR contain a \ca buffer, this is a substance that binds to \ca in proportion to \cca and has the effect of maintaining \cca in a smaller range than would be the case in its absence. The elements of the cell which are represented in Jacobsen's model are shown in \cref{figJacobsenCell}.

The first stage of Jacobsen's test of his model is to see what happens if all of the currents across the cell membrane are switched off and so only the interaction between the cytoplasm and the SR remains. The result (under certain conditions) is an oscillation in cytoplasmic \cca and an inverse oscillation in the SR. The reason that this system oscillates, rather than settling at a single steady-state, is that there is a fast acting positive feedback mechanism that reinforces \ca release from the SR into the cytoplasm once a threshold level of \ca has been reached in the cytoplasm, this mechanism is referred to as Calcium-Induced Calcium Release, or CICR. There is a higher threshold of cytoplasmic \cca beyond which the SR calcium release channel is inhibited rather than activated, but the binding of \ca to the inactivation site is somewhat slower than binding to the activation site. Once enough \ca has been released and enough time has passed to allow the binding of \ca to the inactivation sites on the SR \ca release channels, the channels become inactive and the cytoplasmic \ca is pumped back into the SR, against the concentration gradient, by a calcium-ATPase. An ATPase is a pump that transports ions across a membrane against their natural concentration gradient, so named because they require the energy available from ATP (adenosine-triphosphate) to do so. The restoration of \ca to the SR takes enough time for the unbinding of \ca from the inactivation sites and the cycle can start again with the slow leakage of \ca into the cytoplasm from the SR.

Another feature of Jacobsen's model is that it includes the influence of a substance called inositol-1,4,5-triphosphate, or IP$_3$. This may well be useful in an advanced stage of investigation because it is through the production of IP$_3$ that phenylephrine effects vasoconstriction, so its inclusion in the model may make it possible to directly compare the effects of eNOS inhibition and $\alpha$-adrenergic stimulation (application of phenylephrine).

\subsection{Cornelissen and Carlson}
\label{cornelissenandcarlson}
A model consisting of 10 resistance compartments in series, each with different reactances to flow, pressure, and metabolic factors, is presented by Cornelissen et al. in \cite{Cornelissen02}. The responses of the different generations of vessels were fitted primarily to data from Liao and Kuo \cite{Kuo95,Liao97}. Although the responses are fitted to these data, the functions defining them are not derived from a physical model. The authors tuned the relative strengths of the three mechanisms (myogenic, flow-induced, and metabolic) in the different generations of blood vessels to replicate the tone and pressure distributions seen in the data. Whilst very instructive in the higher level interactions of these mechanisms, the lack of a physical model underlying the response curves limits the usefulness of this approach for the purposes of understanding the processes leading to vessel behaviour.

A very similar  model to that of \cite{Cornelissen02} was developed by Carlson in \cite{Carlson08}. This model comprises of only 7 segments in series and includes the three mechanisms of myogenic response, shear-induced vasodilation, and metabolic influence acting on all of the segments, to varying degrees. Again the modelling of the variation of wall tension due to these factors is non-physical; no mechanical, kinetic, or biochemical mechanisms inside the VSMC or endothelial cells are described. The equations representing the three mechanisms are selected for their ability to re-create experimentally observed behaviour.

\subsection{Koenigsberger}
\label{koenigsberger}
In an investigation into the role of the endothelium on vasomotion \cite{Koenigsberger05}, Koenigsberger et al. model a population of VSMCs coupled to a population of endothelial cells. Their model sought to explain the apparently contradictory experimental data on the role of the endothelium in arterial vasomotion, some of which suggested that vasomotion was abolished by the endothelium \cite{Sell02} or arose in its absence \cite{Sell02,Haddock02,Lamboley03}, some of which suggested that the endothelium was necessary for vasomotion to occur \cite{Gustafsson93,Huang97,Peng01,Okazaki03}. Through their model it was found that the action of the endothelium, in reducing intracellular \cca in the VSMCs, could move \cca either into or out of the range in which whole-cell \cca oscillations are sustainable, depending on the initial concentration of \ca. In other words, because there is only a finite range of \cca in which whole-cell \cca oscillations can be sustained, the lowering of \cca by the endothelium can either bring \cca down into that range, or lower \cca below that range.

In a subsequent paper \cite{Koenigsberger06} Koenigsberger et al. use the same model to investigate the effect of wall stress on the occurrence of vasomotion. They find that wall stress is the principal determinant of intracellular \ca in the VSMC (through the myogenic response) and hence can either induce or abolish vasomotion similarly to the production of cGMP by the endothelium.

\section{Conclusions}
Much has been discovered about the nature and mechanisms of vascular regulation, and of the structure of the arteriole. The role of \ca has been shown to be crucial in modulating smooth muscle tone, and NO has been unveiled as the flow-induced relaxing agent which antagonises the myogenic response. The endothelium has been identified as the source of this NO release. Mathematical models have successfully reproduced many aspects of the behaviour of the components of the arteriole, such as the smooth muscle cell, although no physically based model has yet combined the myogenic response and the influence of NO to successfully reproduce the behaviour observed in \cite{Kuo91}. In fact, several models have mistakenly modelled the myogenic response as a \emph{stretch}-dependent phenomenon, when in reality the observed response of arterioles to persistent increases in intramural pressure could only be produced by a \emph{stress}-dependent mechanism.

The phenomenon of vasomotion has been observed and modelled and some understanding of the conditions under which it occurs has been gained. It has been suggested by mathematical modelling of coupled groups of cells that there is a range of pressures and flow states outside which vasomotion cannot arise.

\chapter{A Model of Vascular Regulation}
\label{chapVascModel}
\section{Introduction}
As stated in \cref{introObjective}, this study is concerned with explaining the lack of impact of eNOS blockade on blood pressure oscillations, presented in \cite{Ide07} and described here in \cref{interactionsNOMyoMech}, which we believe to be linked to cerebral autoregulation. We begin by hypothesising about the lack of apparent change in ABP-CBFV correlation after inhibition of eNOS. Some possibilities are:
\begin{enumerate}
\item NO is not involved in cerebral autoregulation at all, so inhibiting the enzymes that produce it has no effect on the ABP-CBFV correlation.
\item Cerebral autoregulation is not, in fact, reflected by changes in ABP-CBFV correlation.
\item Stimulation of $\alpha$-adrenoreceptors by phenylephrine overrides the influence of NO, so that instead of comparing an NO depleted condition with a normal condition but with matching ABP, what is actually being compared is a case where NO is depleted and a case where NO is abundant but ineffective.
\item There is still enough NO after inhibition of eNOS to regulate the blood flow by the same proportion as before.
\item Some other vasodilatory mechanism takes over from NO release when eNOS is inhibited maintaining the same correlation between ABP and CBFV.
\end{enumerate}

Number 1 seems very unlikely; NO is a vasodilator found in the arterioles of the brain and as such must be either used in autoregulatory mechanisms or compensated for by those mechanisms, either way removing it would have an effect on autoregulation. Number 2 is possible, although again it seems unlikely; we may not be able to predict the actual nature of the change in ABP-CBFV correlation that we would expect to see for any given change in cerebral autoregulation but, as explained in \cref{DeBoer}, we would expect to see a change of some kind. Number 3 is a strong contender, the mechanism by which phenylephrine causes the VSMCs to contract is certainly not independent from that by which NO causes them to relax; it could well be the case that the introduction of phenylephrine inhibits autoregulation just as much as the depletion of NO. Possibility 4 also has potential, and it could be a significant effect either a) because the inhibition of eNOS by L-NMMA is not as complete as assumed and/or the rate of decay of existing NO is lower than thought, both leading to more NO remaining after eNOS blockade than was assumed, or b) the amount of NO remaining after eNOS blockade is small but because the blood vessel is in a contracted state, the amount of NO required to modulate the bloodflow according to the demands of autoregulation is also small, so although there is little NO available, there is more than enough to maintain autoregulation around the new, low radius, operating point. Number 5 is another possibility, although it seems that if any of 1-4 can be shown to be feasible then this explanation should be discarded on the grounds of excessive complexity.

To continue investigations into the viability of proposals 3 and 4 it is clear that a model of the system is required. The model must include the production of NO and factors affecting the rate of that production, e.g., the effect of shear stress on the production of NO in the endothelium; the release and uptake of \ca in the smooth muscle cells; the relationship between the concentrations of the species in the VSMC and the contraction of the same; the effects of each species on the interactions of the others, e.g., the effect of NO on the dephosphorylation of myosin cross bridges \cite{Payne05}.Essentially the model needs to be sufficiently complex that it can eventually provide some insight into the results shown in \cite{Ide07} and hence give us some better understanding of what might be going on. Here, and indeed as should be the case in all mathematical models, we aim to construct the simplest model possible which can shed light on the role of NO and it's interactions with the myogenic response. For this reason some parts of the model will be necessarily complex - to reflect the true action of the VSMC - whilst some parts can very simple. For instance, it is considered necessary to include the kinetics of myosin phosphorylation and attachment, and the subsequent mechanical effects of myosin cross-bridge cycling in order to accurately represent the effects of \ca and NO on the vessel radius - this being the key interaction under investigation. Similarly, the production and transport of NO through the system should be given its proper physical basis in the equations. In contrast, the full complexity of the NO-cGMP pathway need not be represented explicitly - it is sufficient to put in place a reasonable approximation to the mapping that an enzyme mediated pathway produces between NO and cGMP. Nor is the complexity of a full biochemical model of the VSMC (with all the associated ionic currents and channel activations) considered to be justified by this investigation. It is sufficient that \cca is a state of the model and that the overriding influence on \cca is the stress in the vessel wall, as per the myogenic response. Further elaborations on the work presented here may seek deeper insight into the particulars of the processes which we have represented simply, but here we constrain ourselves to the simplest model which can produce insights of value.

\begin{figure}[!h] \centering \includegraphics[scale = 0.8]{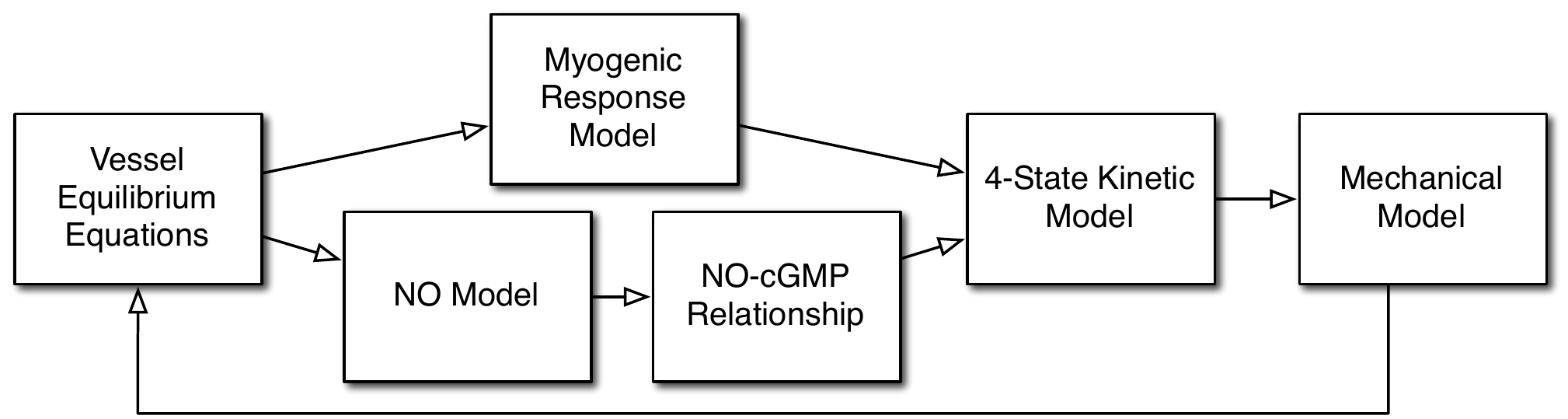} 
\caption{Model schematic; boxes represent sets of equations and lines represent states as they are propagated through the model. ABP is set externally.} \label{figSSoverallflow} \end{figure}

Such a model as described above is presented here, consisting of several sub-models connected by shared states, as shown in \cref{figSSoverallflow}. The NO model calculates the rate of change of NO concentration in the vessel wall dependent on ABP and the given vessel radius. The \cno -- \ccgmp relationship converts \cno into \ccgmp\!\!. The myogenic response model uses ABP and cell length, $\lambda$, to calculate stress in the vessel wall, and hence the rate of change of \cca\!\!. The 4-state kinetic model uses \cca and \ccgmp to calculate the phosphorylation and attachment of the myosin cross-bridges in the VSMC. Finally the mechanical model calculates the rate of change of VSMC length, which is then converted into the rate of change of vessel radius -- thus completing the feedback loop.

\section{Model Equations}
\label{Equations}
\subsection{NO model}
\label{NOeqns}
The NO model has two compartments: the vessel wall, and the blood in the vessel lumen. NO is produced in the vessel wall at a constant rate, $s_w$, supplemented by a shear-stress-dependent production, $k_{\tau} \frac{\hat{\tau}}{A_w} $, where $k_{\tau}$ is a constant and $\hat{\tau}$ is the normalised shear stress at the inner vessel wall surface. The cross-sectional areas of the lumen and the vessel wall are $A_b$ and $A_w$ respectively. The area of the wall, $A_w$, is assumed not to deviate from its value at baseline conditions. This is equivalent to a constant volume assumption for the vessel wall. Diffusion into the blood from the wall is assumed to be proportional to the relative concentration difference between the two compartments with constant of proportionality $h_w$. Blood entering the upstream end of the vessel has an NO concentration of $C_{in}$; as it progresses downstream more NO is accumulated in the blood from the vessel wall. The NO concentration in the vessel wall, $C_w$, is assumed to be constant along the length of the vessel. NO decays in both the wall and the blood with time, although the decay rate in the blood, $r_{db}$, is much higher than that in the wall, $r_{dw}$, due to NO scavenging by haemoglobin. Thus the governing equation for the NO concentration in the blood, $C_b$, at a downstream distance, $x$, from the entrance to the arteriole is:
\beq \diff{C_b(x)}{x} = \frac{1}{Q} \cdot  \left(h_w \left( C_{w_e} - C_b(x) \right) - r_{db} A_b C_b(x) \right) \label{NO1} \eeq

where $Q$ is the volumetric blood flow rate given by Poiseuille's law (\cref{Poiseuille}) and $C_{w_e}$ is the effective NO concentration in the vessel wall, given by:

\beq C_{w_e} = \frac{C_w}{\gamma_{wb}} \eeq

such that when $C_w = \gamma_{wb} C_b$, i.e., $C_{w_e} = C_b$, there will be no transport of NO between the vessel wall and the blood. In the absence of other processes, the equilibrium concentration of NO in the wall would be greater than that in the blood by a factor of $\gamma_{wb}$.

The spatially averaged NO concentration in the blood over a distance $L$ from the upstream end of the vessel is calculated from the solution to \cref{NO1} to give:
\beq \bar{C_b} = \left( \frac{h_w C_{w_e}}{h_w + r_{db} A_b} - C_{in}\right) \cdot \left( 1 - \frac{Q}{\left(h_w + r_{db} A_b \right) L} \cdot \left( 1 - e^{-\left(h_w + r_{db} A_b \right) \frac{L}{Q}} \right) \right)+ C_{in} \label{cbbar} \eeq

The rate of change of \cno in the vessel wall is then calculated using this spatial average concentration, $\bar C_b$. Thus assuming that the net effect of NO transfer to the blood on the NO concentration in the wall can be approximated by using the difference between the spatially homogenous NO concentration in the vessel wall and the spatially averaged concentration in the blood, we can write the following equation for the conservation of mass of NO:
\beq \diff{C_w}{t} = \frac{-h_w}{A_w} \cdot \left( C_{w_e} - \bar{C_b} \right) - r_{dw} C_w + s_w + k_{\tau} \frac{\hat{\tau}}{A_w} \label{dNOdt} \eeq

The normalised shear stress is given by:
\beq \hat{\tau} = \frac{\Delta P}{\Delta P_0} \cdot \frac{r}{r_0} \eeq

\subsection{\conc{NO} -- \conc{cGMP} relationship}
\label{NOcGMP}
The mapping from \cno to \conc{cGMP} assumed here is that the logarithm of \ccgmp is a sigmoidal function of \cno\!\!. This is done so that the range of \cno values from the NO model corresponds with the active range of \ccgmp values from \cite{Lee97} and to reflect the fact that the pathway by which cGMP production is stimulated by NO is limited by the availability of soluble guanylate cyclase in the VSMC \cite{Denninger99}. We define:
\beq p\ccgmp = \log_{10}\!\left( \ccgmp\!\!\right)\eeq

which is given by a sigmoidal function of \cno\!\!:
\beq p\ccgmp =  p\ccgmp\!\!_L + \frac{p\ccgmp\!\!_U - p\ccgmp\!\!_L}{1 + e^{-\frac{\cno - z_{half}}{k_z} }}\eeq

where $p\ccgmp\!\!_L$ and $p\ccgmp\!\!_U$ are constants representing the lower and upper bounds on \ccgmp respectively and $z_{half}$ and $k_z$ are functions of similar bounds on \cno given by:
\beq z_{half} = \cno\!\!_L + \frac{\cno\!\!_U - \cno\!\!_L}{2}\eeq

and:
\beq k_z = \frac{\cno\!\!_L - \cno\!\!_U}{2 \ln{\left( \frac{1}{0.95} - 1 \right)}}\label{sigmoiddef}\eeq

\begin{figure}[!hbtp] \centering \includegraphics[scale = 0.6]{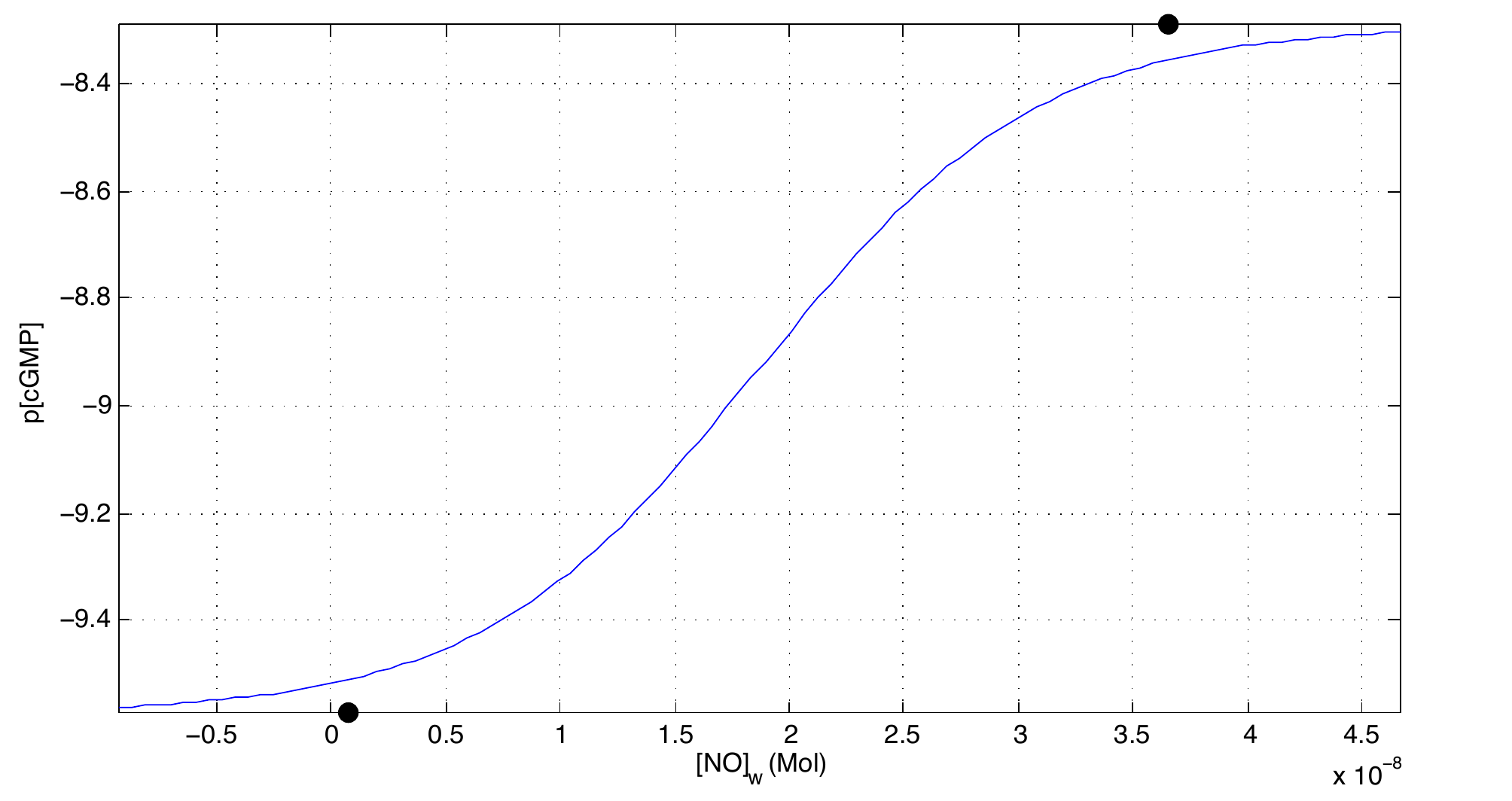} \caption{\cno -- $p$\ccgmp relationship. The coordinates of the two black circles are given by the parameters defining the function:  ($\cno\!\!_L$, $p\ccgmp\!\!_L$) and ($\cno\!\!_U$, $p\ccgmp\!\!_U$)} \label{figcGMPcNO} \end{figure}

The definition of a sigmoid in this fashion may seem obscure at first but actually it allows a better intuitive grasp of the active range of the function; the four parameters defining the relationship are the two saturation limits of the output, and the two input values at which the function reaches 5\% and 95\% transition respectively from the lower output saturation value to the higher. It is from this formulation that \cref{sigmoiddef} is derived; a graphical representation is given in \cref{figcGMPcNO}.

The \cno -- \ccgmp relationship is assumed to reach equilibrium instantly, or at least to be much faster than the rest of the system and so to have negligible dynamics. This assumption is made in the context of an NO model of tuneable speed, thus the overall speed of the feedback between flow and VSMC tone via the NO model is still free in the model as a whole.

\subsection{Myogenic Response Model}
Despite full implementation and testing of the electrochemical models described in §\cref{Models in the Literature} \cite{Yang03a, Abatay, Jacobsen07a, Rowley04}, we decided that a full electrochemical model of the cell was an unnecessary introduction of complexity at an early stage in the investigation. For the initial stage of replicating behaviour in the absence of vasomotion it is only necessary to model behaviour in line with the wall tension hypothesis of the myogenic response. The ability of an electrochemical model to replicate spontaneous vasomotion may be necessary at a later stage, in which case one of the more complex biochemical models may be re-inserted.

For the present implementation then, we have only to define a function relating stress in the vessel wall, $\sigma$, to equilibrium \cca\!\!, $\overline{\ccam}$; for this we will use a sigmoidal function of exactly the same form as that of the \conc{NO} -- \conc{cGMP} relationship of \cref{NOcGMP}:
\beq p\overline{\ccam} =  p\ccam_L + \frac{p\ccam_U - p\ccam_L}{1 + e^{-\frac{\sigma - z_{half}}{k_z} }}\eeq

where, as before,
\beq p\overline{\ccam} = \log_{10}\!\left( \overline{\ccam}\right) \eeq

and in which $z_{half}$ and $k_z$ are functions of the bounds of the active region of $\sigma$ given by:
\beq z_{half} = \sigma_L + \frac{\sigma_U - \sigma_L}{2}\eeq

and:
\beq k_z = \frac{\sigma_L - \sigma_U}{2 \ln{\left( \frac{1}{0.95} - 1 \right)}}\eeq

In this context, $\sigma$ does not strictly signify stress in the vessel wall, rather force per unit vessel length, being defined as:
\beq \sigma = \frac{Pr}{t_w} \eeq

where $P$ is intramural pressure, $r$ is vessel internal radius, and $t_w$ is thickness of the vessel wall. The concentration of \ca is assumed to tend towards its equilibrium value $\overline{\ccam}$ with first order dynamics dictated by the time constant $\tau_{Ca}$:
\beq \diff{\ccam}{t} = \frac{1}{\tau_{Ca}}\left(\overline{\ccam} - \ccam \right) \label{dCadt} \eeq

\subsection{4-state Kinetic Model}
\label{4stateKM}
The 4-state kinetic model from \cite{Hai88a} describes the attachment and phosphorylation of myosin cross-bridges. The proportions of the total number of cross-bridges are: $M$, myosin cross-bridges; $Mp$, phosphorylated cross-bridges; $AMp$, phosphorylated and attached cross-bridges; and $AM$, attached, non-phosphorylated cross-bridges. 
\begin{figure}[!hbtp] \centering \includegraphics[scale = 0.5]{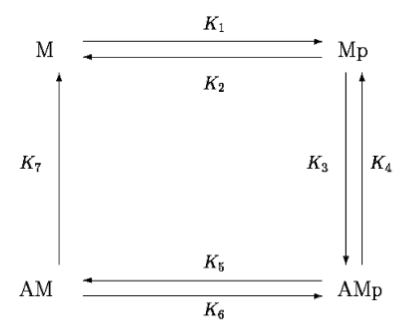} \caption{The four state kinetic system of myosin cross-bridges, adopted from Hai and Murphy, \cite{Hai88a}.} \label{fig4state} \end{figure}
Rate constants, $K$, govern the transitions between states (i.e., phosphorylation/dephosphorylation and attachment/detachment) such that the state derivatives are given by:

\beq \diff{}{t} \left[ \begin{array}{c}
M\\ Mp\\ AMp\\ AM
\end{array} \right]
= \left[ \begin{array}{cccc}
-K_1 & K_2 & 0 & K_7 \\
K_1 & -K_2 -K_3 & K_4 & 0\\
0 & K_3 & -K_4-K_5 & K_6\\
0 & 0 & K_5 & -K_6-K_7
\end{array} \right] \cdot
\left[ \begin{array}{c}
M\\ Mp\\ AMp\\ AM
\end{array} \right] \label{4state} \eeq 

subject to: \beq M + Mp + AMp + AM = 1 \label{4stateCons} \eeq

The phosphorylation and dephosphorylation rate constants are assumed to depend on \cca and \conc{cGMP} in order that the effects of MLCK and MLCP respectively can be simulated. It is via these rate constants that the myogenic response and the NO model exert their influence on the mechanical model, and so on the behaviour of the vessel. The phosphorylation rates are modulated by the concentration of \ca:
\beq K_1 = K_6 = \frac{\ccam^{n_{Ca_1}}}{\ccam^{n_{Ca_1}} + k_{Ca_1}^{n_{Ca_1}}} \label{phos} \eeq

whereas the \emph{de}-phosphorylation rates are negatively modulated by \cca and positively modulated by \conc{cGMP}:
\beq K_2 = K_5 = 0.55 + 2 \cdot \frac{\conc{cGMP}^{n_{cGMP}}}{\conc{cGMP}^{n_{cGMP}} + k_{cGMP}^{n_{cGMP}}} \cdot \frac{k_{Ca_2}^{n_{Ca_2}}}{\ccam^{n_{Ca_2}} + k_{Ca_2}^{n_{Ca_2}}} \label{dephos} \eeq

\subsection{Mechanical Model}
\begin{figure}[!h] \centering \includegraphics[scale = 0.6]{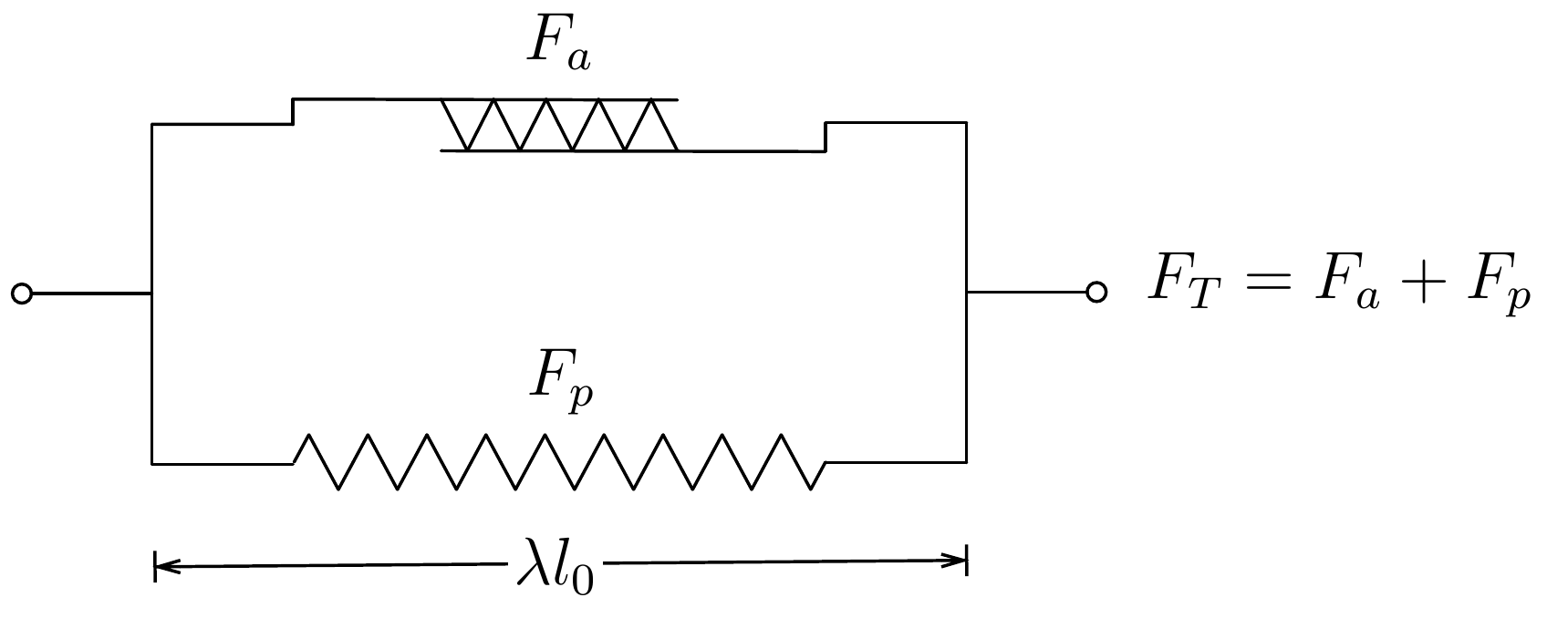} 
\caption{Schematic of the mechanical model showing the active force generating sarcomere (upper) and the passive non-linear spring element (lower).} \label{figmechModel} \end{figure}

The mechanical model is based on those of Yang \cite{Yang03a} and Stalhand \cite{Stalhand08}. The elastic extension of myosin cross bridges has been omitted here as their stiffness is so high that the error introduced by assuming that the cross bridges are rigid is negligible. Thus the model consists of two parallel elements, one active element representing the overlapping actin and myosin filaments, and one passive element, representing the passive stiffness of the cell, \cref{figmechModel}. 

The force in the active element is determined by the viscous friction coefficients $f_1$ and $f_2$ (for phosphorylated and un-phosphorylated attached myosin cross bridges respectively):
\beq F_a = l_0 \left(f_1 AMp \left(\hat v + \dot \lambda \right) + f_2 AM \dot \lambda \right) e^{-\left(\lambda - \lambda_{opt} \right)^2} \label{Fa} \eeq

where normalised cross-bridge cycling speed $ \hat v = v/l_0$ and normalised cell length $\lambda = l_{cell}/l_0$. The passive element takes the form of an exponential spring-like element, giving a passive force:
\beq F_p = q_1 \left( e^{q_2 \left(\lambda - 1\right)} - 1 \right) \label{Fp} \eeq

 The sum of the active and passive forces gives the total force in the cell:
\beq F_T = F_a + F_p \label{FT} \eeq

\begin{figure}[!h] \centering \includegraphics[scale = 0.7]{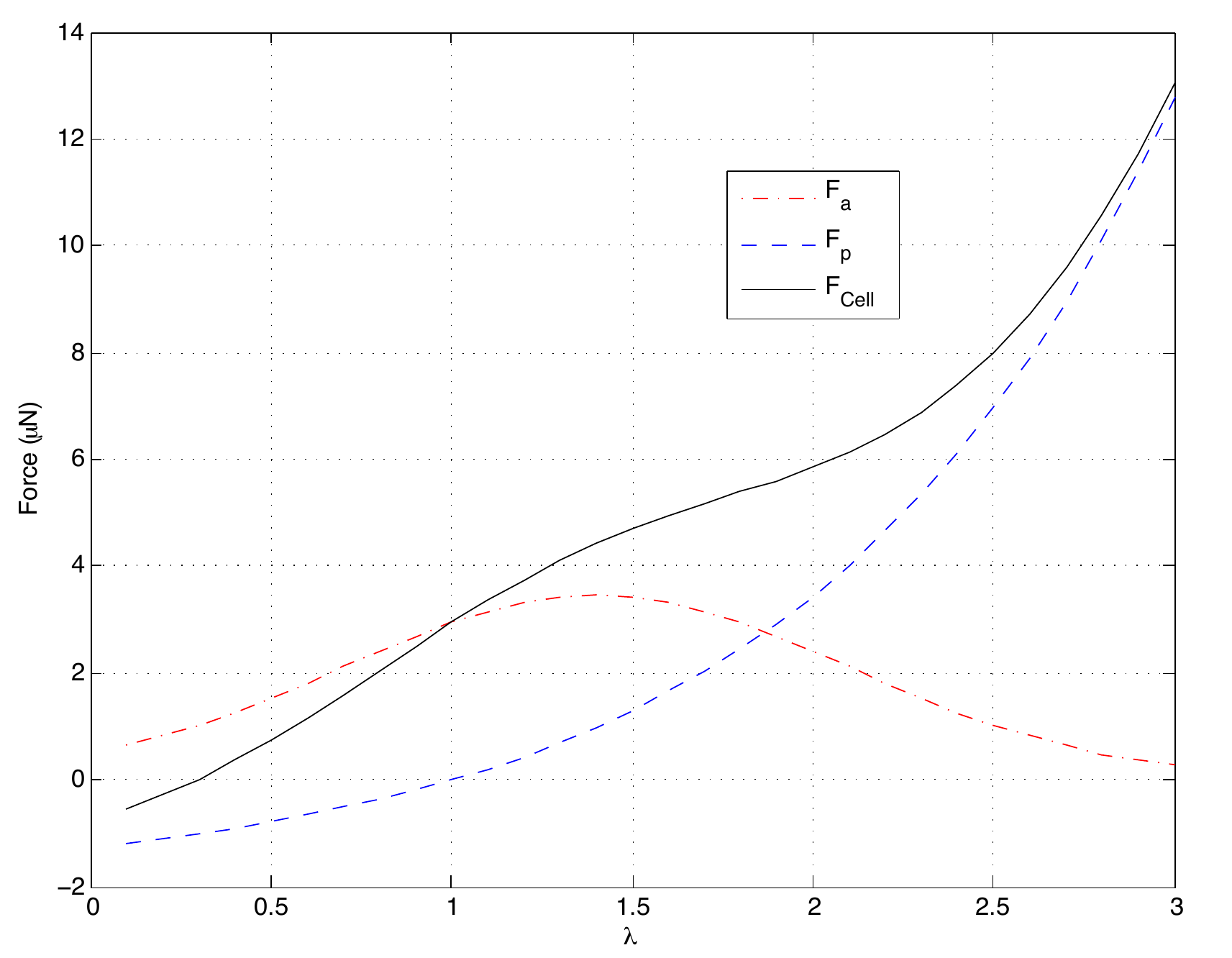} \caption{Contributions of the parallel passive and active elements of the mechanical muscle cell model at high phosphorylation for a range of cell lengths: $F_a$ - red, dot-dashed - force generated in the active element by the friction of cycling (phosphorylated and attached) cross-bridges, \cref{Fa}; $F_p$ - blue, dashed - force generated by the passive element, \cref{Fp}; $F_{Cell}$ - black, solid - total force in the cell, the sum of passive and active contributions, \cref{FMFP}.} \label{figFTbreakdown} \end{figure}

\Cref{figFTbreakdown} shows the contributions of the passive and active sides of the mechanical model to total force in the cell over a range of cell lengths for a fixed phosphorylation state. Note that the peak active force is reached when $\lambda = \lambda_{opt} = 1.4$ here; this is when the overlap between myosin and actin filaments is maximal.

Normalised cell length is related to radius by:
\beq r = \frac{1}{2} \left( \frac{\lambda l_0 n_c}{\pi} - t_w \right) \label{l2r} \eeq
 
 where $n_c$ is the number of VSMCs end-to-end around the circumference of the vessel and $t_w$ is the vessel wall thickness. The expansive force of the blood pressure can be equated with the contractive force of the muscle from \cref{FT}, for a single cell of width $w_c$:
\beq F_T = F_{Cell} = w_c \cdot P \cdot r \label{FMFP} \eeq

where $P$ denotes intramural pressure; the difference in pressure between the blood within the vessel and the tissue surrounding the vessel.

\section{Solution of the Model in the Steady-State}
\label{ssmodel}
The model has been presented as a series of ODEs and algebraic equations. In order to solve for the equilibrium state of the model it is necessary to set the state derivatives to zero and rearrange the equations to be explicit in terms of the equilibrium state values, rather than the state derivatives. In the equilibrium case, the model has inputs: intramural pressure, $P$, and pressure drop along the vessel, $\Delta P$; the outputs are the equilibrium states of the system, principally the radius of the vessel, or rather length of the cell, which is related to the vessel radius by \cref{l2r}. The overall form of the steady-state model is then:
\beq \mathbf{\overline{x}} = \left[ \begin{array}{c}
\ccam \\ C_w \\ \lambda
\end{array} \right]  = f\!\left(P, \Delta P \right) \label{states} \eeq

where $\mathbf{\overline{x}}$ is the state vector in the steady-state case, i.e., $\mathbf{\overline{x}} = \mathbf{x}$ when $\dot{\mathbf{x}} = \mathbf{0}$. The following sections detail the method of solution for each part of the model under equilibrium assumptions.

\subsection{NO}
Let us simplify our notation for brevity and define:
\beq \psi = 1 - \frac{Q}{\left(h_w + r_{db} A_b \right) L} \cdot \left( 1 - e^{-\left(h_w + r_{db} A_b \right) \frac{L}{Q}} \right) \label{psi} \eeq

such that \cref{cbbar} becomes:
\beq \bar{C_b} = \left( \frac{h_w C_{w_e}}{h_w + r_{db} A_b} - C_{in}\right) \cdot \psi + C_{in} \label{cbbarpsi} \eeq

We then rearrange \cref{dNOdt} for $C_w$ after setting $\diff{C_w}{t} = 0$ to get:
\beq C_w = \frac{\frac{h_w}{A_w}\bar{C_b} + s_w + k_\tau\frac{\hat{\tau}}{A_w} }{\frac{h_w}{A_w\gamma_{wb}} + r_{dw}} \eeq

Into which we can substitute \cref{cbbarpsi} and rearrange to yield the equilibrium equation:
\beq C_w = \frac{\frac{h_w}{A_w}C_{in}\left(1 - \psi \right) + s_w + k_{\tau} \frac{\hat{\tau}}{A_w}}{\frac{h_w}{A_w\gamma_{wb}} + r_{dw} - \frac{h_w^2}{A_w\left(h_w + A_br_{db}\right)\gamma_{wb}}\psi} \label{NOSS} \eeq

\subsection{Myogenic Response}
The myogenic response model is trivial to solve in the steady-state, we simply rearrange \cref{dCadt} with $\diff{\ccam}{t} = 0$ to get: 
\beq \ccam = \overline{\ccam} \label{ssCa} \eeq

\subsection{4-state Kinetic}
Let the vector of states in the 4-state kinetic model be denoted by:
\beq \left[M, Mp, AMp, AM \right]^T = \boldsymbol{\alpha} \label{alphadef} \eeq

and its equilibrium value by $\overline{\boldsymbol{\alpha}}$, then setting the derivatives to zero, \cref{4state} becomes:
\beq \mathbf{0}
= \left[ \begin{array}{cccc}
-K_1 & K_2 & 0 & K_7 \\
K_1 & -K_2 -K_3 & K_4 & 0\\
0 & K_3 & -K_4-K_5 & K_6\\
0 & 0 & K_5 & -K_6-K_7
\end{array} \right] \cdot
\overline{\boldsymbol{\alpha}} \label{4stateSS} \eeq 

This matrix, $\mathbf{K}$, is rank-deficient and so non-invertible. To solve for the equilibrium state vector $\overline{\boldsymbol{\alpha}}$ we replace one line of \cref{4stateSS} with \cref{4stateCons}, yielding an invertible matrix and so the equilibrium solution:
\beq \overline{\boldsymbol{\alpha}} = \left[ \begin{array}{cccc}
-K_1 & K_2 & 0 & K_7 \\
K_1 & -K_2 -K_3 & K_4 & 0\\
0 & K_3 & -K_4-K_5 & K_6\\
1 & 1 & 1 & 1
\end{array} \right] ^{-1}\cdot
\left[ \begin{array}{c}
0\\ 0\\ 0\\ 1
\end{array} \right] \eeq

\subsection{Mechanical}
To solve the mechanical model in the steady-state we set $\dot \lambda = 0$ in \cref{Fa}, giving the relation:
\beq F_T = F_p + F_a|_{\dot \lambda = 0} = q_1 \left( e^{q_2 \left(\lambda - 1\right)} - 1 \right) + l_0  f_1  \alpha_3  \hat v  e^{-\left(\lambda - \lambda_{opt} \right)^2} \label{FTSS} \eeq

Here force in the cell is a function only of its current length and the proportion of cycling cross-bridges  ($\alpha_3 = AMp$, as per \cref{alphadef}). The value of $\lambda$ at which $F_T = F_{Cell}$ from \cref{FMFP} is then found by solving:
\begin{align}
0 &= F_p + F_a|_{\dot \lambda = 0}  - F_{Cell}\\
0 &= q_1 \left( e^{q_2 \left(\lambda - 1\right)} - 1 \right) + l_0  f_1  \alpha_3  \hat v  e^{-\left(\lambda - \lambda_{opt} \right)^2} - w_cP\frac{1}{2} \left(\frac{\lambda l_0 n_c}{\pi} - t_w \right) \label{mechSS}
 \end{align}

Unfortunately there is no analytical solution to \cref{mechSS}, instead it must be solved numerically, in this case a modified Newton-Raphson method was used.

\section{Solution of the Dynamic Model}
In the dynamic case the equations of the model remain as they are presented in \cref{Equations}. The driving inputs to the model are $P(t)$ and $\Delta P(t)$, both of which are functions of time. The states of the model are the concentration of \ca in the vessel wall, the concentration of NO in the vessel wall, and the normalized length of the VSMCs which constitute the vessel wall, as in \cref{states}, and so the system of ODEs governing these states, given by \cref{dCadt,dNOdt,Fa} respectively, takes the form:
\beq \dot{\mathbf{x}} = f\!\left(\mathbf{x}, P(t), \Delta P(t) \right) \eeq

This system is solved using either a custom fixed-step solver or one of MATLAB's built-in ODE solvers, depending on which is more appropriate for the specific task.

\section{Assumptions}
\label{Assumptions}
As with any mathematical representation of a physical system, a number of assumptions have been made and the principle ones shall be discussed here, grouped by the part of the model to which they relate.
\subsubsection{NO}
\begin{enumerate}
\item The diffusion of NO is calculated using volume-averaged concentrations.
\item The transport of NO between the blood and the endothelial cells is assumed to be linear, whilst the transport of NO from the endothelium into the VSMCs is assumed to be perfect.
\item The relationship between pressure and flow is assumed to follow Poiseuille's law, despite the presence of red blood cells and other bodies in the blood meaning that it is not a Newtonian fluid.
\item The relationship between \cno and \conc{cGMP} is assumed to be sigmoidal and to equilibrate instantly.
\item The concentration of NO in the blood entering the vessel is assumed to be zero.
\end{enumerate}
\subsubsection{Myogenic}
\begin{enumerate}
\item The myogenic mechanism is assumed to be sensitive only to circumferential stress in the vessel wall, not to any other factors.
\item The myogenic mechanism is assumed to have the sole effect of raising \cca levels.
\item The dynamics of the myogenic response are assumed to be first order.
\end{enumerate}
\subsubsection{Kinetic}
\begin{enumerate}
\item Phosphorylation and dephosphorylation rates of myosin cross-bridges are assumed to be independent of attachment.
\item The dependance of phosphorylation and dephosphorylation rates on \ca and cGMP respectively are both assumed to be sigmoidal.
\end{enumerate}
\subsubsection{Mechanical}
\begin{enumerate}
\item Force from sliding of cycling cross-bridges and latch (attached, non-cycling) bridges is assumed to be linear with sliding velocity; given the complex interaction between myosin and actin in smooth muscle filaments it is highly unlikely that the friction force is simply proportional to the sliding speed.
\item Extracellular pressure is assumed to be constant, despite expansion of the blood vessel to compress extracellular tissue. This assumption will vary in its validity depending on the proximity of one vessel to another.
\item Supply pressure of the blood is assumed not to vary with vessel radius; whilst this is valid for the variation in radius of an individual vessel, if the behaviour of the vessel being modelled is representative of the behaviour throughout the brain then the arterial blood pressure will be affected by changes in peripheral resistance in the brain.
\item Forces in the vessel wall are assumed to be in equilibrium at all times, i.e., there is no inertia associated with changes in vessel radius, what's more, the changes in flow resulting from changes in vessel radius displacing blood volume are not considered.
\end{enumerate}
Most of these assumptions are reasonable approximations, and necessary for the construction of a compact model. Some are more gross than others, and these may be addressed in the future according to how severely they inhibit the ability of the model to replicate the behaviour of the physiological system.

\section{Conclusions}
In this chapter we have presented a new model of the arteriole. This model includes the myogenic response and the flow-induced release of NO from the endothelium. The effects of these mechanisms on the contraction of the VSMC are modelled and thus the feedback loop from vessel radius influencing flow and stress, to flow and stress influencing vessel radius is completed. We have noted the simplifications that are possible when the model is solved in the steady-state, and we have outlined the general form of the model in the dynamic case. Although several aspects of the model as presented are hypothetical approximations, the aim is to identify the nature of the relationships that are necessary to accurately recreate experimental observations. To this end it is thought better to start with simple relationships and complicate them as necessary than to start with great complexity and later attempt to extract that which is essential.

\chapter{Fitting and Analysis of Vascular Regulation Model}
\label{chapArtFit}
\section{Introduction}
Before our model can be used to make useful predictions about the behaviour of the cerebral vasculature, first the parameter values must be determined. This is done here by tuning the parameter values such that the behaviour of the model is as close as possible to some experimental measurements of the behaviour of a real arteriole. In this chapter we will first present the data to which we have chosen to fit the model. We will then describe the fitting method before presenting the results of the model fitting, both static and dynamic. Finally we will analyse the sensitivity of the model to variance in the values of its parameters -- highlighting those parameters which have the greatest influence on the behaviour of the system.
\section{Model Fitting Data}
Ideally quantitative data would be available for all processes in our model, allowing each process to be fitted individually before being compiled into a completely fitted model. Unfortunately very few quantitative data are available for most of the processes included in the model, and those quantitative results which are available often do not come from a common tissue type or experimental condition. Where data are available for low level structures and processes they invariably do not come from humans, especially not the brain. We are forced therefore to use data from other tissues, and indeed other mammals, in the hope that a model built on such data can provide answers to questions concerning the human cerebral vasculature. If we cannot build a useful model on data from other animals then we might revisit the question of applicability of data across species. A key exception to the general lack of low level quantitative data (from any animal) is the work of Lee et al. \cite{Lee97} which does provide quantitative data with which to fit the parameters of the 4-state kinetic model. For this reason the parameters of the 4-state kinetic model will be fitted to the data of \cite{Lee97} before the other parameters of the model are fitted to the data of Kuo at al. \cite{Kuo91} and Falcone et al. \cite{Falcone91}. The results presented in \cite{Kuo91} are particularly suited to our purpose as they include the steady-state responses of arterioles over a range of intramural pressures, both with and without flow; they also include the dynamic responses of the same or similar arterioles to step changes in pressure and flow, thus in theory providing enough information to separate these two effects.

\subsection{Fitting the 4-State Kinetic Model}
\label{LeeFit}
\begin{figure}[!hbtp] \centering \includegraphics[scale = 0.6]{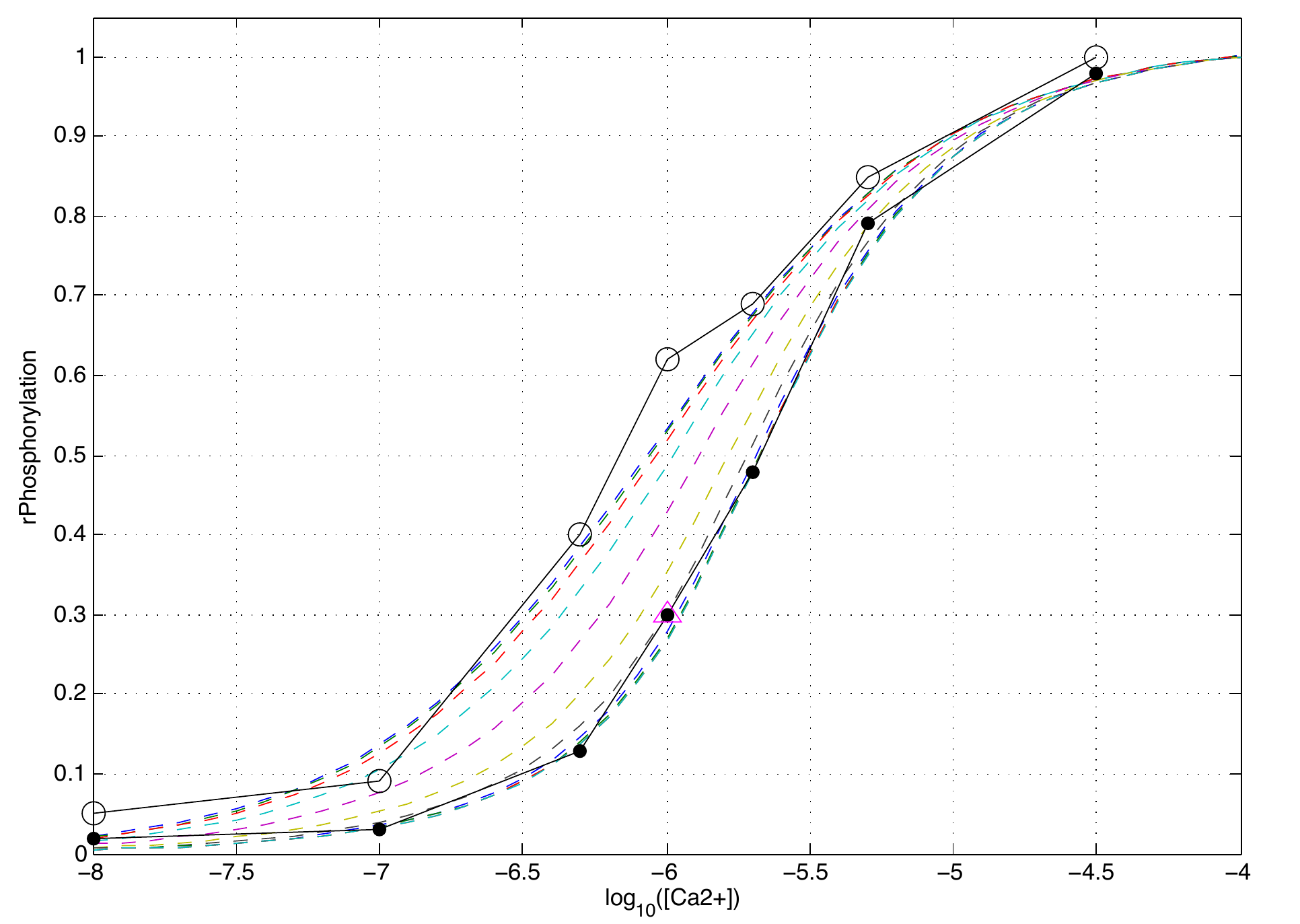} \caption{Proportion of maximum phosphorylation as a function of \cca\!\!. Hollow circles are data from \cite{Lee97} for the control condition, \conc{8Br-cGMP} = $0$M. Filled circles are the data from \cite{Lee97} for the \conc{8Br-cGMP} = $2\mu$M condition. Dashed lines are the values of relative phosphorylation (AMp + Mp) from the 4-state kinetic model for values of \ccgmp between $0.1$nM and $10\mu$M.} \label{figrPhospCa} \end{figure}

Lee et al. \cite{Lee97} use the cGMP analogue 8-bromo-cGMP (8Br-cGMP) to investigate the effect of cGMP on the phosphorylation of myosin light chains (myosin heads) and the force subsequently generated by the VSMC. Although Lee's results were obtained from VSMCs extracted from rabbit femoral arteries, it is  a reasonable working assumption that the biochemistry of MLCK and MLCP is similar in VSMCs found in all mammals; indeed we are forced to make this assumption due to the lack of experimental data relating to isolated muscle cells, or even isolated arterioles, from humans. The phosphorylation and dephosphorylation rates of the 4-state kinetic model have therefore been determined such that the phosphorylation of the model at varying levels of \cca and \ccgmp agrees with the data presented in \cite{Lee97}. The phosphorylation rate, \cref{phos}, depends solely on \cca\!\!. The dephosphorylation rate, \cref{dephos}, depends on \cca and on \ccgmp\!\!. \Cref{figrPhospCa} shows the total phosphorylation of the 4-state kinetic model alongside the data from \cite{Lee97}, to which the sigmoidal functions in \cref{phos,dephos} were fitted; fitting was manual, by inspection.

\begin{figure}[!hbtp] \centering \includegraphics[scale = 0.6]{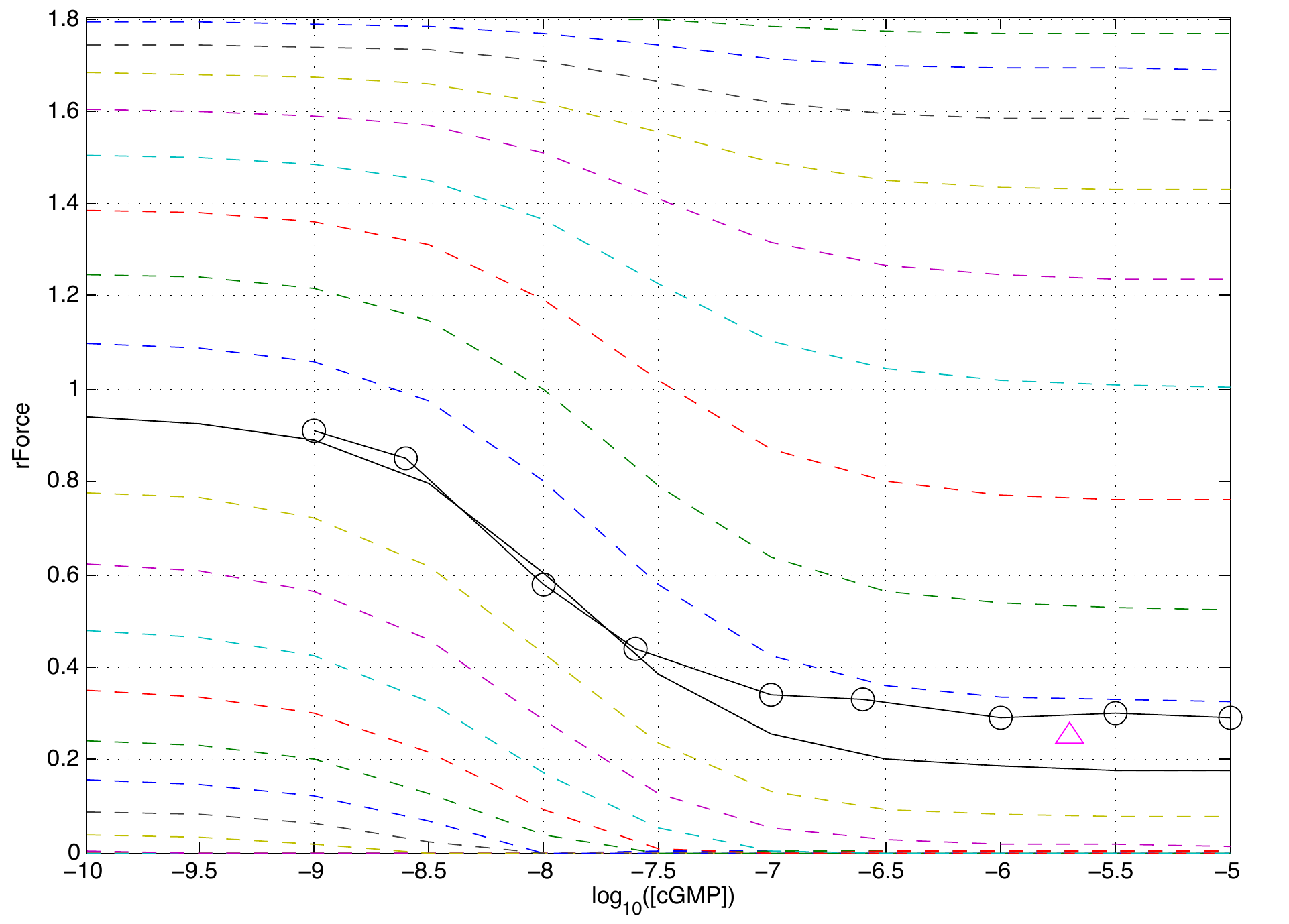} \caption{Relative force developed by the VSMC against \ccgmp\!\!. Circles show data from \cite{Lee97} for  \cca = $1\mu$M, dashed lines show the results of the 4-state kinetic model (converted from phosphorylation to force using the equation given in the caption of Figure~3 of \cite{Lee97}) for values of \cca between 10nM and $100\mu$M. The unmarked solid line represents model behaviour at \cca = $1\mu$M; this is the same condition at which the data were recorded.} \label{figFccGMP} \end{figure}

\Cref{figFccGMP} shows the relative force generated by the cell at varying concentrations of cGMP and a fixed \cca of $1\mu$M. The reason that the fitted sigmoid does not exactly overlay the data is that there is an experimental discrepancy in \cite{Lee97}; the two sets of experimental data shown in \cref{figrPhospCa,figFccGMP} have slightly different values of phosphorylation/force at their common point, \cca = $1\mu$M, \ccgmp = $2\mu$M, shown by the magenta triangle in \cref{figrPhospCa,figFccGMP}, making it impossible to exactly fit both data sets. \Cref{figrPhospCa,figFccGMP} can be thought of as perpendicular planar sections through the same three dimensional surface, with dimensions phosphorylation/force, \cca\!\!, and \ccgmp\!\!.

The initial parameter values of the \cno -- \ccgmp relationship given in \cref{NOcGMP} are also determined in part by the data of \cite{Lee97}, the range of \ccgmp  chosen was between $10^{-9}$M and $10^{-7}$M, to coincide with the region of greatest sensitivity of force to \ccgmp in \cref{figFccGMP}. The values of the 4-state kinetic model parameters are given in \cref{tab4State}: some values have been taken from the original model presentation by Hai and Murphy \cite{Hai88a}, whereas those relating to the effects of \ca and cGMP have been found through fitting outlined above.

\begin{table}[!h] \centering
\footnotesize{
\begin{tabular}{l l r l c}
\hline
Parameter & Description & Value & Units & Source\\
\hline
$K_3$ & phosphorylated attachment rate constant & $0.4$ & 1/s & \cite{Hai88a}\\
$K_4$ & phosphorylated detachment rate constant & $0.1$ & 1/s & \cite{Hai88a}\\
$K_7$ & non-phosphorylated detachment rate constant & $0.01$ & 1/s & \cite{Hai88a}\\
$k_{Ca_1}$ & phosphorylation \ca sensitivity constant & $3\e{-6}$ & M & fit to \cite{Lee97}\\
$n_{Ca_1}$ & phosphorylation \ca sensitivity constant & $0.85$ & -- & fit to \cite{Lee97}\\
$k_{Ca_2}$ & dephosphorylation \ca sensitivity constant & $8\e{-7}$ & M & fit to \cite{Lee97}\\
$n_{Ca_2}$ & dephosphorylation \ca sensitivity constant & $1.5$ & -- & fit to \cite{Lee97}\\
$k_{cGMP}$ & dephosphorylation \conc{cGMP} sensitivity constant & $3\e{-8}$ & M & fit to \cite{Lee97}\\
$n_{cGMP}$ & dephosphorylation \conc{cGMP} sensitivity constant & $1$ & -- & fit to \cite{Lee97}\\
\hline
\end{tabular}}
\caption{4-state kinetic model parameters. Here and elsewhere in this thesis, M stands for molar, or mol/L}
\label{tab4State}
\end{table}

\subsection{Data to Fit the Model Overall}
Here we will present the results of Kuo et al. \cite{Kuo91} and Falcone et al. \cite{Falcone91} and discuss their significance in terms of the fitting of the present model before going into the detail of the fitting process itself in \cref{ssfitting,dynfitting}.

\begin{figure}[hbtp] \centering \includegraphics[scale = 1]{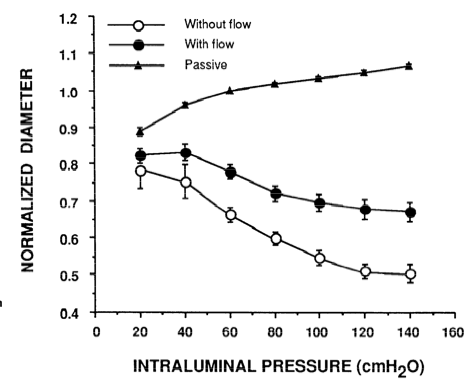} \caption{The steady-state response of an arteriole to changes in intramural pressure (reproduced from fig.~4 in \cite{Kuo91}).} \label{figKuo91Fig4} \end{figure}

\Cref{figKuo91Fig4} is reproduced from fig.~4 in \cite{Kuo91} and shows the equilibrium diameter of an arteriole across a range of intramural pressures in three different cases. The `Passive' case relates to an arteriole bathed in nitroprusside, which prevents \ca from being available to the VSMCs, whereas in the other cases the arteriole is bathed in a \ca containing solution; there is no flow in the passive case. In the `With flow' case $\Delta P = 4$cmH$_{2}$O and the pressure quoted on the \emph{x}-axis is the mean pressure. In the `Without flow' case the pressure at each end of the vessel is equal. Diameters have been normalized to the diameter in the passive case at a pressure of 60cmH$_2$O.

\begin{figure}[hbtp] \centering \includegraphics[scale = 0.8]{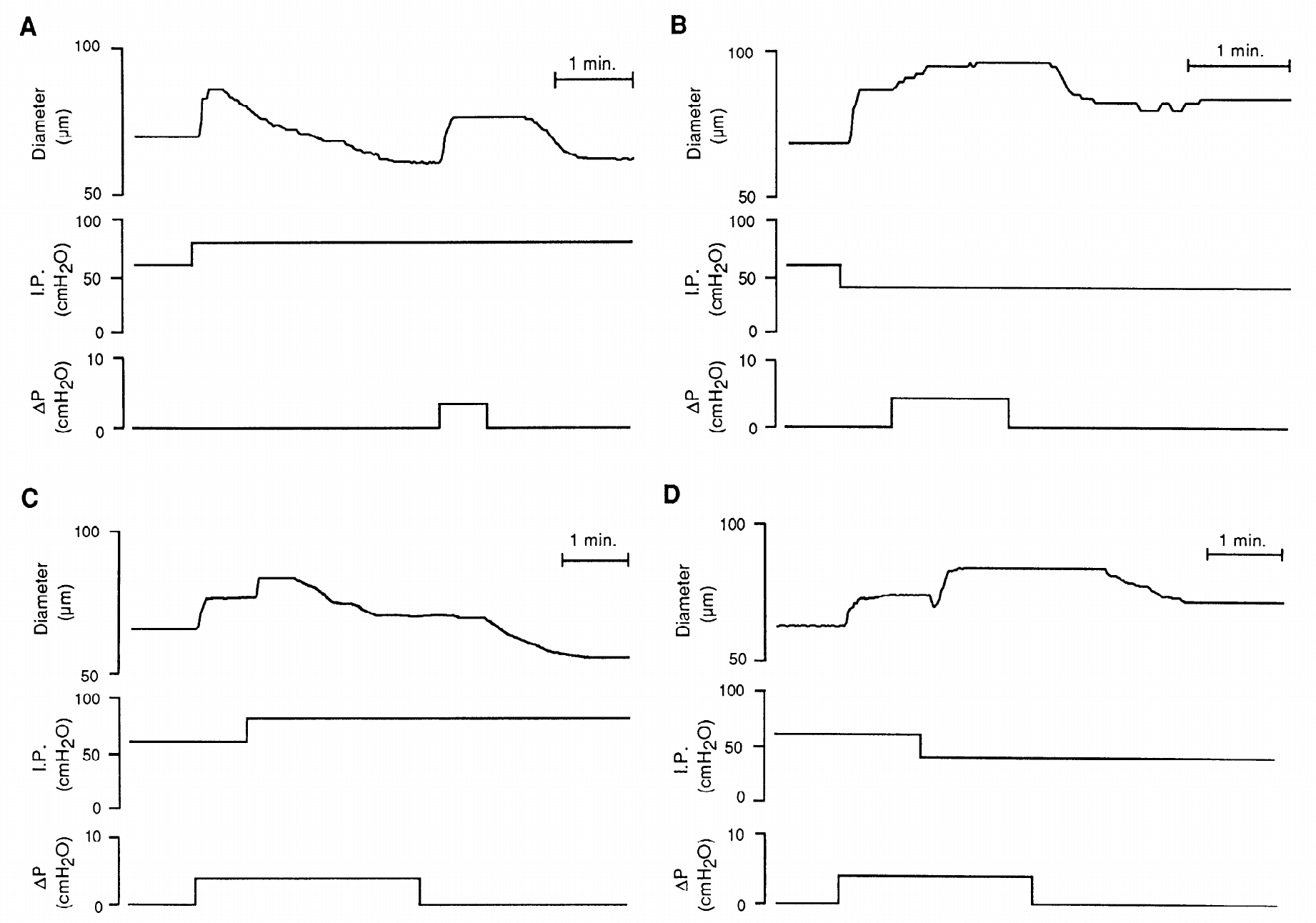} \caption{Four dynamic responses of arterioles to step changes in  $P$ and $\Delta P$ (reproduced from fig.~2 in \cite{Kuo91}). Each of the four subplots represents a different experiment. Each subplot shows the response of vessel diameter (top line) to changing intramural pressure (middle line) and pressure gradient (bottom line).} \label{figKuo91Fig2} \end{figure}

The dynamic responses of arterioles to step changes in $P$ and $\Delta P$ are presented in \cref{figKuo91Fig2}, reproduced from fig.~2 in \cite{Kuo91}. In the first subplot, A, it can be seen that in response to a step increase in intramural pressure the arteriole first passively dilates under the increased stress, then the myogenic response causes an increase in muscular tone and the vessel gradually contracts down to a new steady-state diameter which is smaller than the diameter before the step change in pressure. The second event in this plot is the onset of flow through the vessel, which causes a dilation of the vessel. The flow is short lived however, and the vessel returns to its previous baseline diameter soon (but not immediately) after flow has ceased.

\begin{figure}[!hbtp] \centering \includegraphics[scale = 1]{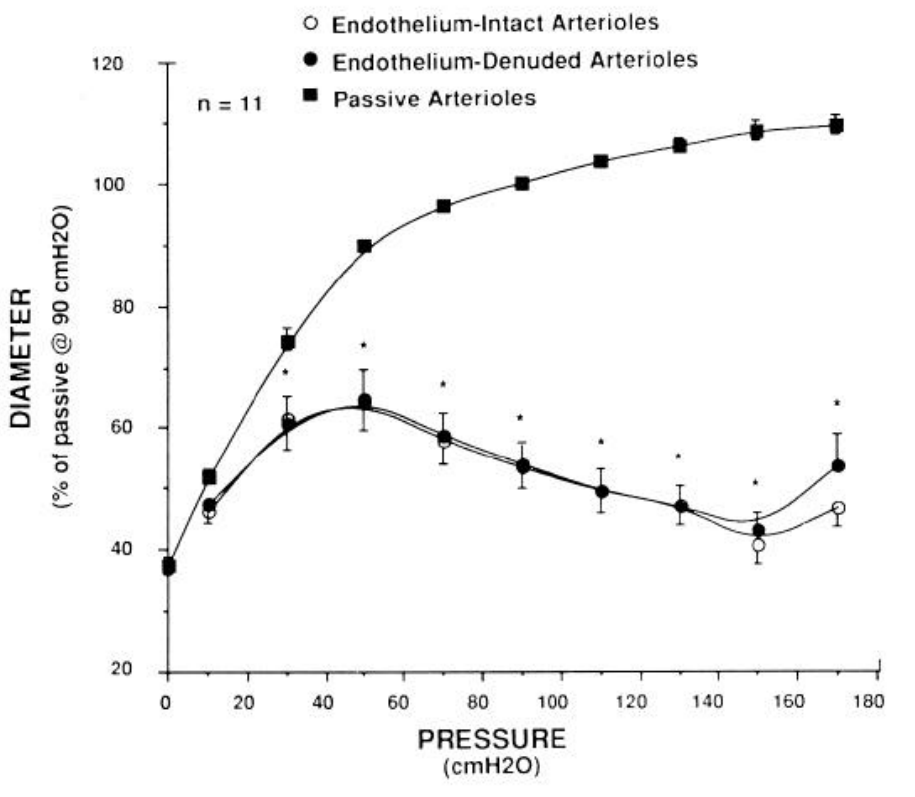} \caption{The steady-state response of an arteriole to changes in intramural pressure (reproduced from fig. 1 in \cite{Falcone91}).} \label{figFalcone91Fig1} \end{figure}

The steady-state results of Falcone et al. are shown in \cref{figFalcone91Fig1}. All of Falcone's experiments were conducted with the same pressure at each end of the isolated arteriole, therefore no flow is present and so we expect the active response to be largely unaffected by the removal of the endothelium (as long as the shear-stress induced production of NO is much greater than the baseline production). This is seen in the results with the exception of the highest pressure where the two active curves do diverge. Similarly to Kuo, Falcone's passive response was achieved through bathing the preparation in a solution devoid of \ca\!\!, although Falcone also introduced adenosine, which is an eNOS activating compound. This additional activation of eNOS is shown to have no effect on the present model in the absence of \ca in \cref{sspassivefit}.

\section{Steady-State Response}
\label{ssfitting}
The first stage of the fitting process for our model is to replicate the steady-state response of the arteriole. For this we will optimize the parameters present in the steady-state equations of \cref{ssmodel} to achieve the closest possible match to the data of \cref{figKuo91Fig4}. It is desirable for the purposes of accuracy to isolate the effects of parameters from each other as far as possible during fitting. To this end we will first optimize those parameters which affect only the passive response of the arteriole (i.e., in the absence of \ca\!\!\!) to fit the passive response curve of \cref{figKuo91Fig4}. In this way we can fit the subset of parameters involved in the passive response without affecting those parameters which appear only in the equations for the active case. Secondly we will fit the active response in the absence of flow, so as to achieve approximately accurate parameter values for the myogenic and mechanical elements of the model without the complication of the flow-induced NO response. Lastly we will use the values already obtained as a starting point to combine the flow and no-flow cases and optimize a combined set of parameters in order to achieve the best fit to both cases simultaneously.

\subsection{Fitting the Passive Response}
\label{sspassivefit}
The passive case shown in \cref{figKuo91Fig4}, where the arteriole was bathed in a solution containing nitroprusside, can be simulated in our model by omitting the myogenic response and setting the concentration of \ca to zero, effectively replacing \cref{ssCa} with $\ccam = 0$. This leads to a solution to the 4-state kinetic model of $\underline{\alpha} = \left[1, 0, 0, 0\right]^T$, which in turn leads to zero active force thanks to the factor of $\alpha_3$ in the active force term of \cref{mechSS}. This reduces the entire steady-state model to a single equality, conveniently isolating the values of $q_1$, $q_2$, and $l_0$ to be optimized by fitting to the passive curve of \cref{figKuo91Fig2}:
\beq 0 = q_1 \left( e^{q_2 \left(\lambda - 1\right)} - 1 \right) - w_cP\frac{1}{2} \left(\frac{\lambda l_0 n_c}{\pi} - t_w \right) \label{passive} \eeq

By extracting the data points from \cref{figKuo91Fig4} and interrogating our model at the same values of $P$, with $\Delta P = 0$ and $\ccam = 0$, we can compute the root-mean-squared (rms) error between our model's predictions of passive response and the data. By formulating the problem such that this error is expressed as a function of the parameter values then we can use an optimization function to find the set of parameter values which minimizes the error between the model predictions and the data. In other words, if the problem is expressed as:
\beq \epsilon = f\!\left(q_1, q_2, l_0\right) \eeq
where $\epsilon$ is the rms error between the data and the model prediction, then we can use an optimizer to find the parameter values which minimize this error, i.e., to find:
\beq \argmin_{q_1, q_2, l_0} f\!\left(q_1, q_2, l_0 \right) \label{minimization} \eeq

or, more generally, we optimize sets of parameters to fit a certain target dataset by finding:
\beq \argmin_\mathbf{p} f\!\left(\mathbf{p}\right) \label{minimizationGeneric} \eeq

where the rms fitting error $\epsilon = f\!\left(\mathbf{p}\right)$, and $\mathbf{p}$ is a vector of the parameter values to be optimized. The optimizer used in this case is the Matlab built-in function \texttt{fmincon} which finds the minimum of the given function subject to constraints. The constraints imposed here are simple bounds on the parameter values. This was done to bound the domain of the optimizer and so to enable it to solve the problem more quickly. It is important when doing this that we ensure that the optimum solution is not actually resting on one of the bounds; if that were the case the arbitrary value of the bound would have affected the solution, rather than just simplifying the optimizer's search for the unconstrained minimum of the objective function.

After an initial optimization, the solution with the model in the form of \cref{passive} was found not adequately to fit the passive data of \cref{figKuo91Fig4}. The rate of stiffening with cell elongation was found to be insufficient to match the curvature of the response curve in the data. To remedy this, another parameter, $q_3$, was introduced to exponentiate the length term in \cref{Fp}, leading to a new equation for passive force, and hence a new expression for the model in the passive case:
\beq 0 = q_1 \left( e^{q_2 \left(\lambda^{q_3}  - 1\right)} - 1 \right) - w_cP\frac{1}{2} \left(\frac{\lambda l_0 n_c}{\pi} - t_w \right) \label{passiveNew} \eeq

\begin{figure}[!hbtp] \centering \includegraphics[scale = 0.8]{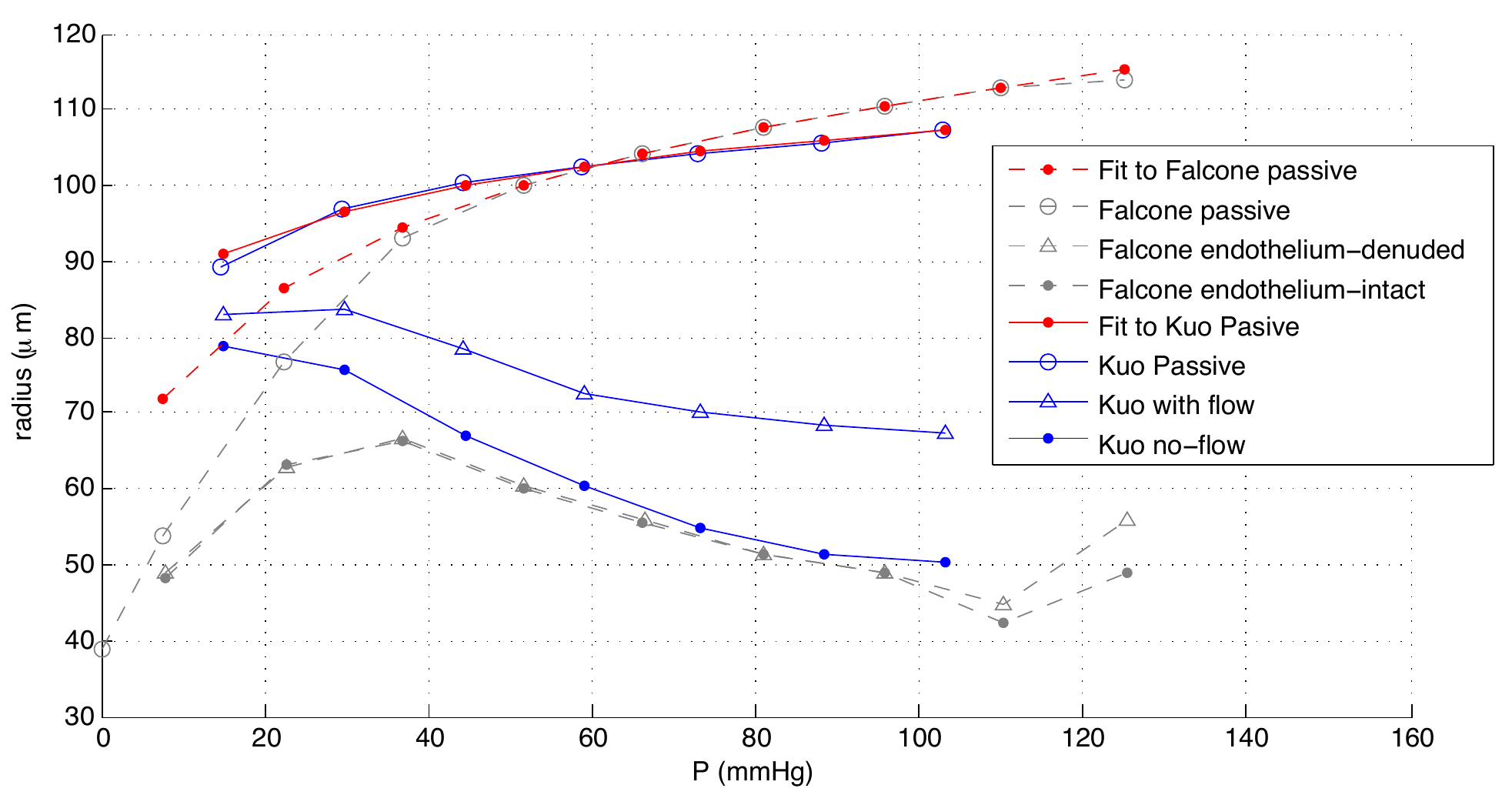} \caption{The steady-state data from \cite{Kuo91} have been extracted and are shown here in blue, the extracted data from \cite{Falcone91} are shown in grey. The red points and line show the fit achieved after optimization of the parameter values $q_{1-3}$ to minimize the rms error between the model and the data, i.e., to achieve the closest fit between the red and uppper blue lines. The normalized diameter of both sets of original data has been transposed to an absolute radius about a set point of $r = 60\mu$m when intramural pressure $P = 60$mmHg in the active, no-flow case from \cite{Kuo91}.} \label{figPassiveFitBoth} \end{figure}

With this modification the optimization was able very closely to fit the passive response of \cref{figKuo91Fig2}. This optimization was then applied to the fitting of the Falcone data with the exception that the results for pressures below 30mmHg were omitted from the fitting. This was done because the function given in \cref{passiveNew} was not capable of fitting this section of the curve and because it is possible that at very low pressures there is another mechanical element which comes into play. The element would be less stiff than that in \cref{passiveNew} but have a hard end-stop when it reached full extension, somewhat like a shorter elastic band in parallel with a longer stiff cord. Given that we are chiefly concerned with the autoregulatory regime, we decided not to model this additional element. The results of the optimized fitting to both Kuo and Falcone data are shown in \cref{figPassiveFitBoth}, together with the original data which have been extracted and re-normalized for comparison.

Of the terms in \cref{passiveNew}, only four are optimized parameters, one is a variable, one is an input, and the rest are physiological parameters which have been reliably experimentally determined and so are treated as fixed. The values of all terms in \cref{passiveNew} after optimal fitting to each set of data are given in \cref{tabPassiveFit}.

\begin{table}[!hbtp] \centering
\footnotesize{
\begin{tabular}{l l r r l c}
\hline
Parameter & Description & Kuo & Falcone & Units & Source\\
\hline
$q_1$ & passive stiffness coefficient & $26.4\e{-9}$ & $61.4\e{-9}$& N & Fitting\\
$q_2$ & passive stiffness coefficient & $8.38$ & $8.27$ & -- & Fitting\\
$q_3$ & passive stiffness coefficient & $1.06$ & $0.619$ & -- & Fitting\\
$l_0$ & relaxed length of cell & $77.5\e{-6}$ & $64.3\e{-6}$ & m & Fitting\\
\hline
$w_c$ & effective width of muscle cell & \multicolumn{2}{c}{$2.67\e{-6}$} & m & \cite{Yang03a}\\
$n_c$ & number of cells around vessel & \multicolumn{2}{c}{$6$} & -- & \cite{Yang03a}\\
$t_w$ & thickness of vessel wall & \multicolumn{2}{c}{$15\e{-6}$} & m & \cite{Yang03a}\\
\hline
\end{tabular}}
\caption{Parameters of the model fitted to the passive (no \ca\!\!) case. As can be seen from \cref{figPassiveFitBoth}, the passive response in the Kuo data is stiffer than that found in the Falcone data; although the coefficient $q_1$ is smaller in the Kuo case, $q_2$ and $q_3$ are larger and these coefficients have a stronger effect.}
\label{tabPassiveFit}
\end{table}

\subsection{Fitting the Active Response}
\label{activefitting}
Having found optimal values for those parameters which are involved in the passive response, we now turn our attention to the active cases, the lower blue and grey lines in \cref{figPassiveFitBoth}. Initially only the no-flow case was considered from the Kuo data; the problem is formulated exactly as for the passive case, \cref{minimization}, except that the parameters given in \cref{tabPassiveFit} are now fixed and the error is a function of all of the other parameters of the model. The purpose of considering first only the no-flow case is to again remove as many of the unknowns as possible whilst finding at least approximate values for a subset of the parameters. When optimizing over all remaining parameters to fit only the no-flow data it is possible (results not shown) to exactly match the model response to the data. However, an exact fit to the no-flow data can only be achieved by setting the parameters such that the NO concentration is extremely low, so low that using these parameter values (optimized on the no-flow case only) for the flow case gives a very poor fit to the flow data. Therefore, once good starting values for the parameters had been found from the fitting to no-flow only, the objective function was changed to equal the sum of the rms errors of the fit to both flow and no-flow cases. With this combined objective function the optimizer was not driven to overfit one case at the expense of the other, but rather to find the best set of parameter values to fit both cases at the same time.

\begin{figure}[hbtp] \centering \includegraphics[scale = 0.8]{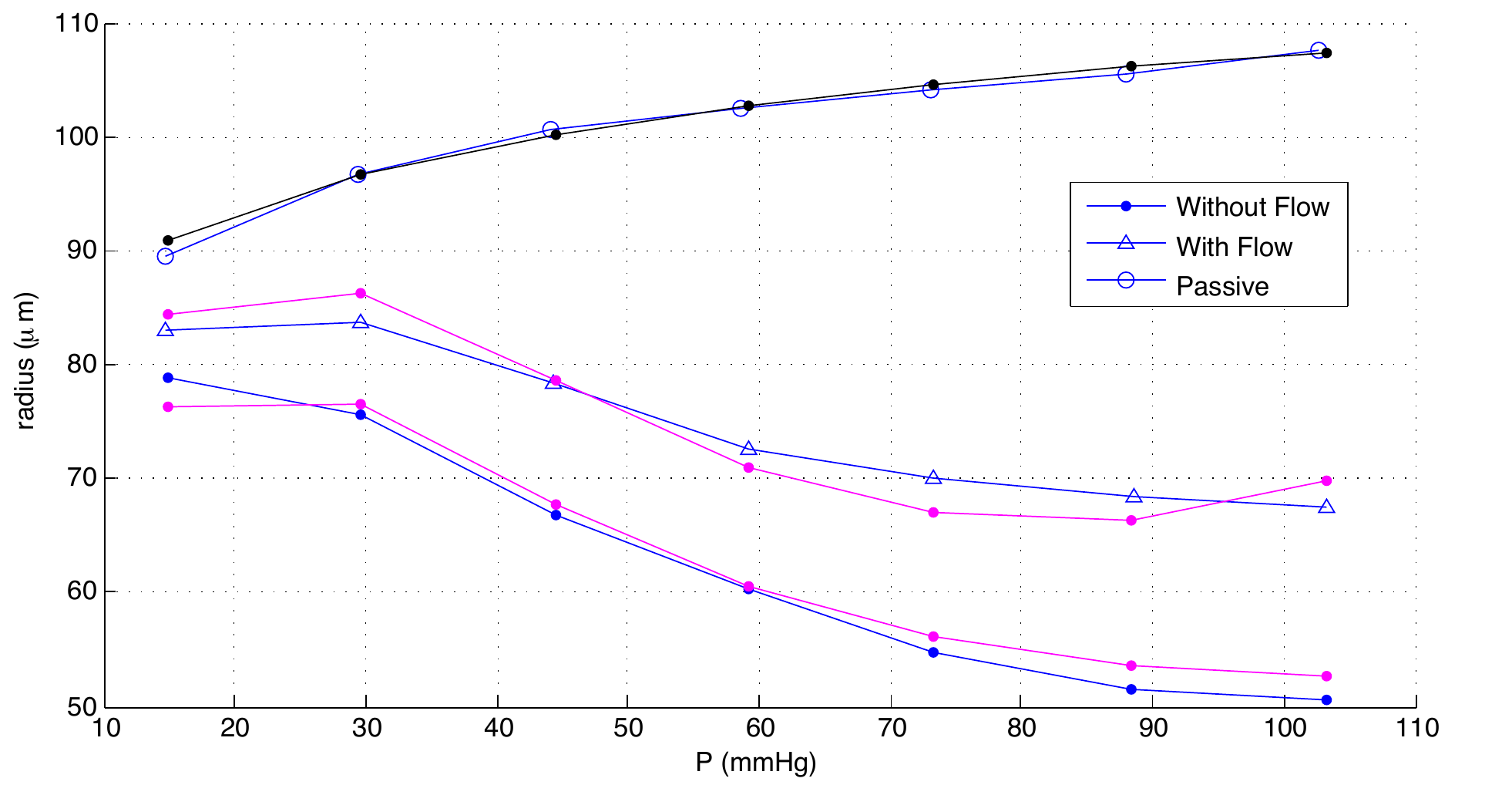} \caption{The steady-state data from Kuo \cite{Kuo91} have been extracted and are shown here in blue. The model responses are shown in black (passive) and magenta (active).} \label{figActiveFitKuo} \end{figure}

\begin{figure}[hbtp] \centering \includegraphics[scale = 0.8]{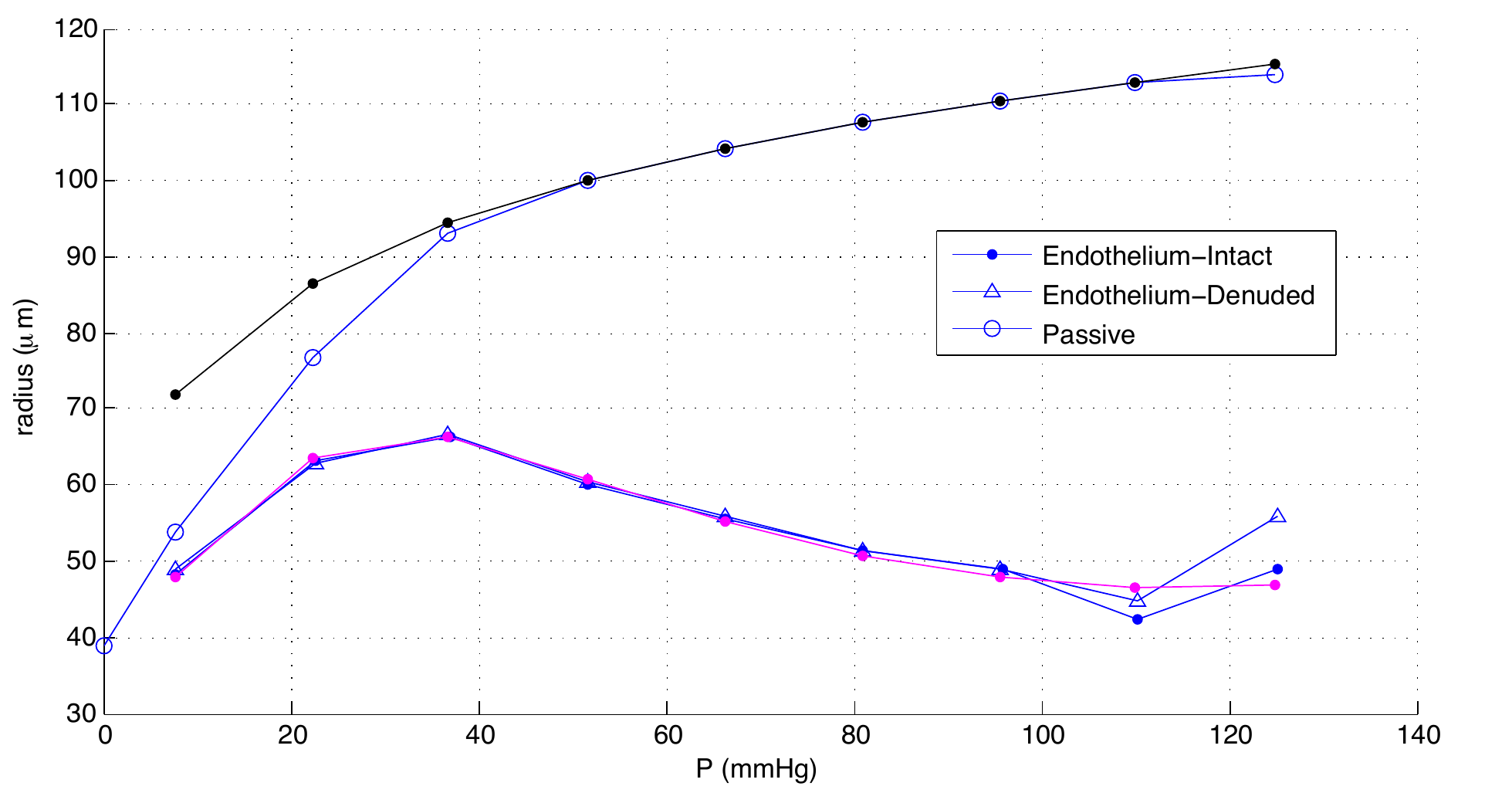} \caption{The steady-state data from Falcone \cite{Falcone91} have been extracted and are shown here in blue. The model responses are shown in black (passive) and magenta (active).} \label{figActiveFitFalcone} \end{figure}

As with the passive fitting, it was found that the model as described above is not sufficiently flexible to adequately fit all of the data. Specifically the reaction of the vessel to the onset of flow was found to be too large, this was caused by the sensitivity of active force to vessel length being too low\footnote{See gradient of muscle force lines at intersection with equilibrium force lines in \cref{figForceRadiusFalcone}.}. To remedy this another slight modification was introduced, this time into the active side of the mechanical model. A variance term was added to the Gaussian describing the amount of overlap between actin and myosin filaments in the cell. The addition of this variance term changes \cref{Fa} to:
\beq F_a = l_0 \left(f_1 AMp \left(\hat v + \dot \lambda \right) + f_2 AM \dot \lambda \right) e^\frac{-\left(\lambda - \lambda_{opt} \right)^2}{2\eta} \label{FaNew} \eeq

and \cref{mechSS} similarly becomes (incorporating the extra parameter, $q_3$, from \cref{passiveNew}):
\beq 0 = q_1 \left( e^{q_2 \left(\lambda^{q_3} - 1\right)} - 1 \right) + l_0  f_1  \alpha_3  \hat v  e^\frac{-\left(\lambda - \lambda_{opt} \right)^2}{2\eta} - w_cP\frac{1}{2} \left(\frac{\lambda l_0 n_c}{\pi} - t_w \right) \eeq

where $\eta$ represents the variance of the distribution of sarcomeres in the VSMC, previously fixed to an implicit value of $1/2$. The results of the fitting after inclusion of this extra term are shown in \cref{figActiveFitKuo,figActiveFitFalcone} and the corresponding parameter values given in \cref{tabActiveFit}.

\begin{table}[!hbtp] \centering
\footnotesize{
\begin{tabular}{l l r r l c}
\hline
Parameter & Description & Kuo & Falcone & Units & Source\\
\hline
$h_w$ & NO wall transport coefficient & $7.61\e{-9}$ & -- & m$^2$/s & Fitting\\
$s_w$ & production rate of NO in vessel wall & $69.2\e{-9}$ & -- & M/s & Fitting\\
$\gamma_{wb}$ & baseline wall:blood \conc{NO} ratio & $7.27$ & -- & --  & Fitting\\
$r_{db}$ & decay rate of NO in blood & $89.5$ & -- & 1/s & Fitting\\
$r_{dw}$ & decay rate of NO in vessel wall & $9.51$ & -- & 1/s & Fitting\\
$k_{\tau}$ & shear stress NO production coefficient & $248\e{-15}$ & N/A & m$^2$M/s & Fitting\\
\hline
$C_{in}$ & upstream (incoming) \conc{NO} in the blood &  \multicolumn{2}{c}{$0$} & M & \cref{Assumptions}\\
$\mu_b$ & viscosity of blood &  \multicolumn{2}{c}{$4.37\e{-3}$} & Ns/m & \cite{Rosenson96}\\
$\Delta P_0$ & baseline arterial blood pressure &  \multicolumn{2}{c}{$60$} & mmHg & Typical\\
$r_0$ & baseline internal vessel radius &  \multicolumn{2}{c}{$60 \e{-6}$} & m &  Typical\\
$L$ & length of arteriole &  \multicolumn{2}{c}{$1\e{-3}$} & m &  Typical\\
\hline
$\conc{NO}_L$ & lower edge of \conc{NO} active region & $780\e{-12}$ & -- & M & Fitting\\
$\conc{NO}_U$ & upper edge of \conc{NO} active region & $36.5\e{-9}$ & -- & M  & Fitting\\
$p\conc{cGMP}_L$ & lower bound of  $p\conc{cGMP}$ in muscle cells & $-9.57$ & -- & -- & Fitting\\
$p\conc{cGMP}_U$ & upper bound of  $p\conc{cGMP}$ in muscle cells  & $-8.29$ & -- & -- & Fitting\\
\hline
$\sigma_L$ & lower edge of $\sigma$ active region & $-8.16\e{3}$ & $11.3\e{3}$ & N/m & Fitting\\
$\sigma_U$ & upper edge of $\sigma$ active region & $52.4\e{3}$ & $47.3\e{3}$ & N/m  & Fitting\\
$p\ccam_L$ & lower bound of $p\ccam$ in muscle cells & $-8.50$ & $-8.84$ & -- & Fitting\\
$p\ccam_U$ & upper bound of $p\ccam$ in muscle cells  & $-6.94$ & $-7.57$ & -- & Fitting\\
\hline
$f_1$ & friction from cycling cross-bridges & $9.34\e{-6}$ & $98.7\e{-6}$ & Ns/m & Fitting\\
$f_2$ & friction from latch bridges & $500\e{-3}$ & N/A & Ns/m & Fitting\\
$\lambda_{opt}$ & normalized cell length at max overlap & $1.49$ & $1.79$ & --  & Fitting\\
$v$ & cross bridge cycling velocity & $9.71$ & $8.99$ & m/s & Fitting\\
$\eta$ & variance of distribution of sarcomeres & $264\e{-3}$& $217\e{-3}$ & -- & Fitting\\
\hline
\end{tabular}}
\caption{Parameter values of the model after optimization to fit active steady-state data of Kuo \cite{Kuo91} and Falcone \cite{Falcone91}. Note that the parameters of the NO model were not re-optimized for the Falcone data as they have negligible effect when no flow is present.}
\label{tabActiveFit}
\end{table}

\subsection{Discussion}
\begin{figure}[hbtp] \centering \includegraphics[scale = 0.8]{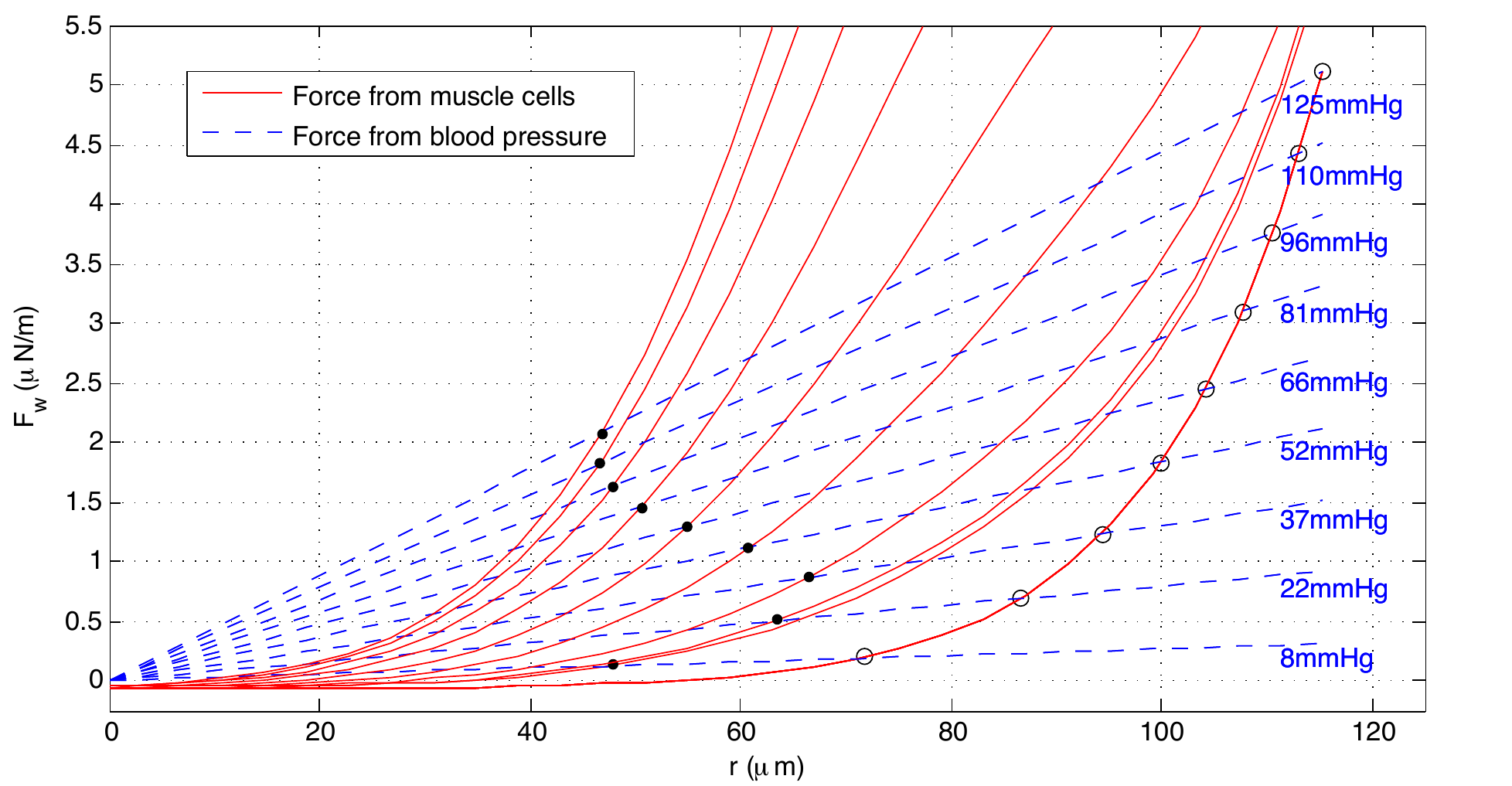} \caption{The relationship between force required to resist the blood pressure, dashed blue line, blue text, and the force available from the muscle cell, solid red. Responses shown are a result of fitting to the Falcone data. Open circles on the right show the equilibrium points of the passive response, solid circles in the centre show the equilibria of the active response -- as intramural pressure increases, the myogenic response makes more \ca available and the resulting phosphorylation of myosin cross-bridges moves the muscle force curve upward and the equilibrium radius is reduced.} \label{figForceRadiusFalcone} \end{figure}

Whilst the magenta lines of the model response in \cref{figActiveFitKuo,figActiveFitFalcone} do not overlay exactly with the blue lines of the experimental results, the form of both sets of curves is the same: both exhibit active autoregulation over the given range of intramural pressures in the presence of \ca and fail to do so in the absence of \ca\!\!. It is clear, from the model's emulation of the Kuo data, that the impact of flow is to significantly reduce the vasoconstriction caused by the myogenic response, exactly as observed in \cite{Davis99}. \Cref{figForceRadiusFalcone} shows the relationship between the forces in the vessel wall and the pressure of the fluid in the vessel for the model as optimized to fit the Falcone data. It is clear from this figure how the myogenic response moves the force curve of the muscle upward and hence the equilibrium radius decreases. \Cref{tabActiveFit} shows that the primary difference between the parameter values for the two fitting cases is in the friction associated with cycling cross-bridges, $f_1$. The lesser passive stiffness of the vessel in the Falcone data, as shown in \cref{figPassiveFitBoth}, necessitates a stronger active component of the VSMC to achieve the same level of autoregulatory response. The sensitivity of the model to the values of its parameters will be further examined in \cref{sensitivity}.

\section{Dynamic Responses}
\label{dynfitting}
\begin{figure}[hbtp] \centering \includegraphics[scale = 0.8]{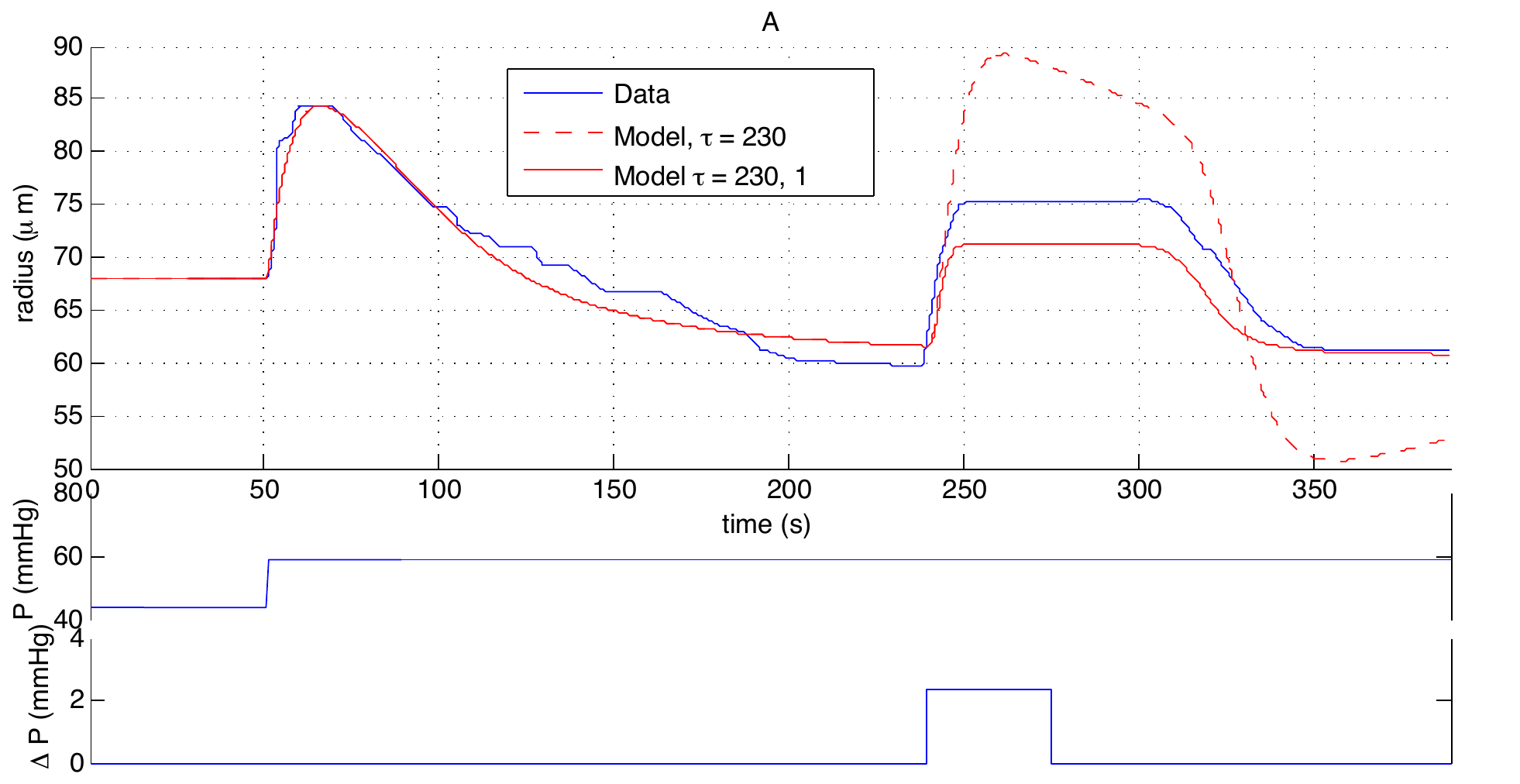} \caption{Data from fig.~2A in \cite{Kuo91} (\cref{figKuo91Fig2}) shown with two model responses: the dashed line is given by $\tau_{Ca} = 230$s throughout, the solid line is given when the value of $\tau_{Ca}$ is reduced from 230s to 1s at the onset of flow. Radius values from data have been normalised to match initial value from the model.} \label{figKuo91Fig2ACombo} \end{figure}

We now turn our attention to the dynamic responses to various steps in pressure and flow presented in fig.~2 of \cite{Kuo91} and reproduced here in \cref{figKuo91Fig2}. In order to simulate a dynamic response with our model, we return to the fully dynamic equations presented in \cref{Equations}. This introduces the additional parameter $\tau_{Ca}$ which will require tuning to achieve good agreement with the data. The first part of \cref{figKuo91Fig2}A is the dynamic response of the system to a step change in pressure under no-flow conditions, this is the ideal experiment against which to tune the value of $\tau_{Ca}$ without the dynamics of the NO model interfering. \Cref{figKuo91Fig2ACombo} shows the data from this experiment, together with two model responses. For the first 239s of this experiment there is no pressure difference between the ends of the vessel and so no flow along it. The initial response of the vessel to the step increase in intramural pressure is to dilate elastically towards the passive equilibrium radius at the new pressure; this occurs between 51 and 60s. The secondary, slower, response is a gradual contraction down to a new active equilibrium radius which is smaller than the initial radius. The speed of the myogenic response is governed by $\tau_{Ca}$ as per \cref{dCadt}; the response shown by the dashed line in \cref{figKuo91Fig2ACombo} was obtained with a value of $\tau_{Ca} = 230$s, chosen by visual inspection. Clearly this value fits the slow response to the pressure step between 60 and 239s quite well, however, the model response to the brief period of flow is distinctly different from the data. The equilibrium radius of the model during the period of flow is actually quite close to the level of the plateau seen in the data, this is not evident in the dashed line of \cref{figKuo91Fig2ACombo} because the myogenic response (with a time constant of 230s) is too slow to achieve the very quick saturation at the new equilibrium level shown in the data, in fact radius has only just started to decrease back towards equilibrium when the flow stimulus is removed. If the time constant for the myogenic response is reduced at the onset of flow then it is possible to closely match the shape of the data. The solid line in \cref{figKuo91Fig2ACombo} is the model response when the value of $\tau_{Ca}$ follows, the lower value again being chosen by visual inspection to match the data:
\beq \tau_{Ca} = \begin{cases}
230\mathrm{s} & \text{if } t < 239,\\
1\mathrm{s} & \text{if } t \ge 239.
\end{cases}
\eeq

\begin{figure}[hbtp] \centering \includegraphics[scale = 0.8]{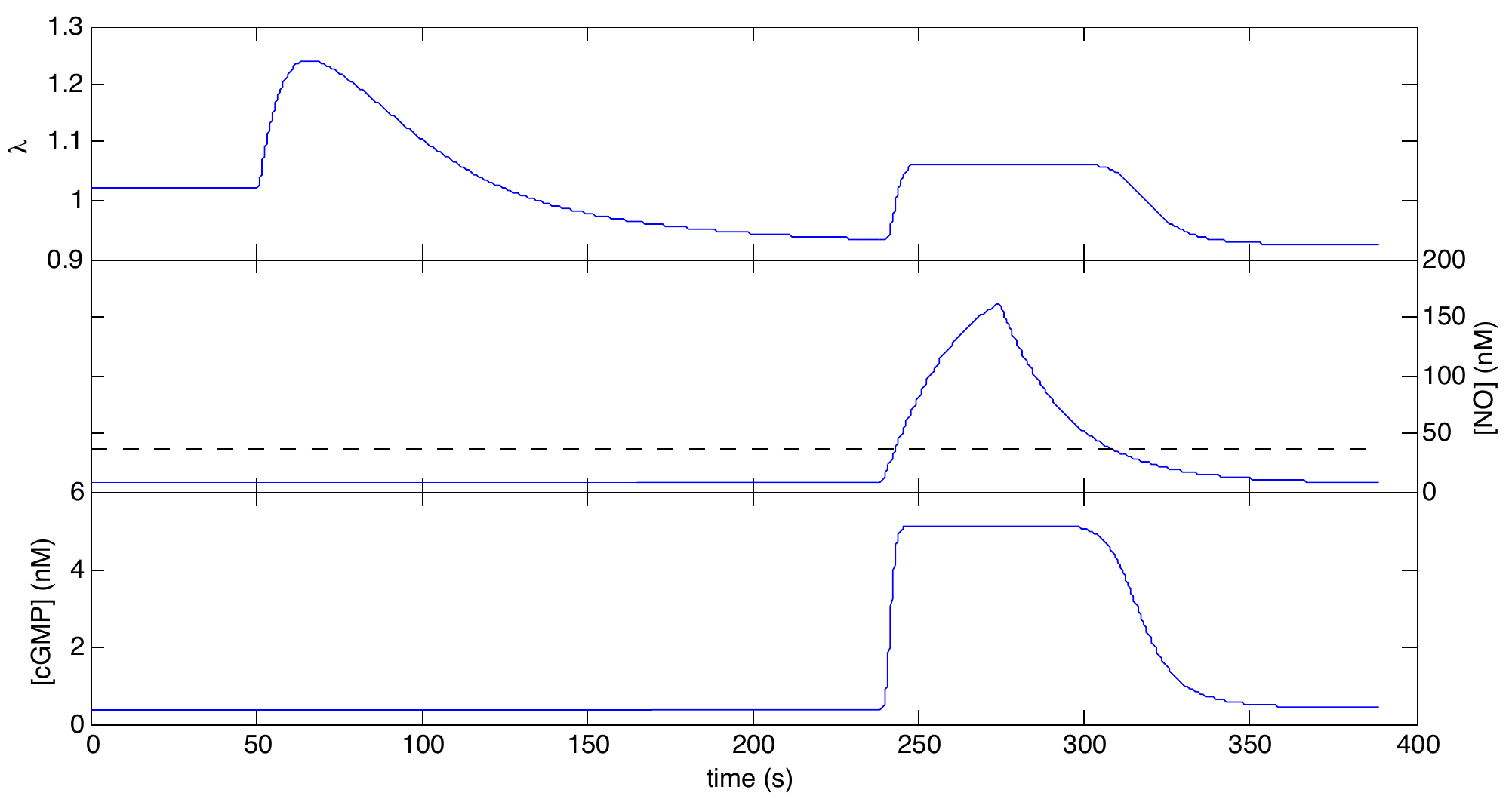} \caption{The states of the model during the experiment shown in \cref{figKuo91Fig2ACombo}. The dashed line in the middle panel shows the level of \cno at which the cGMP concentration is 95\% saturated. The peak of the sawtooth in the NO concentration curve marks the cessation of flow -- its slow return to below the level of the dashed line gives rise to the delay in the response of cGMP, and hence radius.} \label{figKuo91Fig2AStates} \end{figure}

Once the myogenic response is speeded up, the model can closely match the behaviour seen in the data in its response to the onset of flow. There is a significant delay seen in the data between the removal of the flow stimulus and the reaction of the vessel. This is achieved in the model through a combination of the saturation of the NO -- cGMP relationship with the slow response of NO; when flow begins the level of NO in the vessel wall rises rapidly and the NO -- cGMP sigmoid quickly saturates at the maximum level of \ccgmp\!\!, the effect of the cessation of flow is not felt by the model until the NO concentration falls back below the cGMP saturation level and the concentration of cGMP begins to fall, and with it the vessel radius. Because of the exponential sawtooth shape of the NO response there is a significant delay between cessation of flow and the de-saturation of the NO -- cGMP sigmoid. The result is the desired delay between the cessation of flow and the resultant reduction in vessel radius. Although it is not known whether this is the actual mechanism for delay which occurs in reality, it seems likely that the delay in vessel response to cessation of flow is the result of one of the species in the NO pathway reaching saturation whilst the concentration of its progenitor continues to increase -- the slow decay of the progenitor then causing the delay in the reaction of the saturated species. The states of the model throughout the period of this simulation are shown in \cref{figKuo91Fig2AStates}, the dashed line in the middle panel indicates $\conc{NO}_U$, the concentration of NO at which cGMP reaches 95\% of its maximum value. The tuning of this delay was done manually by adjusting the speed of the NO model. This is possible without changing the steady-state behaviour of the model because \cref{NOSS} has a degree of freedom; the values of $h_w$, $r_{dw}$, $s_w$, and $k_{\tau}$ can be scaled by a common factor without affecting the result. These are the coefficients for each of the four terms in \cref{dNOdt}, transport into the blood, decay in the wall, baseline production in the wall, and shear-stress-induced production in the wall. Their scaling thus affects the speed of the model without changing its steady-state behaviour. The common factor by which they were all scaled is 0.005, resulting in the values given in \cref{tabNewNOValues}:

\begin{table}[!hbtp] \centering
\footnotesize{
\begin{tabular}{l l r r l c}
\hline
Parameter & Description & Value & Units & Source\\
\hline
$h_w$ & NO wall transport coefficient & $38.0\e{-12}$ & m$^2$/s & Fitting\\
$r_{dw}$ & decay rate of NO in vessel wall & $47.6\e{-3}$ & 1/s & Fitting\\
$s_w$ & production rate of NO in vessel wall & $346\e{-12}$ & mol/Ls & Fitting\\
$k_{\tau}$ & shear stress NO production coefficient & $1.24\e{-15}$ & m$^2$mol/Ls & Fitting\\
\hline
\end{tabular}}
\caption{Parameter values of the coefficients of \cref{dNOdt} after manual tuning to fit the delay observed in \cref{figKuo91Fig2}A. These are the values in \cref{tabActiveFit} scaled by a common factor of 0.005.}
\label{tabNewNOValues}
\end{table}

\begin{figure}[hbtp] \centering \includegraphics[scale = 0.8]{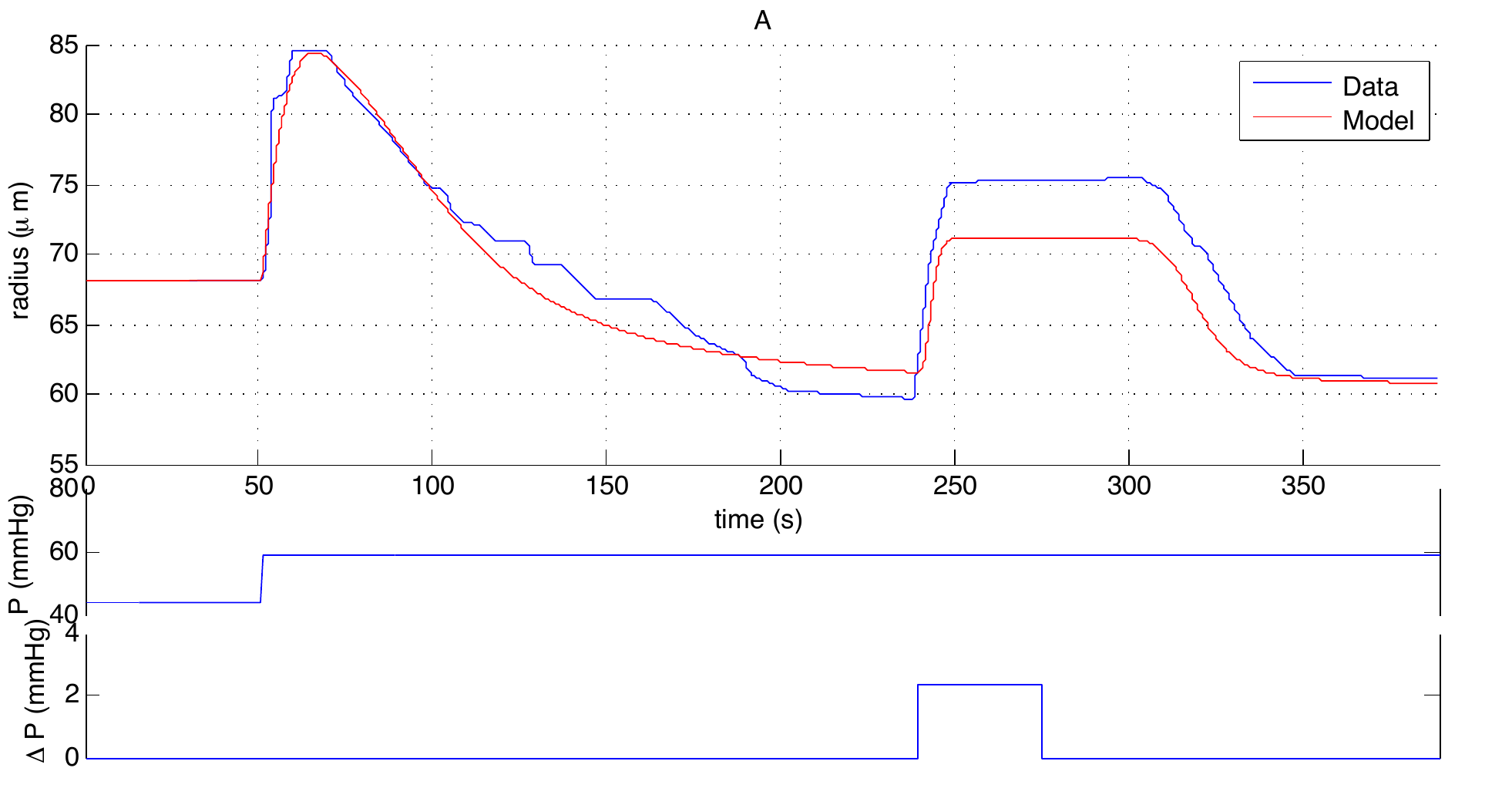} \caption{Data from fig.~2A in \cite{Kuo91} (\cref{figKuo91Fig2}) in blue, shown with the model response in red.} \label{figKuo91Fig2A} \end{figure}

\begin{figure}[hbtp] \centering \includegraphics[scale = 0.8]{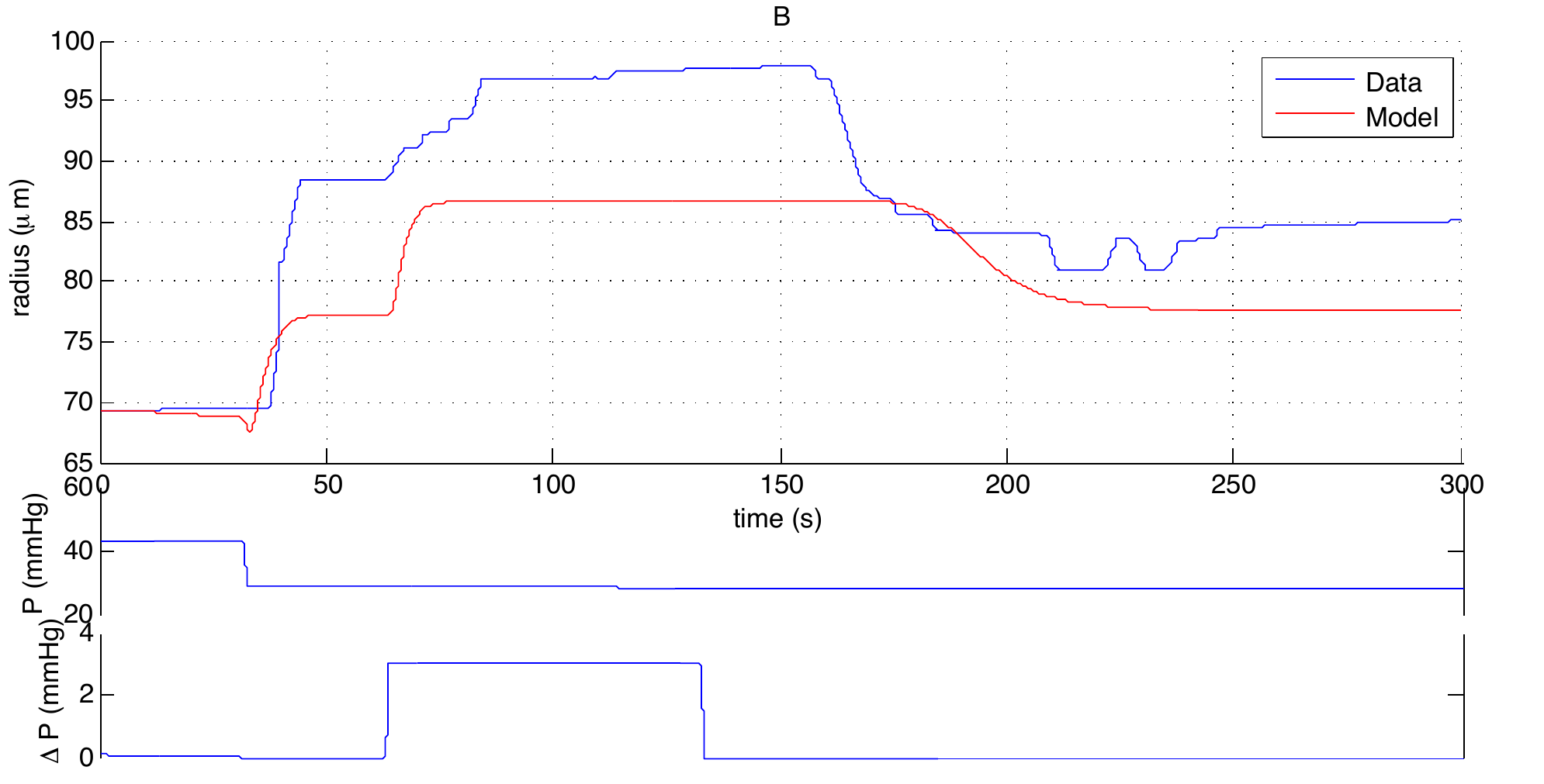} \caption{Data from fig.~2B in \cite{Kuo91} (\cref{figKuo91Fig2}) in blue, shown with the model response in red.} \label{figKuo91Fig2B} \end{figure}

\begin{figure}[hbtp] \centering \includegraphics[scale = 0.8]{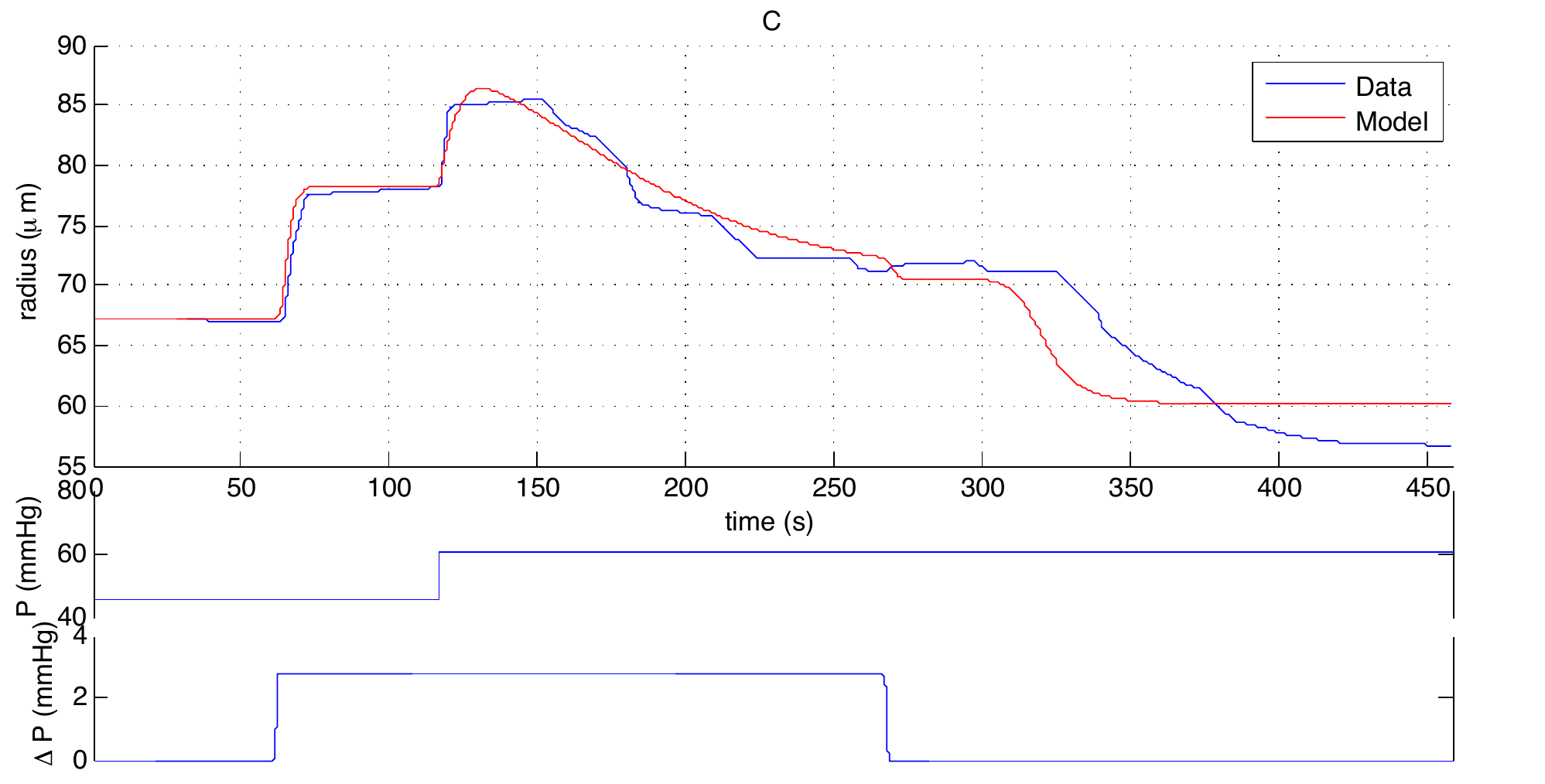} \caption{Data from fig.~2C in \cite{Kuo91} (\cref{figKuo91Fig2}) in blue, shown with the model response in red.} \label{figKuo91Fig2C} \end{figure}

\begin{figure}[hbtp] \centering \includegraphics[scale = 0.8]{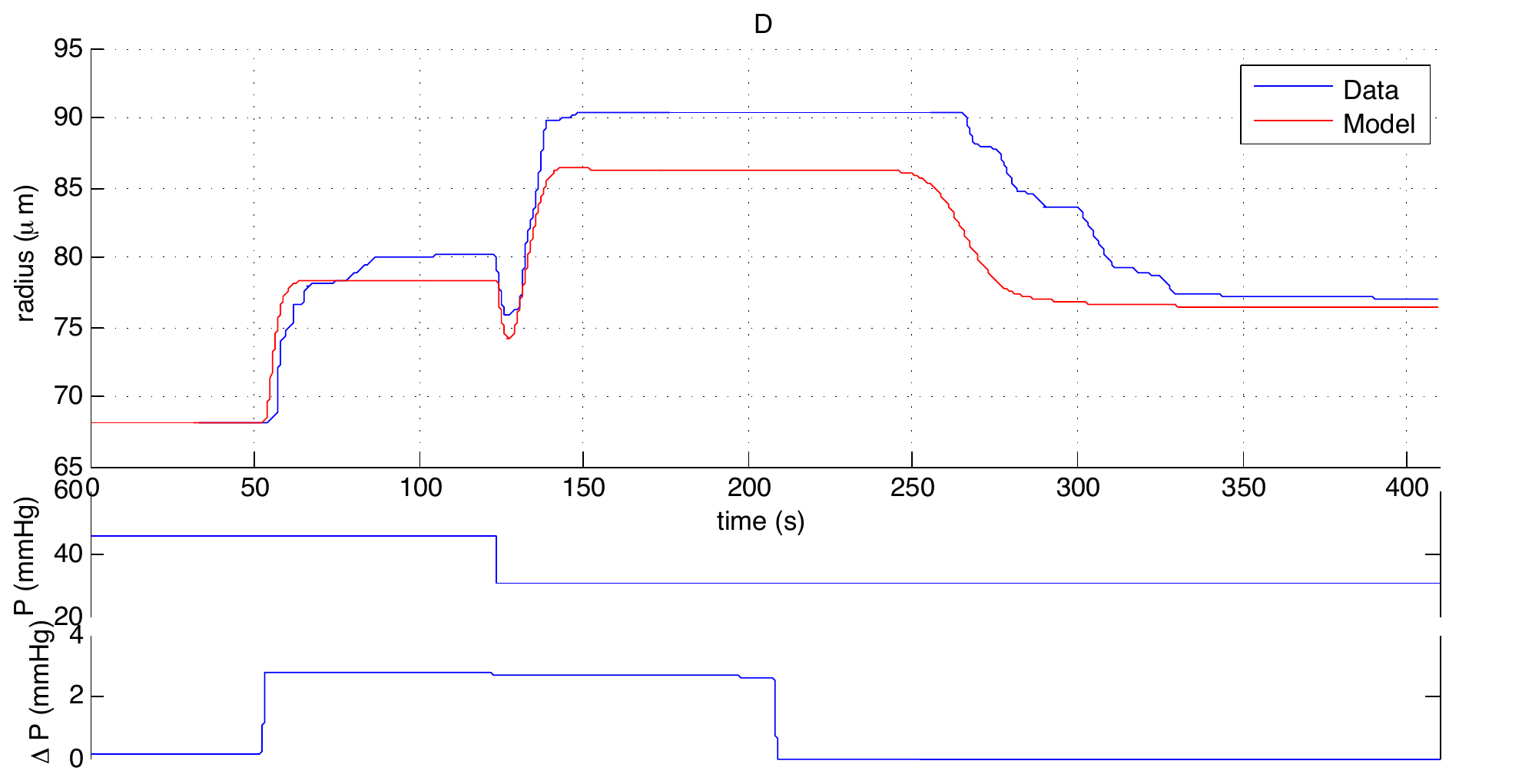} \caption{Data from fig.~2D in \cite{Kuo91} (\cref{figKuo91Fig2}) in blue, shown with the model response in red.} \label{figKuo91Fig2D} \end{figure}

Having found from this first dynamic simulation that a variation in the speed of the myogenic response was necessary to match the data of \cref{figKuo91Fig2}A, we proceeded to apply the same approach to matching the responses of the model to the remaining three dynamic experiments presented in \cite{Kuo91}. In order to maintain a credible ability to hypothesise about the possible physical mechanism for such changes in speed, the time constant was only changed at points in the experiments when the conditions (either $P$ or $\Delta P$) change. The results of all four simulations are shown in \cref{figKuo91Fig2A,figKuo91Fig2B,figKuo91Fig2C,figKuo91Fig2D} and the time constants used are summarised in \cref{tabTauCaVals}. Radius values from the data have been normalised to match the initial radius from the model. The values of these time constants were again chosen manually to provide a close fit to the data, the response to the drop in pressure in \cref{figKuo91Fig2D} requires a time constant of 5s, rather than the previously used 1s, in order to replicated the brief passive contraction seen in the data before the active expansion.


\begin{table}[!hbtp] \centering
\footnotesize{
\begin{tabular}{r r r r r}
\hline
Experiment & Start time (s) & End time (s) & $\tau_{Ca}$ (s)\\
\hline
A & 0 & 239 & 230\\
   & 239 & end & 1\\
\hline
B & 0 & end & 1\\
\hline
C & 0 & 117 & 1\\
& 117 & 267 & 210\\
& 267 & end & 1\\
\hline
D & 0 & 123 & 1\\
& 123 & end & 5\\
\hline
\end{tabular}}
\caption{Values of the myogenic response time constant, along with corresponding epoch start and end times, used to obtain the model simulation results presented in \cref{figKuo91Fig2A,figKuo91Fig2B,figKuo91Fig2C,figKuo91Fig2D}. All values are given in seconds.}
\label{tabTauCaVals}
\end{table}

It is suspected that the experimental data of the second experiment, \cref{figKuo91Fig2B}, contains an error; the instant dilation of the arteriole in response to the drop in pressure at the beginning of the experiment not only exhibits no initial elastic contraction but it goes on to an equilibrium of much greater radius than would be expected given the steady-state results presented in the same paper. It is possible that some kind of experimental or recording error occurred coincident with the drop in pressure which caused the results to be shifted upward from then on.

It should also be noted that the discrepancies between the data and the model responses during periods of constant radius (i.e., plateaus in the traces) are not a feature of the dynamic response per se, but rather a combined feature of the fitting inaccuracies of the steady-state model, \cref{figActiveFitKuo}, and disagreement between the steady-state and dynamic data of \cite{Kuo91}. Shape, relative magnitude, and timescales provide more meaningful comparisons between the data and the model responses than absolute values.

\subsection{Hypothesis/Justification for a Variable \ca Time Constant}
\label{variableTCaJust}
Whilst at first it might seem physiologically unjustifiable to allow tuning of the myogenic response time constant in an ad hoc manner to fit the available data, there is actually a viable physiological mechanism that could cause such an effect. \ca ions can be admitted to the cell cytoplasm from two sources: the extracellular fluid, and the sarcoplasmic reticulum, or SR, see \cref{figcell}. Influx of \ca from the extracellular fluid is relatively slow due to the fact that the concentration of \ca in the extracellular fluid is relatively low. Conversely, the SR contains fluid at a very high concentration of \ca relative to the cytoplasm. Some processes lead to \ca being released from the SR, causing a rapid rise in the cytoplasmic \ca concentration; one such process is calcium-induced calcium release, or CICR, which is thought to be a vital part of the generation of  rhythmic action potentials within VSMCs, leading to vasomotion. So there is actually a plausible physical basis for the duality in \ca response times, in the slow cases the \ca could be coming through the cell membrane from the extracellular fluid, in the fast cases it could be coming from the SR.

Another hypothesis for the physical basis of the apparent need for a varying \ca time constant is suggested when examining the conditions under which the time constant is very large. Only when the intramural pressure is stepped \emph{up} do we see that the time constant must be significantly greater than in all other situations. When intramural pressure suddenly increases the vessel is forced to dilate quickly by the imbalance of forces, this sudden elongation of the VSMCs may well cause phosphorylated attached cross-bridges to be forcibly detached from their actin filaments. If myosin heads so detached took some time to recover and regain attachment, perhaps to dephosphorylate and re-phosphorylate, then this could explain the slowness of the myogenic response in the case of sudden pressure increase. In no other circumstance are VSMCs forcibly elongated, and in no other circumstance is the \ca time constant required to be large.

Looking at the $\tau_{Ca}$ values in \cref{tabTauCaVals} and the experimental conditions of \cref{figKuo91Fig2A,figKuo91Fig2B,figKuo91Fig2C,figKuo91Fig2D}, it can be seen that the only two occasions on which the \ca response is required to be very slow are the only two occasions on which the vessel is reacting to a step increase in intramural pressure. It could be that some aspect of the vessel's initial passive dilation in response to these step increases of pressure causes the influx of \ca to the cytoplasm from the SR to be inhibited, resulting in a much slower than `usual' myogenic response. This is an interesting potential avenue for future experimental investigation.

\section{Sensitivity Analysis}
\label{sensitivity}
So far we have presented the equations that constitute the model, we have described how the parameters of these equations have been optimized to best fit the data, and we have presented the comparison between the model's behaviour and the experimental data for both steady-state and dynamic cases. In order to determine which of the many parameters in the model have the greatest effect, and which have very little effect, we now perform a sensitivity analysis on the model. Whereas the derivatives of the objective function of the optimization with respect to the parameters, $\diff{\epsilon}{\mathbf{p}}$ from \cref{minimizationGeneric}, give the sensitivity of the model output to the parameters at a specific operating point, sensitivity analysis aims to evaluate the mean and variance of these terms throughout the parameter space (i.e., for all sensible values of $\mathbf{p}$). This approach thus gives a better idea of the effects of parameter values over the whole parameter domain of a strongly non-linear model such as ours.

\subsection{Current Methods}
We will look at the case where the output variable of which we are measuring the sensitivity is the rms error between the model radius and the radius data for both the Kuo flow and no-flow cases combined. This was the form of the model for the fitting to the Kuo data described in \cref{activefitting}. In this case the model has 17 variable parameters. In the parlance of sensitivity analysis, the derivative of the model output (rms fitting error) with respect to a parameter is referred to as the \emph{elementary effect} of that parameter. It is the mean and variance of each elementary effect that sensitivity analysis aims to evaluate. Because the model is analytically intractable it is necessary to infer the desired properties of the distribution of elementary effects from the samples of their distribution, rather than to compute them directly. Due to limitations on computational cost it is necessary that this sampling be relatively sparse. For instance, to sample the elementary effect of only one parameter over a grid with 10 divisions in each dimension of the parameter space would take 2.5 billion years, for just one parameter (2 $\times$ $10^{17}$ $\times$ 0.4s). This has led to the development of techniques which aim to sample the distribution of elementary effects with the fewest model evaluations possible.

One group of such techniques is known as Morris Methods; briefly, the domain is gridded evenly, then one vertex is chosen and the model evaluated there, the next evaluation is made by stepping to an adjacent vertex of the grid, the dimension along which this step is taken being chosen at random. A further step is then taken, again along a randomly chosen dimension, providing that a step in that dimension has not already been taken. This stepping continues until the number of steps taken is equal to the number of dimensions, and so the elementary effects of all parameters have been sampled once along the trajectory. For a system with $k$ parameters, this path consists of $k+1$ model evaluations. Many such paths may be generated until a sufficiently accurate approximation to the true distribution of elementary effects has been found. By contrast, random selection of points for each sample of each elementary effect would necessitate $2k$ evaluations of the model, one at each point and one after one parameter had been incremented by the finite difference $\Delta$, hence the great advantage of Morris' method whereby the second evaluation in one dimension serves as the first evaluation in the next dimension.

The separation of trajectories can be improved by using another sampling technique for the selection of a starting point for each trajectory. Such a sampling technique is the latin hypercube method, whereby $n$ samples are taken from a $k$-dimensional space by dividing each dimension into $n$ divisions and ensuring that no two samples occupy the same division in any dimension. Although the points are chosen at random within these criteria, latin hypercube sampling does not guarantee an even, space filling, distribution; the leading diagonal of a matrix, for example, satisfies the criteria of the latin hypercube sampling method. It is common when employing latin hypercube sampling, to generate several distributions and to choose the one with the greatest mean separation between points. Even when the starting points are well distributed, it is possible with a Morris method that two trajectories could become close if the random direction and dimension of each step brings two trajectories together.

In order to resolve the problems of latin hypercube sampling and of Morris trajectories converging, Campolongo et al. \cite{Campolongo07} proposed a scheme whereby a large number of Morris trajectories are generated from random starting points and from that a subset is chosen which has the maximum total distance between the trajectories, where the distance between two trajectories is taken as the sum of the distances from each point in one trajectory to every point in the other trajectory:
\beq d_{ml} = \sum_{i=1}^{k+1}\sum_{j=1}^{k+1} \sqrt{\sum_{z=1}^k\left( X_i^m(z) - X_j^l(z)\right)^2} \label{totaldist} \eeq

where $k$ is the number of parameters and $X_i^m(z)$ is the $z$th coordinate of the $i$th point on the $m$th Morris trajectory. 

\subsection{A Novel Method}
The aim of the computationally intensive trajectory selection given in \cite{Campolongo07} is to achieve an even, space-filling pattern of Morris trajectories; although this is achieved, it is not the correct objective. The distributions being sampled are of derivatives, not of values, so the location of the sample in the $z$th dimension lies midway between the points on the Morris trajectory that are separated by a step in the $z$th dimension. Not only that but when looking purely at elementary effects (as opposed to the combined effects of two or more parameters) the distributions of each elementary effect are independent, so the only relevant distance measure between two trajectories is the sum of the distances between pairs of midpoints of steps in the same dimension. The distances between a step in the $z$th dimension and all the steps in the non-$z$ dimensions in another trajectory are irrelevant. The illusion of optimal space filling is given by the idea of trajectories weaving though the parameter space evenly, but for any given elementary effect, each trajectory contains only one sample point. The reality is then that a lot of effort is devoted to ensuring that $k$ independent sets of points are well separated from each other, for no reason.

The method proposed here is a much simpler one. We generate several sets of $k$-dimensional points using the latin hypercube method, pick the one with the greatest mean separation between points, evaluate the model at these points and at $k$ additional points per starting point, each separated from the starting point by a step in a different dimension. The result is a `constellation' of star-shaped experiments dotted throughout the parameter space. Each elementary effect is sampled on an equally well spaced basis and very little computational effort is required to design experiments, or to select a subset thereof. The total number of model evaluations necessary for a star-shaped experiment is equal to that of a Morris trajectory, but the unnecessary complexity of choosing random directions in random, unexplored dimensions is removed.

The parameter space itself can be defined in many ways; some parameters might have intrinsic bounds but for the rest there is some arbitrary choice of which values should define the limits of the parameter space from which the sensitivity analysis samples are to be drawn. Bounds for the parameters have already been set during the optimization/fitting process, and these bounds can sensibly be used here for consistency with \cref{chapVascModel}. Alternatively we can define the parameter space to extend some fixed percentage either side of the values found to give the best fit. This is perhaps a preferable strategy as the definition of the bounds on each variable ensures that the parameter space is centred on a point where all of the mechanisms of the model are active. Conversely, if we change parameter values too much, we will find ourselves in regions of the model where some sigmoid functions are effectively flat and so the related parameters have no local effect. Here we conduct sensitivity analyses using the latin hypercube `constellation' method detailed above on both an additive, absolute bounds parameter space, and on a multiplicative, relative bounds parameter space.

The result of any sensitivity analysis such as this is a plot of variance, $\sigma$, versus mean, $\mu$, of the sensitivity of the output to variations in each parameter.

\subsection{Additive Results}
The bounds used to define the parameter space in the additive case were those bounds which were used for the optimization of the fitting in \cref{activefitting}, given in \cref{tabBounds}. 

\begin{table}[!hbtp] \centering
\footnotesize{
\begin{tabular}{l r r r l}
\hline
Parameter & Lower Bound & Optimal Value & Upper Bound & Units\\
\hline

$h_w$ & $5\e{-15}$ & $38.0\e{-12}$ & $5\e{-10}$ & m$^2$/s\\
$s_w$ & $0$ & $346\e{-12}$ & $5\e{-7}$ & M/s \\
$\gamma_{wb}$ & 1 & $7.27$ & 100 & -- \\
$r_{db}$ & 0 & $89.5$ & 200 & 1/s \\
$r_{dw}$ & 0 & $47.6\e{-3}$ & $5\e{-2}$ & 1/s \\
$k_{\tau}$ & $1.65\e{-18}$ & $1.24\e{-15}$ & $5\e{-15}$ & m$^2$M/s \\
\hline
$\conc{NO}_L$ & $600\e{-12}$ & $780\e{-12}$ & $32\e{-9}$ &M \\
$\conc{NO}_U$ & $32\e{-9}$ & $36.5\e{-9}$ & $40\e{-9}$ & M \\
$p\conc{cGMP}_L$ & -10 & $-9.57$ & -5 & -- \\
$p\conc{cGMP}_U$ & -10  & $-8.29$ & -5 & -- \\
\hline
$\sigma_L$ & $-4\e{4}$ & $-0.816\e{4}$ & $20\e{4}$ & N/m \\
$\sigma_U$ & 0 & $52.4\e{3}$ & $20\e{4}$ & N/m \\
$p\ccam_L$ & -20 & -8.50 & -6.75 & -- \\
$p\ccam_U$ & -9 & $-6.94$ & -6 & -- \\
\hline
$f_1$ & 0 & $9.34\e{-6}$ & $1\e{-2}$ & Ns/m \\
$\lambda_{opt}$ & 1.3 & 1.49 & 2 & -- \\
$\eta$ & $10\e{-3}$ & $264\e{-3}$& $500\e{-3}$ & -- \\
\hline
\end{tabular}}
\caption{Bounds placed on variables during optimization and used to define the limits of the parameter space for additive sensitivity analysis. Optimal values shown are as per \cref{tabActiveFit}, `Kuo' column. These are the values used as the centre of the multiplicative sensitivity analysis scheme.}
\label{tabBounds}
\end{table}

\begin{figure}[hbtp] \centering \includegraphics[scale = 0.5]{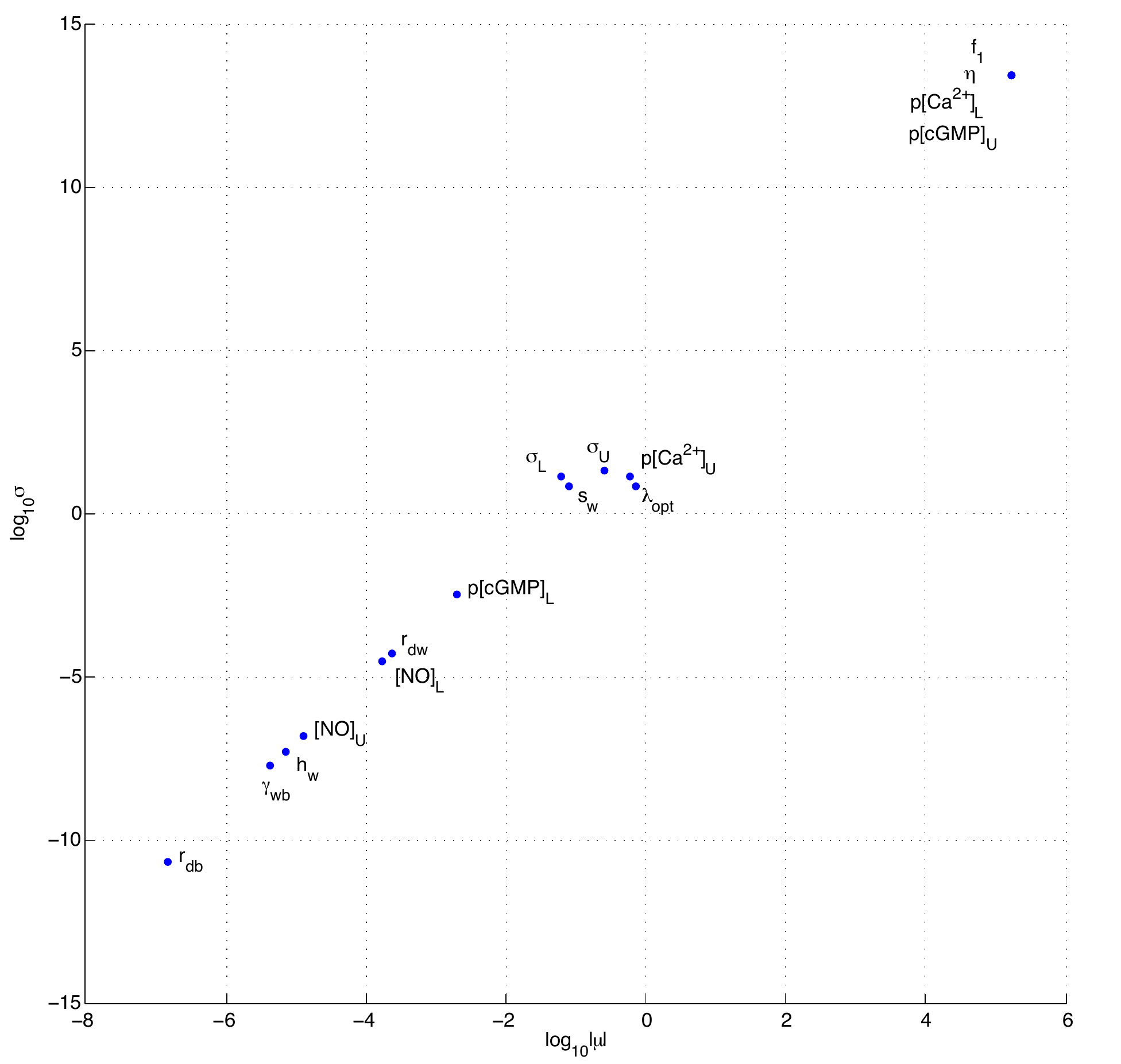} \caption{Results of sensitivity analysis using the additive scheme, variance of elementary effects versus mean magnitude of elementary effects; axes are logarithmic to show detail.} \label{figsensitivityAdditive} \end{figure}

The results of this analysis carried out with 1000 samples and using a finite difference of $1\e{-9}$ are shown in \cref{figsensitivityAdditive}. Aside from the arbitrary choice of bounds on the parameter space, another major problem with performing sensitivity analysis in an additive fashion is that the scale of the parameters scales both the mean and variance of the sensitivities.

\subsection{Multiplicative Results}
\begin{figure}[hbtp] \centering \includegraphics[scale = 0.5]{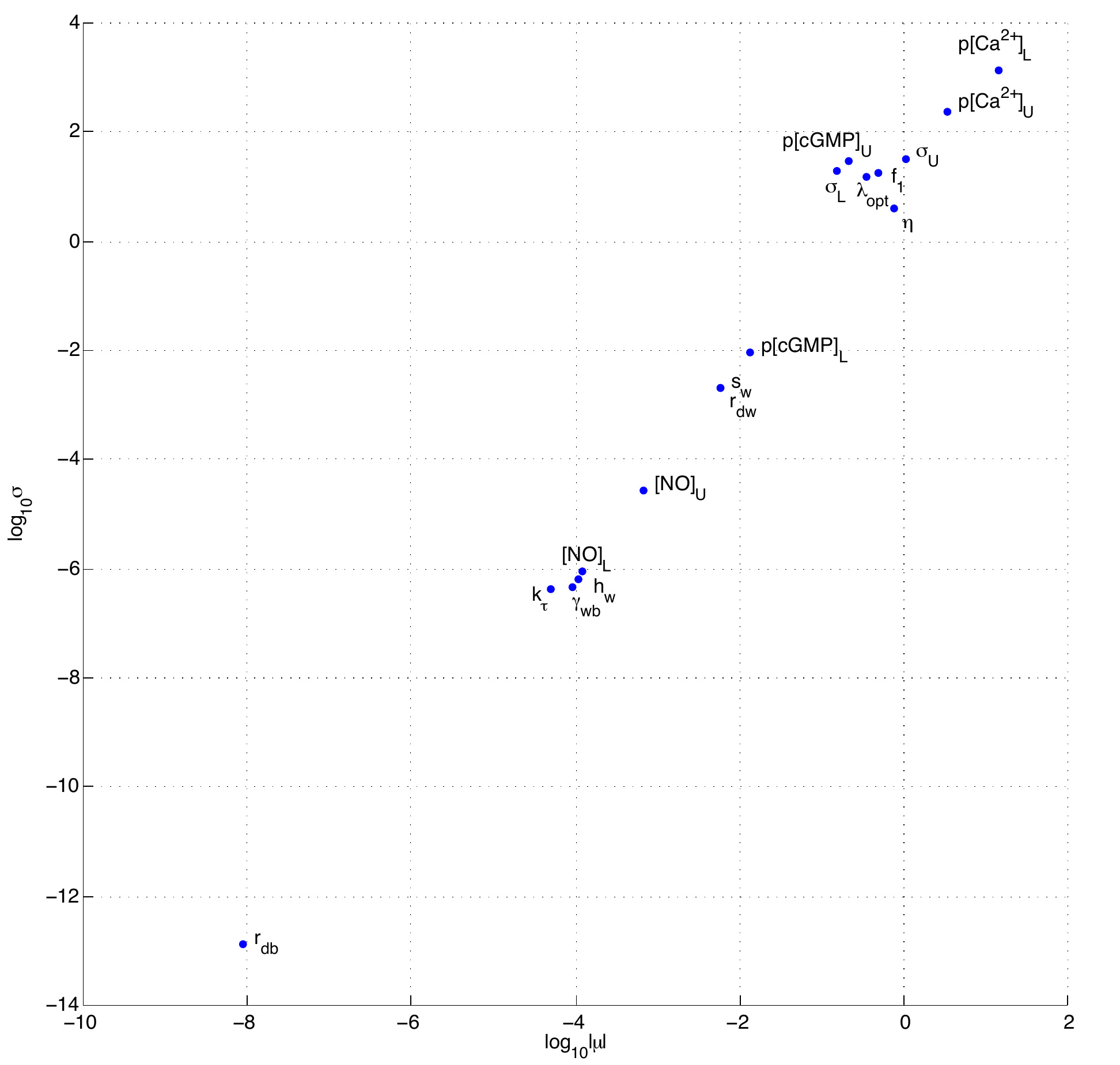} \caption{Results of sensitivity analysis using the multiplicative scheme with a range of $\pm$10\%, variance of elementary effects versus mean magnitude of elementary effects; axes are logarithmic to show detail, sensitivities are dimensionless.} \label{figsensitivityMultiplicative10} \end{figure}

\begin{figure}[hbtp] \centering \includegraphics[scale = 0.5]{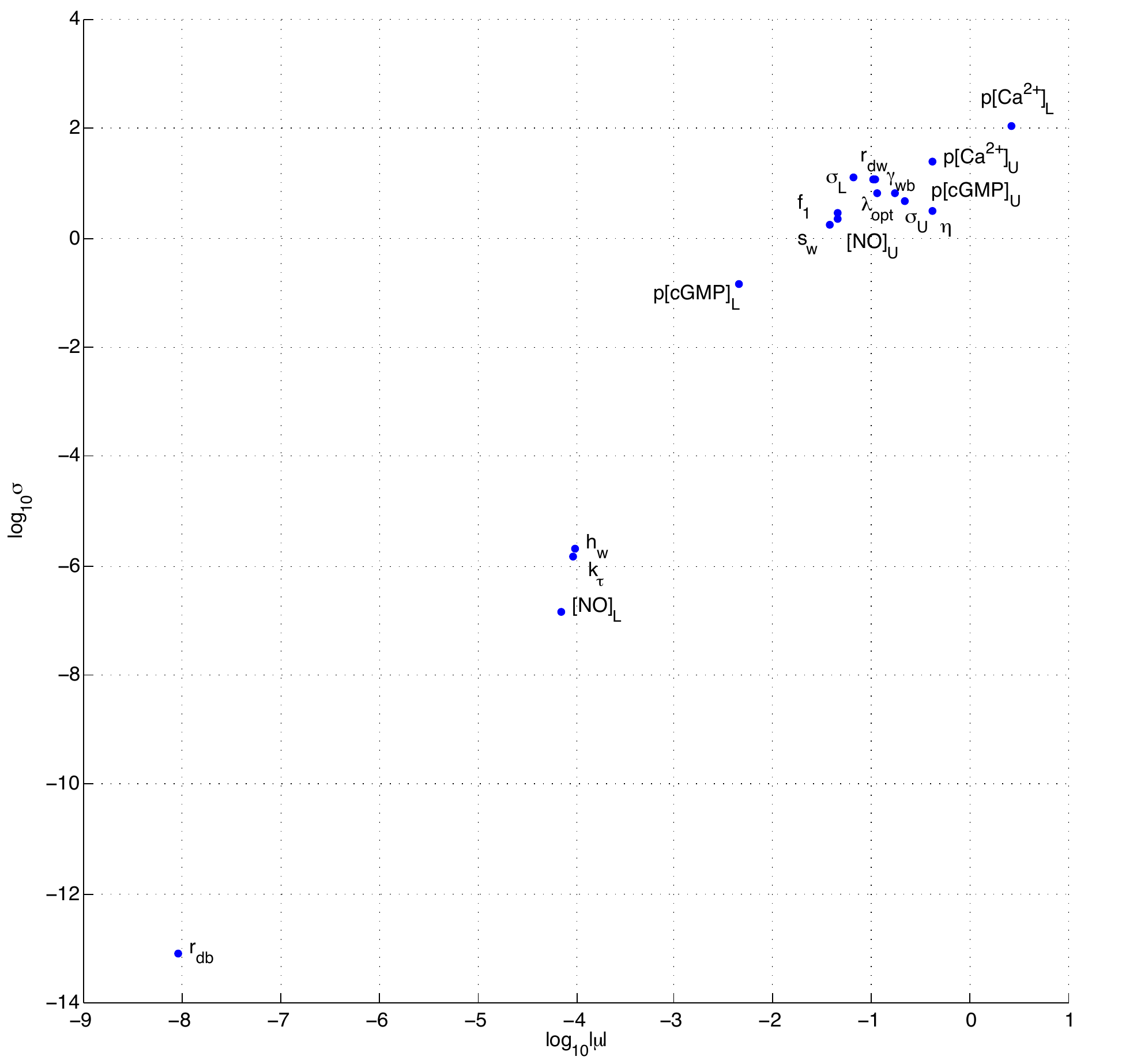} \caption{Results of sensitivity analysis using the multiplicative scheme with a range of $\pm$50\%, variance of elementary effects versus mean magnitude of elementary effects; axes are logarithmic to show detail, sensitivities are dimensionless.} \label{figsensitivityMultiplicative50} \end{figure}

To overcome some of the limitations of the additive scheme, another analysis was carried out, this time with the parameter space being defined to extend $\pm$10\% or $\pm$50\% either side of the optimal values given in \cref{tabBounds}. The finite difference used to evaluate the main effects is also defined as a percentage of the value of the central point, specifically 100 times smaller than the factor defining the width of the sample space. The resulting sensitivities are then normalized by the baseline values of the parameters, thus creating a dimensionless sensitivity which can be thought of as \%/\%, or proportional change in output per proportional change in parameter. To give a concrete example of the procedure for the $\pm$10\% case (in one dimension): if the optimal value of a parameter, having been tuned to fit the data, is 6units, then the parameter space over which samples are taken extends from 5.4units to 6.6units ($\pm$10\%). Let us suppose that one of the samples chosen by the latin hypercube method lies at 5.814units - the main effect for this sample will be found by evaluating the model output both at 5.814units and at 5.1892units, this being an increment of 10\%/100 = 0.1\% of the sample point value. If the output values for the sample point value and the incremented value are $a$ and $b$ respectively, then the dimensionless main effect is given by $\left(\left(b - a\right)/a\right)/0.001$. Although dimensionless, this can be thought of as percentage change in output per percent change in parameter. The results of this scheme are shown in \cref{figsensitivityMultiplicative10,figsensitivityMultiplicative50}.

\subsection{Dynamic Results}
\begin{figure}[hbtp] \centering \includegraphics[scale = 0.5]{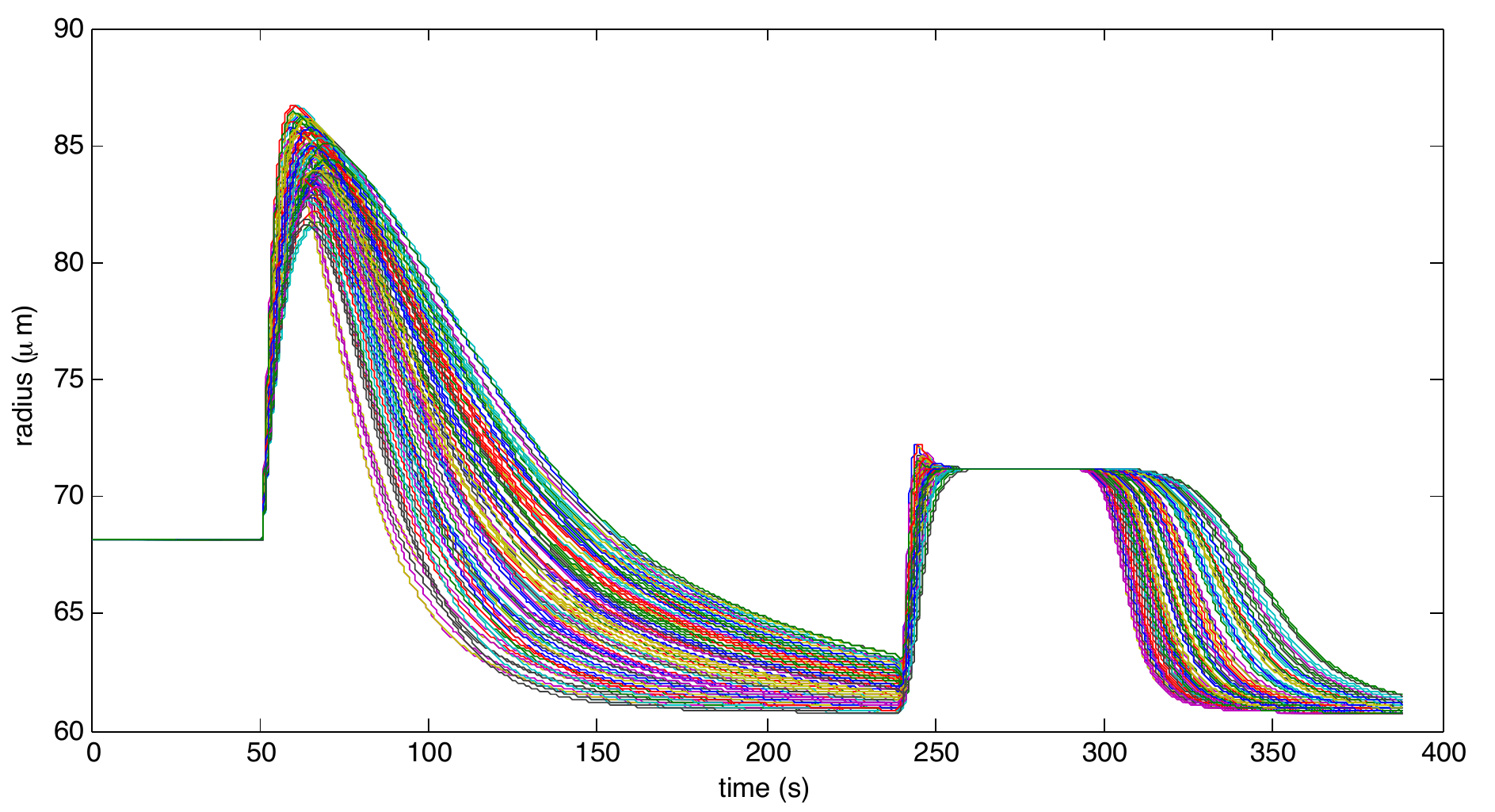} \caption{The 100 different baseline dynamic responses used to evaluate the sensitivity of the dynamic response. Parameters were varied $\pm$50\% around their optimal values.} \label{figdynSensitivity50radius} \end{figure}

\begin{figure}[hbtp] \centering \includegraphics[scale = 0.5]{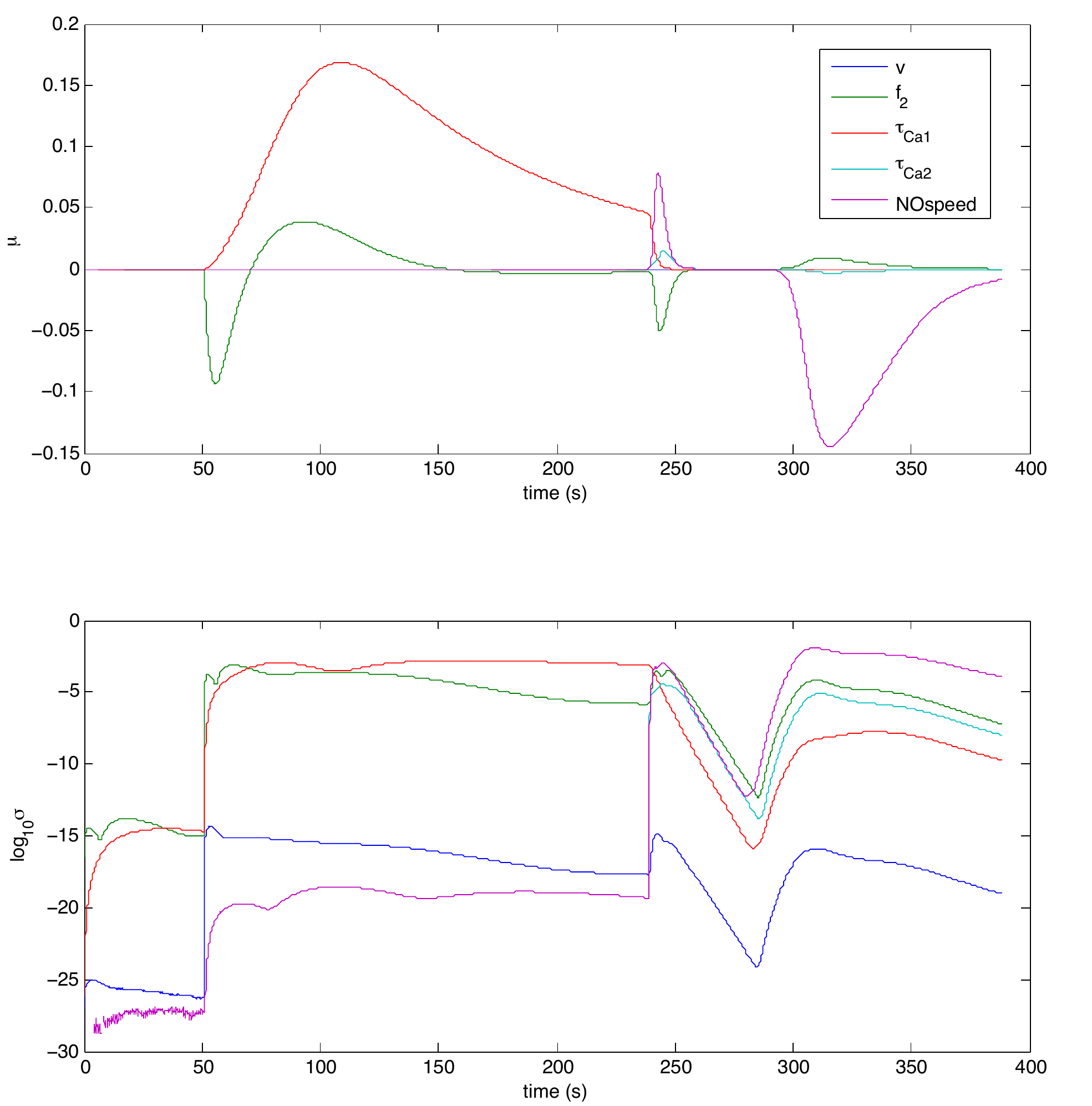} \caption{The mean and variance of the sensitivity of the dynamic response. Variance is shown on a log y-axis to show detail. Sensitivity is dimensionless as the muliplicative scheme yields \%/\% values.} \label{figdynSensitivity50} \end{figure}

We can also apply the same technique for sensitivity analysis in the dynamic case. We measure the sensitivity of the model radius to variations in the dynamic parameters at every time during the response. This analysis was done on the response of \cref{figKuo91Fig2A} over a range of $\pm$50\% either side of the optimal values of the parameters $v$, $f_2$, $\tau_{Ca}$, and $NOspeed$,  where $NOspeed$ is a scaling factor applied to the four coefficients of \cref{dNOdt} as described in \cref{dynfitting}. To ensure that we examine only the sensitivity of the dynamics of the system, without changing the steady-state behaviour, we scale $f_1$ inversely to the scaling applied to $v$ at each experimental location. The values of $\tau_{Ca}$ for the two phases of the response are treated as separate variables, $\tau_{Ca1}$ and $\tau_{Ca2}$. The 100 baseline responses at which the 5 elementary effects were sampled are shown in \cref{figdynSensitivity50radius}. The mean and variance of the sensitivities are shown in \cref{figdynSensitivity50}.

It can be seen from \cref{figdynSensitivity50} to which parameters the response is sensitive in different phases of the response: whilst the vessel is passively dilating in response to the step increase in pressure, it is very sensitive to changes in the passive sliding friction coefficient $f_2$; during the slow active contraction the model is most sensitive to the \ca time constant $\tau_{Ca1}$; sensitivity to $f_2$ returns along with sensitivity to $NOspeed$ as the step in flow causes sudden dilation; finally the delayed drop in radius following the cessation of flow is dominated by the speed of the NO response, as described in \cref{dynfitting}.

\subsection{Discussion}
The two methods, additive and multiplicative, of sensitivity analysis yield quite different results. The additive scheme suggested that four parameters were over 100,000 times more influential than any others, whereas the multiplicative scheme shows that there is not such a clear distinction. Two of the most important parameters in the multiplicative results are the upper and lower levels of \ca concentration in the myogenic response. Given that the myogenic response is the key driver in reducing the vessel radius with increased pressure, this is not surprising. All of the NO system parameters are relatively unimportant, with the exception of the saturation level of \ccgmp\!\!, suggesting that the system may have settled in a region of the parameter space where the effect of NO is binary, i.e., NO is effectively `on' or `off'. This could only happen because the parameters were fitted to data in which there were only two levels of flow.

Because the output of which we have measured the sensitivity is the goodness of fit to experimental data, we can say that we have greatest confidence in the accuracy of the parameters to which the model is most sensitive. In other words, we can be more sure that we have realistic values for the myogenic response parameters than for the NO model parameters. Of course with the caveat that this sensitivity analysis has not included the parameters which were independently fitted to the data of \cite{Lee97} in \cref{LeeFit}. In terms of the robustness of the physical system, our results suggest that a perturbation of the \ca levels would significantly affect the regulation of bloodflow by the arterioles. Conversely an excess of NO would have very little effect on the radius of the vessel without an accompanying rise in \ccgmp\!\!, which would require additional soluble guanylate cyclase (sGC) enzymes and/or additional guanylate triphosphate, the substrate for the production of cGMP by sGC. It might be deduced that those quantities which are limited by the availability of an enzyme are far less able to perturb the system than those which are no so limited. For instance, an excess of NO introduced by some external agent will not affect the equilibrium of a vessel through which blood is flowing (because \ccgmp is already saturated) whereas a similar introduction of \ca would immediately upset the \cca in the cytoplasm of the VSMC (it being able to diffuse across the cell membrane) and so the equilibrium of the vessel. There is a point at which excess \ca no longer has an effect on myosin phosphorylation, \cref{figrPhospCa}, but it is many times the equilibrium level. In this way the vessel is robust to excess of NO in a way that it cannot be robust to an excess of \ca\!\!. Enzymes being large molecules, unable to cross the cell membrane, must be produced by the cells themselves and so some derangement of the cell's biochemistry would be necessary to alter the availability of an enzyme. In this sense enzymes limit the sensitivity of the system to all but one (at a time) of the species which precede them in any pathway. While \cno is below the saturation level, \ccgmp is sensitive only to \cno and not to [GTP] or [Mg$^{2+}$], once NO is in excess, the system is insensitive to \cno and becomes sensitive to either [GTP], [Mg$^{2+}$], or the capacity/availability of the sGC itself\footnote{The production of cGMP by sGC requires activation by NO, catalysation by Mg$^{2+}$, and guanosine tri-phosphate (GTP) as the substrate \cite{Denninger99}.}.

\section{Conclusions}
In this chapter we have presented the data to which we have fitted our model, we have explained the methods by which our model has been optimized to achieve the best fit to the data, and we have presented the results of that optimization. The model closely replicates the behaviour of an isolated arteriole both in the steady-state response to changes in intramural pressure, and in its dynamic responses to step changes in pressure and flow. The necessity of a varying myogenic \ca time constant to fit the dynamic data has been discussed, and possible physical mechanisms which could underlie this process have been outlined. We have developed and implemented a novel technique for sensitivity analysis and it has shown us a possible limitation in the way that our model is tuned. Whilst the fit of the model responses to the data is not perfect, neither is our model very complex. To achieve such a good correlation as we have to both the steady-state and dynamic data with a model containing many simplifications (such as the sigmoids) shows that the operation of the physical system could well be fundamentally similar to that of our model. Although our model reproduces the behaviour of the system at a high level and neglects to explicitly model the very complex biochemistry of the VSMC.

\chapter{A Model of Autoregulation}
\label{chapAutoreg}
\section{Introduction}
In the previous chapter we presented a model of an individual arteriole and showed that its behaviour is representative of that of a real arteriole. In this chapter we integrate that model of a single arteriole into a model of the vasculature of the whole brain. We then examine the effect that our arteriolar model has within a larger vascular system. The integration into a whole brain model also allows comparison of model results with in-vivo data from humans. In-vivo flow data for single arterioles is not available for humans, whereas data for the flow in the middle cerebral artery (MCA) and arterial blood pressure in humans are readily available.

\section{Integration of Arteriole Model Into Existing Whole-Brain Model}
\label{autoregFirst}
\begin{figure}[hbtp] \centering \includegraphics[scale = 4]{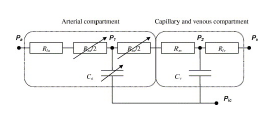} \caption{Schematic of the electrical equivalence model of the cerebral vasculature, reproduced from \cite{Payne06}. $P_a$, systemic arterial pressure; $R_{la}$, resistance of non-regulating arterial compartment; $P_1$, $R_{sa}$, and $C_a$, pressure, resistance and compliance of regulating arterial compartment; $R_{sv}$, resistance of capillary compartment and small veins; $C_v$, venous compliance; $P_2$, venous pressure; $P_v$ and $R_{lv}$, venous pressure and resistance of large veins, respectively; $P_{ic}$, intracranial pressure.}\label{figPayneSchematic} \end{figure}

The model of the cerebral vasculature which we will adapt to include our arteriolar model is taken from \cite{Payne06}. A schematic of this vasculature model is reproduced from \cite{Payne06} in \cref{figPayneSchematic}. The different levels of the vascular tree (large arteries, small arteries, capillaries and small veins, and large veins) are each represented as a resistance in an electrical equivalent circuit. Changes in volume of the arteriolar and venous compartments are accommodated by two capacitors. Because it is the arterioles, or the small arteries, which have the greatest effect in the autoregulation feedback loop, it is the properties of this level of the vascular tree which change in response to haemodynamic and metabolic stimuli. The work presented in \cite{Payne06} is concerned with the effects of metabolic factors, whereas we do not wish to investigate feedback from neural activation, rather we wish to replace the variable components of the arterial compartment with components whose properties are determined by the arteriolar model of \cref{chapVascModel}. In this way we can examine the behaviour of our arteriolar model in the context of the whole brain vasculature and compare the results to experimental data measured in-vivo from humans. 

\begin{figure}[hbtp] \centering \includegraphics[scale = 0.8]{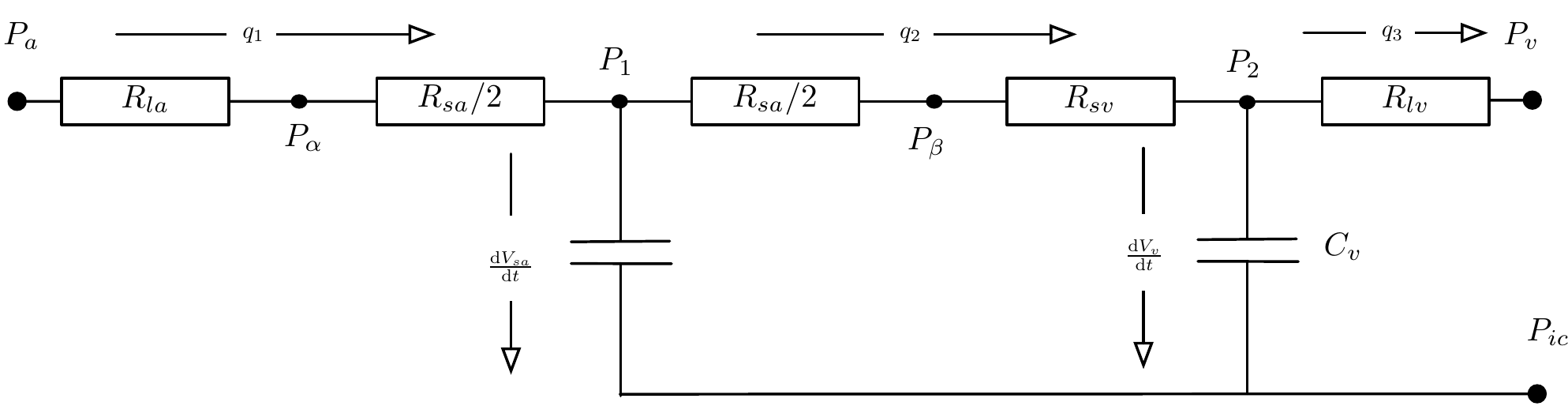} \caption{Schematic of the electrical equivalence model of the cerebral vasculature.}\label{figvasculatureSchematic} \end{figure}

The arteriolar network in \cref{figPayneSchematic} is represented by two resistive elements, each presenting half of the resistance of the whole arteriolar network, and a capacitance, accounting for the blood volume which is stored or dispensed when the arterioles dilate or contract respectively. To begin coupling our vessel model with this model we first define the two intermediate pressures $P_{\alpha}$ and $P_{\beta}$ as the pressures immediately upstream and downstream of the arteriolar level of the vascular tree, see \cref{figvasculatureSchematic}. We can relate the resistance of the vessels to the radius via:
\beq R_{sa} = \frac{8\mu_b L}{\pi r^4} \cdot \frac{1}{n_{sa}} \label{Rsa} \eeq

where $n_{sa}$ is the effective number of parallel arterioles in the vascular tree. The volume of blood in the arterioles at any time is given by:
\beq V_{sa} = \pi r^2 L n_{sa} \label{Vsa} \eeq

from the volume of $n_{sa}$ cylinders of radius $r$, length $L$. This volume is equivalent to the charge on the capacitor in the arterial compartment, and its derivative:
\beq \diff{V_{sa}}{t} = \dot{V}_{sa} = 2\pi r L n_{sa} \diff{r}{t} \label{dVsadt} \eeq

is equivalent to the current down that leg of the circuit, as shown in \cref{figvasculatureSchematic}. These relationships are sufficient to combine the arteriolar model of \cref{chapVascModel} with the vasculature model of \cite{Payne06}. 

\subsection{Solution in the Steady-State}
\label{brainSS}
In the steady-state the flows $\dot{V}_{sa}$ and $\dot{V}_v$ are zero. Conservation of mass then gives:
\beq q_1 = q_2 = q_3 = q = \frac{P_a - P_v}{R_{la} + R_{sa} + R_{sv} + R_{lv}} \label{q} \eeq

We then calculate $P_{\alpha}$ and $P_{\beta}$ from \cref{Poiseuille}:
\beq P_{\alpha} = P_a - q R_{la} \label{Palpha} \eeq
\beq P_{\beta} = P_{\alpha} - q R_{sa} \label{Pbeta} \eeq

The form of the arteriolar model in the steady-state is given in \cref{states}, the inputs for which can be calculated from $P_{\alpha}$ and $P_{\beta}$ as follows (to avoid confusion, here we will refer to the mean intramural pressure as $P_{im}$, in \cref{chapVascModel} this was referred to as $P$):
\beq P_{im} = \frac{P_{\alpha} + P_{\beta}}{2} \label{Pab2Pim} \eeq
\beq \Delta P = P_{\alpha} - P_{\beta} \label{Pab2DP} \eeq

Our arteriolar model can now be used to close the algebraic loop by calculating $r$ from $P_{im}$ and $\Delta P$. The procedure employed for iteratively solving the steady-state of the combined model is as follows (using $k$ to denote iteration number and starting from an initial value of $r_k = r_0 = 60\e{-6}$m):
\begin{enumerate}
\item Calculate $R_{sa}$ from $r_k$, \cref{Rsa}.
\item Calculate $q$ from $R_{sa}$, \cref{q}.
\item Calculate $P_{\alpha}$ and $P_{\beta}$, \cref{Palpha,Pbeta}.
\item Convert to $P_{im}$ and $\Delta P$, \cref{Pab2Pim,Pab2DP}.
\item Use arteriolar model to find $\bar{\lambda}$ and hence $\bar{r}$, \cref{states,l2r}.
\item Update estimated value of $r$ using $r_{k+1} = r_k + \frac{1}{2}\left(\bar{r} - r_k\right)$.
\item Repeat until $r_{k+1} = r_k$ to within some tolerance ($1\e{-15}$m).
\end{enumerate}

This iterative process is itself repeated for different values of $P_a$ to yield the steady-state response of the model over a range of arterial pressures.

\subsection{Solution in the Dynamic Case}
The combined model has four states, the three states of the arteriolar model given in \cref{states} and a fourth state from the vascular model, $V_v$, the volume of blood in the venous compartment. Converting $\lambda$ to $r$ through \cref{l2r}, we can write the form of the combined model as:
\beq \dot{\mathbf{x}} = \diff{}{t}\left[ \begin{array}{c}
\ccam \\ C_w \\ r \\ V_v
\end{array} \right]  = f\!\left(\mathbf{x}, P_a(t) \right) \label{combDyn} \eeq

The derivatives of the first three states are given by the arteriolar model as a function of the states themselves and $P_{\alpha}$ and $P_{\beta}$. The calculation of $P_{\alpha}$ and $P_{\beta}$ depends (in the dynamic case) on the flows $\dot{V}_{sa}$ and $\dot{V}_v$, which in turn depend on the state derivative $\diff{r}{t}$ (\cref{dVsadt}), leaving us with an algebraic loop in the solution of the dynamic system. Similarly to \cref{brainSS}, this can be solved using an iterative approach at each time step, again using $k$ to denote iteration number and starting with an initial value of $\dot{V}_{sa_k} = \dot{V}_{sa_0} = 0$. From conservation of mass in the arterial compartment:
\beq \frac{P_a - P_1}{R_1} = \dot{V}_{sa_k} + \frac{P_1 - P_2}{R_2} \eeq

where $R_1 = R_{la} + \frac{R_{sa}}{2}$ and $R_2 = \frac{R_{sa}}{2} + R_{sv}$. Rearranging for $P_1$ gives:
\beq P_1 = \frac{\frac{P_a}{R_1} - \dot{V}_{sa_k} + \frac{P_2}{R_2}}{\frac{1}{R_1} + \frac{1}{R_2}} \label{P1} \eeq

Equation (5) in \cite{Payne06} gives the venous compliance:
\beq C_v = \frac{1}{k_{ven}\left(P_2 - P_{ic} - P_{v1} \right)} \label{Cv} \eeq

where $k_{ven}$ and $P_{v1}$ are constants. This leads to the following equation for venous volume (eq.~(6) in \cite{Payne06}):
\beq V_v = \frac{1}{k_{ven}} \ln\! \left(P_2 - P_{ic} - P_{v1} \right) + V_{vn} \eeq

where the constant of integration, $V_{vn}$ is chosen to given a suitable baseline venous volume fraction. Thus we can calculate $P_2$ directly from the state $V_v$ by rearranging:
\beq P_2 = e^{\left(V_v - V_{vn}\right)k_{ven}} + P_{ic} + P_{v1} \label{P2} \eeq

We can now calculate $P_{\alpha}$ and $P_{\beta}$ as follows:
\beq P_{\alpha} = P_a - \frac{P_a - P_1}{R_1} R_{la} \eeq
\beq P_{\beta} = P_1 - \frac{P_1 - P_2}{R_2} \frac{R_{sa}}{2} \eeq

and from them use the dynamic form of the arteriolar model to yield $\diff{r}{t}$. We then use \cref{dVsadt} to find $\dot{V}_{sa_{out}}$, the inferred value of net bloodflow into the arteriolar compartment. Finally we update our estimated value and iterate:
\beq \dot{V}_{sa_{k+1}} = \dot{V}_{sa_k} + \frac{1}{2} \left( \dot{V}_{sa_{out}} - \dot{V}_{sa_k} \right) \eeq

Once $\dot{V}_{sa_{k+1}} = \dot{V}_{sa_k}$ to within some tolerance, the derivatives of the first three states have been found from the arteriolar model and the fourth state derivative can be found from conservation of mass in the venous compartment:
\beq \diff{V_v}{t} = \dot{V}_v = \frac{P_1 - P_2}{R_2} - \frac{P_2 - P_v}{R_{lv}} \label{dVvdt} \eeq

With this iterative scheme solving for  $\dot{V}_{sa}$ at each time step, the dynamic system of \cref{combDyn} can be integrated to yield the dynamic response of the combined model.

\section{Fitting and Analysis}
In this section we will present the fitting of the combined model to experimental data and examine the model's behaviour, both static and dynamic, under a variety of conditions.

\subsection{Steady-State Response}
\begin{figure}[hbtp] \centering \includegraphics[scale = 0.6]{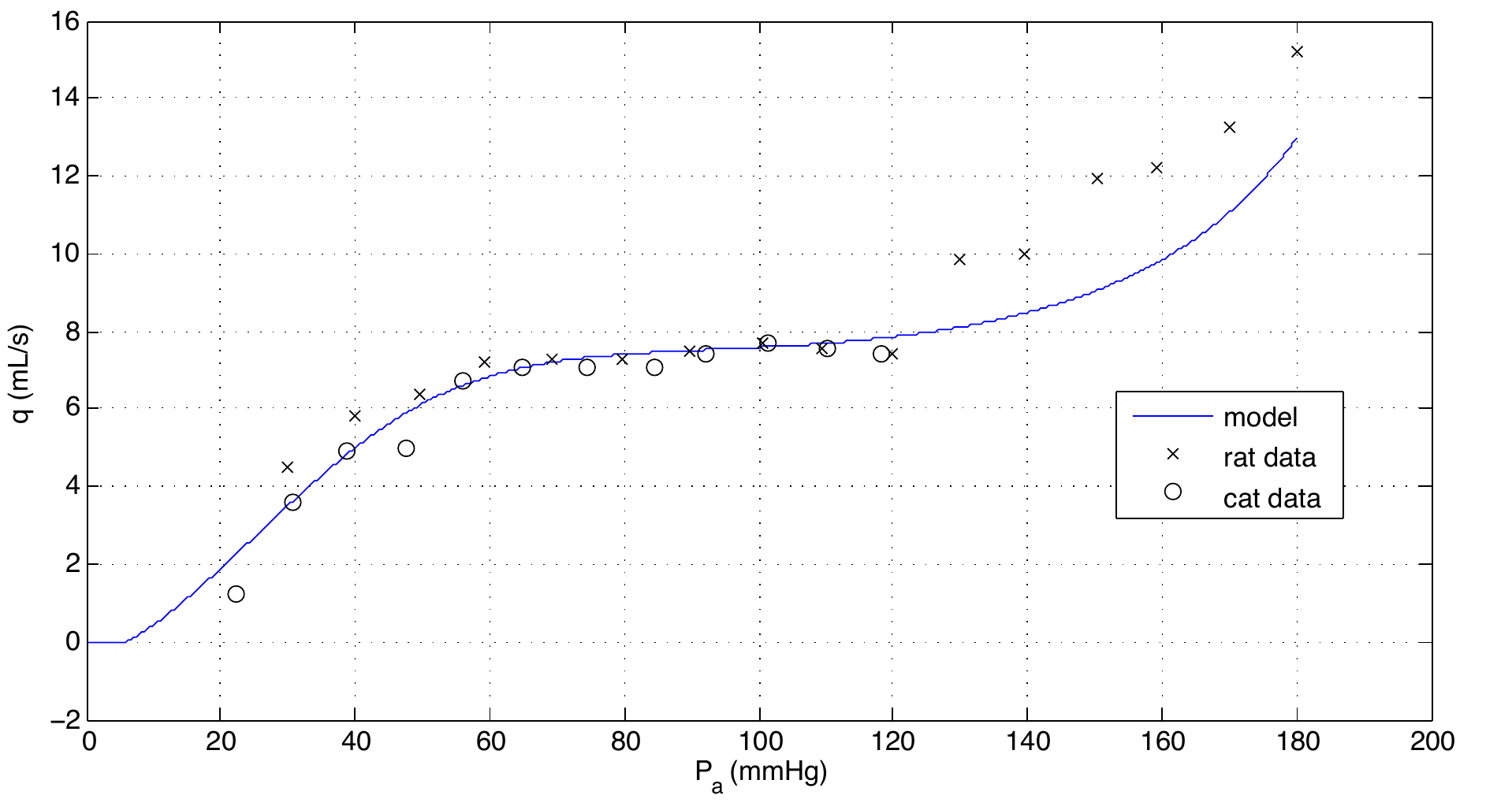} \caption{The steady-state flow/pressure behaviour of the model shown alongside rat and cat CBF data taken from \cite{Ursino98} fig. 5B where they were presented as normalized around 100mmHg.}\label{figcombiSS} \end{figure}

Values of the parameters of the vasculature model were taken from \cite{Payne06}. The only parameter not given in \cite{Payne06} and not already assigned a value in the fitting of the arteriolar model is $n_{sa}$, the effective number of arterioles in parallel in the arteriolar level of the vascular tree. In order to simultaneously achieve a realistic resistance for the arteriolar part of the network and a realistic volume in the arterial compartment, it is necessary to change the value of vessel length, $L$, from that in \cref{chapArtFit}. It can be seen from \cref{Rsa,Vsa} that if $R_{sa}$ and $V_{sa}$ are given, and if $r$ is to be maintained at a value consistent with arteriolar radius, then two degrees of freedom are needed; varying $L$ provides this second degree of freedom (the first being $n_{sa}$). Values of $L$ and $n_{sa}$ were thus found to achieve the correct resistance, matching the flow/pressure data of \cite{Ursino98}, whilst also achieving an arteriolar radius of $\approx 60 \mu$m and an arterial volume approximately one third of that of the venous compartment. \Cref{figcombiSS} shows data extracted from fig. 5B of \cite{Ursino98} alongside the fitted steady-state response of the combined model.

In \cite{Ursino98} the data are presented as percentage deviations from the flow at 100mmHg, we show them here multiplied by the model flow at 100mmHg. It can be seen that the low pressure side of the curve is a very close fit to the data, the high pressure side less so, possibly because it is much harder to collect high pressure data than low pressure data. It should also be noted that in terms of investigation of pathological conditions of bloodflow in the brain, we are much more interested in abnormally low, than abnormally high, values of blood pressure. \Cref{tabcombParams} lists the parameter values used to achieve this fit, parameters not listed take the values presented in \cref{chapVascModel}.

\begin{table}[!h] \centering
\footnotesize{
\begin{tabular}{l l r l c}
\hline
Parameter & Description & Value & Units & Source\\
\hline
$P_v$ & venous pressure & 6 & mmHg & \cite{Payne06}\\
$P_{ic}$ & intracranial pressure & 10 & mmHg & \cite{Payne06}\\
$R_{la}$ & resistance of large arteries & 0.4 & mmHg s/mL & \cite{Payne06}\\
$R_{sv}$ & resistance of small veins (inc. capillaries) & 1.32 & mmHg s/mL & \cite{Payne06}\\
$R_{lv}$ & resistance of large veins & 0.56 & mmHg s/mL & \cite{Payne06}\\
$k_{ven}$ & stiffness coefficient for venous compliance & 0.186 & 1/mL & \cite{Payne06}\\
$P_{v1}$ & pressure offset for venous compliance & -2.25 & mmHg & \cite{Payne06}\\
$V_{vn}$ & offset for venous volume & 28 & mL & \cite{Payne06}\\
\hline
$L$ & characteristic length of arterioles & $50\e{-3}$ & m & Fitting\\
$n_{sa}$ & effective number of arterioles in parallel & 16,000 & -- & Fitting\\
\hline
\end{tabular}}
\caption{Parameters of the combined, whole-brain model.}
\label{tabcombParams}
\end{table}

\begin{figure}[hbtp] \centering \includegraphics[scale = 0.6]{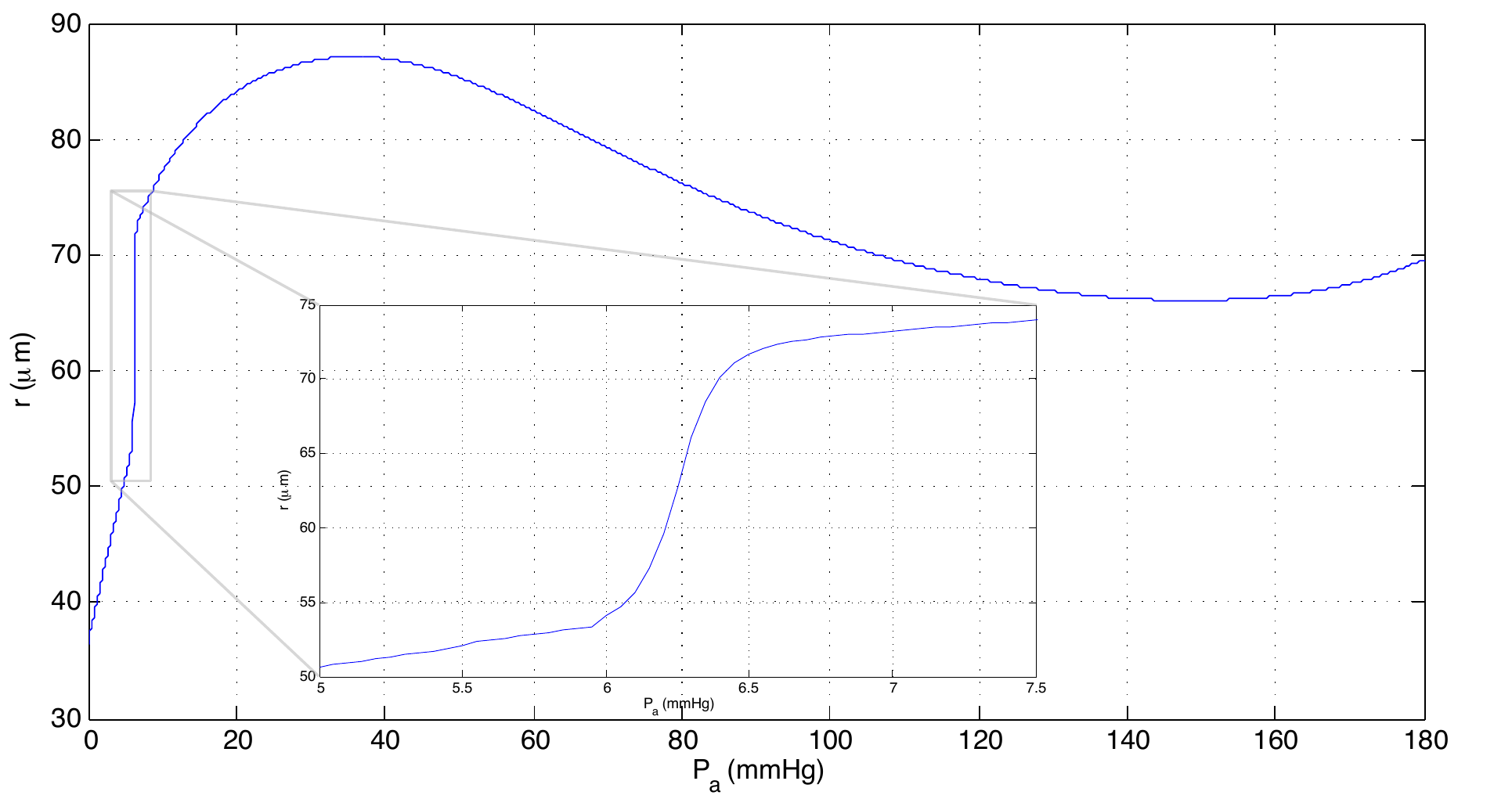} \caption{Steady-state arteriolar radius vs. arterial pressure.}\label{figautoregSSr} \end{figure}

\begin{figure}[hbtp] \centering \includegraphics[scale = 0.6]{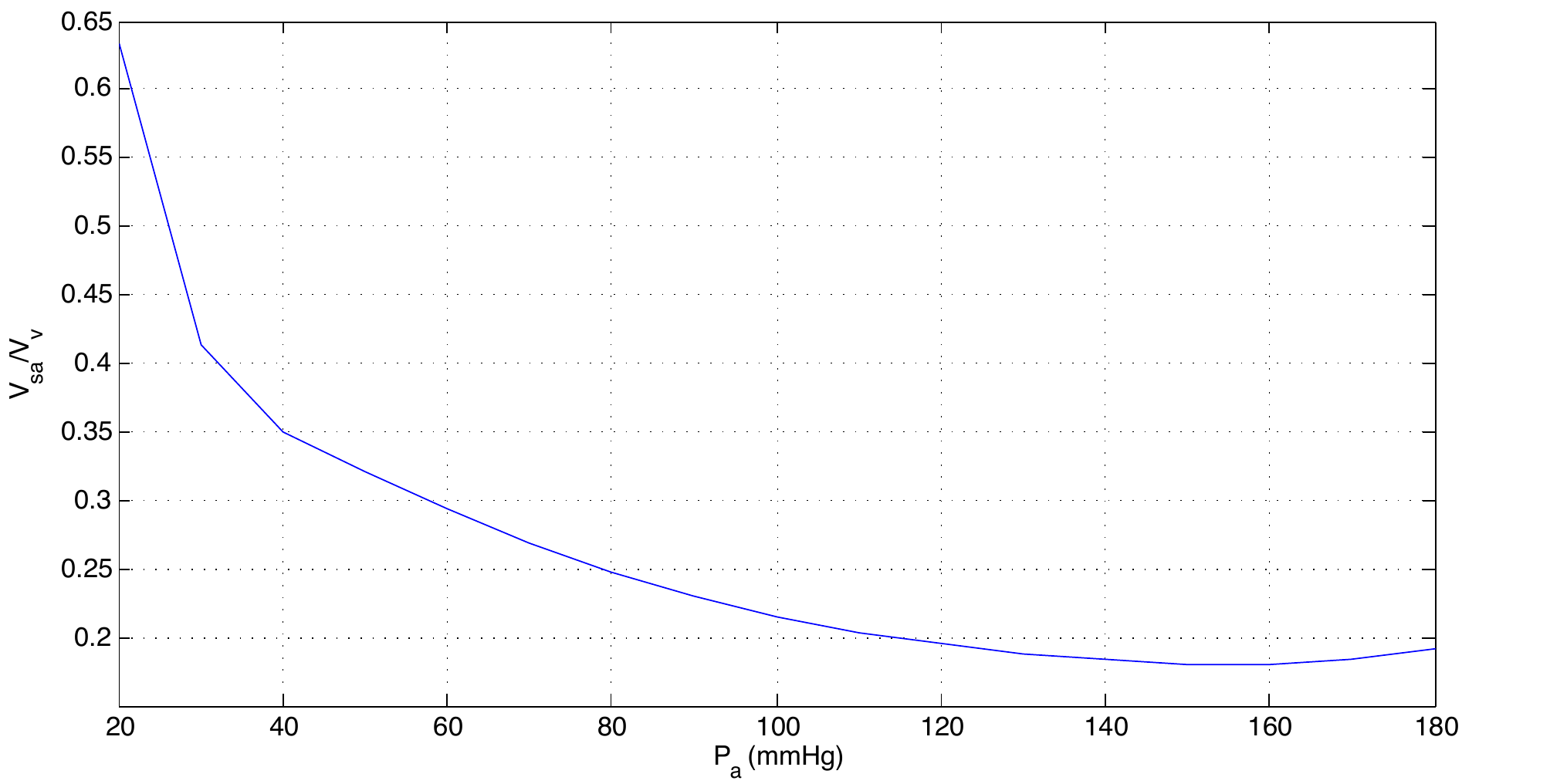} \caption{Steady-state arteriolar volume fraction, $V_{sa}/V_v$ vs. arterial pressure.}\label{figautoregVolFrac} \end{figure}

Note that the values for $L$ and $n_{sa}$ given in \cref{tabcombParams} are not representative of actual vessel length or number, rather they are the values which must be assumed in order to match the volume and resistance of an actual arteriolar network whilst maintaining the gross simplification that the network is a single level of identical arterioles arranged in parallel. The implications of the values that these parameters must take will be discussed further in \cref{autoregDisc}. The steady-state vessel radius which gives rise to this plateau in the pressure/flow relationship of \cref{figcombiSS} is shown in \cref{figautoregSSr} and the arteriolar volume fraction $\frac{V_{sa}}{V_v}$ is shown in \cref{figautoregVolFrac}; note that the $x$-axis shows systemic arterial pressure, $P_a$, whereas the $x$-axes of \cref{figActiveFitKuo,figActiveFitFalcone} show pressure immediately upstream of the arteriole, $P_{\alpha}$ in this combined model. The inset in \cref{figautoregSSr} shows the sigmoidal step in vessel radius as the NO level rises and the cGMP concentration goes from its low value to its saturation value. Arteriolar volume fraction is seen to vary with arteriolar radius squared, with the exception of the lowest pressures where the collapse of venous volume inflates the arteriolar volume in proportion to it.

\subsection{Dynamic Responses}
\begin{figure}[hbtp] \centering \includegraphics[scale = 0.6]{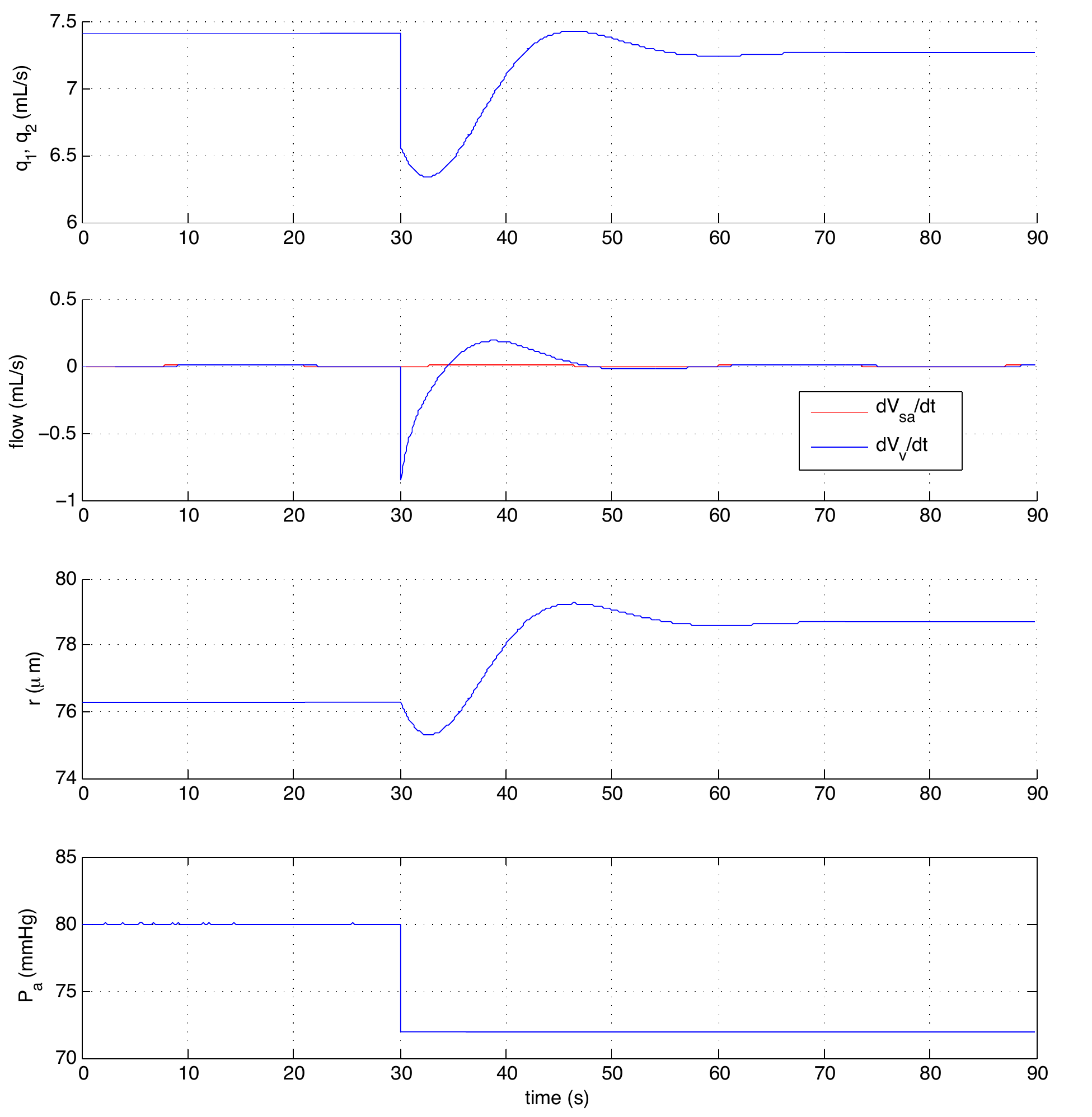} \caption{Response of the dynamic model to a 10\% step decrease in arterial pressure $P_a$. In the top pane, $q_1$ and $q_2$ are both shown but lie exactly atop one another.}\label{figautoregStepDn5} \end{figure}

The dynamic behaviour of the system can be assessed through analysis of its responses to variations in arterial pressure $P_a$. The response of the system to a 10\% step decrease in $P_a$ from 80 to 72mmHg is shown in \cref{figautoregStepDn5} for a $\tau_{Ca}$ value of 5s. It can be seen from the flows $\diff{V_a}{t}$ and $\diff{V_v}{t}$ that the change in volume of the arterial compartment is minimal compared to that of the venous compartment. It is the compliance of the venous compartment which drives the overshoot of vessel radius, and so flow, past the equilibrium value after the initial passive contraction of the vessels; this overshoot is not pronounced when considering an isolated arteriole in the absence of venous compliance (\cref{figKuo91Fig2D}).

\begin{figure}[hbtp] \centering \includegraphics[scale = 0.6]{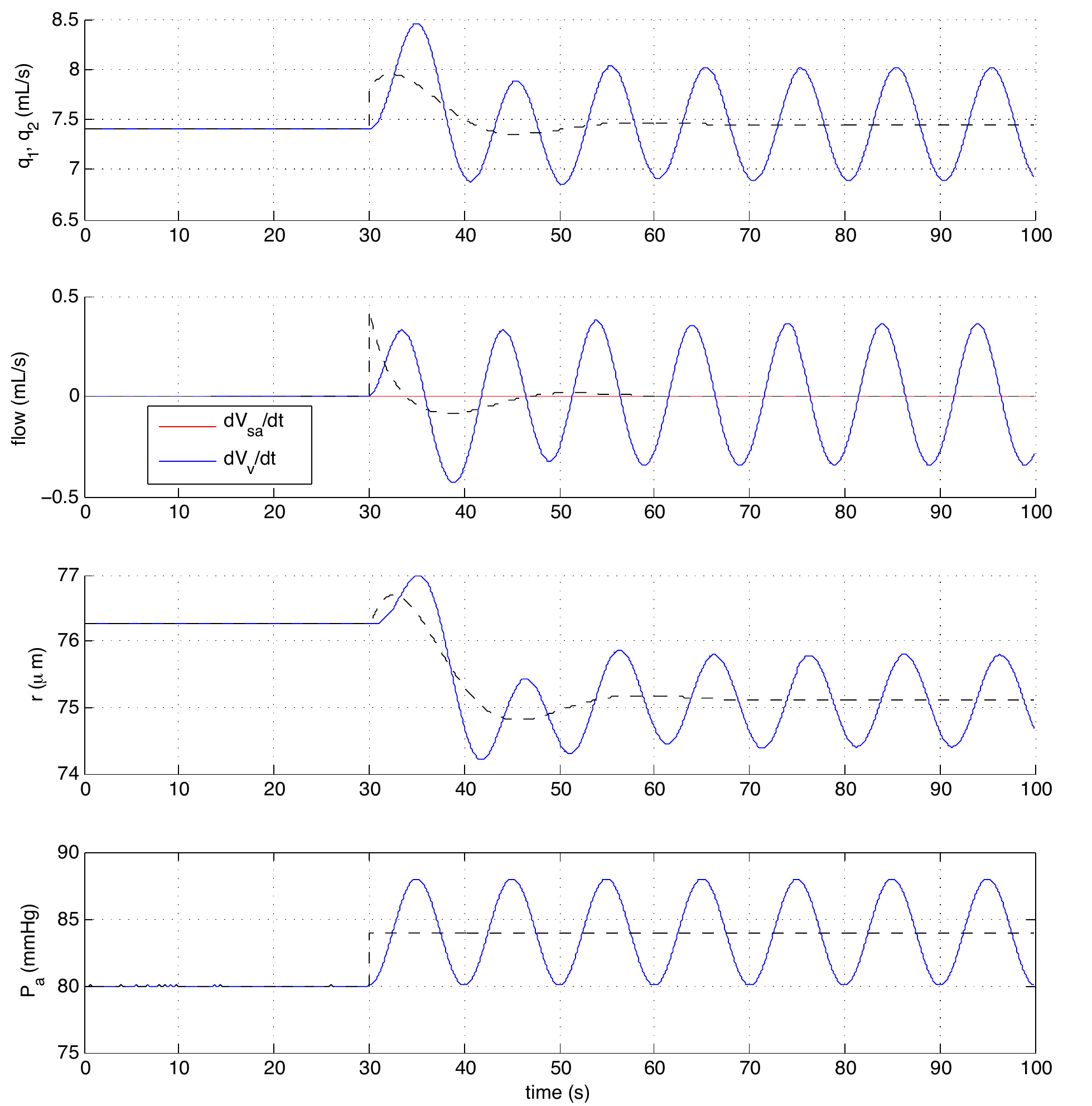} \caption{Response of the dynamic model to the sudden onset of a sinusoidal oscillation in pressure. In the top pane, $q_1$ and $q_2$ are both shown for each case but lie exactly atop one another.}\label{figautoregSine5} \end{figure}

\begin{figure}[hbtp] \centering \includegraphics[scale = 0.6]{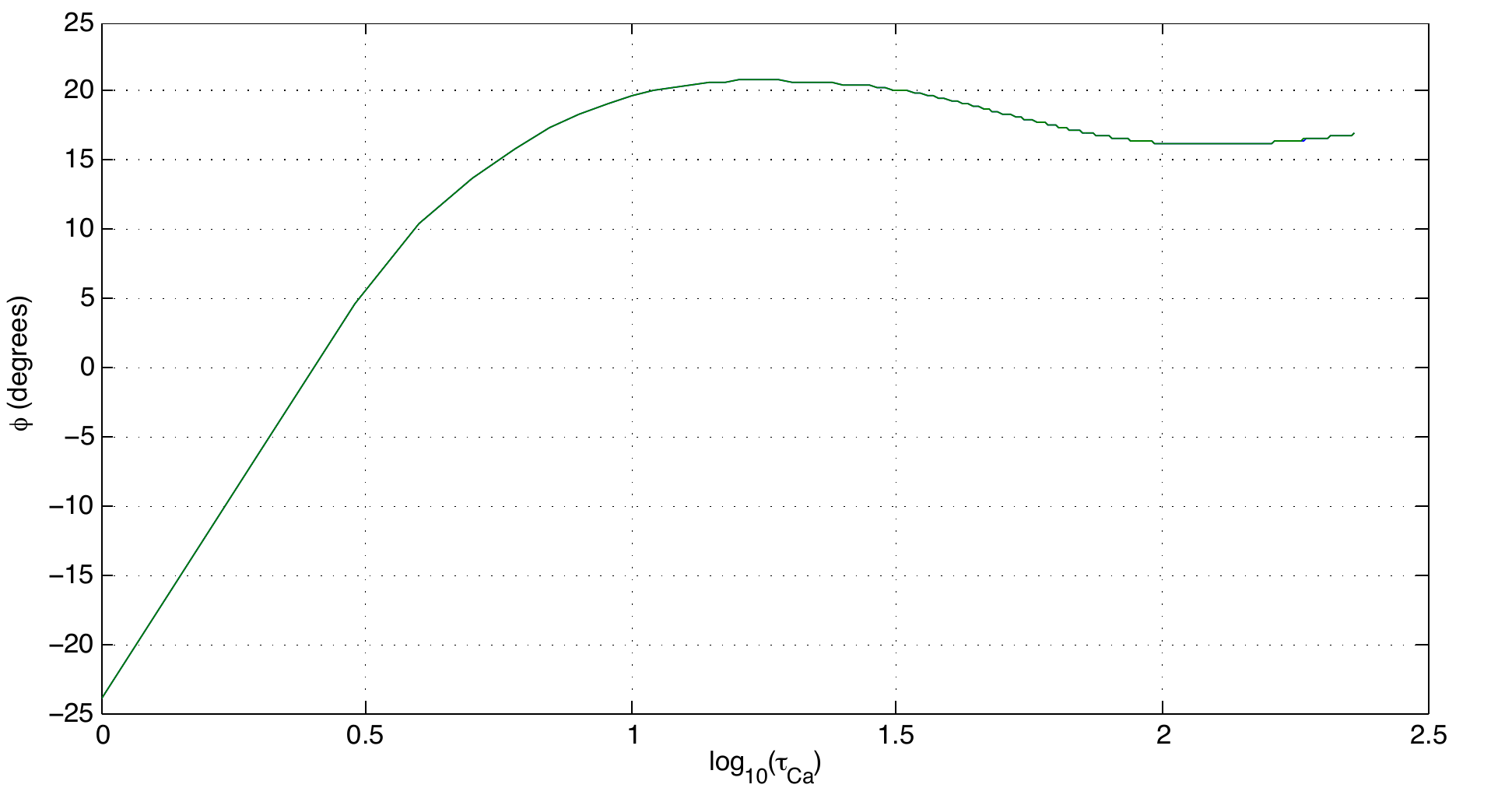} \caption{Phase difference between $P_a$ and $q_{1,2}$ for varying values of $\tau_{Ca}$. Positive $\phi$ indicates flow lagging pressure.}\label{figautoregPhase} \end{figure}

We next examine the response of the combined model to a pressure oscillation at 0.1Hz, as this is the frequency which is most commonly used to characterise autoregulatory influence. \Cref{figautoregSine5} shows the model's response to such an oscillatory driving pressure, again with a value for $\tau_{Ca}$ of 5s. The forcing function (bottom, blue) can be thought of as the superposition of a 4mmHg step increase and a 4mmHg amplitude sine wave. For comparison, the black dashed lines show the response to a 4mmHg step increase in pressure (excluding the unresponsive $\diff{V_{sa}}{t}$). It can be seen that the response is nearly linear in that it is close to a sine wave superposed on the step response. There is however a phase lag between the driving pressure and the flows $q_1$ and $q_2$ (which remain almost identical). This phase lag changes as we change the speed of the myogenic response by varying $\tau_{Ca}$, as shown in \cref{figautoregPhase}.

Altering the speed of the NO system, as described in \cref{dynfitting}, has no effect on the phase lag between pressure and flow. This is to be expected because of the saturation of \ccgmp at relatively low levels of \cno\!\!. The initial (and minimum) value of \cno in the sinusoidal pressure case is over 130 times greater than the \ccgmp saturation level $\conc{NO}_U$.

\subsection{Downstream Transport of NO and the Possibility of NO Signalling}
It has been postulated \cite{Duling87,Beach98} that the regulation of bloodflow in the brain in response to metabolic factors may involve some signalling mechanism between the downstream perfused tissue and the upstream regulating arterioles. Because it responds to flow and diffuses easily through both tissue and blood, it is possible that NO may be the medium for such signalling. In order to evaluate the feasibility of this hypothesis we augment our combined model with equations describing the transport of NO through the venous compartment and the equilibrium of NO concentrations between the tissue surrounding the veins and the blood flowing through them. We begin by assuming that the venous compartment is well-mixed, then conservation of mass gives:
\beq \diff{}{t}\left(C_vV_v\right) = q_2 C_{in,v} - q_3 C_v - r_{db} C_v - h_{wv}\left(C_v - C_T\right) \eeq

where $C_v$ is the NO concentration in the venous blood, $C_T$ is the NO concentration in the tissue surrounding the veins, $h_{wv}$ is the transfer coefficient between the blood and the tissue (across the wall), and $C_{in.v}$ is the NO concentration in the blood entering the venous compartment which can be found by evaluation of the solution to \cref{NO1} at $x = L$:
\beq C_{in,v} = C_b(L) = \left( \frac{h_w C_{w_e}}{h_w + r_{db} A_b} - C_{in} \right) \cdot \left( 1 - e^{-\left(h_w + r_{db} A_b \right) \frac{L}{Q}} \right) + C_{in} \label{cinv} \eeq

where all terms are as defined in \cref{NOeqns} with the exception of $Q$ which here is taken as $(q_1 + q_2)/2$ in order to reconcile the singular flow of the arteriolar model with the split arteriolar compartment of the combined model. As we have seen from \cref{figautoregStepDn5,figautoregSine5}, the two flows are nearly identical so this averaging has a negligible effect.

The tissue compartment can be modelled similarly:
\beq \diff{}{t}\left(C_TV_T\right) = h_{wv}\left(C_v - C_T\right) - r_{dw} C_T \eeq

If we assume that the tissue volume is constant then we can multiply out both equations to yield:
\beq \diff{C_v}{t} = \frac{1}{V_v}\left(q_2 C_{in,v} - q_3 C_v - r_{db} C_v - h_{wv}\left(C_v - C_T\right) -C_v \diff{V_v}{t} \right) \label{dcvdt} \eeq

\beq \diff{C_T}{t} = \frac{1}{V_T}\left( h_{wv}\left(C_v - C_T\right) - r_{dw} C_T \right) \label{dctdt} \eeq

If we assume that the ratio of volumes between the arterial and venous compartments is approximately 1:3 and that the transfer coefficient between the tissue and the blood is proportional to vessel circumference, and hence to the cube root of volume, then $h_{wv}$ should be related to the known value for the arteriolar model, $h_w$, by the relation $h_{wv} = h_w \cdot \sqrt[3]{3}$. If we then assume that the volume of the tissue adjacent to the blood in the venous compartment (which includes the capillary bed) is equal to that of all brain tissue, and furthermore that the blood accounts for 3\% of total brain volume, then we have that $V_T = 97\%/3\% \cdot V_v \cdot 4/3$. This approximation is completed with the assumption of a nominal venous volume of $V_v = 60$mL. Thus we have found values for the two parameters hitherto undefined, albeit very approximate ones, which will suffice for the analysis presented here.

The steady-state solutions to \cref{dcvdt,dctdt}, at which the states of our augmented dynamic model will be initialized, are given by:
\beq C_{v_0} = \frac{q_2 C_{in,v}}{q_3 + r_{db} + h_{wv}\left(1 - \frac{h_{wv}}{h_{wv} + r_{dw}} \right)} \eeq

\beq C_{T_0} = \frac{h_{wv}C_{v_0}}{h_{wv} + r_{dw}} \eeq

Incorporating these two new states into our model we find that for an arterial pressure of 80mmHg, the steady-state value of NO concentration in the venous blood is $1.4\e{-20}$M and that the steady-state concentration in the tissue is $1.6\e{-29}$M. Both of these are extremely low, primarily due to the rapid rate of decay of NO in the blood, $r_{db}$. Even if the decay rate in the blood is reduced to equal that in the tissue/vessel wall, $r_{dw}$, i.e., decreased by a factor of 1900, the initial concentration in the venous blood rises to only $2.7\e{-17}$M, and in the tissue to only $3.1\e{-26}$M. Compare this to the initial level of NO in the arteriole wall, $4.9\e{-6}$M. Maintaining this artificially reduced value of $r_{db}$, even after flow has increased (thus releasing more NO from the arteriolar endothelial cells), in the sinusoidal pressure case of \cref{figautoregSine5} the peak to trough amplitude of the variation in $C_v$ is $\approx 3\e{-17}$M. Given this evidence that the downstream concentrations of NO in the blood are negligible, the possibility of NO being the medium of a signal from venous blood to upstream vessels appears remote. When we consider that deoxyhaemoglobin takes up NO very readily and so realise that the decay rate of NO in venous blood should in fact be higher than that in arterial blood, it becomes evident that NO will not persist long enough in whole blood to transmit signals of detectable magnitude.

An experiment, described in \cite{Kuo91}, where downstream endothelium-denuded vessels react to NO transported from upstream intact vessels may be taken to suggest that this is not the case, and that NO released from arteriolar endothelium can effect the dilation of downstream vessels. However, this experiment was carried out using saline solution as the intraluminal fluid, rather than whole blood, and since the NO scavenging capacity of haemoglobin has such a profound effect on the lifetime of free NO molecules, this experiment does not provide a sound basis for extrapolation to in-vivo processes (this not being the purpose of the experiment in \cite{Kuo91}).

\subsection{Discussion}
\label{autoregDisc}
The integration of the isolated arteriole model of \cref{chapArtFit} into a simple electrical equivalent model of the whole-brain vasculature is not without some inconsistencies. The first derives from the assumption that all the small arteries are identical, and that they are all arranged in parallel. This leads to the misleading values of $L = 50$mm and $n_{sa} = 16,000$, when in fact the mean length of an arteriole of radius 70$\mu$m is closer to the original value of 1mm and we might expect the actual number of arterioles in the brain to be somewhere between 372,000 and 2,790,000\footnote{It is estimated that there are $\approx 186\e{5}$ arterioles in the body \cite{Leondes07}; if we assume arteriolar density goes with mass density then we take 2\% of this number for the brain's proportion of body mass \cite{Clarke99}, whereas if arteriolar density goes with density of oxygen-consumption then we take up to 15\% of this number as the brain's proportion of total cardiac output \cite{anaesthesiaUK}.}. How then do we reconcile the short, low volume arterioles of \cref{chapArtFit} with the volume of the arterial compartment? We may suppose, for instance, that the arteriolar level is composed of two generations of arterioles, the first being large, compliant, and passive, the second being those of our earlier model -- narrow and with an active myogenic response. This arrangement would accommodate the volume required of the arterial compartment, primarily in the first generation of larger arterioles. The resistance presented by the two generations together would be only slightly greater than that of the second generation alone. If we then generalise to a series of multiple generations, each smaller than the last, then we might easily conceive that there is an arrangement which satisfies simultaneously the requirements of number, resistance, and volume. The volume is thus largely attributable to the earlier generations, the resistance largely to the later generations, and the number being greatly increased.

Whilst, as stated above, it may be possible to match certain aspects of a more complex arteriolar network with an equivalent single level of identical arterioles in parallel, there are other properties of such a network which are not well represented. For example, we would not expect the larger arteries to exhibit a strong myogenic response; in fact we would expect the walls of such vessels to behave essentially purely elastically. This is supported by the data shown in fig.~3 of \cite{Payne06} which show arterial blood volume increasing with blood flow, and by implication with arterial pressure. In our single layer representation all arterioles show active contraction with increased blood flow, thus a diminishing arterial blood volume. In order to capture this passive compliance of the arterial compartment, another capacitative element must be added to the equivalent circuit. This addition would also redress the balance between $\diff{V_{sa}}{t}$ and $\diff{V_v}{t}$; currently the former is negligible compared to the latter, causing $q_2$ to follow $q_1$ almost exactly. Inclusion of the passive arterial compliance would partially decouple $q_2$ and $q_1$ and may increase the phase lag between ABP and CBF significantly enough to bring it into line with the $\approx 45^{\circ}$ repeatedly observed in vivo \cite{Diehl95,Birch95,Liu03}.

The change in phase difference between ABP and CBF, shown in \cref{figautoregPhase}, with varying $\tau_{Ca}$ can be explained in terms of the relative speeds of the passive, elastic, response and the active myogenic response of the arteriole. When $\tau_{Ca}$ is large the initial elastic response of the arteriole is large, and persists for some time before the myogenic response compensates and drives the system in the opposite direction, as we saw in \cref{figKuo91Fig2A}. Conversely, when $\tau_{Ca}$ is small and the myogenic response is fast, the initial elastic response of the vessel is quickly reversed by the active myogenic response and so is of much smaller magnitude and duration. With a driving pressure oscillation at 0.1Hz, the important factor for phase lag is the macro response of the vessel over a 10s period -- when $\tau_{Ca}$ is large this is in the passive direction, when $\tau_{Ca}$ is small, this is in the active direction. The phase thus reverses, from lead to lag, when the speed of the active response is just sufficient to balance the elastic reaction of the vessel over the initial few seconds; this occurs at $\tau_{Ca} \approx 2.5$s. Once the myogenic response is slower than the driving oscillations, the phase ceases to change significantly, as seen on the right of \cref{figautoregPhase} where $\tau_{Ca} > 10$s. This sensitivity of ABP-CBF phase difference to the speed of the \ca response suggests that it may be possible to diagnose pathological conditions affecting the release/uptake of \ca by measuring the ABP-CBF phase difference. If, as hypothesised in \cref{variableTCaJust}, the quick response of the \ca system is due to \ca release from the SR, then a pathological blockade of the \ca channels of the SR could lead to an observable phase shift between ABP and CBF, from lead to lag.

The insensitivity of the phase response to changes in the speed of the NO system may be unrealistic. Whilst the effect of flow on equilibrium radius does saturate at higher flow levels, it may be that our system saturates too early, and is thus insensitive to variations in flow to which it should respond. There are some data on the response of equilibrium radius to varying levels of flow in \cite{Kuo91} but they are inconsistent with the other data therein presented. Figure 3 in \cite{Kuo91} presents the variation in vessel diameter with flow for different levels of intramural pressure, however, these data, when plotted alongside those shown in \cref{figKuo91Fig4}, do not follow the same trend, and so it is difficult to integrate them into the fitting procedure used here. The sigmoidal step in \cref{figautoregSSr} shows the range of pressures over which NO transitions to its saturation value. This region of positive feedback between flow and radius, under the right conditions, could lead to a minimum pressure below which the vessel `collapses', or rather contracts fully, to become completely occluded. Whilst this was not observed in either \cite{Kuo91} or \cite{Falcone91}, it is possible that the fluid surrounding the isolated vessels in these experiments was at a lower pressure than that of the cerebral tissue. In their analysis of pressure-flow relationships in the cerebral vasculature Tzeng et al. \cite{Tzeng11} use the value of 33mmHg for the arterial pressure at which vessels collapse and flow stops, taken from in-vivo experiments presented in \cite{Aaslid03}. Our model exhibits no such sudden collapse, although the on-off nature of the NO response does hint at it. Perhaps in reality this positive feedback from the NO system contributes to this phenomenon. In our fitting of the isolated arteriole model in \cref{chapArtFit} we assume an extramural pressure of zero, there is some inconsistency in this vasculature model as we present the arteriolar compliance as reacting against an intracranial pressure of 10mmHg whilst calculating the wall stress for the myogenic response assuming zero extramural pressure. This inconsistency will be most noticeable at the lowest pressures. We should then perhaps not pay too much attention to the behaviour of the model at this extreme end of the pressure range. The effect of this inconsistency will be further reduced when we allow intracranial pressure to vary with the introduction of cranial compliance in \cref{addICCap}.

Whilst the inclusion of the equations for downstream NO transport has shown that there is a signal in the downstream NO concentration which reflects upstream conditions, it has been assumed that this signal is too weak to have an effect on autoregulation upstream. In numerical simulations such as ours it is possible to keep track of such small numbers representing numbers of molecules, and in writing and solving equations for exponential decay we do not make allowances for the true, stochastic nature of such low concentrations. In reality it may be that the downstream NO concentration is truly reduced to zero, or that at such low concentrations other effects which we have not modelled become important. It should also be noted that we have not included a discussion here of the sensitivity of any receptors to low concentrations of NO, the only NO sensitive system we have approached in this text is that of cGMP production. Without further investigation into the dynamics of low-concentration NO in blood and possible highly sensitive receptors for downstream NO it is not possible to rule out NO absolutely as a candidate for upstream metabolic signalling. However, on the basis of these results it seems highly unlikely.

\subsection{Conclusions}
So far in this chapter we have integrated the single arteriole model from \cref{chapVascModel} into a simple electrical equivalence model of the whole brain vasculature from \cite{Payne06}. The equilibrium behaviour of the combined model fits well with experimental data for the variation of cerebral blood flow with arterial pressure. We have examined several dynamic responses of the model and found that a phase difference exists between pressure and flow, that this phase difference is largely caused by the venous compliance, and that its magnitude and direction are dependent on the speed with which the myogenic response overrides the passive response of the arterioles. We have identified several shortcomings of this combined model and suggested possible improvements, namely the inclusion of a passive arterial compliance to the represent the larger arteries and the inclusion of a more complex vascular topology, better representing the branching levels of the network. Lastly we have concluded, through our examination of the downstream transport of NO, that the decay of NO in the blood is such that venous concentrations of NO are extremely low; consequently it is very unlikely that NO is involved as the signalling medium for any large-scale autoregulatory feedback loop.

\section{Inclusion of Passive Arterial Compliance}
\label{artComp}
\begin{figure}[hbtp] \centering \includegraphics[scale = 0.8]{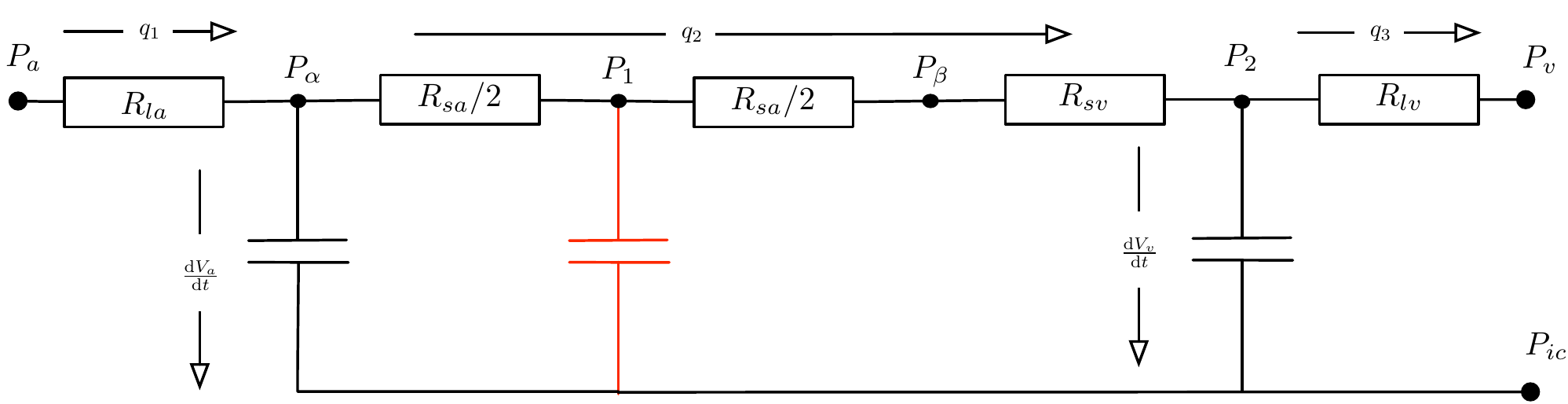} \caption{The modified electrical equivalent model of the vasculature; flow down the red path is neglected.}\label{figvasculatureSchematicModified} \end{figure}

In order to address the inconsistencies referred to in \cref{autoregDisc}, the model of \cref{autoregFirst} has been modified to include a capacitative element representing the passive elastic behaviour of the larger, non-regulating arteries and arterioles. As well as adding this passive arterial compliance, we neglect flow through the capacitative element associated with regulating arterioles, noting from the results above that this flow is negligible. A schematic for this modified model is shown in \cref{figvasculatureSchematicModified}.

The new arterial passive compliance is governed similarly to the venous compliance:
\beq C_a = \frac{1}{k_{art}\left(P_{\alpha} - P_{ic} - P_{a1} \right)} \label{Ca} \eeq

The removal of $\diff{V_{sa}}{t}$ from the model also removes the algebraic loop and so it is no longer necessary to iteratively solve for the states of the arteriolar model at each time step. Values for the parameters introduced in \cref{Ca} were found such that the arterial to venous volume ratio was approximately 1:3 and such that this ratio too increases slightly with increasing arterial pressure over the mid pressure range (as opposed to the previous case where this ratio fell across the mid pressure range, see \cref{figautoregVolFrac}). The value for arterial stiffness was set at three times the stiffness of the venous compartment, reflecting the relative stiffness of arterial walls relative to those of veins. These values are given in \cref{tabautoregMod}: note that the values of $L$ and $n_{sa}$ have both been reduced by a factor of fifty, thus reducing the characteristic length to a more realistic value whilst maintaining overall resistance; again, the value for the number of arterioles is not representative of the value found in a realistic network architecture.

\begin{table}[!h] \centering
\footnotesize{
\begin{tabular}{l l r l c}
\hline
Parameter & Description & Value & Units & Source\\
\hline
$k_{art}$ & stiffness coefficient for arterial compliance & 0.558 & 1/mL & Fitting\\
$P_{a1}$ & pressure offset for venous compliance & -2.25 & mmHg & \cite{Payne06}\\
$V_{an}$ & offset for venous volume & 3 & mL & Fitting\\
\hline
$L$ & characteristic length of arterioles & $1\e{-3}$ & m & Fitting\\
$n_{sa}$ & effective number of arterioles in parallel & 320 & -- & Fitting\\
\hline
\end{tabular}}
\caption{Parameters of the modified whole-brain model.}
\label{tabautoregMod}
\end{table}

\begin{figure}[hbtp] \centering \includegraphics[scale = 0.6]{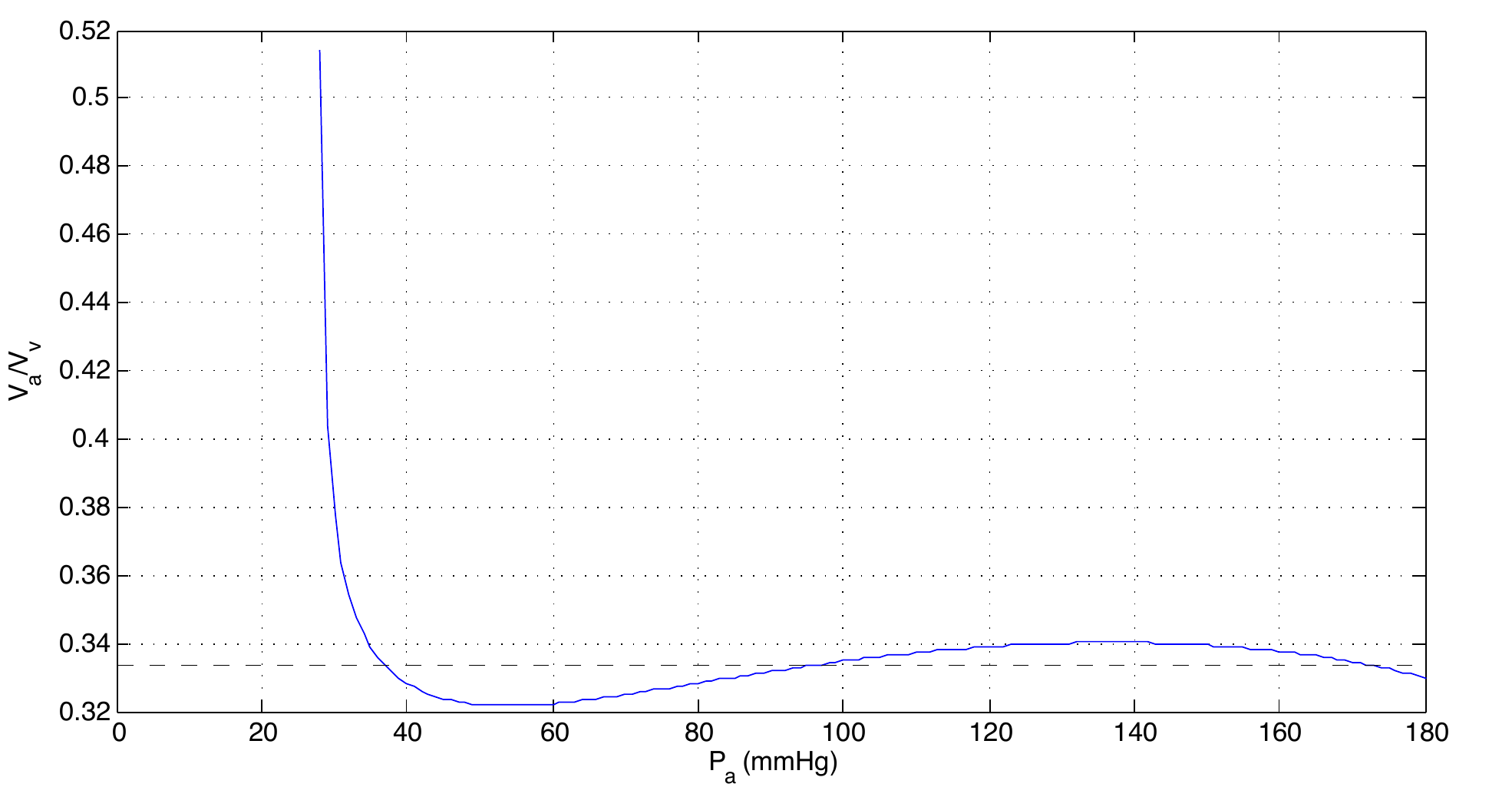} \caption{Arterial to venous volume fraction in the modified whole-brain model, variation with arterial pressure. Dashed line shows $V_a/V_v = 1/3$}\label{figautoregModVolFrac} \end{figure}

\begin{figure}[hbtp] \centering \includegraphics[scale = 0.6]{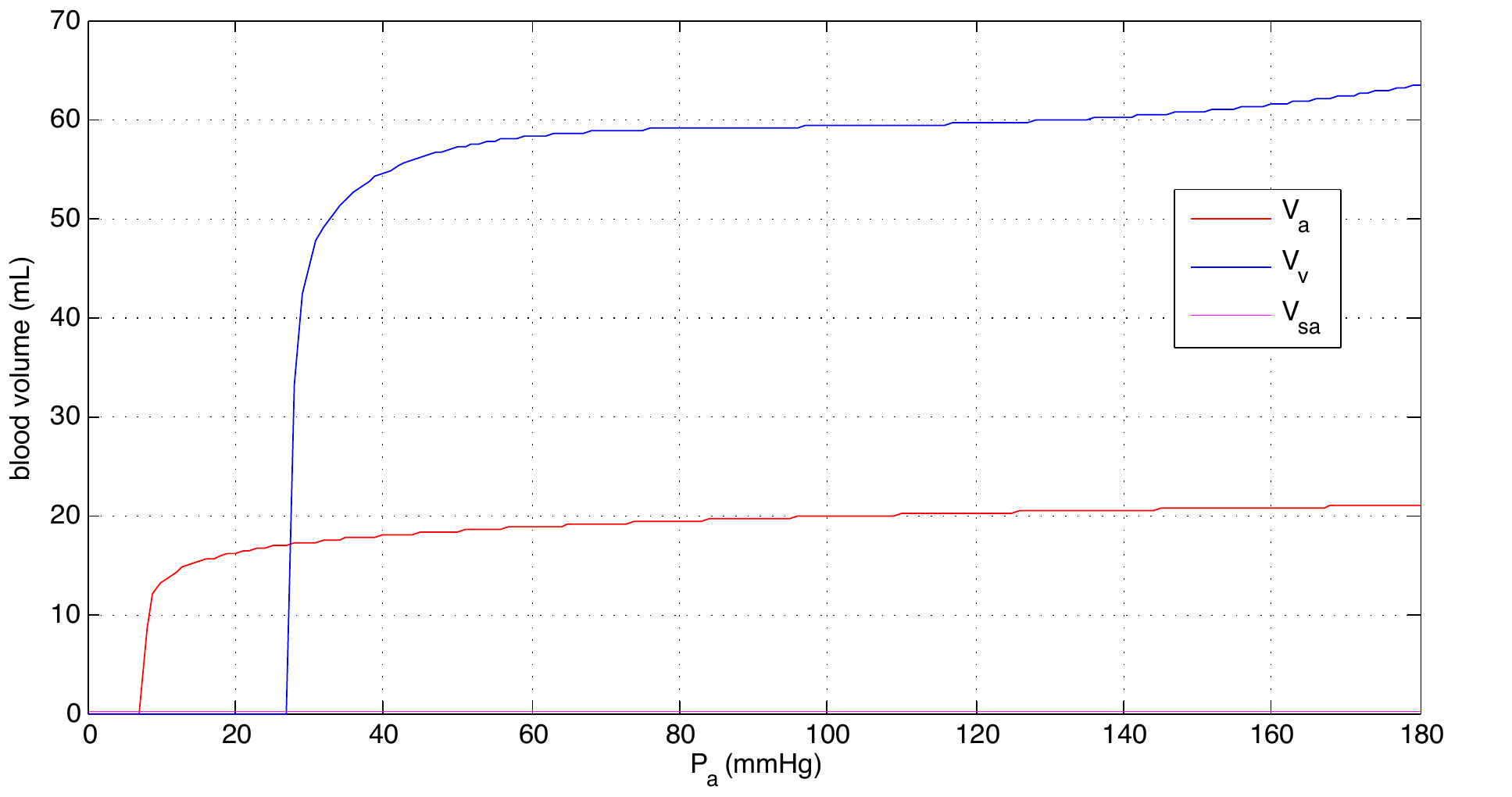} \caption{Variation of the volumes of the veins, arteries and small arteries (reactive arterioles) with arterial pressure.}\label{figautoregModVol} \end{figure}

The steady-state flow response of the modified model is identical to that of the original model; the volume fraction response however is significantly altered and now exhibits a slight upward trend through the central, autoregulatory, region; \cref{figautoregModVolFrac}. It should be noted that whilst the flow related to the change in arteriolar volume ($\diff{V_{sa}}{t}$ in the original model) has been neglected, the actual change in volume of the arterioles has been included in the calculation of arterial volume fraction, i.e., $\diff{V_{sa}}{t} \simeq 0$ but $\Delta V_{sa}$ is not. The discontinuities in the blue curve of \cref{figautoregModVolFrac} are due to the points at which the volume in the arteriolar and venous compartments reach zero, \cref{figautoregModVol} shows the three volumes individually varying with arterial pressure.

\begin{figure}[hbtp] \centering \includegraphics[scale = 0.6]{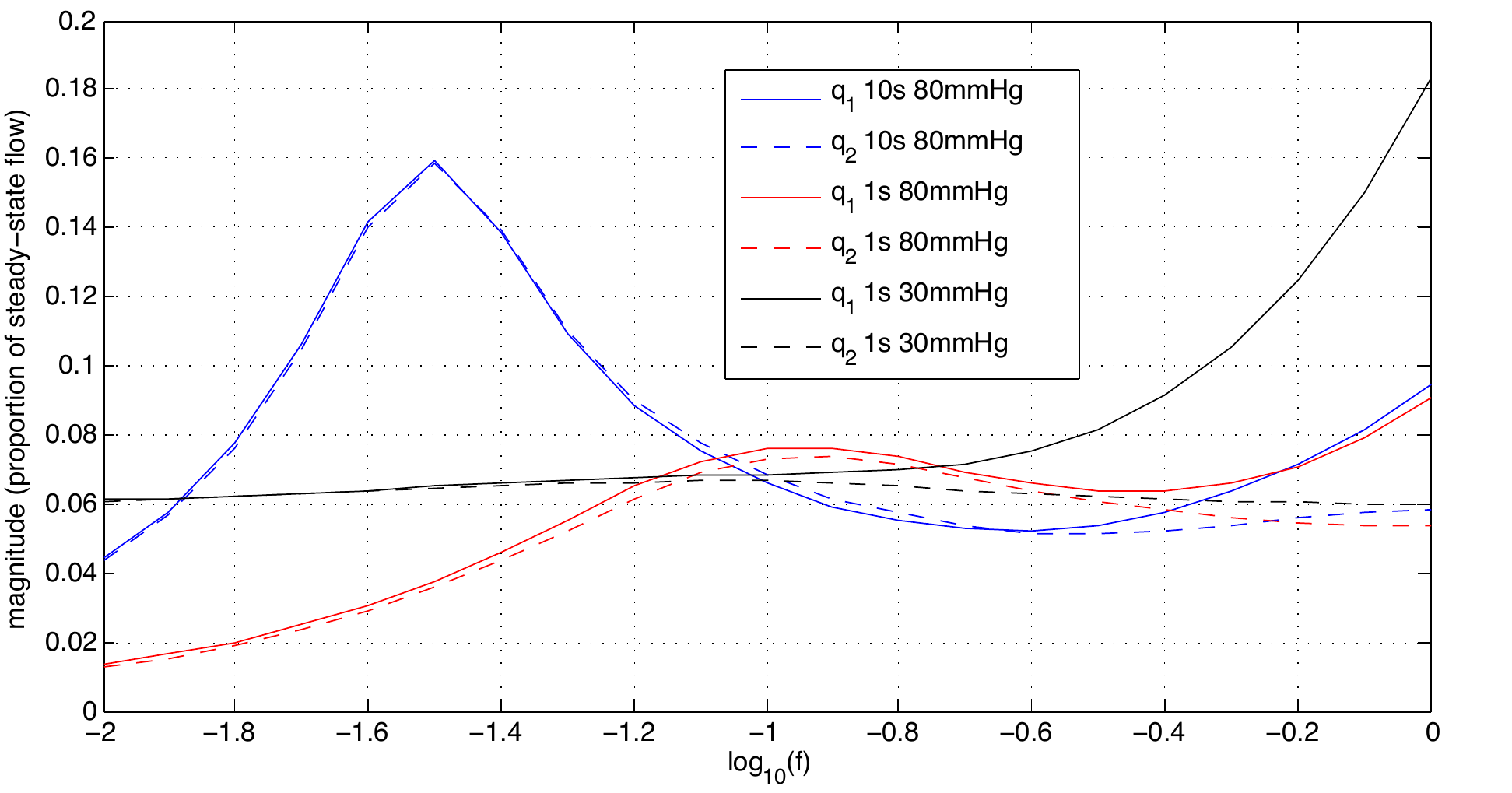} \caption{Magnitude of the response of the model flows to oscillations of various frequencies in arterial pressure; y-axis shows oscillation amplitude relative to steady state value, legend shows flow, $\tau_{Ca}$ value, and baseline pressure for 10\% oscillations in $P_a$.}\label{figautoregFSweepMag} \end{figure}

\begin{figure}[hbtp] \centering \includegraphics[scale = 0.6]{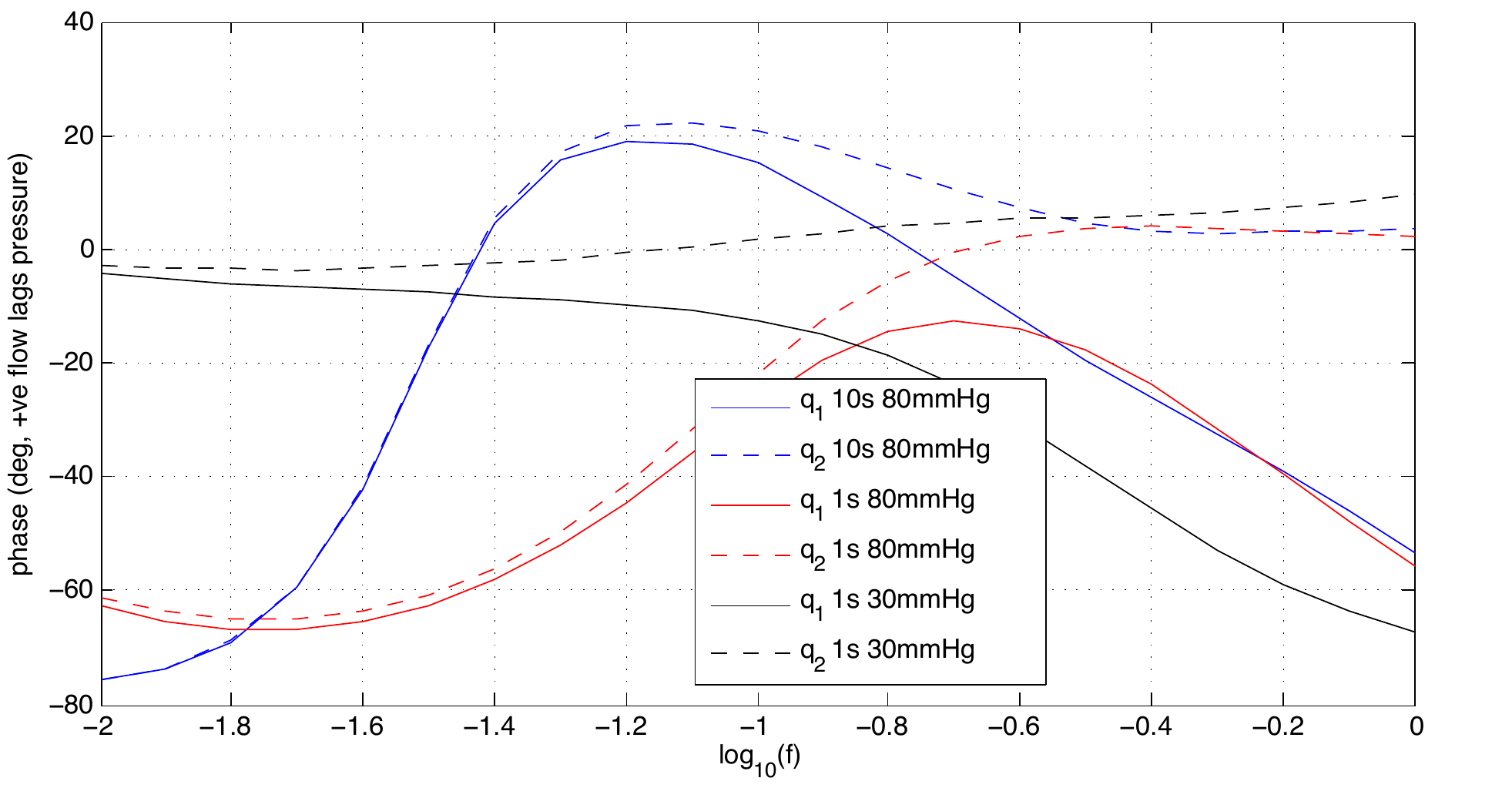} \caption{Phase of the response of the model flows to oscillations of various frequencies in arterial pressure; y-axis shows phase lag of flow relative to driving pressure, legend shows flow, $\tau_{Ca}$ value, and baseline pressure for 10\% oscillations in $P_a$.}\label{figautoregFSweepPhase} \end{figure}

The dynamic behaviour however is markedly different. In the previous model, the flow into the arterial capacitative element, $\diff{V_{sa}}{t}$, was so small that $q_1$ and $q_2$ were effectively identical. The passive compliance introduced in the modified model partially decouples $q_1$ and $q_2$. Now that arterial volume fraction changes realistically with pressure, and that $q_1$ and $q_2$ have been decoupled we can look at the responses of $q_1$ and $q_2$ to variations in the frequency of the driving oscillation in $P_a$. As previously, the forcing function used in this frequency response analysis is a 10\% step in pressure coupled with a 10\% amplitude sine wave. \Cref{figautoregFSweepMag,figautoregFSweepPhase} show the magnitude and phase respectively of the responses of the flows $q_1$ and $q_2$ to various frequencies of oscillation in $P_a$.

The red and blue lines correspond to the cases where $\tau_{Ca} = 1$s and 10s respectively, from a baseline of $P_a = 80$mmHg. There is a clear resonant peak in the red lines at 0.1Hz, interestingly this peak strengthens and moves to a lower frequency when the calcium response is slowed. When the oscillation is applied about a point which is not within the autoregulatory region of the system, $P_a = 30$mmHg, as in the case of the black lines ($\tau_{Ca} = 1$s), the resonance is no longer present. The increasing values of $q_1$ (solid lines) at higher frequencies culminates in another resonance with a magnitude of 1.2 around 60Hz -- however it is not thought that the behaviour of the model at such high frequencies is relevant to the real system, rather  a mathematical feature of our very simplified system architecture.

The phase of the flow relative to the pressure oscillations, \cref{figautoregFSweepPhase}, tells a similar story: when the calcium response is slowed by increasing $\tau_{Ca}$ from 1s to 10s, the maximum phase lag increases, moves to a lower frequency, and the transition from lead to lag is compressed. The response at low pressure is tending towards the response of the system in the absence of autoregulation -- a zero phase shift in $q_2$ and a transition from $0^{\circ}$ to $-90^{\circ}$ for $q_1$ as $f$ increases. This is to be expected when the baseline pressure is so close to the peak of the radius/pressure curve, \cref{figautoregSSr}, where radius is least sensitive to changes in pressure and so closest to presenting a fixed resistance.

\subsection{Conclusions}
The inclusion of a passive compliance in the arterial compartment has led to a more realistic ratio of volumes between the arterial and venous compartments. The elastic dilation of the arteries, in contrast to the active contraction of the arterioles, leads to the volume ratio increasing  slightly with $P_a$; this brings the model into closer agreement with \cite{Payne06} and hence the data of \cite{Lee01}. We have also found that this modified model exhibits a resonant peak and that this peak is found at a frequency around 0.1Hz when the calcium response is fast, being lower and more pronounced as the calcium response slows down. We also find that at lower frequencies, the phase lead of flow is increased relative to the original model and achieves a maximum lead of nearly $70^{\circ}$ at lower frequencies, being nearer $30^{\circ}$ at 0.1Hz. As found in the original model, the phase can invert at frequencies around 0.1Hz when the calcium system is slow. The low pressure case (black line) shows that the resonance disappears at low perfusion pressures, outside the active range of the arterioles. However, we also see that the magnitude of the response at 1Hz is nearly twice as great at low pressure as it is at higher pressure.

\section{Conclusions}
In this chapter we have integrated our well-fitted single arteriole model into a model of the cerebral vasculature. We have shown that a realistic pressure/flow relationship and pressure/volume-ratio can be achieved. Our results suggest that signalling through tissue to upstream vessels is very unlikely to be possible using NO as a medium, due to its rapid decay in the blood stream. By examining the response of the flows in our model to sinusoidal variations in arterial pressure, we have found that the system has a resonant peak, the frequency of which is sensitive to the speed of the myogenic \ca response. We have also found that the phase lag between pressure and flow is sensitive to both frequency and \ca response speed, and that this may have diagnostic applications.

\chapter{Vasomotion and Spontaneous Oscillations}
\label{chapOscill}

\section{Introduction}
Now that we have produced a model of the cerebral vasculature including a well-fitted (although simplified) model of the arteriole, we are in a position to investigate the potential of the model to produce spontaneous oscillations in vascular tone, representative of vasomotion. We aim to gain an insight into the mechanism of vasomotion by examining the requirements for its onset in our model.

\section{Stability Analysis}
Having identified an equilibrium radius for each value of arterial pressure for our modified whole-brain model, we can use a technique known as \emph{stability analysis} to characterise these equilibria -- specifically to identify them as either stable nodes or possibly unstable equilibria or oscillatory attractors. We perform this characterisation by evaluating the Jacobian of the model at the equilibria and calculating the eigenvalues of this matrix. A Jacobian is a matrix of all first order partial derivatives of a vector-valued function; in our case the vector-valued function is the modified whole-brain model with a fixed arterial pressure $P_a$:
\beq \dot{\mathbf{x}} = f \left(\mathbf{x}\right) \eeq

where $\mathbf{x}$ is the vector of state values for the modified model
\beq \mathbf{x} = \left[ \begin{array}{c}
x_1 \\ x_2 \\ x_3 \\ x_4 \\ x_5
\end{array} \right]  = \left[ \begin{array}{c}
\ccam \\ C_w \\ r \\ V_v \\ V_a
\end{array} \right]  \label{autoregStates} \eeq

The Jacobian matrix required for bifurcation analysis of our system is then:
\beq \mathbf{J} = \diff{\dot{\mathbf{x}}}{\mathbf{x}} = \left[ \begin{array}{cccc}
\pdiff{\dot{x}_1}{x_1}  & \pdiff{\dot{x}_1}{x_2} & \ldots & \pdiff{\dot{x}_1}{x_5} \\
\pdiff{\dot{x}_2}{x_1}  & \pdiff{\dot{x}_2}{x_2} & \ldots & \pdiff{\dot{x}_2}{x_5} \\
\vdots & \vdots & \ddots &\vdots \\
\pdiff{\dot{x}_5}{x_1}  & \pdiff{\dot{x}_5}{x_2} & \ldots & \pdiff{\dot{x}_5}{x_5} \\
\end{array} \right] \eeq

The Jacobian tells us how the derivatives of the system will respond to a perturbation in its state. If a small perturbation in the state will cause a bigger change in the derivatives, or a change in the derivatives which pushes the system further in the same direction as the purturbation, then the starting point is unstable. Any small perturbation from such a point will quickly grow into a large perturbation and the system may transition to another equilibrium, oscillate around an equilibrium, or continue in one direction until it hits a limit of some kind. The eigenvalues of a matrix tell us by how much, and in what direction, that matrix amplifies perturbations in various directions (along the eigenvectors). The Jacobian can be compiled simply using finite differences: having evaluated the model at the equilibrium point, we perturb each element, $x_i$, of $\mathbf{x}$ (each state) in turn, recording the change in $\dot{\mathbf{x}}$. We then divide each change in $\dot{\mathbf{x}}$ by the magnitude of the perturbation in $x_i$ to yield $\pdiff{\dot{\mathbf{x}}}{x_i}$, each of which is a column of the Jacobian. Having thus evaluated the Jacobian at an equilibrium point we proceed to calculate the eigenvalues of this matrix, analysis of which reveals the nature of the equilibrium.

In our case we find that all eigenvalues have negative real parts at all pressures, meaning that all equilibria are stable. If we consider the iterative scheme which we used to solve for the equilibria in the first place we realise that all equilibria found by such a scheme must be stable, otherwise the iterative scheme would not have been able to converge upon them. It is interesting to note that the eigenvalues for the highest and lowest pressures, below 40mmHg and above 150mmHg, are all real whereas the equilibria of the middle pressures all have two complex eigenvalues. Real negative eigenvalues indicate an equilibrium to which the system will monotonically decay after a perturbatioin, whereas complex eigenvalues with a negative real part indicate that the system will oscillate about an equilibrium with an exponentially decaying amplitude. Thus we have some indication that although the system is everywhere stable, it is closer to displaying oscillatory behaviour in the regulatory pressure range than outside it. At the point where venous volume reaches its minimum limit (see \cref{figautoregModVol}) the discontinuity in the sensitivity of $\diff{V_v}{t}$ artificially magnifies $\pdiff{\dot{x}_4}{x_4}$, and thus the fourth eigenvector, $\lambda_4$, enormously whilst reducing the other elements on the same row to zero. This is because when one or more of the states (i.e., $V_v$) reaches a hard limit, its derivative is no longer sensitive to perturbations in other states; however the discontinuity between the baseline evaluation of $\dot{x}_4$, which returns zero, and the finite difference evaluation, which returns a negative value, of that state derivative gives the misleading impression that it is extremely sensitive to perturbations in its corresponding state. This is an artefact of the numerical evaluation; it can be shown that if the finite difference step used was in fact negative then the sensitivity would be zero, and the Jacobian matrix would be rank-deficient.

\subsection{Discussion}
The absence of any oscillatory behaviour in our model suggests that NO is not an important factor in vasomotion. When we review the elements of the arteriolar model we find that the absence of oscillatory behaviour is perhaps to be expected: the NO--cGMP system is saturated at all but the very lowest levels of flow, and so plays no part in the dynamics of the system in its normal pressure range; the \ca dynamics of the myogenic response are first order, and as such are over-damped and cannot drive oscillation; and the mechanical model includes viscous friction terms which damp any oscillation that might appear.

The saturation of the NO--cGMP system is of particular interest as it implies that although NO is an important factor in determining the radius of the arteriole, it plays virtually no role whatsoever in the dynamics of the vessel. From \cref{figKuo91Fig2AStates} we can see that the NO--cGMP sigmoid saturates quickly even with a pressure drop along the vessel as low as 2.4mmHg. The arterial pressure in the whole brain model which creates this pressure drop through the arterial component (in steady state) is only 9mmHg, which is itself lower than the 10mmHg at which intracranial pressure is set. It is clear then that unless a vessel is seriously occluded, or bloodflow is otherwise severely disrupted, there is always sufficient flow through the arterioles to saturate the NO--cGMP pathway. Given that sGC (the enzyme which produces cGMP) is the \emph{only} proven receptor for NO \cite{Denninger99}, once this pathway is saturated further increases in \cno have no effect on the vessel whatsoever. Furthermore, as long as the level of NO does not drop below saturation level (which as we have seen would require a very large fall in arterial pressure), the vessel is completely insensitive to oscillations in \cno\!\!. Whilst \cno may vary in response to radius/flow dynamics, there is no feedback element. This has important implications for interpreting the results of \cite{Ide07} -- if NO plays no role in the dynamics of the system except for a steady state dilatory component, then the blockade of eNOS by L-NMMA would be expected to produce exactly the same effect as an $\alpha$-adrenergic agonist such as phenylephrine in terms of the correlation between ABP and CBF.

\begin{figure}[hbtp] \centering \includegraphics[scale = 0.8]{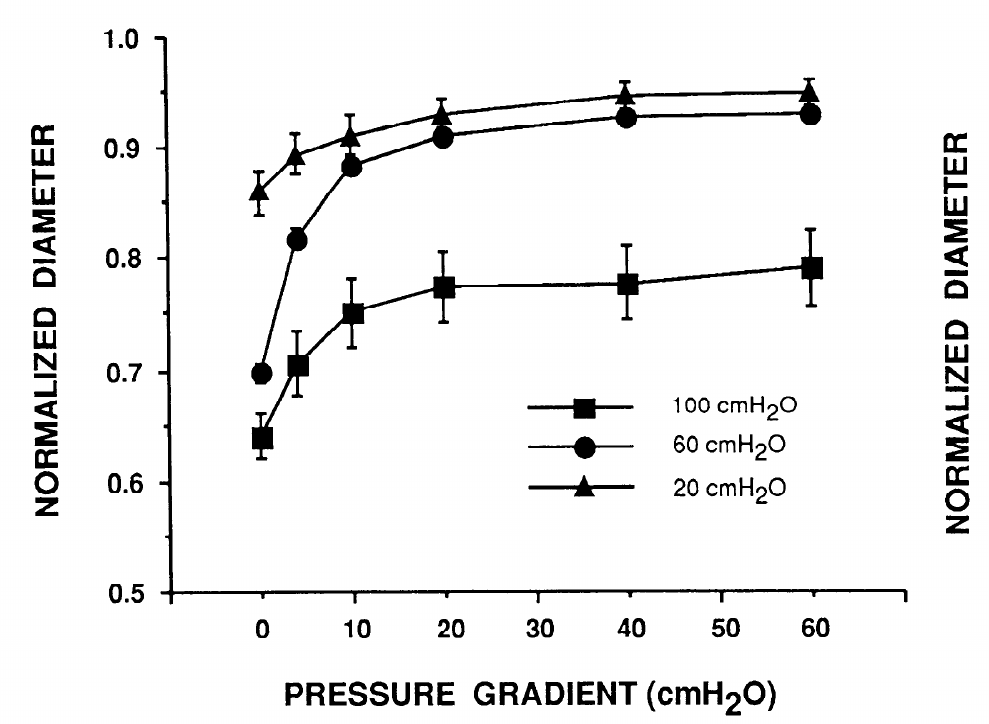} \caption{Response of vessel diameter to pressure drop along vessel. Reproduced from fig.~3 in \cite{Kuo91}.}\label{figKuo91Fig3} \end{figure}

The saturation of the NO pathway is derived from fitting the arteriolar model to the dynamic results of \cite{Kuo91} as presented in \cref{dynfitting}, specifically the delay in vessel response to the cessation of flow. However, there are conflicting results presented in the same paper which suggest that, although the NO pathway does saturate, this does not happen at such low flow levels. \Cref{figKuo91Fig3} (reproduced from fig.~3 in \cite{Kuo91}) presents data which suggest that the vessel continues to show sensitivity to NO (or at least to flow) until the pressure drop along the vessel is approximately 15mmHg, although as mentioned in \cref{autoregDisc}, these results are themselves inconsistent with the data presented in fig.~4 of the same paper and with the dynamic results of \cref{figKuo91Fig2A}.

It could be postulated that the saturation of the flow response manifested in the dynamic experiments was due to the use of saline solution as opposed to whole blood, whereas if whole blood were used the response of \cno within the vessel wall would have been much reduced due to a far greater loss of NO into the blood, thus extending the range in which the vessel was sensitive to flow induced NO. However, the results presented in \cref{figKuo91Fig3} were also obtained using saline solution and not whole blood, so the inconsistency remains unresolved. A better candidate explanation could lie in the time domain: the results presented in \cref{figKuo91Fig3} are steady state results, and the saturation seen in \cref{figKuo91Fig2A} is observed only for about a minute, so it could be the case that there is an initial fast response to NO which reaches saturation quickly, and then a much slower process continues to cause vasodilation until the final steady state radius is reached. Although this explanation allows the possibility of consistency between the dynamic results and the steady state results -- it does not say in \cite{Kuo91} how long it took to reach a steady radius -- it seems like an unlikely scenario. As far as we are aware, there is no experimental evidence which suggests a dual-timescale NO response. Not only that but if the second, slower, phase of the NO response takes $>$ 1min to manifest itself then it makes little difference to our conclusions; the oscillations we are studying are much too fast to be affected by such a slow response.

\section{Replicating Vasomotion}
Having concluded that the current model is incapable of spontaneous oscillations which could represent vasomotion, we will now modify the model with the aim of investigating the possibility of  such oscillations occurring.

\subsection{Adding Intracranial Compliance}
\label{addICCap}
\begin{figure}[hbtp] \centering \includegraphics[scale = 0.8]{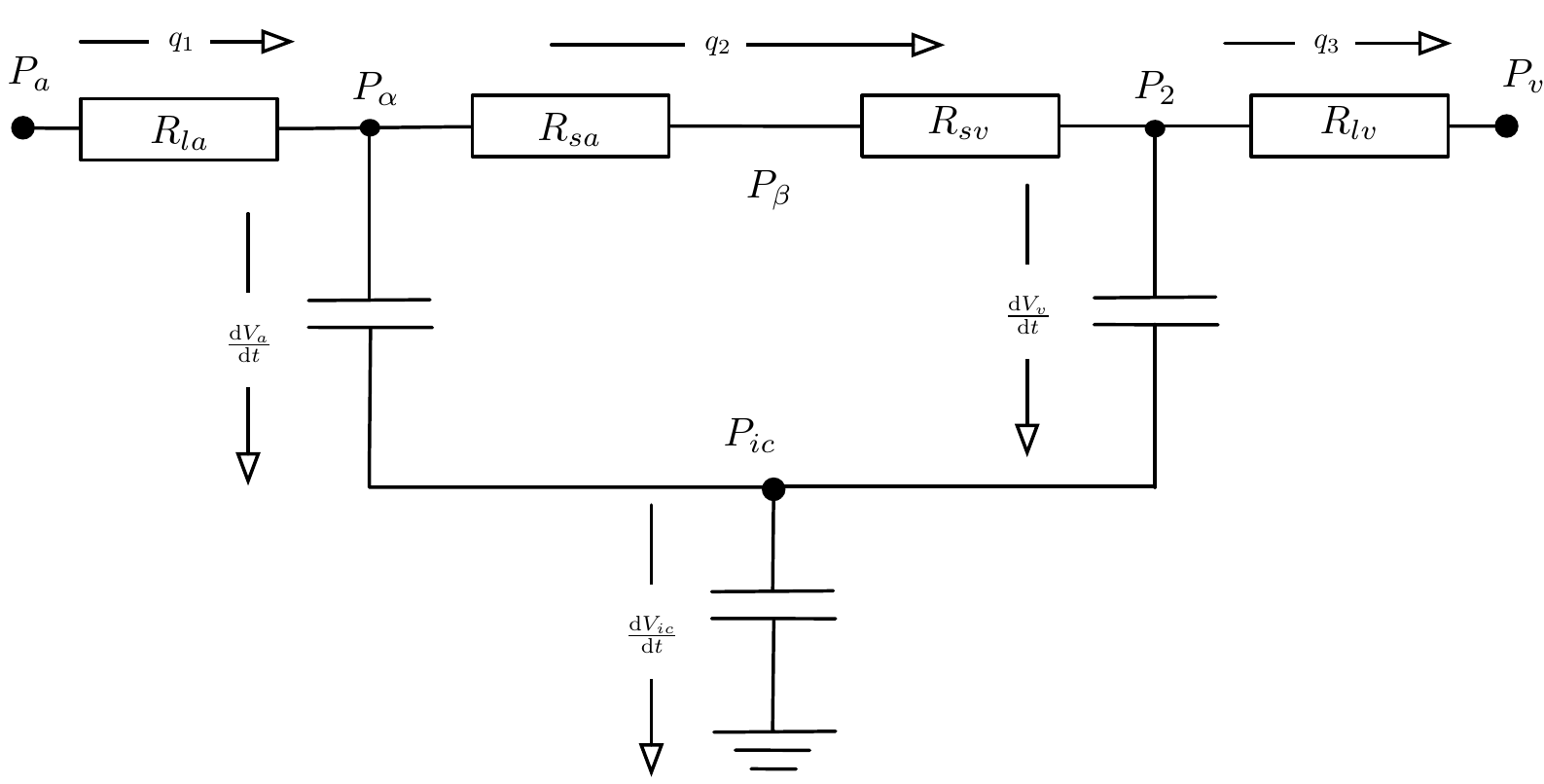} \caption{The vasculature again modified to allow intracranial pressure to vary.}\label{figvasculatureSchematicWithICCap} \end{figure}

Thus far intracranial pressure, $P_{ic}$, has been fixed. If we model cranial compliance and thus allow $P_{ic}$ to vary, we may find a resonance between the blood and intracranial fluid volumes. To this end we introduce a third capacitative element to the model, coupling the intracranial fluid to ground (outside the head), as found in \cite{Peng08} as a modification to the original basis for our model from \cite{Payne06}. \Cref{figvasculatureSchematicWithICCap} shows the schematic of the electrical equivalent circuit.

This extra capacitative element represents the compliance of the cranium and other meninges of the brain to accommodate changes in intracranial fluid pressure. In electrical circuits capacitors are formed of two charge reservoirs (metal plates) separated by an impermeable membrane (the dielectric), in hydraulic circuits capacitance is manifested by two reservoirs of fluid separated by an impermeable membrane. The stiffness of the membrane determines the equivalent capacitance of the coupling, a stiff membrane deflects little and so has a low capacitance, the more flexible the membrane, the more fluid can be accommodated by its deflection. The compliance of the extra element is taken from \cite{Peng08} as:
\beq C_{ic} = \frac{1}{k_{ic}P_{ic}} \eeq

where $k_{ic} = 0.08$/mL \cite{Peng08}. This is similar to the expressions for $C_a$ and $C_v$ but without the pressure offsets. As with previous compliant elements, the expression for volume is:
\beq V_{ic} = \frac{1}{k_{ic}}\ln\left(P_{ic}\right) \label{Vic} \eeq

In this case however the volume does not represent the total volume of intracranial fluid, rather the change in volume of the cerebral cavity caused by the cranium and other meninges deflecting in response to elevated intracranial pressure. If we assume that the volume of cerebrospinal fluid remains constant then any increase in cerebral blood volume must be matched by an equal increase in volume of the cerebral cavity, thus we can write:
\beq V_a + V_v - V_{ic} = V_{b0} \label{Vb0} \eeq

where $V_{b0}$ is a constant representing the cerebral blood volume at which cranial deflection would be zero. This expression completes a set of four equations which can be solved in the steady state condition to yield values for the three volumes and $P_{ic}$ given $P_{\alpha}$, $P_2$, and $V_{b0}$; these equations are:
\beq V_a = \frac{1}{k_{art}}\ln\left(P_\alpha - P_{ic} - P_{a1}\right) + V_{an} \eeq
\beq V_v = \frac{1}{k_{ven}}\ln\left(P_2 - P_{ic} - P_{v1}\right) + V_{vn} \eeq
\beq V_{ic} = \frac{1}{k_{ic}}\ln\left(P_{ic}\right) \eeq
\beq V_a + V_v - V_{ic} = V_{b0} \eeq

$V_{b0}$ can be determined by fixing the intracranial pressure at any particular arterial pressure, in this case we fixed $P_{ic} = 10$mmHg at $P_a = 80$mmHg. Having thus determined the steady state values of all of the volumes at any given arterial pressure, we can then integrate the system as before to reproduce its dynamic behaviour. The extra state, $V_{ic}$, is propagated by calculating $P_{ic}$ from $V_{ic}$ via \cref{Vic}, using this to calculate flows $\diff{V_a}{t}$ and $\diff{V_v}{t}$, and then rearranging and differentiating \cref{Vb0} to give:
\beq \diff{V_{ic}}{t} = \diff{V_a}{t} + \diff{V_v}{t} \eeq

\begin{figure}[hbtp] \centering \includegraphics[scale = 0.6]{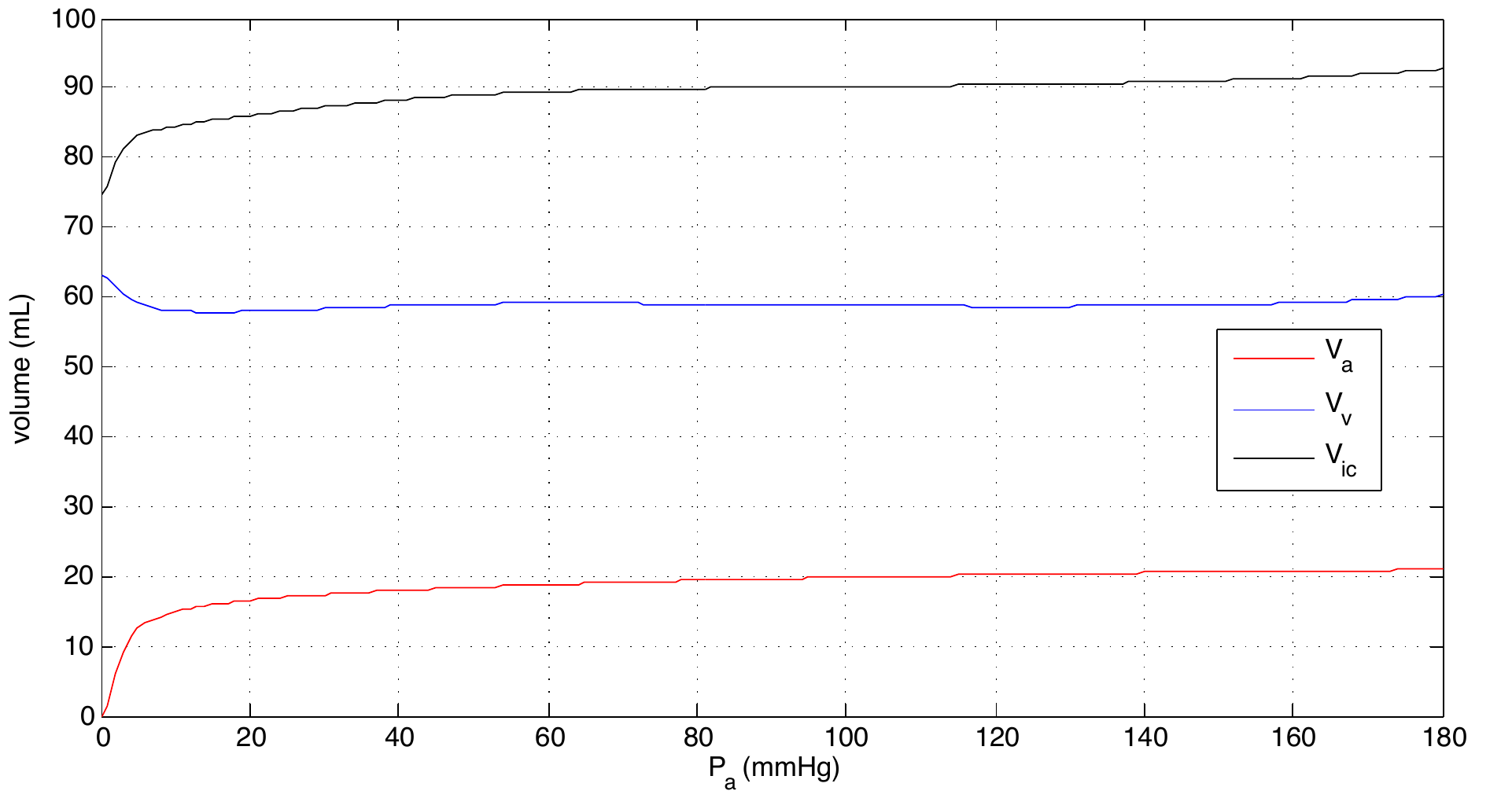} \caption{Arterial, venous, and intracranial volumes with varying arterial pressure.}\label{figautoregICCapVol} \end{figure}

The variation of steady state volumes with arterial pressure is shown in \cref{figautoregICCapVol}. Note that because $P_{ic}$ is no longer fixed, and is allowed to vary alongside blood pressure, the blood compartments no longer collapse at some non-zero arterial pressure, as they are seen to do in the previous iteration of the model (\cref{figautoregModVol}).

Despite this additional dynamic element, the model still exhibits no spontaneous oscillation at any arterial pressure. In fact its dynamic behaviour is very similar to that of the previous iteration of the model. We therefore consider additional reactive components, modelling more and more elements of the biological system, until such point as we develop spontaneous oscillations under some conditions.

\subsection{Adding Flow Inertia}
\label{addInertia}
Thus far the mass, and hence inertia, of the flow has been neglected, allowing instantaneous changes in flow velocity. In reality of course the blood flow does have mass and so takes a finite time to accelerate and decelerate in response to changes in driving pressure. We now add into the model the inertia of the blood in the small arteries (arterioles), starting from Newton's first law:
\beq f = ma \eeq

we can write expressions for each of these quantities in terms of current parameters of our model. The net force acting on the blood in the arterioles is equal to the pressure difference between the upstream and downstream ends multiplied by the area upon which it acts:
\beq f = \Delta P \pi r^2 n_{sa}\eeq

The mass of the blood in the arterioles is given by:
\beq m = \rho_b L \pi r^2 n_{sa} \eeq

where $\rho_b$ is the density of blood.  The acceleration of that mass can be expressed as:
\beq a = \diff{v_{blood}}{t} = \frac{1}{\pi r^2 n_{sa}}\diff{q_2}{t} \eeq

In combination then we have that:
\beq \Delta P \pi r^2 n_{sa} = \rho_b L\pi r^2 n_{sa} \frac{1}{\pi r^2 n_{sa}}\diff{q_2}{t} \eeq

or, rearranging:
\beq \Delta P =  \frac{\rho_b L}{\pi r^2 n_{sa}}\diff{q_2}{t} \label{inertia} \eeq

\begin{figure}[hbtp] \centering \includegraphics[scale = 0.8]{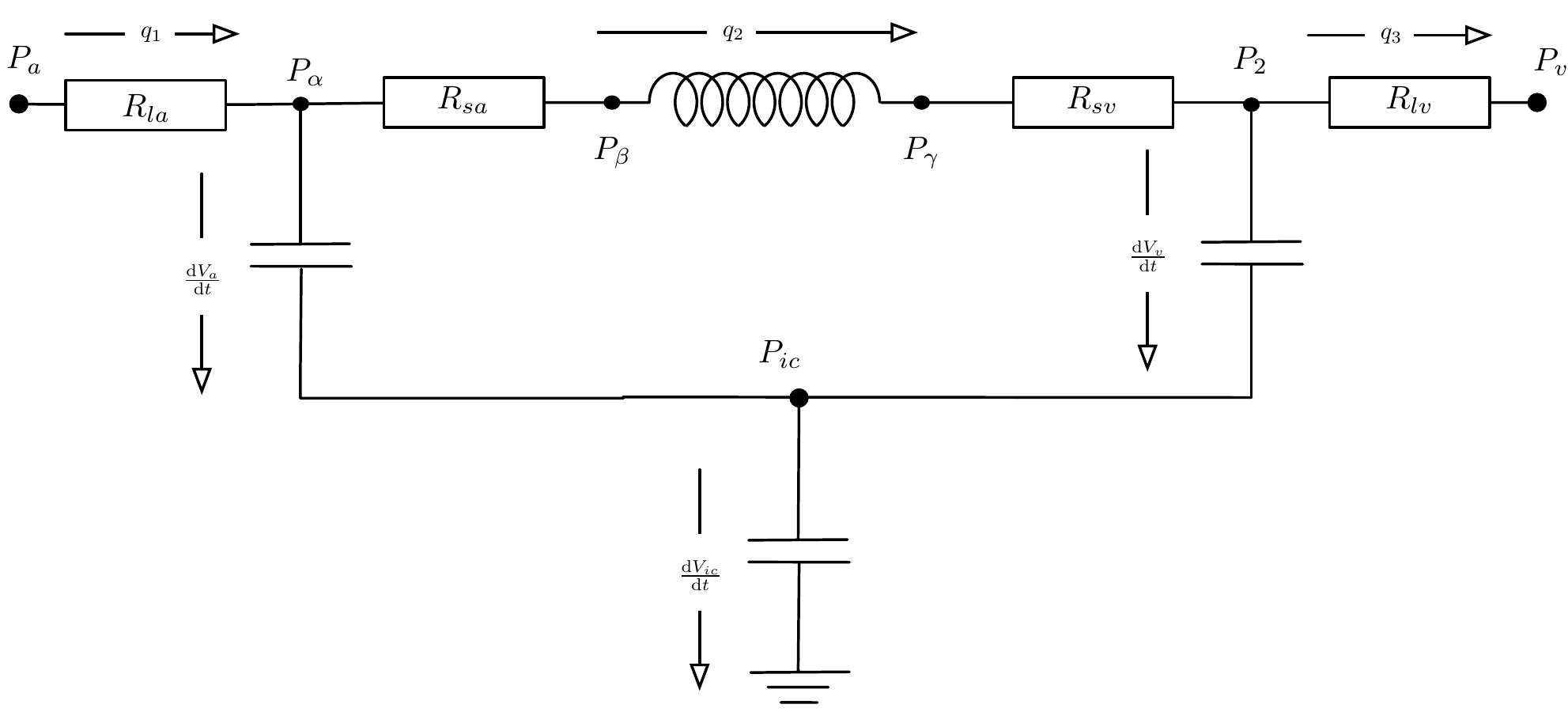} \caption{The vasculature further modified to include the inertia of the bloodflow.}\label{figvasculatureSchematicWithInertia} \end{figure}

This is the hydraulic equivalent of an inductor. \Cref{figvasculatureSchematicWithInertia} shows the revised electrical equivalence circuit including this inductor. The pressure difference $\Delta P$ is equal to $P_\beta - P_\gamma$, and the parameters $r$, $L$, and $n_{sa}$ are taken from the arteriolar model of \cref{chapArtFit}. The density of blood, $\rho_b$, is taken as 1.06kg/L \cite{Cutnell98}.

The addition of this inductance introduces another state, $q_2$ into the model, the derivative of which is given by \cref{inertia}. Steady state behaviour does not change and other values are calculated much as before.

Although the addition of this inertia changes the phase behaviour of the system (by partially decoupling pressure and flow) the stability of the system is unaltered and no spontaneous oscillations are observed at any values of $P_a$ and $\tau_{Ca}$. Considering the literature on vasomotion, there is a clear consensus that vasomotive oscillations are associated with \ca oscillations in the VSMCs, therefore it is to the \ca dynamics of the system that we now turn.

\subsection{Making \ca Dynamics 2nd Order}
The dynamics of the \ca concentration in the VSMC have so far been determined by \cref{dCadt}, a first order asymptotic approach to the steady state equilibrium value. In an analysis of oscillations in vascular networks \cite{Ursino96} Ursino et al. make reference to both a static and a rate-dependent element of the myogenic response, as modelled in a previous paper \cite{Ursino92}. Whilst we do not replicate their model here we note the inclusion of a rate-dependent term in the myogenic response in contrast to our existing model which has no such term. To investigate the possible effect of a rate-dependent term we modify our \ca system to a simple second order response:
\beq \diff{^2\ccam}{t^2} + 2\zeta\omega_n\diff{\ccam}{t} + \omega_n^2\ccam = G_{dc}\omega_n^2\overline{\ccam} \label{dCadt2} \eeq

In our case the DC-gain, $G_{dc}$, is unity since at equilibrium $\ccam = \overline{\ccam}$. We can then write the new state equation:
\beq \diff{^2\ccam}{t^2} = \omega_n^2\left(\overline{\ccam} - \ccam \right) - 2\zeta\omega_n\diff{\ccam}{t} \eeq

The addition of the second derivative adds an extra state to the arteriole model in the form of $\diff{\ccam}{t}$. This appears both as a state and a state derivative. The arteriole model now has the form:
\beq \mathbf{\dot{x}} = \left[ \begin{array}{c}
\dot{x}_1 \\ \ddot{x}_1\\ \dot{x}_2 \\ \dot{x}_3
\end{array} \right]  =  \diff{}{t}\left[ \begin{array}{c}
\ccam \\ \diff{\ccam}{t} \\ C_w \\ \lambda
\end{array} \right]  = f\left(\mathbf{x}\right) = f\left(\left[ \begin{array}{c}
x_1 \\ \dot{x}_1\\ x_2 \\ x_3
\end{array} \right] \right),  \label{2ndOartStates} \eeq

and the whole brain model, which has gained several states since its inception, now has the state vector:
\beq \mathbf{x} = \left[ \begin{array}{c}
\ccam \\ C_w \\ r \\ V_v \\ V_a \\ V_{ic} \\ q_2 \\ \diff{\ccam}{t}
\end{array} \right]  \eeq

\begin{figure}[hbtp] \centering \includegraphics[scale = 0.8]{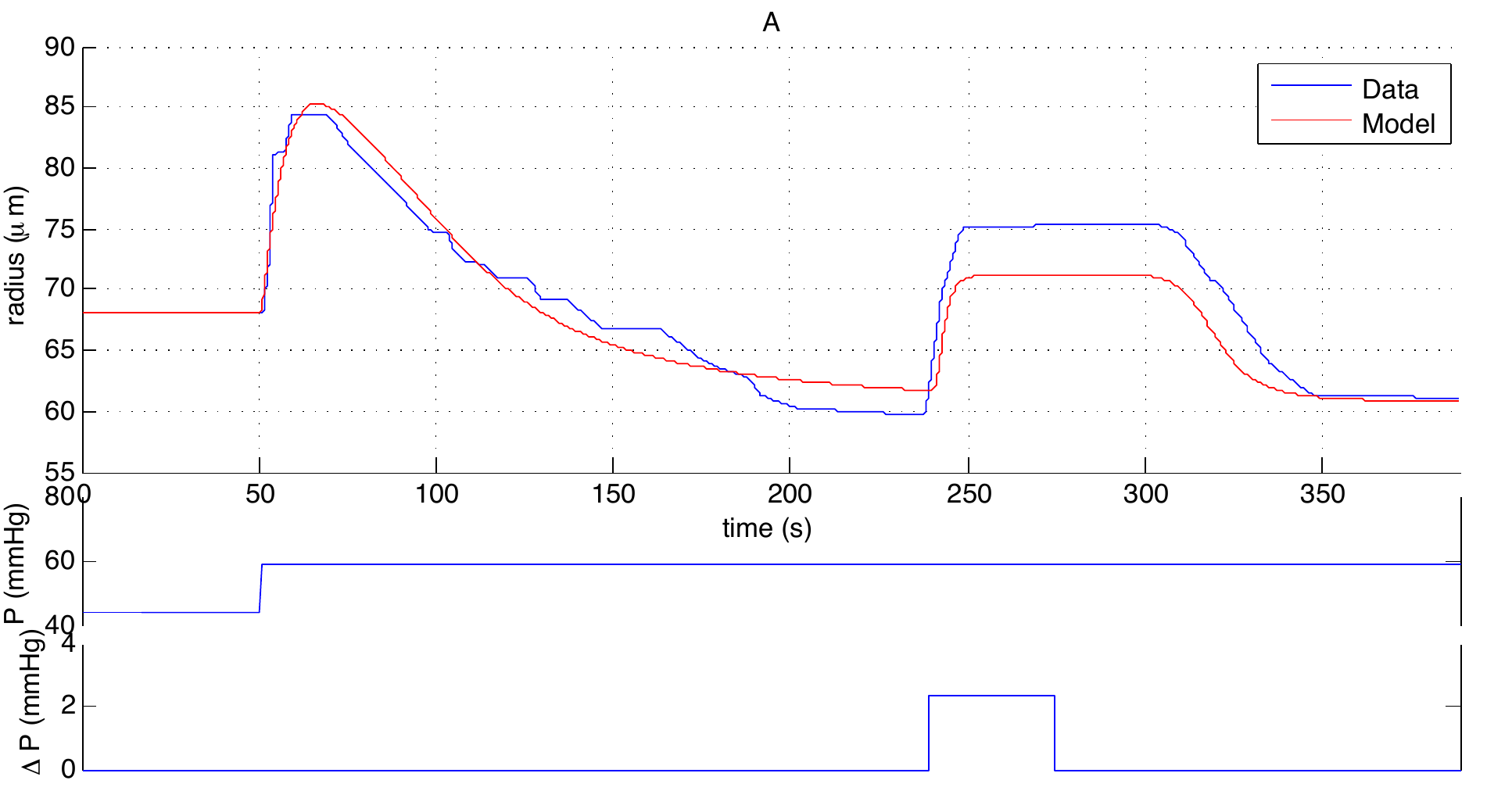} \caption{Fit of model with 2nd order \ca dynamics to the data of \cite{Kuo91} fig. 2A.}\label{fig2ndOCaKuo91Fig2A} \end{figure}

\begin{figure}[hbtp] \centering \includegraphics[scale = 0.8]{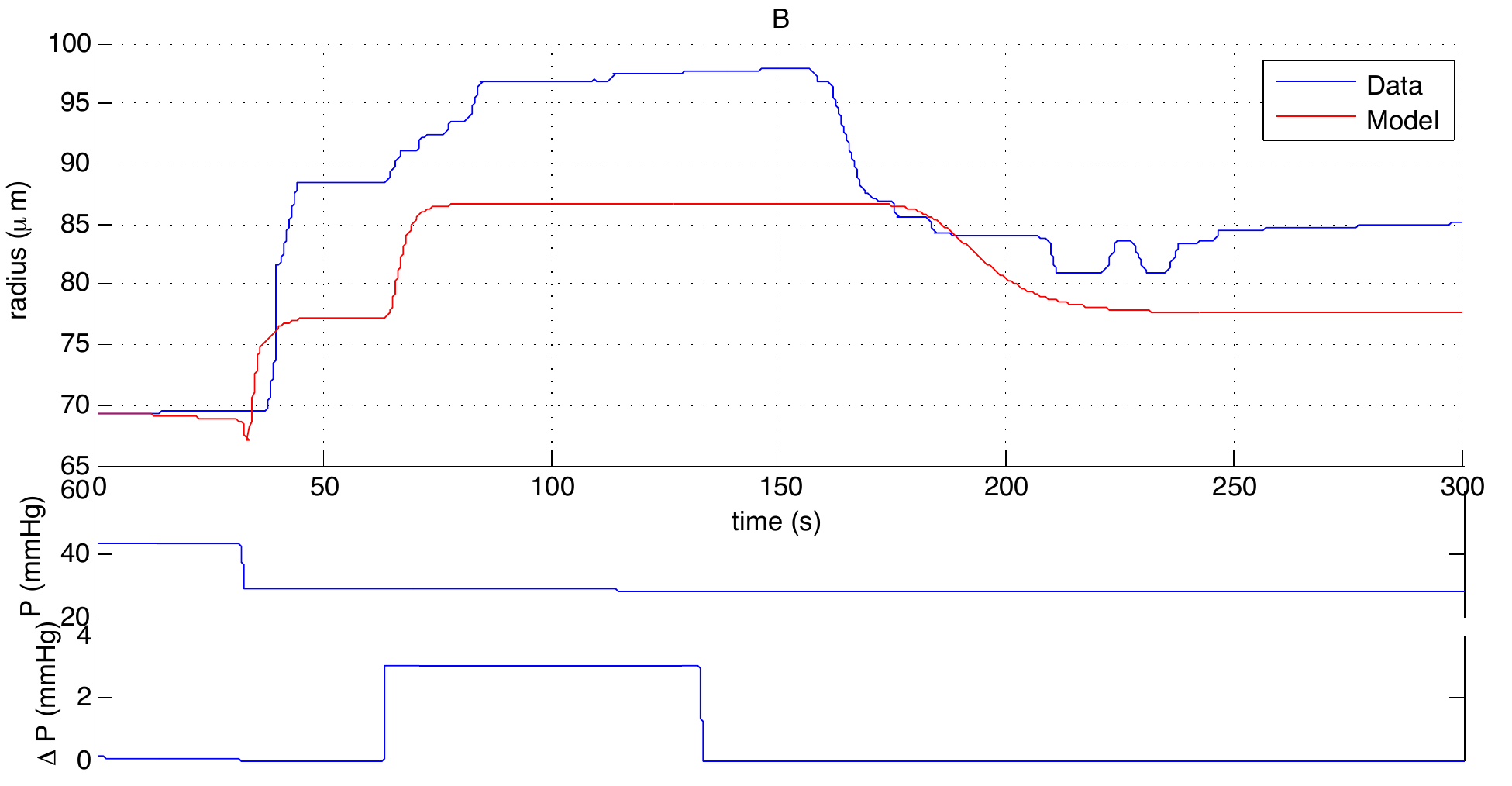} \caption{Fit of model with 2nd order \ca dynamics to the data of \cite{Kuo91} fig. 2B.}\label{fig2ndOCaKuo91Fig2B} \end{figure}

\begin{figure}[hbtp] \centering \includegraphics[scale = 0.8]{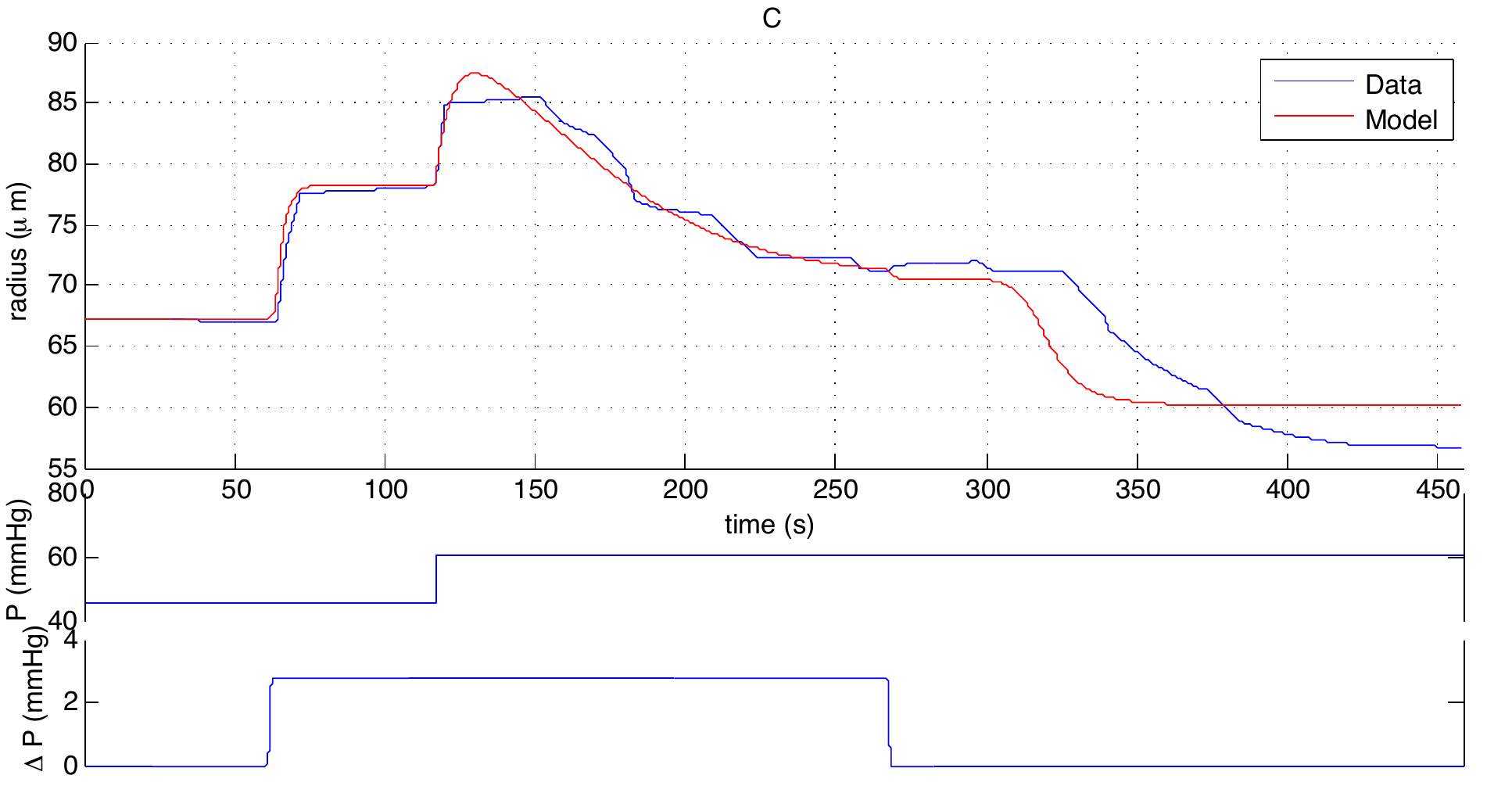} \caption{Fit of model with 2nd order \ca dynamics to the data of \cite{Kuo91} fig. 2C.}\label{fig2ndOCaKuo91Fig2C} \end{figure}

\begin{figure}[hbtp] \centering \includegraphics[scale = 0.8]{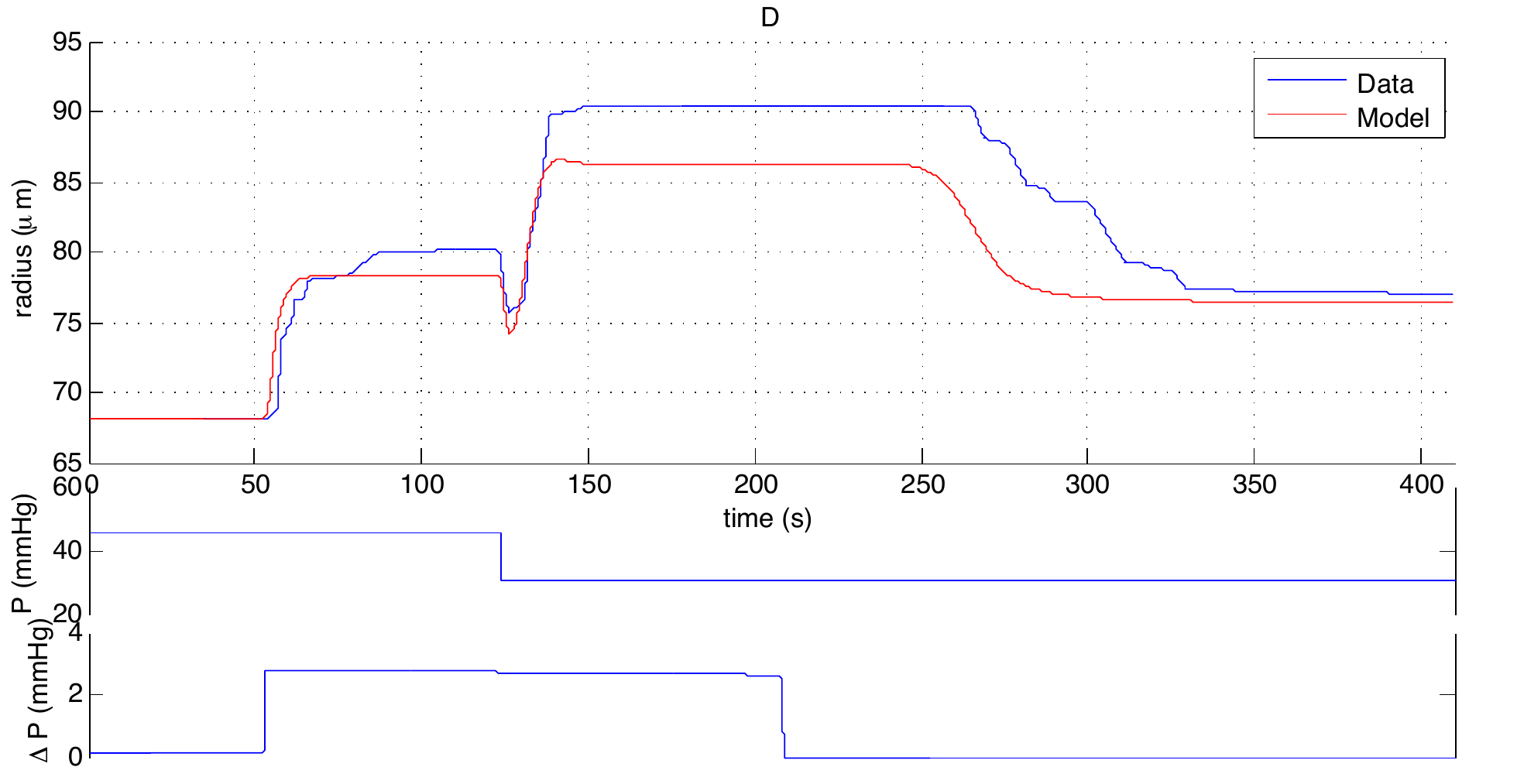} \caption{Fit of model with 2nd order \ca dynamics to the data of \cite{Kuo91} fig. 2D.}\label{fig2ndOCaKuo91Fig2D} \end{figure}

Unlike the addition of the vasculature elements of \cref{artComp,addICCap,addInertia}, this change in the \ca dynamics affects the behaviour of the isolated arteriole model, albeit only the dynamic behaviour and not the steady state results. In order then to preserve the validity of this model we must re-tune the arteriole model with the new \ca dynamics to the dynamic results of \cite{Kuo91}. Whereas previously we tuned the value of $\tau_{Ca}$ for each epoch of each experiment, \cref{tabTauCaVals}, we have now replaced the explicit time constant $\tau_{Ca}$ with the undamped natural frequency $\omega_n$ and the damping factor $\zeta$. It is these then that we must tune in order to regain the fit previously obtained to the dynamic arteriole experiments of \cite{Kuo91}. \Cref{fig2ndOCaKuo91Fig2A,fig2ndOCaKuo91Fig2B,fig2ndOCaKuo91Fig2C,fig2ndOCaKuo91Fig2D} show the fit that was obtained by this tuning process (which was carried out manually as before), the corresponding values of $\omega_n$ and $\zeta$ for each epoch are given in \cref{tabOmegaZeta}.


\begin{table}[!hbtp] \centering
\footnotesize{
\begin{tabular}{r r r r r r}
\hline
Experiment & Start time (s) & End time (s) & $\omega_n$ (rad/s) & $\zeta$ ()\\
\hline
A & 0 & 239 & 0.04 & 5.0\\
   & 239 & end & 1.50 & 0.5\\
\hline
B & 0 & end & 1.50 & 0.5\\
\hline
C & 0 & 117 & 1.50 & 0.5\\
& 117 & 267 & 0.15 & 12.0\\
& 267 & end & 1.50 & 0.5\\
\hline
D & 0 & 123 & 1.50 & 0.5\\
& 123 & end & 0.75 & 1.5\\
\hline
\end{tabular}}
\caption{Values of the myogenic response natural frequencies and damping factors, along with corresponding epoch start and end times, used to obtain the model simulation results presented in \cref{fig2ndOCaKuo91Fig2A,fig2ndOCaKuo91Fig2B,fig2ndOCaKuo91Fig2C,fig2ndOCaKuo91Fig2D}.}
\label{tabOmegaZeta}
\end{table}

The variation in values for $\omega_n$ and $\zeta$ over the different epochs of the experiments is similar to that of the first order system, given in \cref{tabTauCaVals} with the most common fast response being given by $\omega_n = 1.5$ and $\zeta = 0.5$. A natural frequency of 1.5rad/s is equivalent to approximately 0.24Hz. It should be noted that the values presented in \cref{tabOmegaZeta} are not necessarily unique in their fit of the data; previously a similar fit was achieved by tuning only one parameter, now that we have two to tune to the same data, we cannot be sure that there do not exist other combinations of $\omega_n$ and $\zeta$ which would produce the same (or better) results. We have however shown that the new second-order system is at least as capable as the previous, first-order, system of replicating experimental behaviour. It should also be noted that all of the discussion of \cref{variableTCaJust} applies to this new system just as it did to the old. With this new \ca dynamic, we are aiming not to replicate exactly the \ca behaviour of the cell, only to allow the system a little freedom in that dimension, allowing us to explore the possible implications of a more complex \ca dynamic.

\section{Analysis of Instability}
\label{analInst}
\begin{figure}[hbtp] \centering \includegraphics[scale = 0.6]{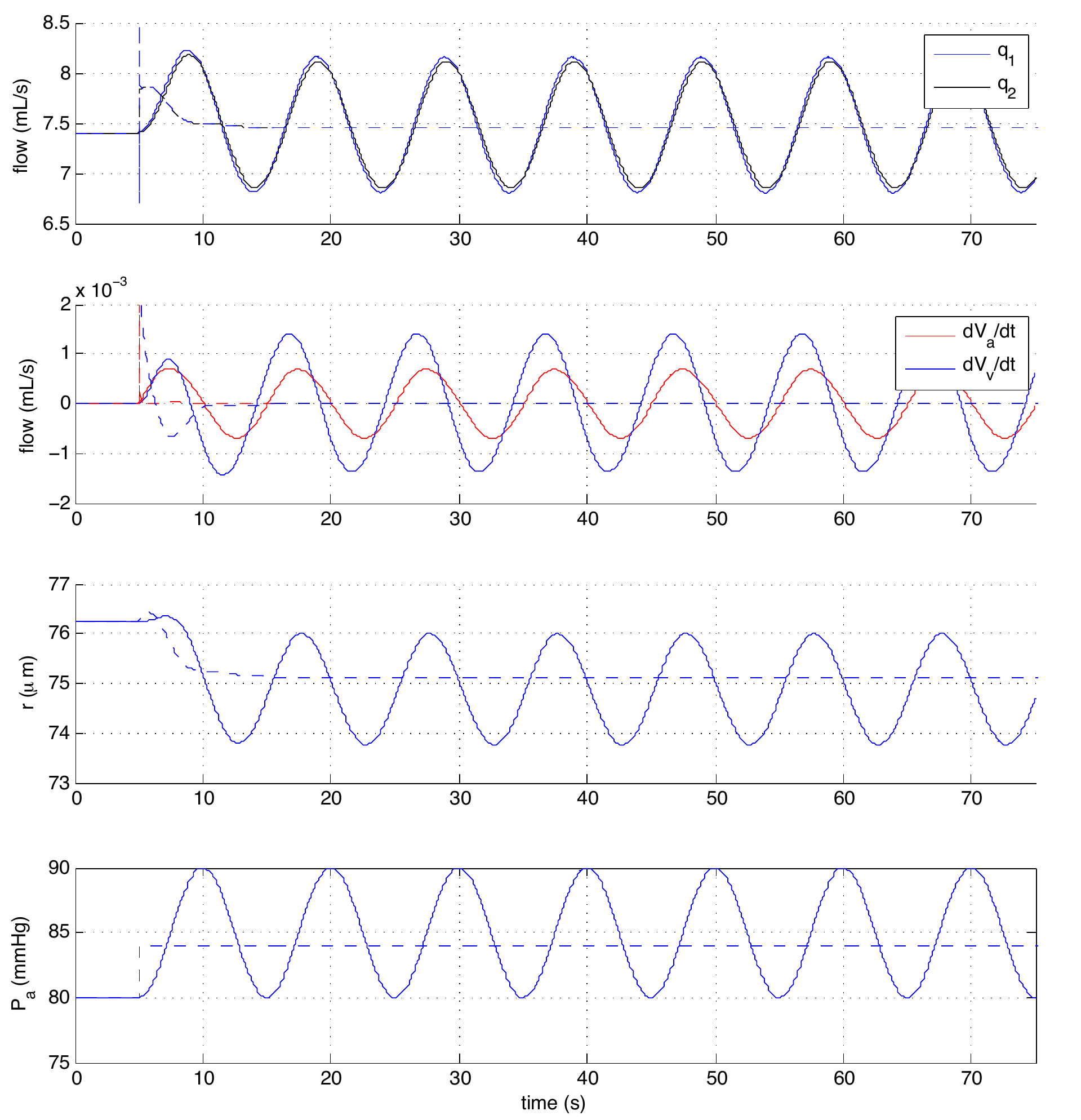} \caption{Response of vasculature model to sinusoidal arterial pressure variations.}\label{fig2ndOCaSine} \end{figure}

The response of the system to the step onset of oscillatory arterial pressure (as in \cref{figautoregSine5}) is shown in \cref{fig2ndOCaSine} and is similar to that of the first iteration of the system. The response is a combination of a straightforward sinusoidal response with a phase lead of flow w.r.t. pressure of 35$^\circ$ and a damped step response to the increase in mean pressure. The dotted lines show the isolated step response while the solid lines show the response to the combined step and sine-wave onset.

\begin{figure}[hbtp] \centering \includegraphics[scale = 0.8]{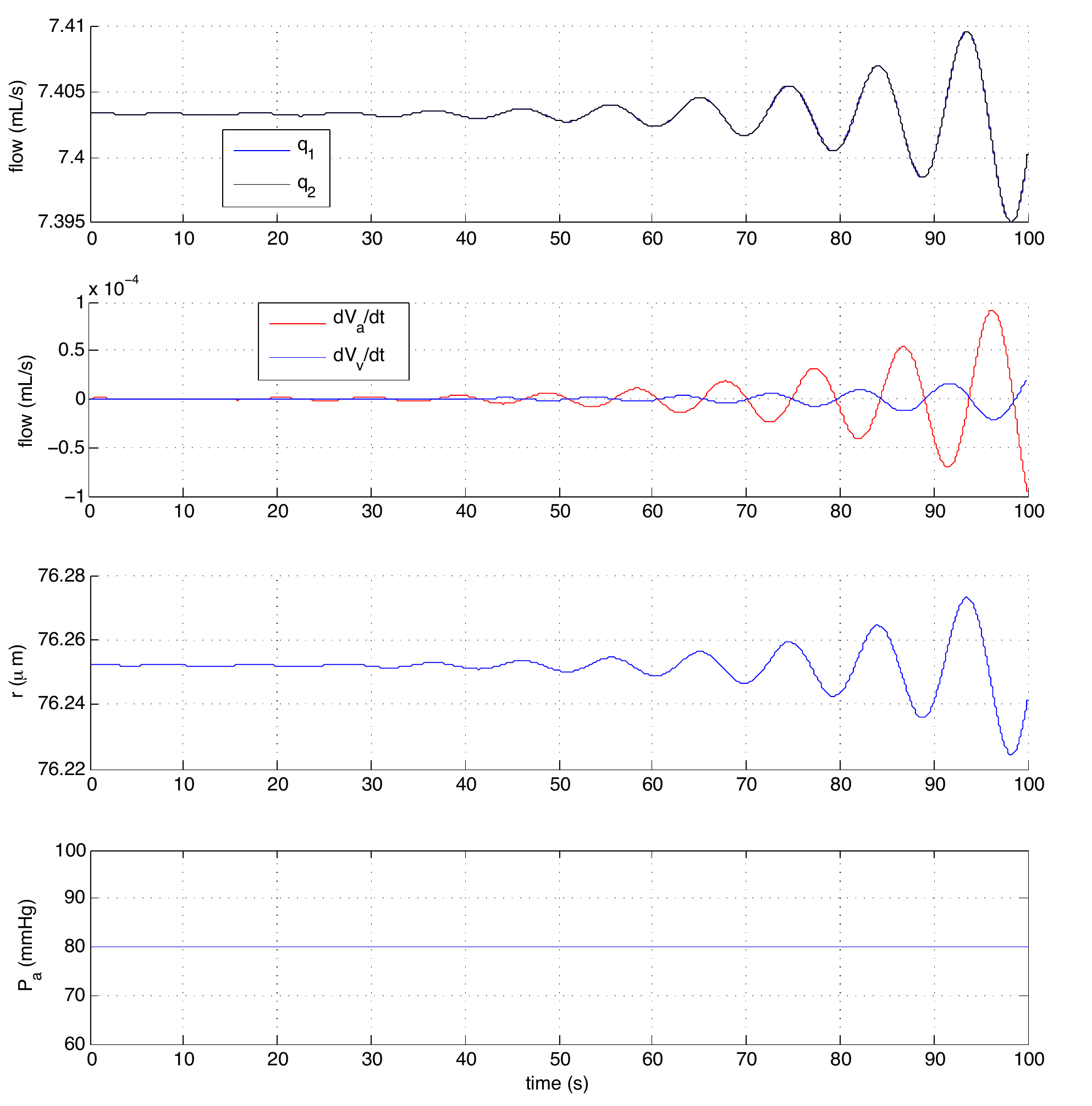} \caption{Initial response of vasculature model to constant arterial pressure when damping is reduced ($\zeta = 0.1$).}\label{fig2ndOCaGrow} \end{figure}

It is not necessarily discouraging that the modified system does not exhibit instability in this response: vasomotion is only observed in vivo under strenuous conditions, such as those of hypo-perfusion due to haemorrhage \cite{Schmidt92} or introduction of an adrenergic stimulant \cite{Colantuoni84}. Whilst our myogenic response model is not physiologically realistic enough to directly model what effects these conditions might have on the biochemistry of the VSMC, we can adjust the parameters of our \ca dynamic (\cref{dCadt2}) and so observe the range of possible effects. \Cref{fig2ndOCaGrow} shows the response of the system to a constant pressure of 80mmHg with $\omega_n = \pi/5$ and $\zeta = 0.1$ ($\frac{\pi}{5}$rad/s = 0.1Hz).

\begin{figure}[hbtp] \centering \includegraphics[scale = 0.8]{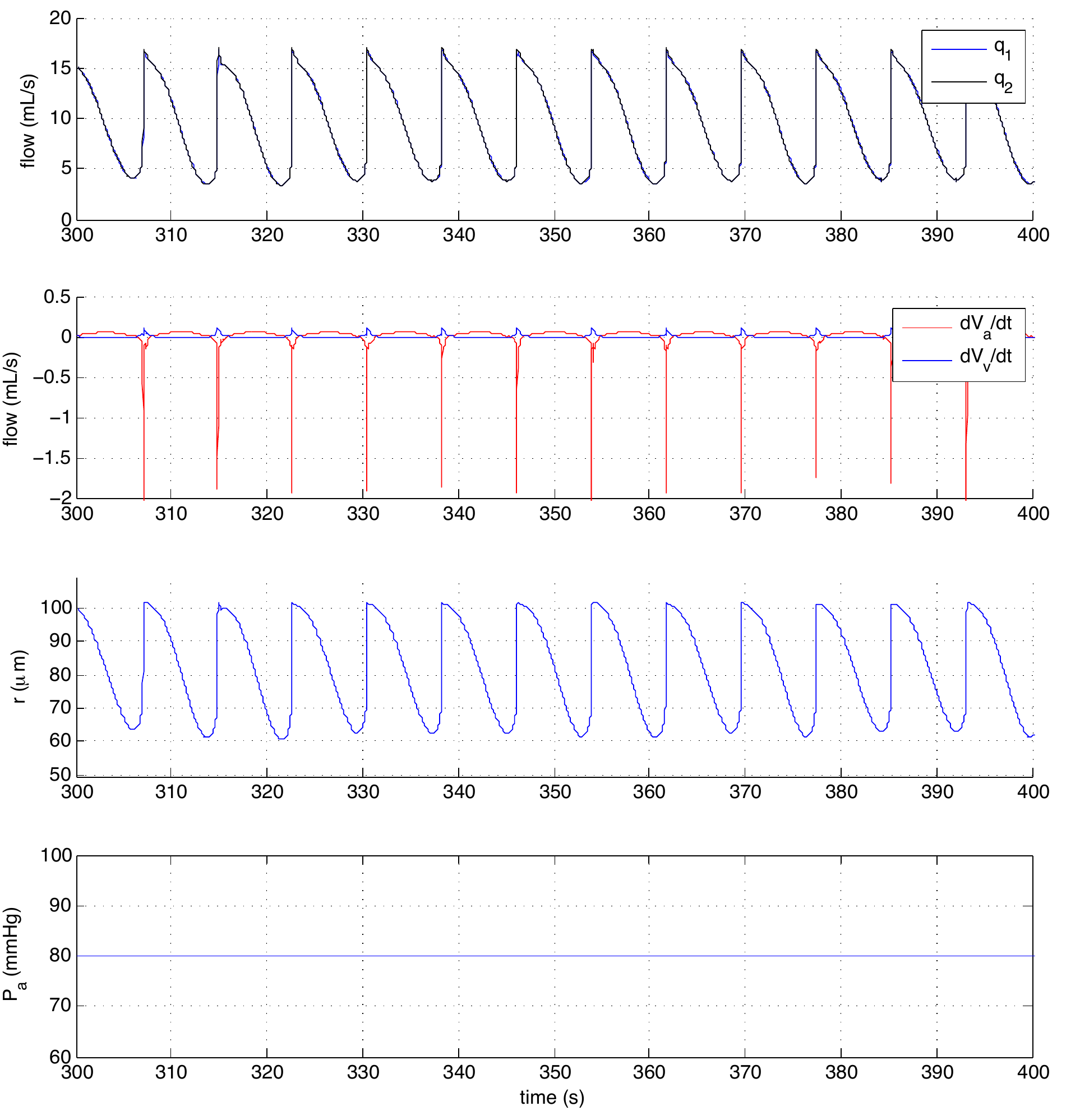} \caption{Response of vasculature model to constant arterial pressure when damping is reduced ($\zeta = 0.1$) -- once limit cycle has been reached.}\label{fig2ndOCaOscill} \end{figure}

\begin{figure}[hbtp] \centering \includegraphics[scale = 0.8]{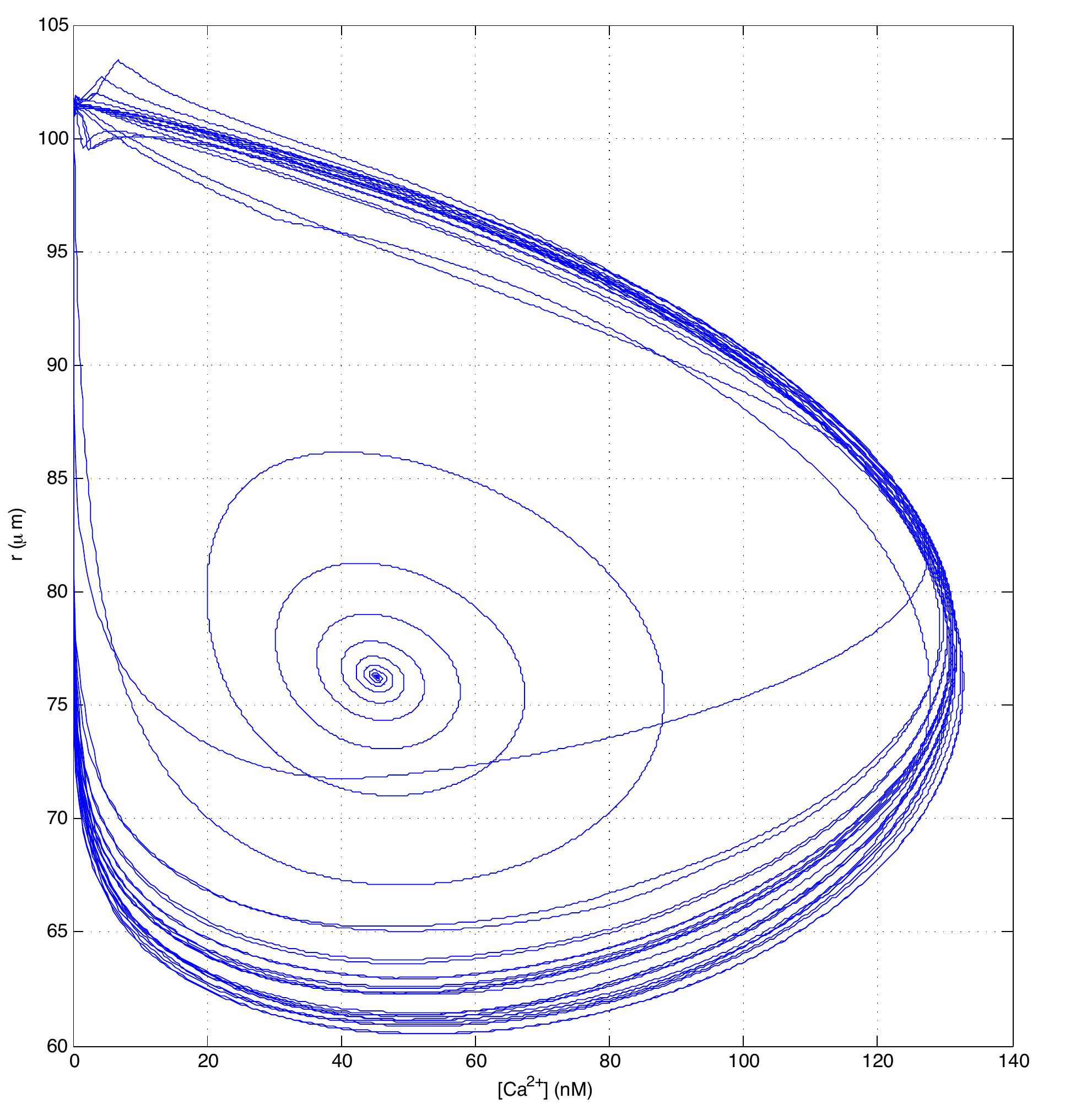} \caption{Initial 400s of the trajectory of the system through \cca\!-$r$ state space under reduced damping ($\zeta = 0.1$).}\label{fig2ndOCaPhase} \end{figure}

The system can be seen to diverge from its initial equilibrium and to oscillate with increasing amplitude, albeit still very low amplitude at 100s. As with any numerically solved system, the initial equilibrium was not exact, rather it was correct to within $1\e{-15}$m, thus no perturbation was needed in order to precipitate these oscillations. If we integrate the system for a longer period we see that the growing oscillation settles into a periodic limit cycle. \Cref{fig2ndOCaOscill} shows the same trajectory as \cref{fig2ndOCaGrow} after it has settled into this limit cycle. \Cref{fig2ndOCaPhase} shows the the first 1000s of the evolution in state-space, specifically \cca\!-$r$ space. The spiral out from the point near the centre corresponds to the slowly growing oscillations seen in \cref{fig2ndOCaGrow} whilst the heavily traversed ovoid paths correspond to the `settled' limit cycle behaviour seen in \cref{fig2ndOCaOscill}. Note that this limit-cycle behaviour is fundamentally different from the sinusoidal response previously observed in the case of sinusoidal variation in arterial pressure -- in the case of a limit cycle there is a bound on the amplitude of any oscillation, but each circuit of the state-space is different from the last.

 As can be seen from the flattening of the trajectories in on the left hand side of \cref{fig2ndOCaPhase}, the cycle is limited by the impossibility of a negative \ca concentration. Once the concentration of \ca has reached zero\footnote{Actually marginally above zero to prevent numerical singularities in the integration.} the vessel quickly dilates to the passive radius of \cref{figPassiveFitBoth}, once this radius is reached the stress in the vessel wall has increased sufficiently to cause \cca to rise again, and so cause the contraction of the vessel which will in turn lower the stress in the wall and cause the return of \cca to near zero, completing the cycle. The cycle is driven to continue by the interaction of the slowness of vessel contraction and the `momentum' of the \ca response\footnote{During integration, $\diff{\ccam}{t}$ is set to zero the instant that \cca reaches its lower limit, effectively losing all `momentum'.}.

\subsection{Sensitivity to $\zeta$}
\begin{figure}[hbtp] \centering \includegraphics[scale = 0.8]{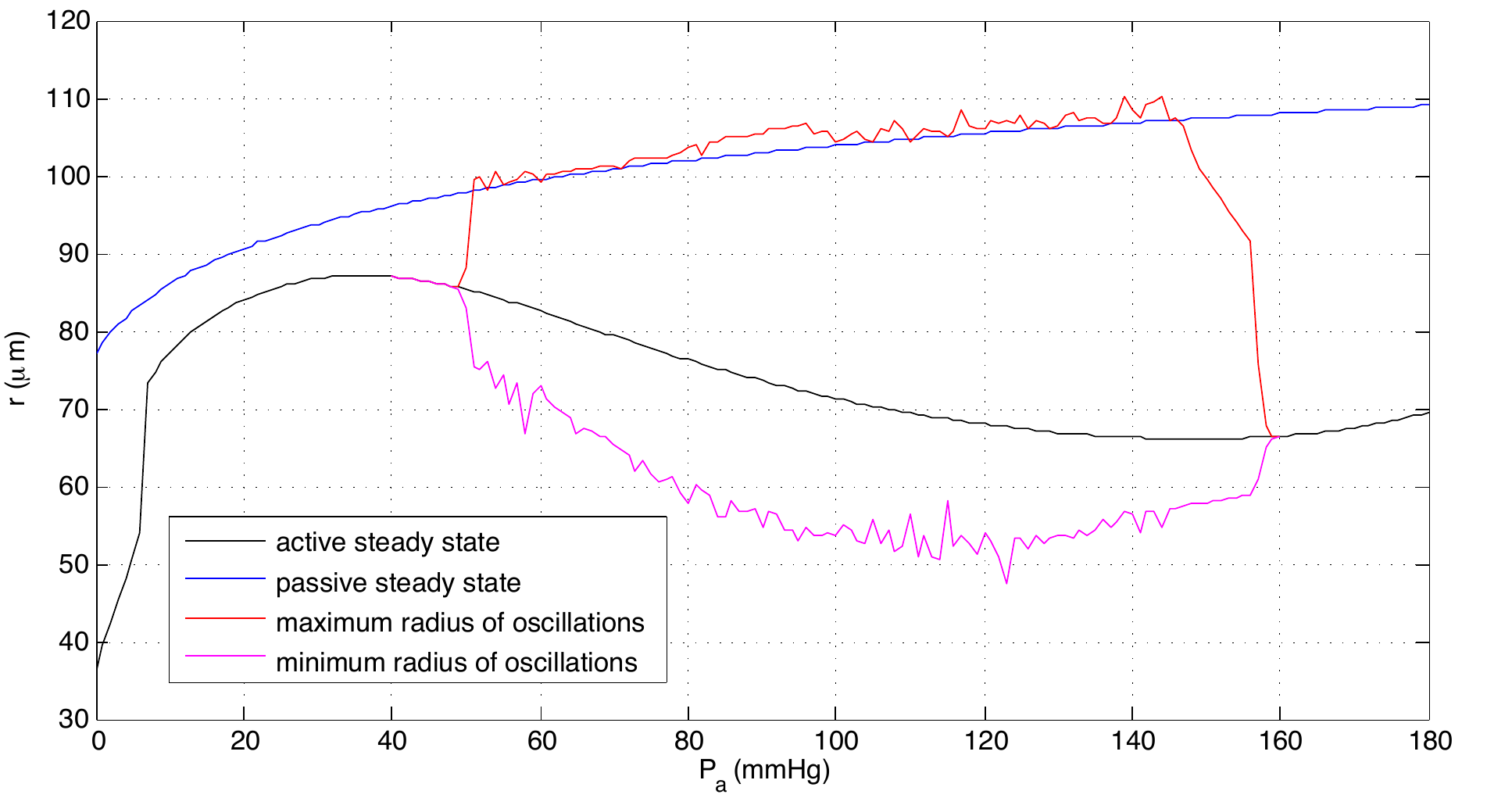} \caption{Variation of oscillatory range with arterial pressure. Maxima and minima taken over first 1000s of integration starting at equilibrium. Arterial pressure was constant throughout each integration ($\zeta = 0.1$).}\label{figzetaSingle0p1} \end{figure}

The oscillatory results so far presented have been at an arterial pressure of 80mmHg and with a natural frequency and damping factor of 0.1Hz and 0.1 respectively. We know that at some combinations of these parameters the system is stable, so it is naturally of interest to define the boundary between stability and instability in the $P_a$--$\omega_n$--$\zeta$ space. Firstly we look at the variation in the limits of vessel radius during oscillation at a fixed value of $\zeta$ over the complete range of arterial pressure. \Cref{figzetaSingle0p1} shows the steady state radii of the vessel under both active and passive ($\ccam = 0$) conditions along with the maximum and minimum radii of the vessel under spontaneous oscillation at the given arterial pressure. The nature of limit cycles is such that the maximum value achieved in the 1000s integration time of the simulations that went into this graph are not necessarily consistent, hence the `noise' in the envelope of oscillatory range.

\begin{figure}[hbtp] \centering \includegraphics[scale = 0.8]{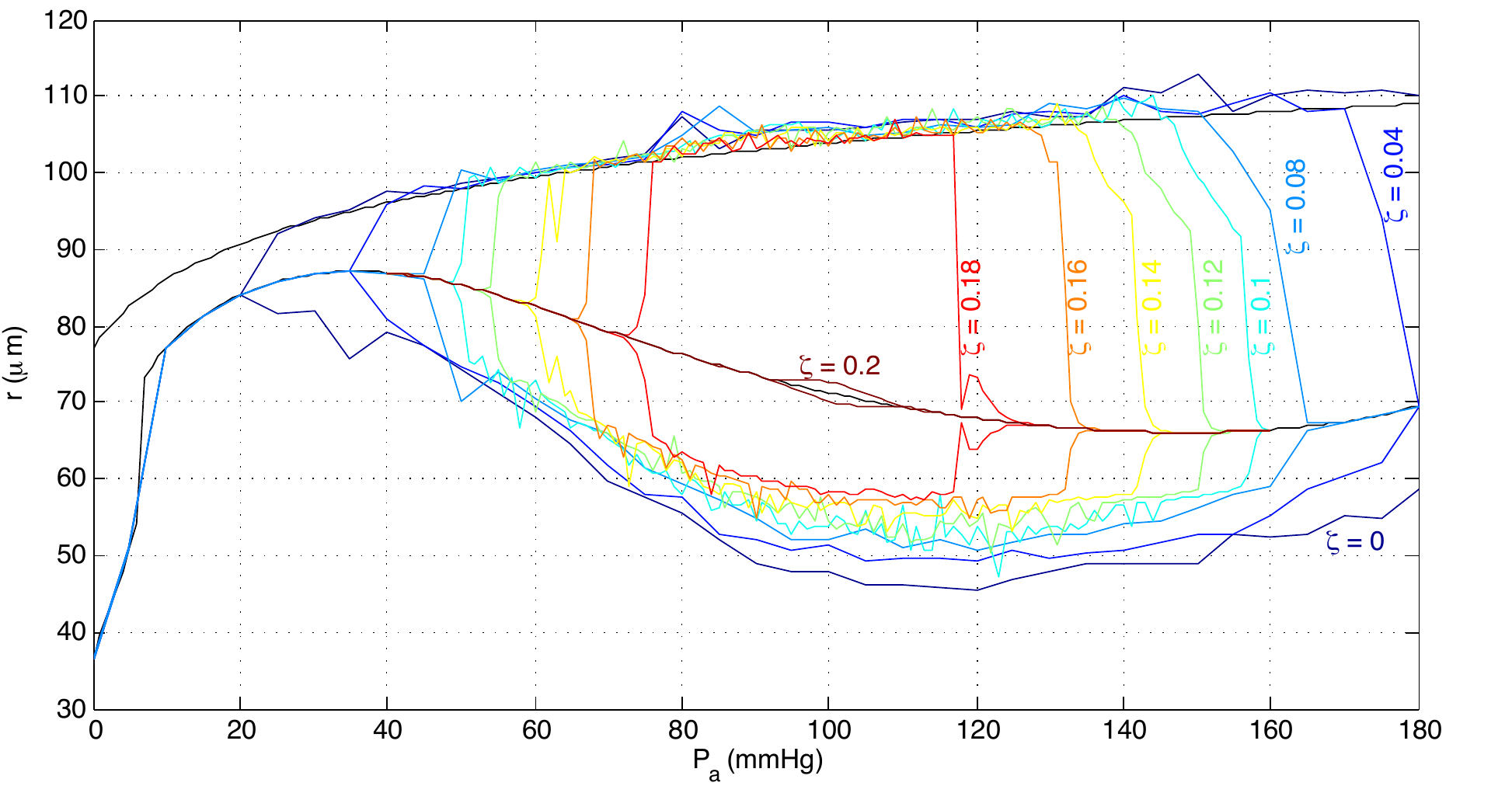} \caption{Variation of oscillatory range with arterial pressure for various values of $\zeta$. Maxima and minima taken over first 1000s of integration starting at equilibrium. Arterial pressure was constant throughout each integration ($\omega_n = 0.1$Hz).}\label{figzetaVariations} \end{figure}

We can see that with a damping factor of 0.1 the system is only unstable in the autoregulatory range of arterial pressure. It is also obvious that the passive radius acts as an effective upper bound on oscillatory radius; given that \cca is limited to be non-negative, and the passive equilibrium is achieved at zero \cca, this is to be expected. Any dilation of the vessel beyond the passive equilibrium radius for the given arterial pressure must be due to a transient pressure fluctuation increasing the mean pressure in the arteriolar compartment above the steady state value. As damping is reduced, the range of arterial pressures at which spontaneous oscillations occur increases, the instability spreading to higher and lower arterial pressures. Conversely, as damping factor is increased the  range of arterial pressures in which oscillation occurs is narrowed. This is shown in \cref{figzetaVariations} where the envelope of oscillation has been plotted for a range of values of $\zeta$.

The steeper region of the steady state active radius curve, where autoregulation is strongest, seems to be able to sustain spontaneous oscillations at higher damping factors than the less steep regions. However there are pressures at which the equilibrium response of radius is to increase with increasing pressure, i.e., outside the autoregulatory range, which do in fact exhibit spontaneous oscillations when damping is reduced far enough.

\subsection{Sensitivity to $\omega_n$}
\begin{figure}[hbtp] \centering \includegraphics[scale = 0.8]{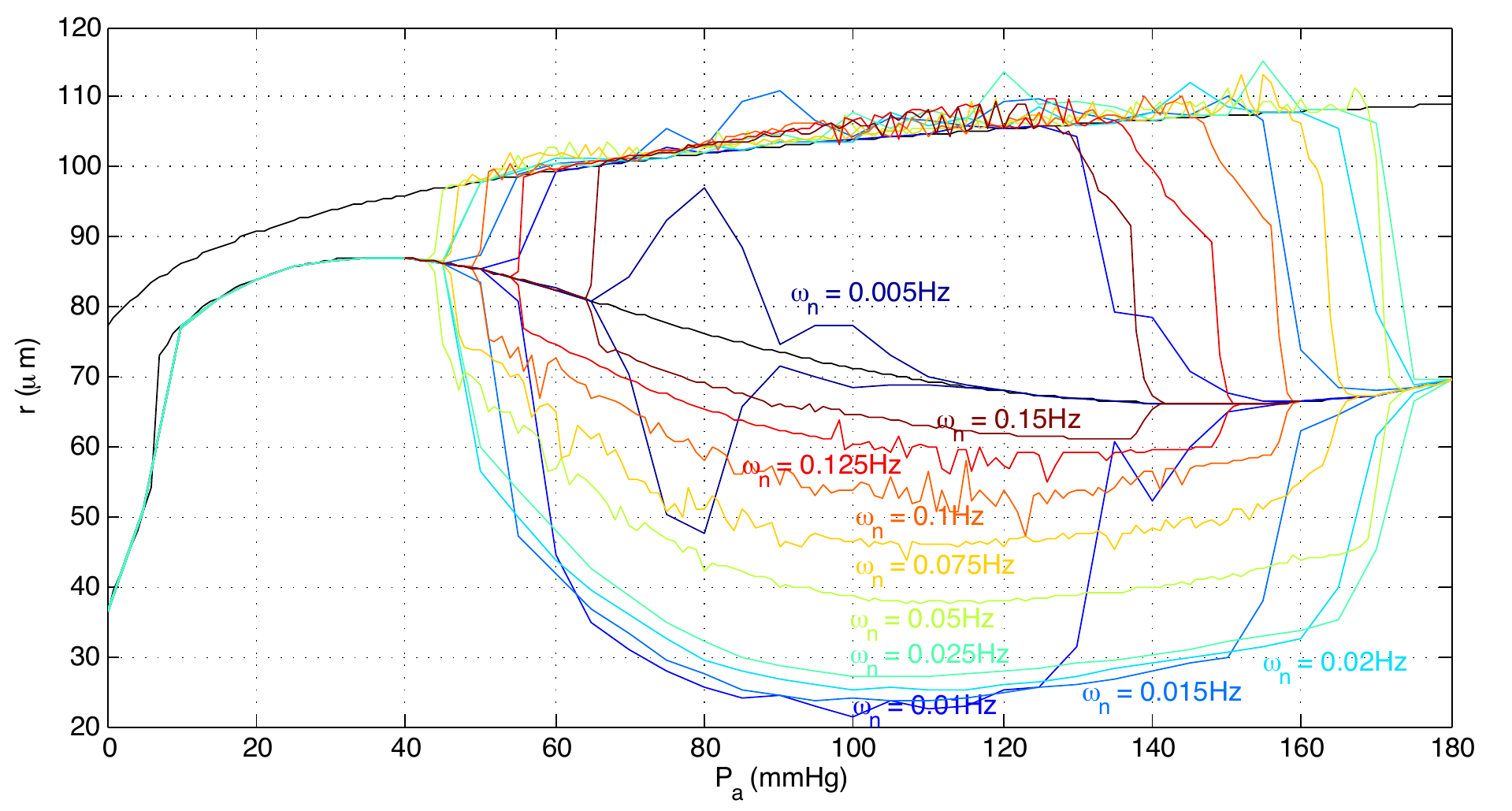} \caption{Variation of oscillatory range with arterial pressure for various values of $\omega_n$. Maxima and minima taken over first 1000s of integration starting at equilibrium. Arterial pressure was constant throughout each integration ($\zeta = 0.1$).}\label{figomegaVariations} \end{figure}

A similar set of simulations reveals that the pressure region over which oscillation occurs varies not only with $\zeta$ but also with $\omega_n$. \Cref{figomegaVariations} shows this variation of oscillatory range, and amplitude, with arterial pressure and undamped natural frequency, $\omega_n$. The pressure range over which oscillation is observed is greatest at a frequency of 0.025Hz -- an oscillatory period of 40s. At frequencies of 0.2Hz and higher no oscillation is seen, similarly at frequencies below 0.005Hz the oscillatory pressure range narrows to nothing. As well as this change in oscillatory pressure range with frequency, we can observe a decrease in minimum radius during the oscillation. At 0.01Hz, 100mmHg, the minimum radius seems to reach a minimum, with the vessel oscillating between $r = 105\mu$m and $r = 22\mu$m. It is possible that 0.025Hz is the frequency at which the \ca dynamics cause the greatest resonance with the speeds of the mechanical model, hence the greatest range of pressures over which oscillation is observed, but lower frequencies afford the mechanical model more time to respond to the elevated levels of \ca and so achieve a smaller minimum radius (greater contraction). The apparent decrease in oscillation magnitude at 0.005Hz may be illusory, at such low frequencies it is possible that the 1000s integration time is not sufficient for the oscillation to fully develop.

\subsection{Role of Non-regulating Vasculature}
\begin{figure}[hbtp] \centering \includegraphics[scale = 0.8]{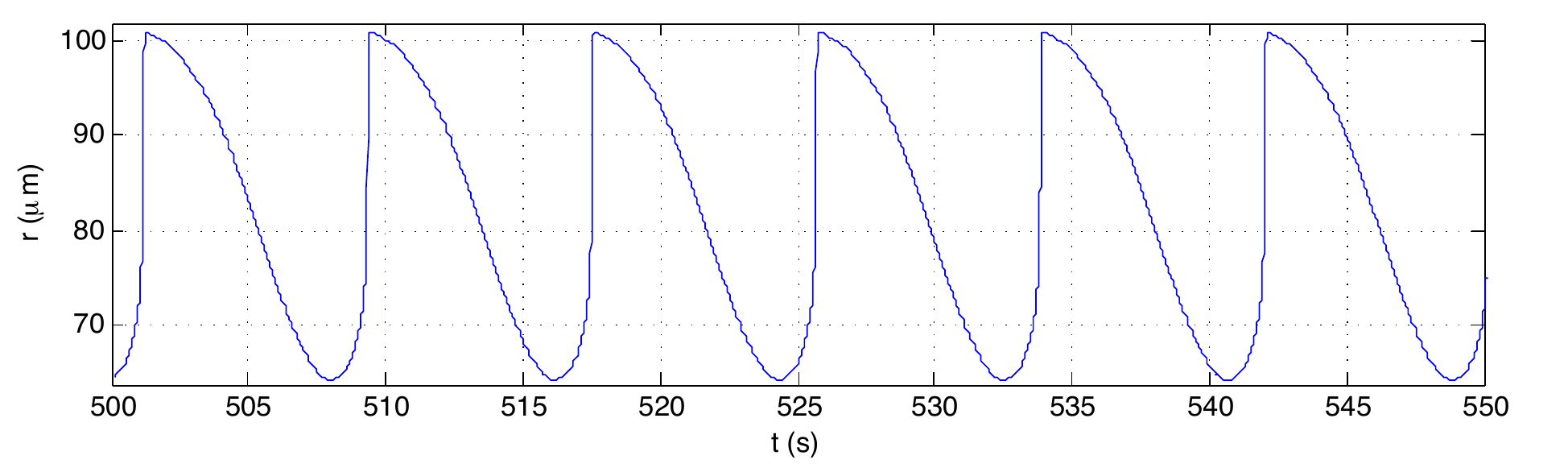} \caption{Oscillation in radius of an isolated arteriole. $P = 48.5$mmHg, $\Delta P = 75$mmHg, $\omega_n = 0.1$Hz, $\zeta = 0.1$.}\label{figsingleOscill} \end{figure}

Spontaneous oscillatory behaviour has been discovered only after incorporation of the arteriole model into the whole-brain vasculature model, and subsequent to the modification of this vasculature model in search of vasomotion like oscillations. However, since the final modification (which made oscillations possible) was to the \ca dynamics within the arteriole model, it is worth investigating the possibility that the isolated arteriole model may now be capable of vasomotion-like oscillations in and of itself, without the peripheral elements of the vasculature model. To this end we set $\omega_n = 0.1$Hz and $\zeta = 0.1$ (as in the cases of \cref{fig2ndOCaGrow,fig2ndOCaOscill,fig2ndOCaPhase}) and we fix the upstream and downstream pressures to match those of the vasculature model ($P_\alpha$, $P_\beta$) with an arterial pressure of 80mmHg. \Cref{figsingleOscill} shows the resulting oscillation in radius of this single, isolated arteriole.

\begin{figure}[hbtp] \centering \includegraphics[scale = 0.8]{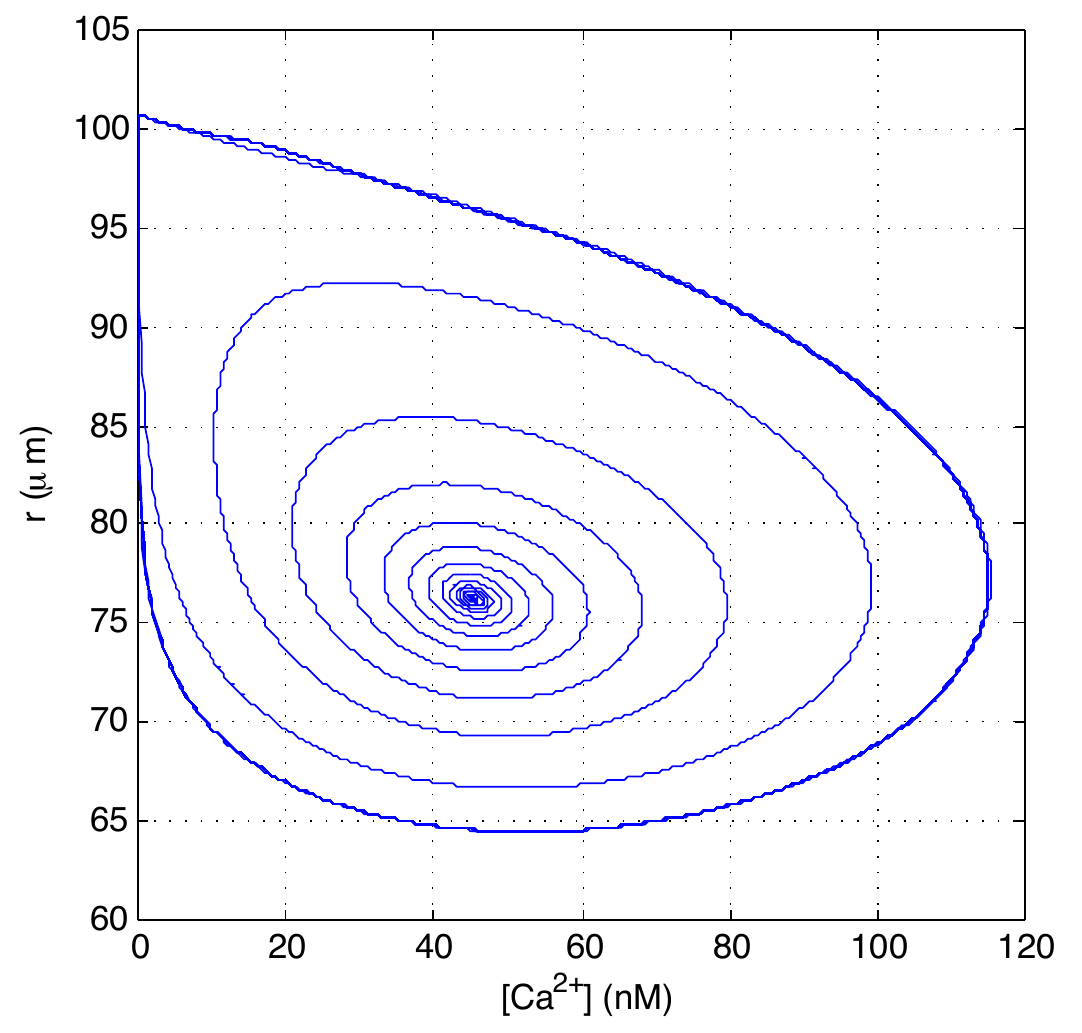} \caption{Oscillation in radius of an isolated arteriole shown in the \cca\!-$r$ state space. $P = 48.5$mmHg, $\Delta P = 75$mmHg, $\omega_n = 0.1$Hz, $\zeta = 0.1$.}\label{figsinglePhase} \end{figure}

This oscillation is similar in character to that of the whole-brain model, we can see from the\cca\!-$r$ phase plane shown in \cref{figsinglePhase} that the trajectory is a similar shape to that of \cref{fig2ndOCaPhase}, however, it is much more regular than in the case of the whole-brain model. The oscillations exhibited by this single arteriole are in fact so regular that it is difficult to say whether any irregularity is due to the nature of the system or the inaccuracies of its integration. Either way the result is that the oscillatory regime into which the single arteriole settles is considerably more regular and predictable than when the arteriole is coupled to the whole-brain vasculature model.

It appears then that the irregularity in the oscillations of the whole-brain model are borne of variations in pressure upstream and downstream of the arteriolar bed, because we can see from the results above that if the arteriole experiences constant pressures at both upstream and downstream ends then it will oscillate in a regular fashion. The irregularity in the whole-brain oscillations can then be thought of as a result of interference between the frequency of the driving arteriolar oscillation and the natural frequencies of the surrounding vasculature model.

\subsection{Effect on Flow}
\begin{figure}[hbtp] \centering \includegraphics[scale = 0.8]{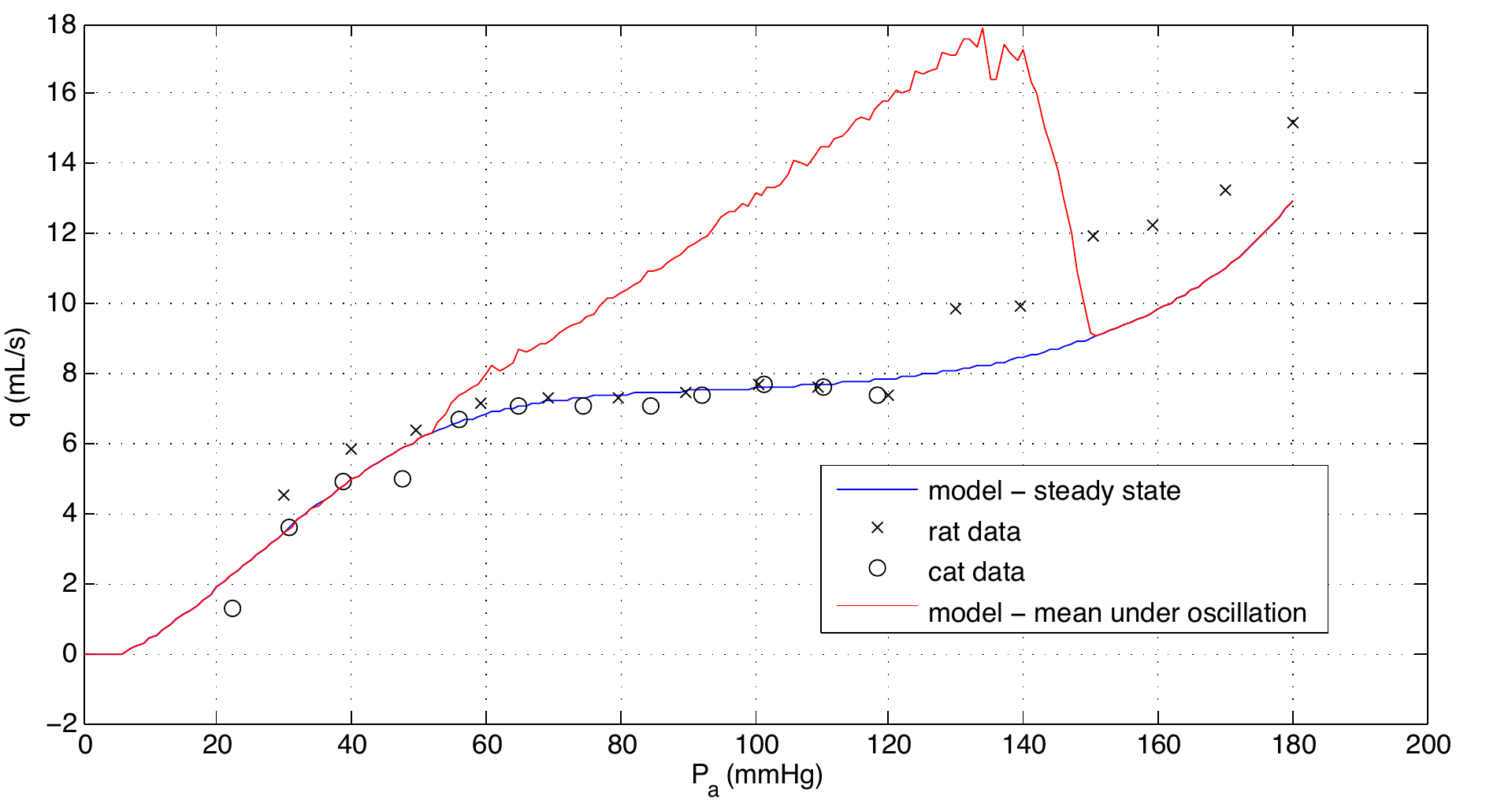} \caption{Effect of oscillation on average flow. $\omega_n = 0.125$Hz, $\zeta = 0.1$.}\label{figflowWithOscillation} \end{figure}

So far we have examined primarily the sensitivity of the range of vessel radius under oscillation to the parameters of the \ca system. A prominent hypothesis for the evolutionary benefit (to avoid using `purpose') of vasomotion is that it increases oxygen perfusion in pathological conditions, where the tissue would otherwise become hypoxic. In \cite{Hapuarachchi10} it is found that symmetrical oscillations in radius about its steady state value can increase oxygen perfusion into the surrounding tissue. In our results however, we find that the oscillation has a very non-symmetrical nature at 0.1Hz; from a steady state radius of 77$\mu$m, the oscillation causes the vessel to dilate as far as 103$\mu$m but only to contract to 60$\mu$m. At higher frequencies this bias towards dilation becomes even more pronounced (\cref{figomegaVariations}). The effect that this has on the response of flow to arterial pressure is pronounced. \Cref{figflowWithOscillation} shows the steady state response of the vasculature model in blue (at higher damping), and the response under the oscillatory regime (reduced damping) in red. The flows shown for the oscillatory regime are the time-averaged flows once oscillations have settled into the fully developed limit cycle. The greatly increased flows under oscillation are partly the result of an asymmetric oscillation in radius, and partly the result of the fourth power of radius in the Poiseuille equation (\cref{Poiseuille}).

\subsection{Comparison to Vasomotion}
We have found that large oscillations in vessel radius are possible with second-order \ca dynamic only if damping factor is below 0.2, whereas in fitting the new \ca dynamics to the dynamic vessel response data of \cite{Kuo91} we required damping factors of 0.5 and higher. As mentioned in \cref{analInst}, this discrepancy is perhaps to be expected given that vasomotion is not often observed under normal physiological conditions. The physiological analogy to our system would be that under normal physiological conditions damping factor is high but in conditions under which vasomotion is observed (hypo-perfusion or adrenergic stimulation) damping factor is reduced and oscillations ensue. We should be careful here not to suggest that our second order model is the correct one, or that work should be carried out to identify the cause of this reduction in `damping factor' as such. Rather a more general view should be taken; we should note that self sustaining oscillations in \ca are not necessary for large oscillations in radius to occur, some amount of damping of the \ca system (such that the \ca system in isolation would quickly settle to a stable equilibrium) is tolerable, more so in the more myogenically active regions of the pressure range. We should note also that some change in the character of the \ca response is likely to take place between normal physiological conditions and those in which vasomotion is observed.

The true nature of \ca dynamics within the VSMC is probably much closer to those found in \cite{Jacobsen07a}, where the cell membrane channels and sarcoplasmic reticulum are represented in enough detail to allow CICR to drive self-sustaining \ca oscillations in the cell, under some conditions\footnote{Stimulation by IP$_3$, which is part of the adrenergic stimulation pathway. Source: implementation and refinement of the model of \cite{Jacobsen07a} in communication with the author (Jens Jacobsen), results not presented here.}. Our simple second order \ca dynamic is useful because it sets a lower bound on the propensity of the \ca concentration in the cytoplasm to oscillate which is necessary to allow large, whole-vessel radius oscillations.

\section{Discussion}
If we reflect upon the components of the model we may be surprised to find any circumstances under which self-sustained oscillations might occur: the \ca dynamics, although less damped than when fitted to the Kuo data, are positively damped such that without some external forcing the \ca oscillations would decay quickly; changes in the length of the mechanical muscle model are damped proportionally to the number of attached cross bridges; and the NO system, although it does exhibit positive feedback at very low pressures, is all but inactive in the mid-range of arterial pressure. With these damping elements dissipating energy during any state transition, there must be some mechanism putting the energy back into the system to propagate the oscillations which we have observed. This mechanism is the contraction of the vessel against the pressure of the blood within it, driven by the cycling of phosphorylated cross bridges to cause contraction of the VSMC. The energy thus expended doing work on the fluid is then released into the mechanical damping elements as the vessel dilates back towards its passive radius. This is only possible because we have implicitly assumed in our 4-state kinetic model (\cref{4stateKM}) that there is always enough ATP present to phosphorylate myosin heads to whatever degree is permitted by the \ca\!\!\!/NO balance. The fact that the oscillation can withstand the greatest degree of damping when arterial pressure places the system at the regions of greatest negative gradient on the radius/pressure curve (most active autoregulation) is consistent with this view. The centre of the autoregulatory range, and so the centre of the myogenic response curve, is the point at which changes in stress cause the greatest change in \ca concentrations. Accordingly it is also the point at which the feedback between stress in the vessel wall and energy usage in vessel contraction is strongest. It is this energy uptake from ATP and its use in vessel contraction which drives the oscillation against the various damping elements removing energy from the system.

We can see from \cref{figzetaVariations} that the oscillation can exist outside the range of pressures within which the response of radius to increased pressure is negative. At zero damping for instance the oscillation extends to arterial pressures as low as 25mmHg, where the vessel in its steady state dilates slightly to positive changes in pressure. This is however not in contradiction to what has already been said about the contractive energy use of the VSMC driving oscillation, for while the net change in radius at these low pressures is positive, it is not \emph{as positive} as it would have been in the absence of the myogenic response -- some phosphorylation of muscle has taken place even if it wasn't enough to cause a net contraction. We can compare the gradients of the passive and active steady state equilibrium curves of \cref{figzetaVariations} and see that the point at which the oscillations become sustainable is the point at which the gradient of the active curve diverges from that of the passive curve (i.e., the first point at which the active curve becomes less steep than the passive curve). It may be highlighted that at the high pressure end the active curve is steeper than that of the passive curve and yet oscillations persist, this is not contradictory either; the active curve may be steeper than the passive curve at their respective equilibrium radii but if the stiffness of the passive curve at the radius of the active curve were to be plotted it would produce a far steeper line than that of the active response. The stiffness of the passive response is exponentially increasing with extension, such that the gradients of the two lines are only meaningfully comparable where they are close in radius, as is the case at the low end of the pressure range.

\section{Conclusions}
The capability of exhibiting self-sustaining oscillations in radius under constant external conditions ($P$ and $\Delta P$) is an important feature of our arteriole model. Whilst the addition of intracranial compliance and flow inertia undoubtedly make our vasculature model more realistic, and so possibly more useful for future comparisons with in-vivo data, we have shown that they are not necessary for vasomotion-like oscillations in arteriolar radius. The sufficiency of a damped \ca dynamic to produce oscillations in radius suggests that the self-sustaining \ca fluctuations of CICR are not necessary (although certainly sufficient) to instigate vasomotion. In fact vasomotion may be initiated by some influence on the VSMC which causes reduced damping of \ca fluctuations, perhaps by upsetting the buffering of \ca in the cytoplasm, and the resulting feedback from the mechanical system, via the myogenic response, could initiate whole-cell \ca waves and CICR. Thus our model shows that self-sustaining \ca waves in the VSMC are not a prerequisite for vasomotion, in fact they may be a consequence of it. The only prerequisites for vasomotion in our model are an active myogenic response and a lightly damped \ca dynamic.

We have also found that the upper bound on vessel radius during vasomotion-like oscillations is that of the passive equilibrium. In most cases this means that the upper bound on radius will be further from the steady state equilibrium than the lower bound, and so the mean radius greater under oscillation than in the steady state. Flow is thus much greater during vasomotion than in the steady state equilibrium at the same pressure, more so than if the oscillation were symmetric around the equilibrium radius. This is consistent with the view that vasomotion is instigated by the sympathetic nervous system as a response to hypo-perfusion because it increases mean blood flow and so perfusion of the distressed tissue.

\chapter{Conclusions and Future Work}
\label{chapConcs}
In this chapter we will summarise the results of our work before discussing its implications and limitations. We will conclude with a discussion of future work which could build on that presented here and provide further insight into the regulation of blood flow in cerebral arterioles.

\section{Summary of Results}
In \cref{chapArtFit} we found, through optimization, a set of parameters for the equations presented in \cref{chapVascModel} which provide a very good fit to both the static and dynamic data found in the experimental literature. We saw that the combined models of the mechanics and biochemistry of the arteriole can create an active response to increased pressure through the stress-induced release of \ca in the VSMC, and that the flow-induced production of NO in the endothelium is antagonistic to this contraction. In the dynamic cases we found that a fixed time constant in the \ca myogenic response model is not sufficient to reproduce the results observed experimentally. Rather than introduce extra complexity at such an early stage by trying to explicitly model the changes in \ca speed we simply found the variation in time constant necessary to fit the data and hypothesised about possible mechanisms which could replicate these changes innately. It was also noted at this point that the level of flow at which \cno reaches saturation is necessarily very low, leading to an almost binary NO state. Although further data are available on the response of vessel radius to variations in flow, it is not consistent with radius-pressure data and so we decided not to attempt a combined fitting of these datasets.

On finding limitations with current methods of sensitivity analysis we contrived a novel implementation of the technique and used it to analyse our model. Consistent with our observations on NO saturation, the sensitivity analysis revealed that the key parameters were those influencing the myogenic response, and not the NO system.

After integrating our arteriole model into a representation of the vasculature of the whole brain we found that we could match the pressure-flow data from rats and cats found in \cite{Ursino98}. After some modification we also reproduced the approximate arterial to venous volume ratio and its variation with ABP at the end of \cref{chapAutoreg}. The frequency response of this combined arteriolar and non-regulating vasculature model was found to have a resonant peak close to 0.1Hz and to exhibit a phase difference between ABP and flow similar to that observed in vivo. The variation of this phase difference with \ca time constant is significant and the possibility of this being used as a diagnostic measure was discussed. We also used this model to investigate the potential for the involvement of NO in the proposed mechanism of chemical signalling from downstream to upstream through the brain tissue. It was found that the decay rate of NO in the blood is far too high for any detectable traces of NO to remain in the venous blood.

In \cref{chapOscill} the possibility of replicating oscillations in vessel radius characteristic of vasomotion was investigated. We first used bifurcation analysis to confirm what we had already suspected from running simulations; that the model as it was would not produce spontaneous oscillations. After adding two more dynamic elements to the non-regulating vasculature part of the model without causing oscillatory behaviour, we changed the dynamics of the \ca equilibrium in the arteriolar model to second-order. This being a change that affected previous dynamic results, we first found parameters of the new dynamic equation which create as good a fit to the dynamic experiments of \cite{Kuo91} as did the previous first-order parameter $\tau_{Ca}$. This was done and the resulting system (with the parameters as fitted to the Kuo data) was stable, however, if the damping factor of the new second-order \ca response was reduced, spontaneous oscillations resulted. Having mapped the extent and range of these oscillations over the pressure-damping-frequency space we examined the stability of the arteriole model in isolation and found that it exhibited vasomotion-like oscillations in the absence of the rest of the vasculature model. We also found that the mean radius under oscillation was higher than the equilibrium radius at which it would settle if damping were higher.

\section{Discussion and Conclusions}
\subsection{The Role of NO}
Our fitting to the flow and no-flow data of \cite{Kuo91} in \cref{chapArtFit} shows that NO does play a significant role in determining the radius of the arteriole. However, because the effect of NO saturates at such low levels of flow, the change in radius due to NO is a constant offset at all physiological levels of flow. The NO pathway has virtually no effect on the dynamic behaviour of the system, either in response to an oscillatory arterial pressure or as part of the spontaneous oscillations the vessel has been found to exhibit, as long as flow remains above a very low minimum level. Because the saturation of the NO pathway occurs due to the exhaustion of cGMP, and not the exhaustion of NO itself, there are interesting delays in the response to the complete cessation of flow, as seen in \cref{chapArtFit}. These delays could serve as a useful hysteresis to allow the brief cessation of flow without closing the arteriole down and possibly exacerbating pathological hypo-perfusion.

\subsection{The Role of \ca}
Whilst we may have found that the role of NO is somewhat less interesting than anticipated, we have found the converse to be true of \ca\!\!. Firstly we established that a variable \ca time constant was necessary to replicate the dynamic isolated vessel experiments of \cite{Kuo91}. We then found that the \ca dynamic itself was critical to the possibility of spontaneous oscillations. The fact that some `momentum' or hysteresis must be associated with the \ca concentration before the system becomes unstable is interesting in two ways; firstly that the system of inductances and capacitances coupled with a very non-linear arteriole model does not oscillate otherwise is notable, the arteriole model must provide significant damping and be everywhere stable; secondly it is very interesting that \emph{all} that is necessary is an underdamped \ca dynamic -- much work has focussed on the mechanism by which whole-cell calcium waves (caused by CICR) in one VSMC might become coordinated with the surrounding cells and so initiate vasomotion. We have found that feedback between the mechanical and biochemical aspects of the myogenic response can amplify an arbitrarily small deviation from equilibrium into a limit cycle oscillation. Although we have implicitly assumed coordination throughout all VSMCs in our model, we have not put in place an electrochemical model of the cell which is capable of generating whole-cell calcium waves in and of itself. Our view of vasomotion in light of these results is that although whole-cell calcium waves are a very likely \emph{result} of vasomotion (given a sufficiently able electrochemical model), they are in no way a necessary precursor to vasomotion.

\subsection{Interpretation of Ide Results}
Let us return to our hypotheses from \cref{chapVascModel}, and to the question which instigated these investigations: we sought an explanation as to why the ABP-CBFV correlation was not different between the two conditions of eNOS blockade and stimulation with phenylephrine. We can now verify, at least partly, two of our hypotheses:

Number 1 was that NO is not involved in cerebral autoregulation at all; whilst NO is indeed involved in autoregulation its effect is that of a DC component. The negligible role that NO plays in the dynamics of autoregulation suggests that eNOS blockade might well have no measurable effect on ABP-CBF correlation. The vasoconstriction and consequent rise in ABP caused by eNOS blockade must move the equilibrium of the system to a new operating point, about which its dynamics may differ. We might therefore expect eNOS blockade to alter ABP-CBFV correlation in comparison to the normal ABP baseline. However, when the comparison is made to a condition with the same ABP, we can expect to see no discernible difference in the ABP-CBFV correlation.

Hypothesis number 3 was that the two conditions being compared were in fact two different methods of achieving the same thing; in one case NO was absent, in the other it was ineffective. We can also partially substantiate this hypothesis from our fitting of the 4-state kinetic model of \cref{LeeFit}. The relative phosphorylation curve shown in \cref{figrPhospCa} shows that cGMP only has influence over the mid-range of \ca concentrations. Where \cca is greater than $10^{-4.5}$ the dotted lines of differing \ccgmp converge. It is quite possible (although we have not investigated this explicitly here) that the action of phenylephrine in releasing \ca from the SR via IP$_3$ raises the concentration of \ca above the level at which cGMP has an effect on phosphorylation of myosin heads.

\section{Future Work}
The model developed here will doubtless/hopefully be valuable in analysing and predicting the behaviour of arterioles and in interpreting experimental results of future experiments. In terms of its future development, it is clear that the first part which should be `upgraded' is the myogenic response model. The inclusion of a more complete electrochemical model of the VSMC, such as that of Jacobsen or Stalhand \cite{Jacobsen07a,Stalhand08}, would greatly increase the ability of this model to replicate observed behaviour and to further elucidate the interactions between flow, pressure, \ca\!\!, and NO. The stress-sensitivity critical for an effective model of the myogenic response would probably take the form of a stress-sensitive \ca channel in the cell membrane, although a stress sensitive channel for a species which then stimulated \ca release from the SR may also be viable. Ideally such a model would be innately capable of changing the speed with which the \ca concentration reaches equilibrium, thus fitting the dynamic results of \cite{Kuo91} without, as we did, resorting to manual manipulation of parameters over different epochs.

The second most promising development of the model would seem to require some experimental results exposing in more detail the relationship between flow and NO production/vasodilation. As mentioned previously, Kuo \cite{Kuo91} presents some such results but the inconsistency between them and the primary dataset makes it hard to integrate them into any model. New investigation and clarification of this issue would enable more precise validation of the NO model and greater confidence in the conclusions that we have drawn from it. Of course if such experiments could be carried out with whole-blood, or with surrounding tissue, or even in vivo, then so much the better.

\bibliographystyle{acm}

\begin{thebibliography}{10}

\bibitem{headway}
Website the brain injury association.
\newblock
  http://www.headway.org.uk/Hypoxic-anoxic-brain-injury.aspx?gclid=COSchd7056oCFUkf4QodqlPv7Q,
  2009.

\bibitem{PISC}
Progress in improving stroke care.
\newblock Tech. rep., National Audit Office, 2010.

\bibitem{anaesthesiaUK}
Website of {A}naesthesia {UK}.
\newblock http://www.frca.co.uk/article.aspx?articleid=249, 2011.

\bibitem{ONS}
Alzheimer's and dementia deaths 2001 to 2010.
\newblock Tech. rep., Office for National Statistics, 2012.

\bibitem{demUK}
Statistics.
\newblock Tech. rep., The Alzheimer's Society, Sept 2014.

\bibitem{Aalkjaer11}
{\sc Aalkj{\ae}r, C.r, C., Boedtkjer, D., and Matchkov, V.}
\newblock {{V}asomotion - what is currently thought?}
\newblock {\em Acta Physiol (Oxf) 202}, 3 (Jul 2011), 253--269.

\bibitem{Aaslid03}
{\sc Aaslid, R., Lash, S.~R., Bardy, G.~H., Gild, W.~H., and Newell, D.~W.}
\newblock {{D}ynamic pressure--flow velocity relationships in the human
  cerebral circulation}.
\newblock {\em Stroke 34}, 7 (Jul 2003), 1645--1649.

\bibitem{Abatay}
{\sc Abatay, H.}
\newblock {\em Analysis of the role of nitric oxide synthase in the control of
  blood flow}.
\newblock PhD thesis, University of Oxford, 2010.

\bibitem{Bayliss02}
{\sc Bayliss, W.~M.}
\newblock {{O}n the local reactions of the arterial wall to changes of internal
  pressure}.
\newblock {\em J. Physiol. (Lond.) 28}, 3 (May 1902), 220--231.

\bibitem{Beach98}
{\sc Beach, J.~M., McGahren, E.~D., and Duling, B.~R.}
\newblock {{C}apillaries and arterioles are electrically coupled in hamster
  cheek pouch}.
\newblock {\em Am. J. Physiol. 275}, 4 Pt 2 (Oct 1998), H1489--1496.

\bibitem{Birch95}
{\sc Birch, A.~A., Dirnhuber, M.~J., Hartley-Davies, R., Iannotti, F., and
  Neil-Dwyer, G.}
\newblock {{A}ssessment of autoregulation by means of periodic changes in blood
  pressure}.
\newblock {\em Stroke 26}, 5 (May 1995), 834--837.

\bibitem{Buerk01}
{\sc Buerk, D.~G.}
\newblock {{C}an we model nitric oxide biotransport? {A} survey of mathematical
  models for a simple diatomic molecule with surprisingly complex biological
  activities}.
\newblock {\em Annu Rev Biomed Eng 3\/} (2001), 109--143.

\bibitem{Buga91}
{\sc Buga, G.~M., Gold, M.~E., Fukuto, J.~M., and Ignarro, L.~J.}
\newblock {{S}hear stress-induced release of nitric oxide from endothelial
  cells grown on beads}.
\newblock {\em Hypertension 17}, 2 (Feb 1991), 187--193.

\bibitem{Campolongo07}
{\sc Campolongo, F., Cariboni, J., and Saltelli, A.}
\newblock An effective screening design for sensitivity analysis of large
  models.
\newblock {\em Environmental Modelling \& Software 22}, 10 (2007), 1509 --
  1518.
\newblock Modelling, computer-assisted simulations, and mapping of dangerous
  phenomena for hazard assessment.

\bibitem{Carlson08}
{\sc Carlson, B.~E., Arciero, J.~C., and Secomb, T.~W.}
\newblock Theoretical model of blood flow autoregulation: roles of myogenic,
  shear-dependent, and metabolic responses.
\newblock {\em American Journal of Physiology-Heart and Circulatory Physiology
  295}, 4 (2008), H1572--H1579.

\bibitem{Caravajal00}
{\sc Carvajal, J.~A., Germain, A.~M., Huidobro-Toro, J.~P., and Weiner, C.~P.}
\newblock {{M}olecular mechanism of c{G}{M}{P}-mediated smooth muscle
  relaxation}.
\newblock {\em J. Cell. Physiol. 184}, 3 (Sep 2000), 409--420.

\bibitem{Clarke99}
{\sc Clarke, D.~D., Sokoloff, L., et~al.}
\newblock Circulation and energy metabolism of the brain.
\newblock {\em Basic neurochemistry: molecular, cellular and medical aspects
  6\/} (1999), 637--669.

\bibitem{Colantuoni84}
{\sc Colantuoni, A., Bertuglia, S., and Intaglietta, M.}
\newblock {{T}he effects of alpha- or beta-adrenergic receptor agonists and
  antagonists and calcium entry blockers on the spontaneous vasomotion}.
\newblock {\em Microvasc. Res. 28}, 2 (Sep 1984), 143--158.

\bibitem{Cornelissen02}
{\sc Cornelissen, A.~J., Dankelman, J., VanBavel, E., and Spaan, J.~A.}
\newblock Balance between myogenic, flow-dependent, and metabolic flow control
  in coronary arterial tree: a model study.
\newblock {\em American Journal of Physiology-Heart and Circulatory Physiology
  282}, 6 (2002), H2224--H2237.

\bibitem{Cutnell98}
{\sc Cutnell, J., and Johnson, K.}
\newblock {\em Physics, Fourth Edition}.
\newblock Wiley, 1998.

\bibitem{Davis99}
{\sc Davis, M., and Hill, M.}
\newblock Signaling mechanisms underlying the vascular myogenic response.
\newblock {\em Physiological Reviews 79\/} (1999), 387--413.

\bibitem{DeBoer}
{\sc deBoer, R.~W., Karemaker, J.~M., and Strackee, J.}
\newblock {{H}emodynamic fluctuations and baroreflex sensitivity in humans: a
  beat-to-beat model}.
\newblock {\em Am. J. Physiol. 253}, 3 Pt 2 (Sep 1987), H680--689.

\bibitem{Denninger99}
{\sc Denninger, J.~W., and Marletta, M.~A.}
\newblock {{G}uanylate cyclase and the .{N}{O}/c{G}{M}{P} signaling pathway}.
\newblock {\em Biochim. Biophys. Acta 1411}, 2-3 (May 1999), 334--350.

\bibitem{Diehl95}
{\sc Diehl, R.~R., Linden, D., Lucke, D., and Berlit, P.}
\newblock {{P}hase relationship between cerebral blood flow velocity and blood
  pressure. {A} clinical test of autoregulation}.
\newblock {\em Stroke 26}, 10 (Oct 1995), 1801--1804.

\bibitem{Duling87}
{\sc Duling, B.~R., Hogan, R.~D., Langille, B.~L., Lelkes, P., Segal, S.~S.,
  Vatner, S.~F., Weigelt, H., and Young, M.~A.}
\newblock {{V}asomotor control: functional hyperemia and beyond}.
\newblock {\em Fed. Proc. 46}, 2 (Feb 1987), 251--263.

\bibitem{Falcone91}
{\sc Falcone, J., Davis, M., and Meininger, G.}
\newblock Endothelial independence of myogenic response in isolated skeletal
  muscle arterioles.
\newblock {\em Am. J. Physiol. Heart Circ. Physiol. 260\/} (1991), H130--5.

\bibitem{Gokina96}
{\sc Gokina, N.~I., Bevan, R.~D., Walters, C.~L., and Bevan, J.~A.}
\newblock Electrical activity underlying rhythmic contraction in human pial
  arteries.
\newblock {\em Circulation research 78}, 1 (1996), 148--153.

\bibitem{Gonzalez94}
{\sc Gonzalez-Fernandez, J.~M., and Ermentrout, B.}
\newblock {{O}n the origin and dynamics of the vasomotion of small arteries}.
\newblock {\em Math Biosci 119}, 2 (Feb 1994), 127--167.

\bibitem{Gustafsson93}
{\sc GUSTAFSSON, H., MULVANY, M.~J., and NILSSON, H.}
\newblock Rhythmic contractions of isolated small arteries from rat: influence
  of the endothelium.
\newblock {\em Acta Physiologica Scandinavica 148}, 2 (1993), 153--163.

\bibitem{Haddock02}
{\sc Haddock, R., Hirst, G., and Hill, C.}
\newblock Voltage independence of vasomotion in isolated irideal arterioles of
  the rat.
\newblock {\em The Journal of physiology 540}, 1 (2002), 219--229.

\bibitem{Hai88a}
{\sc Hai, C.~M., and Murphy, R.~A.}
\newblock Cross-bridge phosphorylation and regulation of latch state in smooth
  muscle.
\newblock {\em Am J Physiol 254}, 1 Pt 1 (Jan 1988), C99--106.

\bibitem{Hapuarachchi10}
{\sc Hapuarachchi, T., Park, C.~S., and Payne, S.}
\newblock {{Q}uantification of the effects of vasomotion on mass transport to
  tissue from axisymmetric blood vessels}.
\newblock {\em J. Theor. Biol. 264}, 2 (May 2010), 553--559.

\bibitem{Harder87}
{\sc Harder, D.}
\newblock Pressure-induced myogenic activation of cat cerebral arteries is
  dependent on intact endothelium.
\newblock {\em Circ. Res. 60\/} (1987), 102--7.

\bibitem{Harder89}
{\sc Harder, D., Sanchez-Ferrer, C., Kauser, K., Stekiel, W., and Rubayani, G.}
\newblock Pressure releases a transferable endothelial contractile factor in
  cat cerebral arteries.
\newblock {\em Circ. Res. 65\/} (1989), 193--8.

\bibitem{Harris91}
{\sc Harris, D.~E., and Warshaw, D.~M.}
\newblock {{L}ength vs. active force relationship in single isolated smooth
  muscle cells}.
\newblock {\em Am. J. Physiol. 260}, 5 Pt 1 (May 1991), C1104--1112.

\bibitem{Herlihy73}
{\sc Herlihy, J.~T., and Murphy, R.~A.}
\newblock {{L}ength-tension relationship of smooth muscle of the hog carotid
  artery}.
\newblock {\em Circ. Res. 33}, 3 (Sep 1973), 275--283.

\bibitem{Hill99}
{\sc Hill, C.~E., Eade, J., and Sandow, S.~L.}
\newblock Mechanisms underlying spontaneous rhythmical contractions in irideal
  arterioles of the rat.
\newblock {\em J Physiol 521 Pt 2\/} (Dec 1999), 507--516.

\bibitem{HH52}
{\sc Hodgkin, A.~L., and Huxley, A.~F.}
\newblock {{A} quantitative description of membrane current and its application
  to conduction and excitation in nerve}.
\newblock {\em J. Physiol. (Lond.) 117}, 4 (Aug 1952), 500--544.

\bibitem{Huang97}
{\sc Huang, Y., and Cheung, K.-K.}
\newblock Endothelium-dependent rhythmic contractions induced by cyclopiazonic
  acid in rat mesenteric artery.
\newblock {\em European journal of pharmacology 332}, 2 (1997), 167--172.

\bibitem{Ide07}
{\sc Ide, K., Worthley, M., Anderson, T., and Poulin, M.~J.}
\newblock {{E}ffects of the nitric oxide synthase inhibitor {L}-{N}{M}{M}{A} on
  cerebrovascular and cardiovascular responses to hypoxia and hypercapnia in
  humans}.
\newblock {\em J. Physiol. (Lond.) 584}, Pt 1 (Oct 2007), 321--332.

\bibitem{Jacobsen07a}
{\sc Jacobsen, J.~C., Aalkjaer, C., Nilsson, H., Matchkov, V.~V., Freiberg, J.,
  and Holstein-Rathlou, N.~H.}
\newblock {{A}ctivation of a c{G}{M}{P}-sensitive calcium-dependent chloride
  channel may cause transition from calcium waves to whole cell oscillations in
  smooth muscle cells}.
\newblock {\em Am. J. Physiol. Heart Circ. Physiol. 293}, 1 (Jul 2007),
  H215--228.

\bibitem{Katusic87}
{\sc Katusic, Z., Shepherd, J., and Vanhoutte, P.}
\newblock Endothelium-dependent contraction to stretch in canine basilar
  arteries.
\newblock {\em Am. J. Physiol. 252\/} (1987), H671--3.

\bibitem{Koenigsberger05}
{\sc Koenigsberger, M., Sauser, R., B{\'e}ny, J.-L., and Meister, J.-J.}
\newblock Role of the endothelium on arterial vasomotion.
\newblock {\em Biophysical journal 88}, 6 (2005), 3845--3854.

\bibitem{Koenigsberger06}
{\sc Koenigsberger, M., Sauser, R., B{\'e}ny, J.-L., and Meister, J.-J.}
\newblock Effects of arterial wall stress on vasomotion.
\newblock {\em Biophysical journal 91}, 5 (2006), 1663--1674.

\bibitem{Kuo91}
{\sc Kuo, L., Chilian, W., and Davis, M.}
\newblock Interaction of pressure- and flow-induced responses in porcine
  coronary resistance vessels.
\newblock {\em Am. J. Physiol. Heart Circ. Physiol. 261\/} (1991), H1706--15.

\bibitem{Kuo95}
{\sc Kuo, L., Davis, M.~J., and Chilian, W.~M.}
\newblock Longitudinal gradients for endothelium-dependent and-independent
  vascular responses in the coronary microcirculation.
\newblock {\em Circulation 92}, 3 (1995), 518--525.

\bibitem{Lacy95}
{\sc Lacy, C.~R., Contracta, R.~J., Robbins, M.~L., Tannenbaum, A.~K., Moreyra,
  A.~E., Chelton, S., and Kostis, J.~B.}
\newblock Coronary vasoconstriction induced by mental stress (simulated public
  speaking).
\newblock {\em The American Journal of Cardiology 75}, 7 (1995), 503 -- 505.

\bibitem{Lamb89}
{\sc Lamb, F., and Webb, R.}
\newblock Regenerative electrical activity and arterial contraction in
  hypertensive rats.
\newblock {\em Hypertension 13}, 1 (1989), 70--76.

\bibitem{Lamboley03}
{\sc Lamboley, M., Schuster, A., B{\'e}ny, J.-L., and Meister, J.-J.}
\newblock Recruitment of smooth muscle cells and arterial vasomotion.
\newblock {\em American Journal of Physiology-Heart and Circulatory Physiology
  285}, 2 (2003), H562--H569.

\bibitem{Lee97}
{\sc Lee, M.~R., Li, L., and Kitazawa, T.}
\newblock {{C}yclic {G}{M}{P} causes {C}a2+ desensitization in vascular smooth
  muscle by activating the myosin light chain phosphatase}.
\newblock {\em J. Biol. Chem. 272}, 8 (Feb 1997), 5063--5068.

\bibitem{Lee01}
{\sc Lee, S.~P., Duong, T.~Q., Yang, G., Iadecola, C., and Kim, S.~G.}
\newblock {{R}elative changes of cerebral arterial and venous blood volumes
  during increased cerebral blood flow: implications for {B}{O}{L}{D}
  f{M}{R}{I}}.
\newblock {\em Magn Reson Med 45}, 5 (May 2001), 791--800.

\bibitem{Leondes07}
{\sc Leondes, C.}
\newblock {\em Biomechanical systems technology}.
\newblock World Scientific, Hackensack, NJ, 2007.

\bibitem{Liao97}
{\sc Liao, J.~C., and Kuo, L.}
\newblock Interaction between adenosine and flow-induced dilation in coronary
  microvascular network.
\newblock {\em American Journal of Physiology-Heart and Circulatory Physiology
  41}, 4 (1997), H1571.

\bibitem{Liu03}
{\sc Liu, Y., Birch, A.~A., and Allen, R.}
\newblock {{D}ynamic cerebral autoregulation assessment using an {A}{R}{X}
  model: comparative study using step response and phase shift analysis}.
\newblock {\em Med Eng Phys 25}, 8 (Oct 2003), 647--653.

\bibitem{Malpas02}
{\sc Malpas, S.~C.}
\newblock {{N}eural influences on cardiovascular variability: possibilities and
  pitfalls}.
\newblock {\em Am. J. Physiol. Heart Circ. Physiol. 282}, 1 (Jan 2002), 6--20.

\bibitem{Martin86}
{\sc Martin, M., Browner, W., Hulley, S., Kuller, L., and Wentworth, D.}
\newblock Serum cholesterol, blood pressure, and mortality: Implications from a
  cohort of 361,662 men.
\newblock {\em The Lancet 328}, 8513 (1986), 933 -- 936.
\newblock Originally published as Volume 2, Issue 8513.

\bibitem{Meininger87}
{\sc Meininger, G., Mack, C., Fehr, K., and Bohlen, H.}
\newblock Myogenic vasoregulation overrides local metabolic control in resting
  rat skeletal muscle.
\newblock {\em Circ. Res. 60\/} (1987), 861--870.

\bibitem{Meyer88}
{\sc Meyer, J.~U., Borgstrom, P., Lindbom, L., and Intaglietta, M.}
\newblock {{V}asomotion patterns in skeletal muscle arterioles during changes
  in arterial pressure}.
\newblock {\em Microvasc. Res. 35}, 2 (Mar 1988), 193--203.

\bibitem{nernst}
{\sc Nernst, W.}
\newblock {\em Theoretische Chemie vom Standpunkte der Avogadro'schen Regel und
  der Thermodynamik}.
\newblock 1893.

\bibitem{Nilsson03}
{\sc Nilsson, H., and Aalkjaer, C.}
\newblock {{V}asomotion: mechanisms and physiological importance}.
\newblock {\em Mol. Interv. 3}, 2 (Mar 2003), 79--89.

\bibitem{Okazaki03}
{\sc Okazaki, K., Seki, S., Kanaya, N., Hattori, J.-i., Tohse, N., and Namiki,
  A.}
\newblock Role of endothelium-derived hyperpolarizing factor in
  phenylephrine-induced oscillatory vasomotion in rat small mesenteric artery.
\newblock {\em Anesthesiology 98}, 5 (2003), 1164--1171.

\bibitem{Payne06}
{\sc Payne, S.}
\newblock A model of the interaction between autoregulation and neural
  activation in the brain.
\newblock {\em Mathematical Biosciences 204}, 2 (2006), 260 -- 281.

\bibitem{Payne05}
{\sc Payne, S., Morris, H., and Rowley, A.}
\newblock {{A} combined haemodynamic and biochemical model of cerebral
  autoregulation}.
\newblock {\em Conf Proc IEEE Eng Med Biol Soc 3\/} (2005), 2295--2298.

\bibitem{Peng01}
{\sc Peng, H., Matchkov, V., Ivarsen, A., Aalkj{\ae}r, C., and Nilsson, H.}
\newblock Hypothesis for the initiation of vasomotion.
\newblock {\em Circulation research 88}, 8 (2001), 810--815.

\bibitem{Peng08}
{\sc Peng, T.}
\newblock {\em Signal processing and analysis of cerebral autoregulation}.
\newblock PhD thesis, University of Oxford, April 2008.

\bibitem{Rashatwar87}
{\sc Rashatwar, S.~S., Cornwell, T.~L., and Lincoln, T.~M.}
\newblock {{E}ffects of 8-bromo-c{G}{M}{P} on {C}a2+ levels in vascular smooth
  muscle cells: possible regulation of {C}a2+-{A}{T}{P}ase by
  c{G}{M}{P}-dependent protein kinase}.
\newblock {\em Proc. Natl. Acad. Sci. U.S.A. 84}, 16 (Aug 1987), 5685--5689.

\bibitem{Rees90}
{\sc Rees, D.~D., Palmer, R.~M., Schulz, R., Hodson, H.~F., and Moncada, S.}
\newblock {{C}haracterization of three inhibitors of endothelial nitric oxide
  synthase in vitro and in vivo}.
\newblock {\em Br. J. Pharmacol. 101}, 3 (Nov 1990), 746--752.

\bibitem{Rosenson96}
{\sc Rosenson, R.~S., McCormick, A., and Uretz, E.~F.}
\newblock {{D}istribution of blood viscosity values and biochemical correlates
  in healthy adults}.
\newblock {\em Clin. Chem. 42}, 8 Pt 1 (Aug 1996), 1189--1195.

\bibitem{Rowley04}
{\sc Rowley, A., and Payne, S.~J.}
\newblock A {C}ompact {P}hysiological {M}odel of {C}erebral {B}lood {F}low
  {C}ontrol {F}or {U}se {I}n {C}linical {A}utoregulation {S}tudies.
\newblock Tech. rep., Oxford University, Mar. 2004.

\bibitem{Rubanyi88}
{\sc Rubanyi, G.}
\newblock Endothelium-dependent pressure-induced contraction of isolated canine
  carotid arteries.
\newblock {\em Am. J. Physiol. 255\/} (1988), H783--8.

\bibitem{Rucker00}
{\sc Rucker, M., Strobel, O., Vollmar, B., Roesken, F., and Menger, M.~D.}
\newblock {{V}asomotion in critically perfused muscle protects adjacent tissues
  from capillary perfusion failure}.
\newblock {\em Am. J. Physiol. Heart Circ. Physiol. 279}, 2 (Aug 2000),
  H550--558.

\bibitem{Schmidt92}
{\sc Schmidt, J.~A., Intaglietta, M., and Borgstrom, P.}
\newblock {{P}eriodic hemodynamics in skeletal muscle during local arterial
  pressure reduction}.
\newblock {\em J. Appl. Physiol. 73}, 3 (Sep 1992), 1077--1083.

\bibitem{Sell02}
{\sc Sell, M., Boldt, W., and Markwardt, F.}
\newblock Desynchronising effect of the endothelium on intracellular ca< sup>
  2+</sup> concentration dynamics in vascular smooth muscle cells of rat
  mesenteric arteries.
\newblock {\em Cell calcium 32}, 3 (2002), 105--120.

\bibitem{Slaaf87}
{\sc Slaaf, D.~W., Tangelder, G.~J., Teirlinck, H.~C., and Reneman, R.~S.}
\newblock {{A}rteriolar vasomotion and arterial pressure reduction in rabbit
  tenuissimus muscle}.
\newblock {\em Microvasc. Res. 33}, 1 (Jan 1987), 71--80.

\bibitem{Stalhand08}
{\sc Stalhand, J., Klarbring, A., and Holzapfel, G.~A.}
\newblock {{S}mooth muscle contraction: mechanochemical formulation for
  homogeneous finite strains}.
\newblock {\em Prog. Biophys. Mol. Biol. 96}, 1-3 (2008), 465--481.

\bibitem{strokestats}
{\sc Townsend, N., Wickramasinghe, K., Bhatnagar, P., Smolina, K., Nichols, M.,
  Leal, J., Luengo-Fernandez, R., and Rayner, M.}
\newblock Coronary heart disease statistics.
\newblock Tech. rep., British Heart Foundation, 2012.

\bibitem{Tzeng11}
{\sc Tzeng, Y.~C., Chan, G.~S., Willie, C.~K., and Ainslie, P.~N.}
\newblock {{D}eterminants of human cerebral pressure-flow velocity
  relationships: new insights from vascular modelling and {C}a{\^A}²{\^a}º
  channel blockade}.
\newblock {\em J. Physiol. (Lond.) 589}, Pt 13 (Jul 2011), 3263--3274.

\bibitem{Ursino96}
{\sc Ursino, M., Cavalcanti, S., Bertuglia, S., and Colantuoni, A.}
\newblock {{T}heoretical analysis of complex oscillations in multibranched
  microvascular networks}.
\newblock {\em Microvasc. Res. 51}, 2 (Mar 1996), 229--249.

\bibitem{Ursino92}
{\sc Ursino, M., and Fabbri, G.}
\newblock {{R}ole of the myogenic mechanism in the genesis of microvascular
  oscillations (vasomotion): analysis with a mathematical model}.
\newblock {\em Microvasc. Res. 43}, 2 (Mar 1992), 156--177.

\bibitem{Ursino98}
{\sc Ursino, M., and Lodi, C.~A.}
\newblock {{I}nteraction among autoregulation, {C}{O}2 reactivity, and
  intracranial pressure: a mathematical model}.
\newblock {\em Am. J. Physiol. 274}, 5 Pt 2 (May 1998), H1715--1728.

\bibitem{VanBavel94}
{\sc VanBavel, E., and Mulvany, M.}
\newblock Role of wall tension in the vasoconstrictor response of cannulated
  rat mesenteric small arteries.
\newblock {\em J. Physiol 477\/} (1994), 103--15.

\bibitem{Yang03a}
{\sc Yang, J., Clark, J.~W., Bryan, R.~M., and Robertson, C.}
\newblock {{T}he myogenic response in isolated rat cerebrovascular arteries:
  smooth muscle cell model}.
\newblock {\em Med Eng Phys 25}, 8 (Oct 2003), 691--709.

\end{thebibliography}

\end{document}